\documentclass{article}

\usepackage{arxiv}

\usepackage[cp1252]{inputenc}
\usepackage{textcomp}
\usepackage{amsmath}
\usepackage{amssymb}
\usepackage{graphicx}
\usepackage{subfigure}
\usepackage{tikz}
\usepackage{array}
\usepackage[abs]{overpic}
\usepackage{color}
\usepackage{soul}
\usepackage{amsthm}
\usepackage{units}
\usepackage{pifont}
\usepackage{upgreek}

\usepackage[ruled]{algorithm2e}
\makeatletter
\renewcommand{\@algocf@capt@plain}{above}% formerly {bottom}

\usepackage{booktabs}       % professional-quality tables
\usepackage{amsfonts}       % blackboard math symbols
\usepackage{nicefrac}       % compact symbols for 1/2, etc.
\usepackage{microtype}      % microtypography
\usepackage{lipsum}
\usepackage{graphicx}
\graphicspath{{figures/}}

\usepackage[square,numbers]{natbib}
\usepackage{textcomp}
\usepackage{amsmath}
\usepackage{amssymb}
\usepackage{graphicx, subfigure}
\usepackage{array}
\usepackage[abs]{overpic}
\usepackage{tabularx} % in the preamble
\usepackage{multirow}
\usepackage{hyperref}

\makeatletter

\usepackage{color}%% to highlight a text with yellow color and using a command \hl{highlighted text}
\usepackage{soul}
\usepackage{amsthm}

\graphicspath{{figures/}}

\newtheorem{theorem}{Theorem}

\newtheorem{remark}[theorem]{Remark}

\newcommand{\bs}{\boldsymbol}

\title{Asymptotic Homogenization in the Determination of \\ Effective Intrinsic Magnetic Properties of Composites}

\author{
 Celal~Soyarslan\\
  Chair of Nonlinear Solid Mechanics \\ University of Twente, The Netherlands \\
  \texttt{c.soyarslan@utwente.nl} \\ %% examples of more authors
   \And
 Jos~Havinga\\
  Chair of Nonlinear Solid Mechanics \\ University of Twente, The Netherlands \\
  \texttt{jos.havinga@utwente.nl} \\ %% examples of more
  \And
Leon~Abelmann \\
 MESA+ Institute for Nanotechnology\\ University of Twente, The Netherlands\\
 \texttt{l.abelmann@utwente.nl}
  \And
 Ton~van~den~Boogaard\\
  Chair of Nonlinear Solid Mechanics \\ University of Twente, The Netherlands \\
  \texttt{a.h.vandenboogaard@utwente.nl} \\ %% examples of more
}

\begin{document}
\maketitle

\begin{abstract}
We present a computational framework for two-scale asymptotic homogenization to determine the intrinsic magnetic permeability of composites. To this end, considering linear magnetostatics, both vector and scalar potential formulations are used. Our homogenization algorithm for solving the cell problem is based on \emph{the displacement method} presented in Lukkassen et al. 1995, Composites Engineering, 5(5),  519-531. We propose using the meridional eccentricity of the permeability tensor ellipsoid as an anisotropy index quantifying the degree of directionality in the linear magnetic response.  As application problems, 2D regular and random microstructures with overlapping and nonoverlapping monodisperse disks, all of which are periodic, are considered.  We show that, for the vanishing corrector function, the derived effective magnetic permeability tensor gives the (lower) Reuss and (upper) Voigt bounds with the vector and scalar  potential formulations, respectively. Our results with periodic boundary conditions show an excellent agreement with analytical solutions for regular composites, whereas, for random heterogeneous materials, their convergence with volume element size is fast. Predictions for material systems with monodisperse overlapping disks for a given inclusion volume fraction provide the highest magnetic permeability with the most increased inclusion interaction. In contrast, the disk arrangements in regular square lattices result in the lowest magnetic permeability and inadequate inclusion interaction. Such differences are beyond the reach of the isotropic  effective medium theories, which use only the phase volume fraction and shape as mere statistical microstructural descriptors.
\end{abstract}

% keywords can be removed

\section{Introduction}
Modern manufacturing processes are becoming increasingly flexible and reactive to changing process conditions \cite{ALLWOOD2016573}. In the case of metal manufacturing processes such as machining, forming, and additive manufacturing, the material microstructure is a crucial factor for process performance, as it determines the local mechanical response of the material. At the same time, the microstructure evolves during production under thermo-mechanical loading. To enable the transition to Industry 4.0, there is a strong need for the development of in-line microstructural observers \cite{CAO2019605}.

A promising technique in this regard is eddy current measurement. This method is used for many purposes, from inferring geometrical properties of the composites, e.g., coating thickness \cite{Meng2021}, film thickness \cite{Wang2014} to size and location of defects \cite{Ghoni2014}. Several studies have shown that the electromagnetic sensor response can be related to mechanical properties such as formability, hardness, or anisotropy through extensive experimental calibration \cite{HEINGARTNER2013,ALAMOS2021}. However, a better understanding of the relations between the electromagnetic and mechanical response of the materials is required for a more effective application of these techniques. Several recent studies have been devoted to the modeling of electromagnetic measurements systems to infer certain statistical moments of the material microstructure using in-line measurement data \cite{YIN20062317,Yinetal2009,Yinetal2017,ZHOU201431,HAO2010305,HALDANE20061761}.

In the low-frequency regimes, what affects the inductive sensor output is mainly the effective magnetic permeability of the material. In contrast, effective electrical conductivity becomes important at frequencies where eddy current losses become significant \cite{HALDANE20061761}. For most alloy steels, the constituents are distinguished by their  magnetic permeabilities, whereas the property contrast in their electrical conductivities (or resistivities) are relatively small \cite{ZHOU201431}. An example is the two-phase ferritic-austenitic steel in which (ferromagnetic) ferrite and (paramagnetic) austenite has a magnetic permeability contrast of around $\mu_{\mathrm{Fe}}/\mu_{\mathrm{Aus}}\simeq1000$. In contrast, their electrical
resistivities are comparable within a measurement uncertainty of 7\% \cite{Voronenko1997}. In the determination of the effective  magnetic permeability, a micro-to-macro transformation of the constituent phase properties is required.

Analytical methods developed to this end constitute various effective-medium theories such as the Maxwell, self-consistent, and differential effective-medium approximations. The (arithmetic) Voigt and the (harmonic) Reuss averages form analytical bounds in this respect. These approximations rely on the simplest microstructural descriptors, such as phase volume fraction and shape. While most effective-medium approximations mixture rules yield high accuracy for low-contrast material properties, high-contrast physical properties, and cases with intricate phase interactions, e.g., due to phase percolation, these mixture rules lose their accuracy.

A full-field numerical computation that accounts for fine microstructural details is required to remedy this.
Such computational methods using field averages on material volumes are modeled either as embeddings into free space with uniform external magnetic fields or as isolated closed-magnetic circuits \cite{HilzingerRodewald2013} subjected to various boundary conditions. In embedded models, due to demagnetizing fields, the resultant predictions underestimate the intrinsic magnetic permeability \cite{HAO2010305}. The use of isolated microstructures with appropriate boundary conditions gives accurate results in determining the intrinsic magnetic properties of the material. In this context, besides the boundary conditions proposed in \cite{HAO2010305}, those satisfying Hill-Mandel conditions are discussed in, e.g., \cite{Garboczi1998, ChatzigeorgiouJaviliSteinmann2014,JAVILI20134197, OstojaStarzewski2019}.

In this work, we present a computational method in first-order two-scale asymptotic homogenization to determine the tensorial magnetic permeability of composites. Limiting ourselves to linear magnetostatics, our derivation of the cell problem and its solution follows the displacement method of Lukassen et al. \cite{LUKKASSEN1995519}, which was initially introduced for an elastomechanical problem. In our formulation of the uncoupled nondimensional magnetostatics equations, we use scalar and vector potentials where the corrector function is identical-type to the applied magnetic potential. We use both uniform Dirichlet and periodic boundary conditions. Inspired by Lam\'{e}'s stress ellipsoid and considering that both the stress and the permeability tensors are of second-order, we propose the use of meridional eccentricity in quantifying the anisotropy in the effective magnetic permeability of the composites.

The developed computational framework is applied to determine the effective properties of regular and random material systems in two dimensions. The motivation behind this choice is two-fold: The first motivation is computational tractability. Secondly, with appropriate changes of variables and parameters, this permits a unified treatment of both scalar and vector potentials through a generic partial differential equation, which becomes handy in applying Dirichlet-based periodic and uniform boundary conditions. Our results are compared with the available analytical solutions and predictions of the selected effective-medium theories and rigorous bounds. In light of our numerical computations, we proposed a simple three-parameter approximation for the effective permeability of overlapping random disk systems.

\section{Theory}
\subsection{A word on notation}

Let us denote by $s(\bs{y},t)$ a scalar field and by  $\bs{C}(\bs{y},t)$ and $\bs{D}(\bs{y},t)$ two vector fields, distributed over the domain represented by material points $\bs{y}$ at time $t$. Considering three-dimensional space and using Einstein's summation convention, we write  $\bs{C}=C_i\bs{e}_i$ and  $\bs{y}=y_i\bs{e}_i$ where $\bs{e}_i$ for $i=1,2,3$ represent Cartesian basis vectors and $C_i$ and $y_i$ the  associated vector components. The gradient of a scalar field, divergence, and curl of vector fields are respectively defined as follows
\begin{align*}
\bs{\nabla}_{\bs{y}}\, s :=\dfrac{\partial s(\bs{y})}{\partial y_{i}} \boldsymbol{e}_{i}\,,\quad
\mathrm{div}_{\bs{y}}\,\bs{C}:=\frac{\partial
\bs{C}(\bs{y}) }{\partial y_{i}} \cdot \bs{e}_{i}\,,\quad
\mathrm{curl}_{\bs{y}}\,\bs{C}:=
-\dfrac{\partial \bs{C}(\bs{y})}{\partial y_{i}}\times \bs{e}_{i}=
\epsilon_{kji}\frac{\partial C_{i}}{\partial y_{j}}\bs{e}_{k}\,.
\end{align*}
Here, $\cdot$ and $\times$ denote single-contraction (scalar) and cross products with $\bs{C}\cdot\bs{D}=C_iD_i$ and  $\bs{C}\times\bs{D}=\epsilon_{ijk}C_jD_k\bs{e}_i$. $\epsilon_{ijk}$ is the  Levi-Civita symbol, such that
$\epsilon_{ijk}$ is $1/-1$ for even/odd permutations of  $(i, j, k)$ and 0 for repeated indices. $\delta_{ij}$ is the Kronecker delta with
$\delta_{ij}=1$ if $i=j$ and $\delta_{ij}=0$ otherwise.

\subsection{Nondimensional Magnetostatics}
Let $\overline{\boldsymbol{H}}$, $\overline{\boldsymbol{B}}$ and $\overline{^%
\mathrm{M} \boldsymbol{x}}$ denote the dimensional magnetic field, magnetic induction
field and the position of the particle, respectively, and, $\boldsymbol{H}$, $\boldsymbol{B}$ and $^%
\mathrm{M} \boldsymbol{x}=\overline{^%
\mathrm{M} \boldsymbol{x}}/^\mathrm{M}x_{\mathrm{ref}}$ their nondimensionalized counterparts, with $^\mathrm{M}x_{\mathrm{ref}}$ representing macroscopic a reference scale length\footnote{For more details we refer the reader to \ref{s:nondimensionalization}.}.
Amp\`{e}re-Maxwell law for steady state conditions links the nondimensional free current density $\boldsymbol{J}$ to $\boldsymbol{H}$ with
\begin{equation}
\mathrm{curl}_{^\mathrm{M}\boldsymbol{x}} \boldsymbol{H} = \boldsymbol{J}\,.
\label{E:freecurrent_NONDIM}
\end{equation}
Gauss's law for magnetism which asserts the absence of magnetic monopoles implying a solenoidal magnetic induction field $\boldsymbol{B}$ by
\begin{equation}
\text{div}_{{^\mathrm{M}\boldsymbol{x}}}{\boldsymbol{B}}=0\,.
\label{E:dimequlibriumNEU1_NONDIM}
\end{equation}
Assumption of linear magnetostatics leads to the following constitutive relation between  ${\boldsymbol{B}}$ and ${\boldsymbol{H}}$
\begin{equation}
{\boldsymbol{B}}={\boldsymbol{\mu}}\cdot{\boldsymbol{H}}
\text{ and }
{\boldsymbol{H}}={\boldsymbol{\beta}}\cdot{\boldsymbol{B}}\,, \label{E:dimconst_NONDIM}
\end{equation}
where, ${\boldsymbol{\mu}}={\boldsymbol{\beta}}^{-1}$ is the nondimensional symmetric second-order magnetic permeability tensor, where ${\mu}_{ij}={\mu}_{ji}$ and ${\beta}_{ij}={\beta}_{ji}$ for $i,j=1,2,3$.

In the formulation of magnetostatics, there exist two possibilities. The first one is the more general vector potential formulation which links the magnetic induction field to a magnetic vector potential $\boldsymbol{A}$ with
\begin{equation}
{\boldsymbol{B}}=\mathrm{curl}_{{^\mathrm{M}\boldsymbol{x}}}{\boldsymbol{A}}\,.
\label{E:vecpotA_NONDIM}
\end{equation}
Alternatively, for vanishing free currents, i.e., ${\boldsymbol{J}}\to \boldsymbol{0}$, using tensor calculus in view of Eq.\ \eqref{E:freecurrent_NONDIM} allows one postulate a magnetic scalar potential ${\varrho}$ from which the magnetic field $\boldsymbol H$ is derived
\begin{equation}
{\boldsymbol{H}}=-\boldsymbol{\nabla}_{{^\mathrm{M}\boldsymbol{x}}} {\varrho}\,.
\label{E:scalarpotvarrho_NONDIM}
\end{equation}
With this, a scalar potential formulation is developed. In the following derivations, for the sake of completeness, we pursue both vector and scalar potential formulations.
\subsection{Asymptotic Homogenization}
The asymptotic homogenization method used here is adapted from
\emph{the displacement method} presented in \cite{LUKKASSEN1995519} in a mechanics context. Similar developments can be found
in, e.g., \cite{EIDEL2018332,Fish2014,SOYARSLAN2019103098,Maetal2010}.

\begin{figure*}[htb!]
% \vspace{5pt}
% trim=left bottom right top, clip
\centering
\begin{tikzpicture}
\coordinate[] (P1) at (2.5,0.);
\coordinate[] (P2) at (5.5,0.);
\coordinate[] (c1o) at (-3.5,-1.5);
\coordinate[] (c1x) at (-2.5,-1.5);
\coordinate[] (c1y) at (-3.5,-0.5);
\coordinate[] (c2o) at (4.5,-1.5);
\coordinate[] (c2x) at (5.5,-1.5);
\coordinate[] (c2y) at (4.5,-0.5);
    \node[inner sep=0pt] (B11) at (0,0){\includegraphics[height=0.25\textwidth,
trim=1238 324 1595 27, clip]{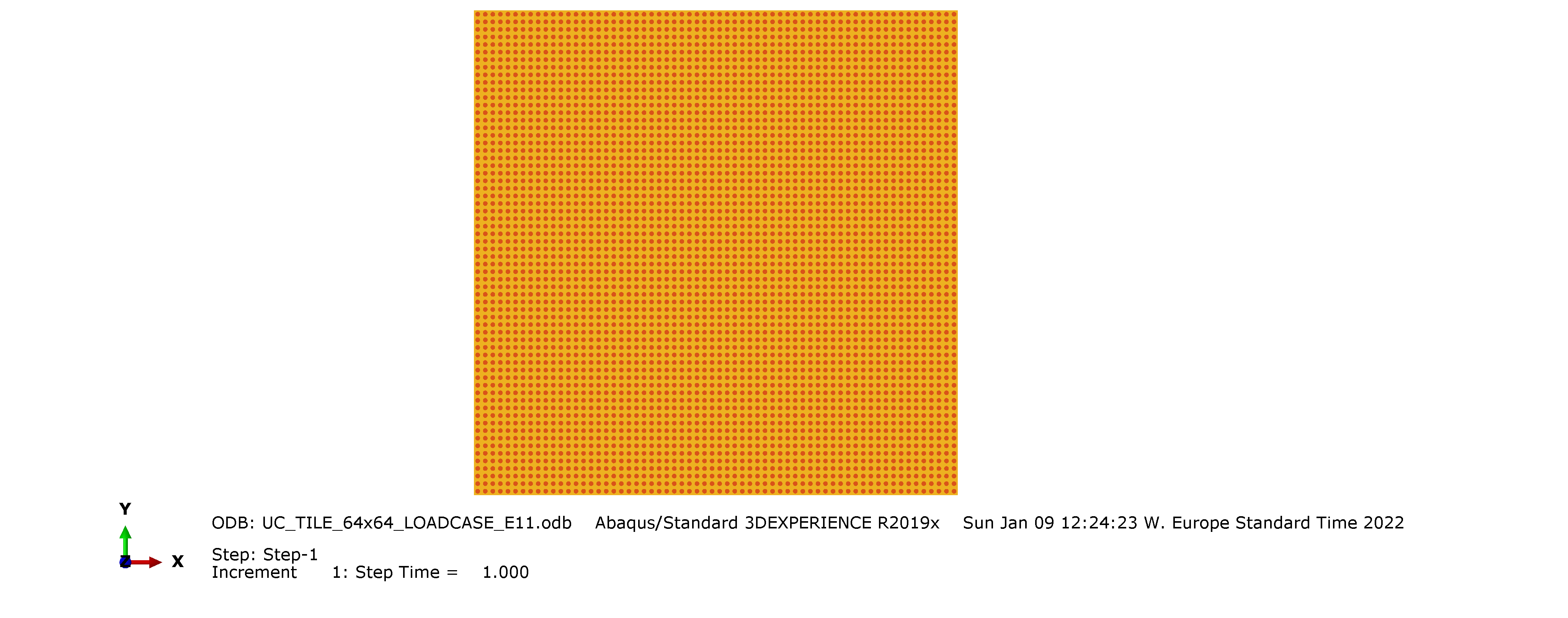}};
%    \node (A) at (0,3.0) {\scriptsize $\mu_\mathrm{i}/\mu_\mathrm{m}=250/1$};
    \node[inner sep=0pt] (B11) at (8,0){\includegraphics[height=0.25\textwidth,
trim=1238 324 1595 27, clip]{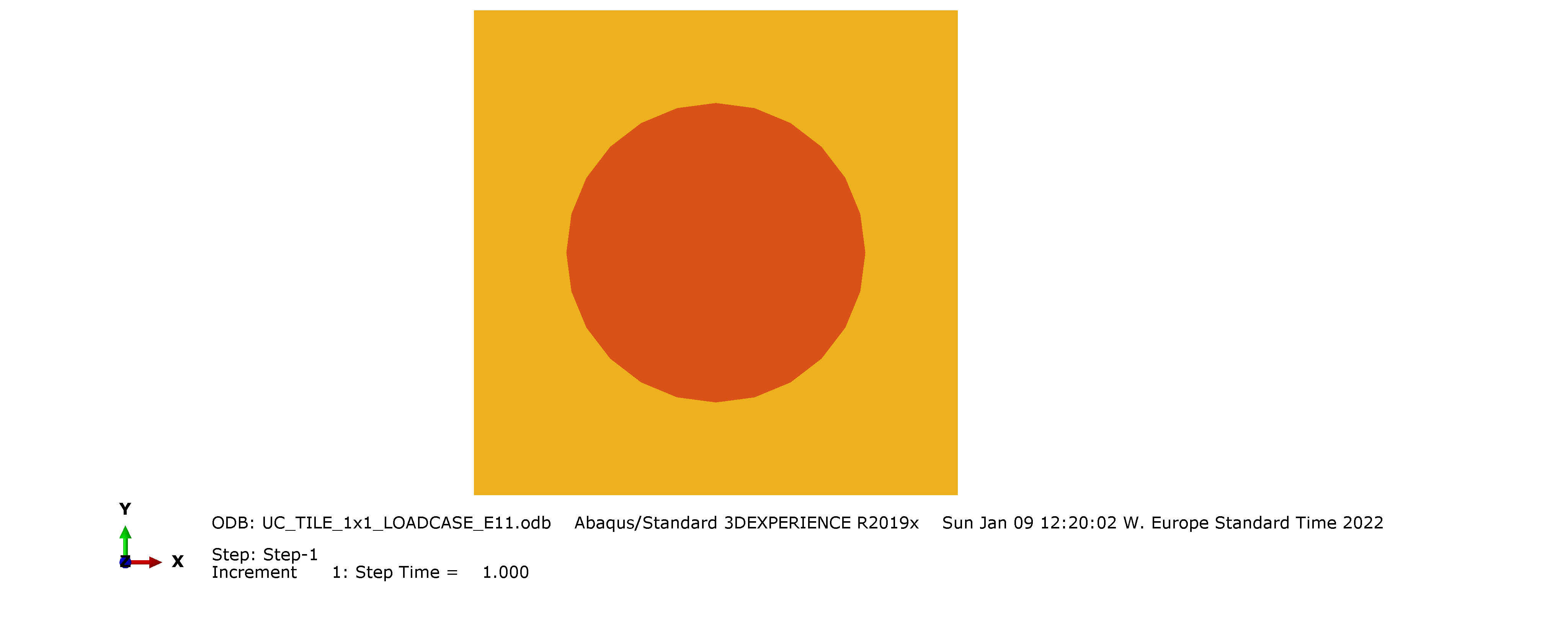}};
%    \node (B) at (3.0,3.0) {\scriptsize $\mu_\mathrm{i}/\mu_\mathrm{m}=250/1$};
\draw[black, ->] (P1) -- node[above] { $\times 1/\epsilon$} (P2);
\draw[black, ->] (c1o) -- node[below right] { $^\mathrm{M}x_1$} (c1x);
\draw[black, ->] (c1o) -- node[above left] { $^\mathrm{M}x_2$} (c1y);
\draw[black, ->] (c2o) -- node[below right] { $x_1$} (c2x);
\draw[black, ->] (c2o) -- node[above left] { $x_2$} (c2y);
\node (v) at (0,2.5) {$\mathcal{B}$};
\node (v) at (8,2.5) {$\mathcal{V}$};
\node (v) at (6.5,1.25) {$\mathcal{V}_1$};
\node (v) at (8,0.0) {$\mathcal{V}_2$};
\end{tikzpicture}
\caption{Schematic description of macroscopic and microscopic domains, respectively denoted by $\mathcal B$ and $\mathcal V$,     and associated coordinates $^\mathrm{M}\boldsymbol x$ and $\boldsymbol x$ in asymptotic homogenization for a periodic composite material. $\mathcal V$ corresponds to the periodic unit cell composed of two constituent domains $\mathcal V_1$ and $\mathcal V_2$. The scale separation parameter $0<\epsilon\ll 1$ controls the fineness of the microstructure and links  domain coordinates $^\mathrm{M}\boldsymbol x$ and $\boldsymbol x$ with
$\boldsymbol x=$$^\mathrm{M}\boldsymbol x/\epsilon$.}
\label{F:schematics_asymptotic_homogenization}
\end{figure*}
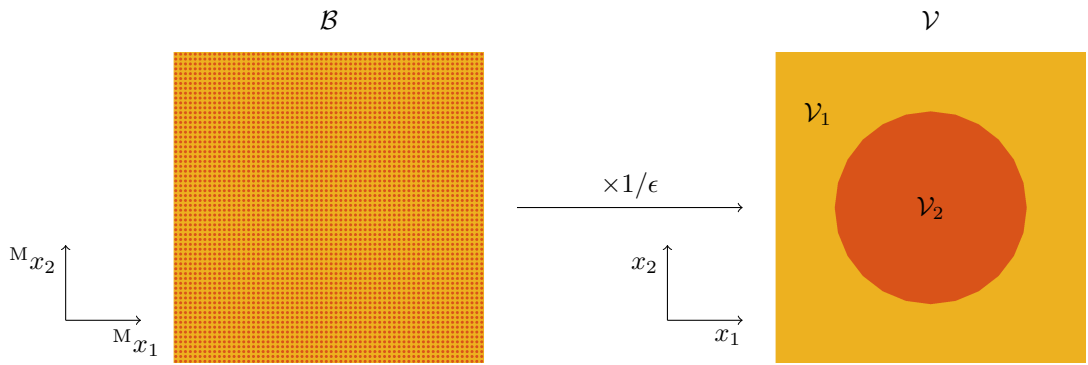

Let $\overline{\mathcal{V}}$ denote a periodic unit cell of size $x_{\mathrm{ref}}\times x_{\mathrm{ref}}\times x_{\mathrm{ref}}$ whose tiling fills the composite material domain. Although the conclusion is not affected if we have more constituent phases, we consider two linearly magnetic phases which occupy unit cell subdomains $\overline{\mathcal{V}}_1$, $\overline{\mathcal{V}}_2$ where $\overline{\mathcal{V}}_1\cup\overline{\mathcal{V}}_2=\overline{\mathcal{V}}$. To characterize the scale of the unit cell, another nondimensional
coordinate $\boldsymbol{x}$ can be generated with
$\boldsymbol{x}= \overline{^\mathrm{M} \boldsymbol{x}}/x_{\mathrm{ref}}\in \mathcal{V}$ where $\mathcal{V}$ is the nondimensional unit cell of size $1\times 1\times 1$ with
unit cell subdomains $\mathcal{V}_1$, $\mathcal{V}_2$ where $\mathcal{V}_1\cup\mathcal{V}_2=\mathcal{V}$. It is convenient to relate $\boldsymbol{x}$ to the previously generated nondimensional coordinate
$^\mathrm{M} \boldsymbol{x}$ with $\boldsymbol{x}= {^\mathrm{M} \boldsymbol{x}}/\epsilon$, using the notation $\epsilon=x_{\mathrm{ref}}/{^\mathrm{M}x_{\mathrm{ref}}}$. Here, $0<\epsilon \ll 1$ controls the fineness (scale separation) of the unit cell, see Fig.\ \ref{F:schematics_asymptotic_homogenization}. If $\epsilon$ is small, the result is a fine microstructure with  rapidly oscillating property functions. Let $\boldsymbol{F}^\epsilon(
$$^\mathrm{M}\boldsymbol{x})=\boldsymbol{F}($$^\mathrm{M}\boldsymbol{x},%
\boldsymbol{x})$ denote a generic tensor-valued function on the composite domain $\mathcal{B}^\epsilon=\mathcal B \times\mathcal V$, see, e.g., Ref. \cite{FISH20012215}, which is $\mathcal{V%
}-$periodic in $\boldsymbol{x}$ over the unit cell $\mathcal{V}$ and which
features slow (with $^\mathrm{M}\boldsymbol{x}$) and fast (with $\boldsymbol{x}
$) variations. It follows that
\begin{align}
\label{E:dispstrainstress1} \boldsymbol{\nabla}_{^\mathrm{M}\boldsymbol{x}}\, \boldsymbol{F}^\epsilon(^%
\mathrm{M}\boldsymbol{x}) & = \boldsymbol{\nabla}_{^\mathrm{M}\boldsymbol{x}}\,
\boldsymbol{F}(^\mathrm{M}\boldsymbol{x},\boldsymbol{x})+\dfrac{1}{\epsilon}%
\boldsymbol{\nabla}_{\boldsymbol{x}}\, \boldsymbol{F}(^\mathrm{M}\boldsymbol{x}%
,\boldsymbol{x})\,, \\
\label{E:dispstrainstress2}
\text{div}_{^\mathrm{M}\boldsymbol{x}}\, \boldsymbol{F}^\epsilon(^\mathrm{M}%
\boldsymbol{x})& = \text{div}_{^\mathrm{M}\boldsymbol{x}}\, \boldsymbol{F}(^%
\mathrm{M}\boldsymbol{x},\boldsymbol{x})+\dfrac{1}{\epsilon}\text{div}_{%
\boldsymbol{x}}\, \boldsymbol{F}(^\mathrm{M}\boldsymbol{x},\boldsymbol{x})\,,
\\
\text{curl}_{^\mathrm{M}\boldsymbol{x}}\, \boldsymbol{F}^\epsilon(^\mathrm{M}%
\boldsymbol{x})& = \text{curl}_{^\mathrm{M}\boldsymbol{x}}\, \boldsymbol{F}(^%
\mathrm{M}\boldsymbol{x},\boldsymbol{x})+\dfrac{1}{\epsilon}\text{curl}_{%
\boldsymbol{x}}\, \boldsymbol{F}(^\mathrm{M}\boldsymbol{x},\boldsymbol{x})\,,
\label{E:dispstrainstress3}
\end{align}
where sub-indices ${^\textrm{M}\bs{x}}$ and  ${\boldsymbol{x}}$ denote derivatives with respect to  the macroscopic and microscopic coordinates $^\textrm{M}\boldsymbol{x}$ and $\bs{x}$, respectively.

In the following, we derive expressions with vector and scalar potential formulations of asymptotic two-scale homogenization in magnetostatics in parallel. To this end, we start by assuming that the magnetic vector and scalar potentials are multiscale functions of the form $%
\boldsymbol{A}^{\epsilon }(^{\mathrm{M}}\boldsymbol{x})=\boldsymbol{A}(^{\text{%
M}}\boldsymbol{x},\boldsymbol{x})$ and $\varrho^\epsilon(^\mathrm{M}\boldsymbol{x})=\varrho(^\mathrm{M}\boldsymbol{x},\boldsymbol{x})$, respectively. Then, we apply
two-scale asymptotic expansion to the nondimensional potentials as the primary solution variables on $\mathcal{B}^\epsilon$,
\begin{align}
\boldsymbol{A}^{\epsilon }(^{\mathrm{M}}\boldsymbol{x})&=\boldsymbol{A}^{(0)}(^{%
\mathrm{M}}\boldsymbol{x},\boldsymbol{x})+\epsilon \boldsymbol{A}^{(1)}(^{%
\mathrm{M}}\boldsymbol{x},\boldsymbol{x})+\epsilon ^{2}\boldsymbol{A}^{(2)}(^{%
\mathrm{M}}\ \boldsymbol{x},\boldsymbol{x})+\mathcal{O}(\epsilon ^{3})\,,\label{E:ASYMPEXPVEC}\\
\varrho ^{\epsilon }(^{\mathrm{M}}\boldsymbol{x})&=\varrho ^{(0)}(^{\mathrm{M}}%
\boldsymbol{x},\boldsymbol{x})+\epsilon \varrho ^{(1)}(^{\mathrm{M}}%
\boldsymbol{x},\boldsymbol{x})+\epsilon ^{2}\varrho ^{(2)}(^{\mathrm{M}}\
\boldsymbol{x},\boldsymbol{x})+\mathcal{O}(\epsilon ^{3})\,,
\label{E:ASYMPEXPSCALAR}
\end{align}
where both $\boldsymbol{A}^{(i)}$ and $\varrho ^{(i)}$ are $\mathcal{V}-$periodic functions in $%
\boldsymbol{x}$. The leading orders $\boldsymbol{A}^{(0)}$ and $\varrho ^{(0)}$ respectively correspond to the homogenized magnetic vector and scalar potential fields, and as shown below, they only depend on the macroscopic coordinates $^{\mathrm{M}}%
\boldsymbol{x}$, i.e., they are constant over the unit cell domain.

We now write the multiscale forms of the differential equations governing both formulations. Considering the vector potential formulation, assuming that the source term depends only on the slow variable $\boldsymbol{J}^{\epsilon }(^{\mathrm{M}}%
\boldsymbol{x})\to \bs {J}(^{\mathrm{M}}%
\boldsymbol{x})$ \cite{Torquato2002}, the solution must fulfill the multiscale version of Eq.\ \eqref{E:freecurrent_NONDIM} at the composite domain
$\mathcal{B}^\epsilon$ with fine-scale microstructural features. Similarly, a multiscale version of Gauss' law given in Eq.\ \eqref{E:dimequlibriumNEU1_NONDIM} sets the stage for further developments. These two conditions can be respectively written as
\begin{align}
\text{curl}_{^{\mathrm{M}}\boldsymbol{x}}\boldsymbol{H}^{\epsilon }(^{\mathrm{M}}%
\boldsymbol{x})& =\bs{J}(^{\mathrm{M}}%
\boldsymbol{x})\quad\text{%
in }\mathcal{B}^\epsilon\,,
\label{E:multiscale_curlHzero}\\
\text{div}_{^{\mathrm{M}}\boldsymbol{x}}\boldsymbol{B}^{\epsilon }(^{\mathrm{M}}%
\boldsymbol{x})&=0\quad\text{%
in }\mathcal{B}^\epsilon\,.
\label{E:multiscale_divBzero}
\end{align}
Eqs.\ \eqref{E:multiscale_curlHzero} and \eqref{E:multiscale_divBzero}  are  subjected to prescribed Dirichlet and Neumann boundary conditions. For vector potential formulation, these respectively read $\bs A^\epsilon=$%
$\widetilde{\bs A^\epsilon}$ on $\partial \mathcal{B}_{\bs A}^\epsilon$ and $\boldsymbol{H^\epsilon}\times \boldsymbol{n}^\epsilon=\widetilde{\bs t^\epsilon
}$ on $\partial \mathcal{B}_{\bs {t}}^\epsilon$, where $\boldsymbol{n}^\epsilon$ is the surface unit normal.
In the same manner, we obtain $\varrho^\epsilon=\widetilde{\varrho}%
^\epsilon$ on $\partial \mathcal{B}^\epsilon_{\varrho}$ and $\boldsymbol{B}%
^\epsilon\cdot \boldsymbol{n}^\epsilon=\widetilde{t}^\epsilon$ on $\partial
\mathcal{B}^\epsilon_{t}$, respectively, for scalar potential formulation. For the domain boundaries, the following conditions apply for vector potential formulation $\partial\mathcal{B}^\epsilon_{\bs A}
\bigcap \partial \mathcal{B}^\epsilon_{\bs t}=\emptyset$ and $\partial\mathcal{B}%
^\epsilon_{\bs A} \bigcup \partial \mathcal{B}^\epsilon_{\bs t}=\partial \mathcal{B}^\epsilon$. Similarly, we have $\partial\mathcal{B}^\epsilon_{\varrho}
\bigcap \partial \mathcal{B}^\epsilon_{t}=\emptyset$ and $\partial\mathcal{B}%
^\epsilon_{\varrho} \bigcup \partial \mathcal{B}^\epsilon_{t}=\partial
\mathcal{B}^\epsilon$ for scalar potential formulation.

Applying the multiscale argument to the magnetostatic constitutive relation, we obtain
\begin{equation}
\boldsymbol{B}^\epsilon=\boldsymbol{\mu}^\epsilon\cdot\boldsymbol{H}%
^\epsilon
\text{ and }\boldsymbol{H}^\epsilon=\boldsymbol{\beta}^\epsilon\cdot\boldsymbol{B}%
^\epsilon \quad\text{in }
\mathcal{B}^\epsilon\,.  \label{E:equlibriumNEU2}
\end{equation}
In general, $\boldsymbol{\mu}^\epsilon=\boldsymbol{\mu}(^\mathrm{M}\boldsymbol{x},\boldsymbol{x})=[\boldsymbol{\beta}^\epsilon]^{-1}$ is  $\mathcal{V}-$periodic in $\boldsymbol{x}$ over the unit cell $\mathcal{V}$ and shows slow as well
as fast variations. For simplicity, we assume $\boldsymbol{\mu}^\epsilon=%
\boldsymbol{\mu}(\boldsymbol{x})$ and hence $\boldsymbol{\beta}^\epsilon=%
\boldsymbol{\beta}(\boldsymbol{x})$ and, thus, disregard its slow variation.

Substituting expansions \eqref{E:ASYMPEXPVEC} and \eqref{E:ASYMPEXPSCALAR} into the following multiscale versions of  Eqs.\ \eqref{E:vecpotA_NONDIM}  and \eqref{E:scalarpotvarrho_NONDIM}, respectively,
\begin{align}
\boldsymbol{B}^{\epsilon }(^{\mathrm{M}}\boldsymbol{x}) =\text{curl}_{^{\text{%
M}}\boldsymbol{x}}\,\boldsymbol{A}^{\epsilon }(^{\mathrm{M}}\boldsymbol{x})\,,
\label{E:multiscaleBfromA}\\
\boldsymbol{H}^{\epsilon }(^{\mathrm{M}}\boldsymbol{x}) =
-\boldsymbol{\nabla }_{^{\mathrm{M}}\boldsymbol{x}}\varrho^{\epsilon }(^{\mathrm{M}}\boldsymbol{x})\,,
\label{E:multiscaleHfromvarrho}
\end{align}
we obtain the following expansions for the nondimensional magnetic fields $\boldsymbol{B}^{\epsilon }$ and $\boldsymbol{H}^{\epsilon }$, respectively
\begin{align}
\boldsymbol{B}^{\epsilon }(^{\mathrm{M}}\boldsymbol{x})& =\dfrac{1}{\epsilon }%
\boldsymbol{B}^{(-1)}(^{\mathrm{M}}\boldsymbol{x},\boldsymbol{x})+\boldsymbol{B%
}^{(0)}(^{\mathrm{M}}\boldsymbol{x},\boldsymbol{x})+\epsilon \,\boldsymbol{B}%
^{(1)}(^{\mathrm{M}}\boldsymbol{x},\boldsymbol{x})+\epsilon ^{2}\,\boldsymbol{B%
}^{(2)}(^{\mathrm{M}}\boldsymbol{x},\boldsymbol{x})+\mathcal{O}(\epsilon
^{3})\,, \label{E:ASYMPEXP_B_VEC}\\
\boldsymbol{H}^{\epsilon }(^{\mathrm{M}}\boldsymbol{x})& =\dfrac{1}{\epsilon }%
\boldsymbol{H}^{(-1)}(^{\mathrm{M}}\boldsymbol{x},\boldsymbol{x})+\boldsymbol{H%
}^{(0)}(^{\mathrm{M}}\boldsymbol{x},\boldsymbol{x})+\epsilon \,\boldsymbol{H}%
^{(1)}(^{\mathrm{M}}\boldsymbol{x},\boldsymbol{x})+\epsilon ^{2}\,\boldsymbol{H%
}^{(2)}(^{\mathrm{M}}\boldsymbol{x},\boldsymbol{x})+\mathcal{O}(\epsilon
^{3})\,.  \label{E:ASYMPEXP_H_VEC}
\end{align}
Using Eqs.\ \eqref{E:dispstrainstress1} and \eqref{E:dispstrainstress3}, the following definitions hold for the expressions making up  Eqs.\ \eqref{E:ASYMPEXP_B_VEC} and \eqref{E:ASYMPEXP_H_VEC}, respectively
\begin{align}
\boldsymbol{B}^{(-1)}&=\mathrm{curl}_{\boldsymbol{x}}\boldsymbol{A}%
^{(0)}\quad \text{and}\quad \boldsymbol{B}^{(k)}=\boldsymbol{B}_{\langle ^{%
\mathrm{M}}\boldsymbol{x}\rangle }^{(k)}+\boldsymbol{B}_{\langle \boldsymbol{x}%
\rangle }^{(k+1)}\quad \text{for }k\geq 0\,,
\label{E:Bs}\\
\boldsymbol{H}^{(-1)}&=-\boldsymbol{\nabla }_{\boldsymbol{x}}\varrho
^{(0)}\quad \text{and}\quad \boldsymbol{H}^{(k)}=\boldsymbol{H}_{\langle ^{\mathrm{M}}\boldsymbol{x}\rangle }^{(k)}+\boldsymbol{H}_{\langle \boldsymbol{x}%
\rangle }^{(k+1)}\quad \text{for }k\geq 0\,,
\label{E:Hs}
\end{align}
where
\begin{align}
\boldsymbol{B}_{\langle ^{\mathrm{M}}\boldsymbol{x}\rangle }^{(k)}&=\mathrm{curl%
}_{^{\mathrm{M}}\boldsymbol{x}}\boldsymbol{A}^{(k)}\quad \text{and}\quad
\boldsymbol{B}_{\langle \boldsymbol{x}\rangle }^{(k)}=\mathrm{curl}_{%
\boldsymbol{x}}\boldsymbol{A}^{(k)}\,,
\label{E:Bkdef}\\
\boldsymbol{H}_{\langle ^{\mathrm{M}}\boldsymbol{x}\rangle }^{(k)}&=-%
\boldsymbol{\nabla }_{^{\mathrm{M}}\boldsymbol{x}}\varrho ^{(k)}\quad \text{%
and}\quad \boldsymbol{H}_{\langle \boldsymbol{x}\rangle }^{(k)}=-\boldsymbol{%
\nabla }_{\boldsymbol{x}}\varrho ^{(k)}\,.
\label{E:Hkdef}
\end{align}
We now proceed by using the multiscale constitutive relation given   in Eq.\ \eqref{E:equlibriumNEU2} in relating the quantities $\boldsymbol{B}^{(k)}$
and $\boldsymbol{H}^{(k)}$
\begin{equation}
\boldsymbol{H}^{(k)}=\boldsymbol{\beta }\cdot \boldsymbol{B}^{(k)}\text{ and }
\boldsymbol{B}^{(k)}=\boldsymbol{\mu }\cdot \boldsymbol{H}^{(k)}
\quad
\text{for }k\geq -1\,.
\label{E:constitutive_VEC}
\end{equation}
Substituting the asymptotic expansions of the magnetic fields into the Eqs.\ \eqref{E:multiscale_curlHzero} and \eqref{E:multiscale_divBzero} for vector and scalar potential formulations, respectively, and comparing the coefficients of different powers in  $\epsilon$ we obtain the following expressions
\begin{align}
\mathcal{O}(\epsilon ^{-2}):\quad & \boldsymbol{0}=\text{curl}_{\boldsymbol{x}}%
\boldsymbol{H}^{(-1)}\,,
& 0&=\text{div}_{\boldsymbol{x}}\boldsymbol{%
B}^{(-1)}\,, \label{E:equlibriumNEU1EXPa} \\
\mathcal{O}(\epsilon ^{-1}):\quad & \boldsymbol{0}=\text{curl}_{^{\mathrm{M}}%
\boldsymbol{x}}\,\boldsymbol{H}^{(-1)}+\text{curl}_{\boldsymbol{x}}\,%
\boldsymbol{H}^{(0)}\,,
& 0&=\text{div}_{^{\mathrm{M}}\boldsymbol{x}%
}\,\boldsymbol{B}^{(-1)}+\text{div}_{\boldsymbol{x}}\,\boldsymbol{B}^{(0)}\,,
\label{E:equlibriumNEU1EXPb} \\
\mathcal{O}(\epsilon ^{0}):\quad & \bs{J}=\text{curl}_{^{\mathrm{M}}%
\boldsymbol{x}}\,\boldsymbol{H}^{(0)}+\text{curl}_{\boldsymbol{x}}\,%
\boldsymbol{H}^{(1)}\,.
& 0 &=\text{div}_{^{\mathrm{M}}\boldsymbol{x}}\,%
\boldsymbol{B}^{(0)}+\text{div}_{\boldsymbol{x}}\,\boldsymbol{B}^{(1)}\,.
\label{E:equlibriumNEU1EXPc}
\end{align}
Since $\boldsymbol{\beta}$ (and thus $\boldsymbol{\mu }$)  depends only on the microscopic variables, one can
rewrite Eqs.\ \eqref{E:equlibriumNEU1EXPa}, for vector and scalar potential formulations, respectively, using Eqs.\ (\ref{E:Bs}-\ref{E:constitutive_VEC}) as follows
\begin{align}
\text{curl}_{\boldsymbol{x}}\left( \boldsymbol{\beta }\cdot \text{curl}_{%
\boldsymbol{x}}\boldsymbol{A}^{(0)}\right)&=\bs 0\,,\\
\text{div}_{\boldsymbol{x}}\left( -\boldsymbol{\mu }\cdot \boldsymbol{\nabla }%
_{\boldsymbol{x}}\varrho ^{(0)}\right)&=0\,.
\end{align}
This implies that both $\boldsymbol{A}^{(0)}$ and $\varrho ^{(0)}$ are functions of only macroscopic coordinates such that $\boldsymbol{A}^{(0)}=\boldsymbol{A}^{(0)}(^{\mathrm{M}}
\boldsymbol{x})$ and $\varrho ^{(0)}=\varrho ^{(0)}(^{\mathrm{M}}\boldsymbol{x}%
)$, respectively. On account of this remark and Eqs.\ \eqref{E:Bs} and \eqref{E:Hs}, we have $\boldsymbol{H}^{(-1)}=\boldsymbol{0}$ and $\boldsymbol{B}^{(-1)}=\boldsymbol{0}$. In consideration of Eqs.\ \eqref{E:Bkdef} and \eqref{E:Hkdef} $\boldsymbol{B}%
_{\langle ^{\mathrm{M}}\boldsymbol{x}\rangle }^{(0)}(^{\mathrm{M}}\boldsymbol{x})
$ and $\boldsymbol{H}_{\langle ^{\text{M%
}}\boldsymbol{x}\rangle }^{(0)}(^{\mathrm{M}}\boldsymbol{x})$ are also independent of $\bs x$, respectively. Therefore, $\boldsymbol{H}^{(0)}$ and $\boldsymbol{B}%
^{(0)}$ are the leading orders of the magnetic fields.
We now seek macroscopic
equations for $\boldsymbol{A}^{(0)}$ and $\varrho ^{(0)}$ in  terms of the magnetic fields $\boldsymbol{B}_{\langle ^{\mathrm{M}}%
\boldsymbol{x}\rangle }^{(0)}$ and  $\boldsymbol{H}_{\langle ^{\mathrm{M}}\boldsymbol{%
x}\rangle }^{(0)}$, respectively. Using Eqs.\ \eqref{E:constitutive_VEC} and \eqref{E:equlibriumNEU1EXPb}  it follows that  $\text{curl}_{\boldsymbol{x}}\,%
\boldsymbol{H}^{(0)}=\boldsymbol{0}$ and $\text{div}_{\boldsymbol{x}}\,%
\boldsymbol{B}^{(0)}=0$. On account of Eqs.\ \eqref{E:Bs} and \eqref{E:Hs}, we conclude that
\begin{align}
\text{curl}_{\boldsymbol{x}}\,(\boldsymbol{
\beta }\cdot \boldsymbol{B}_{\langle \boldsymbol{x}\rangle
}^{(1)})&=-\text{curl}_{\boldsymbol{x}}\,(
\boldsymbol{\beta }\cdot \boldsymbol{B}_{\langle ^{\mathrm{M}}%
\boldsymbol{x}\rangle }^{(0)})\,,
\label{E:curlcondition}\\
\text{div}_{\boldsymbol{x}}\,(\boldsymbol{\mu }\cdot \boldsymbol{H}_{\langle
\boldsymbol{x}\rangle }^{(1)})&=-\text{div}_{\boldsymbol{x}}\,(\boldsymbol{%
\mu }\cdot \boldsymbol{H}_{\langle ^{\mathrm{M}}\boldsymbol{x}\rangle
}^{(0)})\,.
\label{E:divcondition}
\end{align}
According to the above remarks and letting $\bs\chi ^{m}(\boldsymbol{x})$ and $\chi ^{m}(\boldsymbol{x})$ denote vector- and scalar-valued $\mathcal{V}-$periodic $C^{0}$ continuous corrector functions for $m=1,2,3$, we
apply separation of variables and consider solutions of the form $%
\boldsymbol{A}^{(1)}(^{\mathrm{M}}\boldsymbol{x},\boldsymbol{x})=\bs\chi ^{m}(%
\boldsymbol{x})B_{{\langle ^{\mathrm{M}}\boldsymbol{x}\rangle },m}^{(0)}$ and $\varrho ^{(1)}(^{%
\mathrm{M}}\boldsymbol{x},\boldsymbol{x})=-\chi ^{m}(\boldsymbol{x})H_{{%
\langle ^{\mathrm{M}}\boldsymbol{x}\rangle },m}^{(0)}$,  respectively, to obtain
\begin{align}
\boldsymbol{B}_{\langle \boldsymbol{x}\rangle }^{(1)}&=\mathrm{curl}_{%
\boldsymbol{x}}\bs\chi ^{m}B_{{\langle ^{\mathrm{M}}\boldsymbol{x}\rangle }%
,m}^{(0)}\,,
\label{E:ansatzVEC} \\
\boldsymbol{H}_{\langle \boldsymbol{x}\rangle }^{(1)}&=\boldsymbol{\nabla }_{%
\boldsymbol{x}}\chi ^{m}H_{{\langle ^{\mathrm{M}}\boldsymbol{x}\rangle }%
,m}^{(0)}\,.
\label{E:ansatzSCALAR}
\end{align}
$\mathcal{V}-$periodicity of $\bs\chi ^{m}(\boldsymbol{x})$ and $\chi ^{m}(\boldsymbol{x})$ implies for $m=1,2,3$ that
\begin{align}
\int_{\mathcal{V}}\ \mathrm{curl}_{\boldsymbol{x}}\bs\chi ^{m}\mathrm{d}V=%
\boldsymbol{0}\,\text{ and }
\int_{\mathcal{V}}\ \boldsymbol{\nabla }_{\boldsymbol{x}}\chi ^{m}\mathrm{d}%
V=\boldsymbol{0}\,.
\label{E:vanishingintegralSCALAR}
\end{align}
Substituting Eq.\ \eqref{E:ansatzVEC} into Eq.\ \eqref{E:curlcondition}, and
Eq.\ \eqref{E:ansatzSCALAR} into Eq.\ \eqref{E:divcondition} and using the symmetry of $\bs{\beta}$ and $\bs{\mu}$ give the following  cell problems, which, together with the locally periodic boundary conditions, constitute
linear boundary value problems for $\bs\chi ^{m}(\boldsymbol{x})$ and $\chi ^{m}(\boldsymbol{x})$, respectively
\begin{align}
-\dfrac{\partial}{\partial x_j}\,\left[
\beta_{in}\epsilon_{nrs}\dfrac{\partial \chi_s ^{m}}{\partial x_r}\right]&=
\dfrac{\partial \beta_{im}}{\partial x_j}\quad \text{in}\quad \mathcal{V}=
\mathcal{V}_{1}\cup \mathcal{V}_{2}\,,
\label{E:cellPROB2VEC}\\
-\dfrac{\partial }{\partial x_{j}}\left[ \mu _{jl}\dfrac{\partial \chi ^{m}(%
\boldsymbol{x})}{\partial x_{l}}\right]&=\dfrac{\partial \mu _{jm}}{\partial
x_{j}}\quad \text{in}\quad \mathcal{V}=\mathcal{V}_{1}\cup \mathcal{V}_{2}\,.
\label{E:cellPROBSCALAR}
\end{align}
Using the magnetic field influence
functions $\mathcal{E}_{k}^{m}(\boldsymbol{x})$ and $\mathcal{F}_{k}^{m}(\boldsymbol{x})$ associated with the vector and scalar potential formulations, respectively, we can rewrite the expressions   $\boldsymbol{B}^{(0)}(^{\mathrm{M}}\boldsymbol{x},\boldsymbol{x})=\boldsymbol{B}_{\langle ^{\mathrm{M}}\boldsymbol{x}\rangle }^{(0)}+\boldsymbol{B}_{\langle
\boldsymbol{x}\rangle }^{(1)}$
and
$%
\boldsymbol{H}^{(0)}(^{\mathrm{M}}\boldsymbol{x},\boldsymbol{x})=\boldsymbol{H}%
_{\langle ^{\mathrm{M}}\boldsymbol{x}\rangle }^{(0)}+\boldsymbol{H}_{\langle
\boldsymbol{x}\rangle }^{(1)}$, in separated forms
\begin{align}
B_{k}^{(0)}&=\mathcal{E}_{k}^{m}(\boldsymbol{x})B_{{\langle ^{\mathrm{M}}%
\boldsymbol{x}\rangle },m}^{(0)}\quad \text{with}\quad \mathcal{E}_{k}^{m}(%
\boldsymbol{x})=\delta _{km}
+\epsilon_{kjn}\dfrac{\partial \chi_n^{m}(\boldsymbol{x})}{%
\partial x_{j}}\,,
\label{E:cellPROBVEC}\\
H_{k}^{(0)}&=\mathcal{F}_{k}^{m}(\boldsymbol{x})H_{{\langle ^{\mathrm{M}}%
\boldsymbol{x}\rangle },m}^{(0)}\quad \text{with}\quad \mathcal{F}_{k}^{m}(%
\boldsymbol{x})=\delta _{km}+\dfrac{\partial \chi ^{m}(\boldsymbol{x})}{%
\partial x_{k}}\,.
\label{E:cellPROBSCALAR2}
\end{align}
Using Eqs.\ \eqref{E:constitutive_VEC}
following conjugate forms may be obtained
\begin{align}
H_{i}^{(0)}&=\mathcal{S}_{i}^{m}(\boldsymbol{x})B_{{\langle ^{\mathrm{M}}%
\boldsymbol{x}\rangle },m}^{(0)}\quad \text{with}\quad \mathcal{S}_{i}^{m}(%
\boldsymbol{x})=\beta _{ik}(\boldsymbol{x})\mathcal{E}_{k}^{m}(\boldsymbol{x}%
)=\beta _{im}+\beta _{ik}\epsilon_{kjn}\dfrac{\partial \chi_n ^{m}(\boldsymbol{x})}{\partial
x_{j}}\,,
\label{E:cellPROBH}\\
B_{i}^{(0)}&=\mathcal{T}_{i}^{m}(\boldsymbol{x})H_{{\langle ^{\mathrm{M}}%
\boldsymbol{x}\rangle },m}^{(0)}\quad \text{with}\quad \mathcal{T}_{i}^{m}(%
\boldsymbol{x})=\mu _{ik}(\boldsymbol{x})\mathcal{F}_{k}^{m}(\boldsymbol{x}%
)=\mu _{im}+\mu _{ik}\dfrac{\partial \chi ^{m}(\boldsymbol{x})}{\partial
x_{k}}\,,
\label{E:cellPROBB}
\end{align}
where $\mathcal{S}_{k}^{m}(\boldsymbol{x})$ and $\mathcal{T}_{k}^{m}(\boldsymbol{x})$ are magnetic influence functions conjugate to  $\mathcal{E}_{k}^{m}(\boldsymbol{x})$ and $\mathcal{F}_{k}^{m}(\boldsymbol{x})$, respectively. We now apply the divergence theorem. The  periodicity of $\boldsymbol{H}^{(1)}$ and $\boldsymbol{B}^{(1)}$ in  $\mathcal{V}$ yields
$\int_{%
\mathcal{V}}\text{curl}_{\boldsymbol{x}}\,\boldsymbol{H}^{(1)}\mathrm{d}%
V=-\int_{\mathcal{\partial V}}\,\boldsymbol{H}^{(1)}\times \boldsymbol{n}\,%
\mathrm{d}A=\bs 0$ and $\int_{%
\mathcal{V}}\text{div}_{\boldsymbol{x}}\,\boldsymbol{B}^{(1)}\mathrm{d}%
V=\int_{\mathcal{\partial V}}\,\boldsymbol{B}^{(1)}\cdot \boldsymbol{n}\,%
\mathrm{d}A=0$, respectively. Consequently, Eqs.\ \eqref{E:equlibriumNEU1EXPc} can be integrated over the unit cell to give the following homogenized counterparts of Eqs.\ \eqref{E:multiscale_curlHzero} and \eqref{E:multiscale_divBzero}, respectively
\begin{align}
\dfrac{1}{|\mathcal{V}|}\int_{\mathcal{V}}\left[ \text{curl}_{^{\mathrm{M}}%
\boldsymbol{x}}\,\boldsymbol{H}^{(0)}+\text{curl}_{\boldsymbol{x}}\,%
\boldsymbol{H}^{(1)}\right] \mathrm{d}V&=\text{curl}_{^{\mathrm{M}}\boldsymbol{x}}\,^{\mathrm{M}}%
\boldsymbol{H}=\bs J\,,\\
\dfrac{1}{|\mathcal{V}|}\int_{\mathcal{V}}\left[ \text{div}_{^{\mathrm{M}}%
\boldsymbol{x}}\,\boldsymbol{B}^{(0)}+\text{div}_{\boldsymbol{x}}\,%
\boldsymbol{B}^{(1)}\right] \mathrm{d}V&=\text{div}_{^{\mathrm{M}}\boldsymbol{x}}\,^{\mathrm{M}}%
\boldsymbol{B}=0\,.
\end{align}
Here, $^{\mathrm{M}}\boldsymbol{H}$ and $^{\mathrm{M}}\boldsymbol{B}$ denote volume averaged (macroscopic) magnetic fields with
\begin{equation}
^{\mathrm{M}}\boldsymbol{H}=\dfrac{1}{|\mathcal{V}|}\,\int_{\mathcal{V}}%
\boldsymbol{H}^{(0)}\mathrm{d}V\text{ and }
^{\mathrm{M}}\boldsymbol{B}=\dfrac{1}{|\mathcal{V}|}\,\int_{\mathcal{V}}%
\boldsymbol{B}^{(0)}\mathrm{d}V\,.
\label{E:HhomogenizedVEC}
\end{equation}
Considering vector potential formulation, on account of Eqs.~\eqref{E:vanishingintegralSCALAR} and \eqref{E:cellPROBVEC},
we have $^{\mathrm{M}}\boldsymbol{B}=\boldsymbol{B}_{{\langle ^{\mathrm{M}}%
\boldsymbol{x}\rangle }}^{(0)}$.
Analogously, Eqs.~\eqref{E:vanishingintegralSCALAR} and \eqref{E:cellPROBSCALAR2} yield
$^{\mathrm{M}}\boldsymbol{H}=\boldsymbol{H}_{{\langle ^{\mathrm{M}}%
\boldsymbol{x}\rangle }}^{(0)}$, for scalar potential formulation. As a consequence, substitutions of Eqs.\  \eqref{E:cellPROBH}
and \eqref{E:cellPROBB}
into the right-hand side of Eqs.\ \eqref{E:HhomogenizedVEC}
gives the following coarse-scale complementary constitutive equations, respectively
\begin{equation}
^{\mathrm{M}}\boldsymbol{H}=\boldsymbol{\beta }^{\star }\cdot \,^{\mathrm{M}}%
\boldsymbol{B}\text{ and }
^{\mathrm{M}}\boldsymbol{B}=\boldsymbol{\mu }^{\star }\cdot \,^{\mathrm{M}}%
\boldsymbol{H}\,.
\label{E:equlibriumNEU21VEC}
\end{equation}
Here, $\boldsymbol{\mu}^{\star}$ is the nondimensional macroscopic magnetic permeability tensor and $\boldsymbol{\beta }^{\star}$ is its inverse which possess symmetries of $\boldsymbol{\mu}$ and $\boldsymbol{\beta}$, respectively. Derivations of  $\boldsymbol{\beta }^{\star}$ and $\boldsymbol{\mu}^{\star}$ as a result of vector and scalar potential formulations, respectively, result in
\begin{align}
\beta_{im}^{\star }&=\dfrac{1}{|\mathcal{V}|}\,\int_{\mathcal{V}}\mathcal{S}%
_{i}^{m}(\boldsymbol{x})\mathrm{d}V=\dfrac{1}{|\mathcal{V}|}\,\int_{\mathcal{V}}\beta_{ik}\left[%
\delta_{km}+
\epsilon_{kjn}\dfrac{\partial \chi_n ^{m}(\boldsymbol{x})}{\partial
x_{j}}
\right]
\mathrm{d}V\,, \label{E:invPERMhomogenizedVEC} \\
\mu _{im}^{\star }&=\dfrac{1}{|\mathcal{V}|}\,\int_{\mathcal{V}}\mathcal{T}%
_{i}^{m}(\boldsymbol{x})\mathrm{d}V=\dfrac{1}{|\mathcal{V}|}\,\int_{\mathcal{V}}\mu_{ik}\left[%
\delta_{km}+ \dfrac{\partial \chi^{m}(\boldsymbol{x})}{\partial x_k}\right]
\mathrm{d}V\,.
\label{E:PERMhomogenizedSCALAR}
\end{align}
\begin{remark} Considering vector potential formulation and letting $\bs\chi^{m}(\boldsymbol{x})\to\bs 0$ in Eq.\ \eqref{E:invPERMhomogenizedVEC} yields the (lower) Reuss bound for $\boldsymbol{\mu}^\star=[
\boldsymbol{\beta}^\star]^{-1}$ with
\begin{align*}
\mu^{\star,-1}_{im}
\to \dfrac{1}{|\mathcal{V}|}\,\int_{\mathcal{V}}\mu^{-1}_{im} \mathrm{d}V\,.
\end{align*}
Likewise, turning to the case of scalar potential formulation and letting $\chi^{m}(\boldsymbol{x})\to0$ in Eq.\ \eqref{E:PERMhomogenizedSCALAR} gives
the (upper) Voigt bound  for $\boldsymbol{\mu}^\star$ with
\begin{align*}
\mu^\star_{im} \to \dfrac{1}{|\mathcal{V}|}\,\int_{\mathcal{V}}\mu_{im} \mathrm{%
d}V \,.
\end{align*}
\end{remark}
The macroscale boundary value
at $\mathcal{B}$ concerns with finding the
coarse-scale magnetic vector and scalar potentials, that is $^\mathrm{M}\bs A=\bs A^{(0)}(^\mathrm{M}\boldsymbol{x})$ and  $^\mathrm{M}\varrho=\varrho^{(0)}(^%
\mathrm{M}\boldsymbol{x})$, which respectively satisfy the following differential equations
\begin{align}
\boldsymbol{J}&=\text{curl}_{^\mathrm{M}\boldsymbol{x}}\, ^\mathrm{M}\boldsymbol{H}%
\quad\text{in }\mathcal{B}\,,  \label{E:equlibriumNEU1MACROVEC}\\
\boldsymbol{0}&=\text{div}_{^\mathrm{M}\boldsymbol{x}}\, ^\mathrm{M}\boldsymbol{B}%
\quad\text{in }\mathcal{B}\,,
\label{E:equlibriumNEU1MACRO}
\end{align}
as well as the imposed boundary conditions. Analogical to our indications for the composite domain, the Dirichlet and Neumann boundary conditions for the vector potential  formulation constitute $^\mathrm{M}\bs A=$%
$^\mathrm{M}\widetilde{\bs A}$ on $\partial \mathcal{B}_{\bs A}$ and $^%
\mathrm{M}\boldsymbol{H}\times $$^\mathrm{M}\boldsymbol{n}=$$^\mathrm{M}\widetilde{\bs  t
}$ on $\partial \mathcal{B}_{{\bs t}}$, respectively. Here, $^\mathrm{M}\boldsymbol{n}$
is the unit normal to the surface at the macroscale. Likewise, $^\mathrm{M}\varrho=$%
$^\mathrm{M}\widetilde{\varrho}$ on $\partial \mathcal{B}_{\varrho}$ and $^%
\mathrm{M}\boldsymbol{B}\cdot $$^\mathrm{M}\boldsymbol{n}=$$^\mathrm{M}\widetilde{t%
}$ on $\partial \mathcal{B}_{{t}}$ respectively constitute the Dirichlet and Neumann boundary conditions associated with the scalar potential formulation. The boundary parts satisfy $\partial \mathcal{B}_{{\bs A}} \bigcap
\partial \mathcal{B}_{\bs t}=\emptyset$, $\partial \mathcal{B}_{{\bs A}}
\bigcup \partial \mathcal{B}_{\bs t}=\partial \mathcal{B}$,  $\partial \mathcal{B}_{{\varrho}} \bigcap
\partial \mathcal{B}_{t}=\emptyset$ and $\partial \mathcal{B}_{{\varrho}}
\bigcup \partial \mathcal{B}_{t}=\partial \mathcal{B}$.
\subsubsection{Computation of Components through Dirichlet Boundary Conditions}
The cell problems given in Eqs.\ \eqref{E:cellPROB2VEC} and \eqref{E:cellPROBSCALAR}, respectively given for  vector and scalar potential formulations, can be solved for $\bs\chi ^{m}(\boldsymbol{x})$ and $\chi ^{m}(\boldsymbol{x})$ by applying Neumann boundary conditions corresponding to the right-hand-side source term at the phase boundaries, see, e.g., \cite{LUKKASSEN1995519}. Algorithmically, this construction, adapted from \emph{the displacement method} of \cite{LUKKASSEN1995519}, uses only Dirichlet boundary condition associations at the cell boundary and does not  require the detection of phase  boundary surfaces for the imposition of boundary conditions. In what follows, we give details of our implementation. For simplicity of notation, from now on,  we denote the nondimensional cell-scale fields $\boldsymbol{B}^{(0)}$ and $\boldsymbol{H}^{(0)}$ with $\boldsymbol{B}$ and
$\boldsymbol{H}$, respectively.

For each of the magnetic vector and scalar potential formulations, we respectively seek three solutions in the corresponding \emph{total} magnetic potential denoted by $\bs\psi^{m}(\boldsymbol{x})$ and $\psi^{m}(\boldsymbol{x})$ indexed with $m$. For $m=1,2,3$, $\bs\psi^{m}(\boldsymbol{x})$ and $\psi^{m}(\boldsymbol{x})$ have the following additively decomposed forms, respectively
\begin{align}
\bs\psi^{m}(\boldsymbol{x})=\bs\chi^{m}(%
\boldsymbol{x})+\bs\varpi^{m}(\boldsymbol{x})\quad&\text{with}\quad\bs\varpi^{m}(\boldsymbol{x})=[1/2][\bs e_m \times \bs x]\,,\label{E:total_soln_VEC}\\
\psi^{m}(\boldsymbol{x})=-\chi^{m}(\boldsymbol{x})-\varpi^{m}(\boldsymbol{x})
\quad&\text{with}\quad\varpi^{m}(\boldsymbol{x})=x_m\,.
\label{E:total_soln_SCALAR}
\end{align}
Here, $\bs\chi ^{m}(\boldsymbol{x})$ and $\chi ^{m}(\boldsymbol{x})$ are the unknown periodic fluctuation fields whereas, as indicated above,  $\bs\varpi^{m}(\boldsymbol{x})$ and $\varpi^{m}(\boldsymbol{x})$ are known\footnote{The known part is imposed at the control nodes, e.g., the periodic unit cell corners. In this case, $\bs\chi ^{m}(\boldsymbol{x})$ and $\chi ^{m}(\boldsymbol{x})$ vanish at the periodic unit cell corners. Alternatively, the control nodes can be selected over a reference unit cube. In this case, the uniqueness of the fluctuation field can be provided by fixing it at an arbitrary node in the unit cell domain.}. Under the above definitions and using along with Eqs.\
\eqref{E:cellPROBVEC} and \eqref{E:cellPROBSCALAR2} for vector and scalar potential formulations, we  respectively obtain\footnote{Application of averaging given in Eqs.\ \eqref{E:HhomogenizedVEC} to Eqs.\ \eqref{E:BkmVEC} and \eqref{E:Hkm} with the use of
Eqs.\ \eqref{E:vanishingintegralSCALAR} give the following unit vector representations of the
macroscopic magnetic fields $^\mathrm{M}%
\boldsymbol{B}^m=\boldsymbol{e}_m$ and $^\mathrm{M}%
\boldsymbol{H}^m=\boldsymbol{e}_m$ for vector and scalar potential formulations, respectively
\begin{align*}
[^\mathrm{M}\boldsymbol{B}^1]=%
\begin{bmatrix}
1 \\
0 \\
0%
\end{bmatrix}%
\,,\quad [^\mathrm{M}\boldsymbol{B}^2]=%
\begin{bmatrix}
0 \\
1 \\
0%
\end{bmatrix}%
\,,\quad [^\mathrm{M}\boldsymbol{B}^3]=%
\begin{bmatrix}
0 \\
0 \\
1%
\end{bmatrix}%
\,,\text{ and }
[^\mathrm{M}\boldsymbol{H}^1]&=%
\begin{bmatrix}
1 \\
0 \\
0%
\end{bmatrix}%
\,,\quad [^\mathrm{M}\boldsymbol{H}^2]=%
\begin{bmatrix}
0 \\
1 \\
0%
\end{bmatrix}%
\,,\quad [^\mathrm{M}\boldsymbol{H}^3]=%
\begin{bmatrix}
0 \\
0 \\
1%
\end{bmatrix}\,.
\end{align*}}
\begin{align}
B_{k}^m=B_{k}(\bs\psi_{\varrho}^{m})&=\mathcal{E}_{k}^{m}(\boldsymbol{x})\,,
\label{E:BkmVEC}\\
H_{k}^m=H_{k}(\psi_{\varrho}^{m})&=\mathcal{F}_{k}^{m}(\boldsymbol{x})\,.
\label{E:Hkm}
\end{align}
If we apply Eqs.\ \eqref{E:cellPROBH} and  \eqref{E:cellPROBB}, respectively, the conjugate
magnetic fields $\boldsymbol{H}^m$ and $\boldsymbol{B}^m$  are obtained
\begin{align}
H_{i}^m&=H_{i}(\bs\psi^{m})=\mathcal{S}_{i}^{m}(\boldsymbol{x})\,, \label{E:HkmVEC}\\
B_{i}^m&=B_{i}(\psi^{m})=\mathcal{T}_{i}^{m}(\boldsymbol{x})\,.  \label{E:Bkm}
\end{align}
Averaging over the unit cell, corresponding homogenized magnetic fields $^\mathrm{M}\boldsymbol{H}^m$ and $^\mathrm{M}\boldsymbol{B}^m$ are obtained
\begin{align}
^\mathrm{M}{H}_i^m&=\dfrac{1}{|\mathcal{V}|}\,\int_{\mathcal{V}%
}H_{i}(\bs\psi^{m})=\dfrac{1}{|\mathcal{V}|}\,\int_{\mathcal{V}}\mathcal{S}%
_{i}^{m}(\boldsymbol{x}) \mathrm{d}V\,,  \label{E:solnVEC}\\
^\mathrm{M}{B}_i^m&=\dfrac{1}{|\mathcal{V}|}\,\int_{\mathcal{V}%
}B_{i}(\psi^{m})=\dfrac{1}{|\mathcal{V}|}\,\int_{\mathcal{V}}\mathcal{T}%
_{i}^{m}(\boldsymbol{x}) \mathrm{d}V\,.  \label{E:soln}
\end{align}
Comparison of above and Eqs.\ \eqref{E:invPERMhomogenizedVEC} and \eqref{E:PERMhomogenizedSCALAR} yields the following expressions in matrix form for the inverse permeability and permeability tensors computed using vector and scalar potential formulations, respectively
\begin{align}
[\boldsymbol{\beta}^\star]&=[^\mathrm{M}\boldsymbol{H}^1 | ^\mathrm{M}%
\boldsymbol{H}^2 | ^\mathrm{M}\boldsymbol{H}^3] \,, \label{E:MHkmVEC}\\
[\boldsymbol{\mu}^\star]&=[^\mathrm{M}\boldsymbol{B}^1 | ^\mathrm{M}%
\boldsymbol{B}^2 | ^\mathrm{M}\boldsymbol{B}^3] \,.  \label{E:MHkmSCALAR}
\end{align}
This is to say that, for the vector potential formulation, the component of the macroscopic magnetic permeability tensor $%
\boldsymbol{\beta}^\star$ with the indices $km$ corresponds to the homogenized
magnetic induction tensor component $^\mathrm{M}{H}_{k}$ due to \emph{total}
magnetic vector potential influence function $\bs \psi^{m}$ computed for the
imposed macroscopic magnetic field at unit cell vertices with $\bs\varpi^{m}$. The same reasoning applies to the case of the scalar potential formulation; however, this time,  the components of the macroscopic magnetic permeability tensor $\boldsymbol{\mu}^\star$ are inferred. Although, for the most general case, the computation of the six linearly independent macroscopic constitutive constants making up the  permeability tensor requires three load cases given in Eqs.\ \eqref{E:total_soln_VEC} and \eqref{E:total_soln_SCALAR} for $m=1,2,3$, once the symmetry elements of the point group of the underlying composite geometry, which should be included in the physical property symmetry elements with Neumann's principle \cite{voigt1910lehrbuch, nye1985physical}, allow, less number of tests can suffice. For instance, considering two-dimensional tensor formalism, the magnetic response of microstructures possessing cubic symmetry is isotropic; in consequence, their magnetic permeability tensor is spherically symmetric. Therefore, only one test can determine the corresponding nonzero main-diagonal element. The algorithmic treatment of the homogenization procedure for vector and scalar potential formulations are summarized in Algorithms\ \ref{algo:solution_vector} and \ref{algo:solution_scalar}, respectively.

\begin{algorithm}[ht!]
\label{A:1}
\SetAlgoLined
\caption{FE-based computational periodic homogenization for vector potential formulation. Macroscopic loading is applied with PBC to result in $^\mathrm{M}\boldsymbol{B}^m\gets\boldsymbol e_m$. The notations $\boldsymbol\psi_{+}^{m}=\boldsymbol\psi^{m}(\boldsymbol{x}_{+})$ and $\boldsymbol\psi_{-}^{m}=\boldsymbol\psi^{m}(\boldsymbol{x}_{-})$ apply where $[ \boldsymbol x_+ - \boldsymbol x_- ]$ are position vectors of periodically located nodes on the surface $\partial \mathcal V$.}%
    \KwIn{%
         Periodically discretized FE model of the cell $\mathcal V$ with identified $[\boldsymbol{x}_+,\boldsymbol{x}_-]$;
         }%

         {\bf Initialization:} for an arbitrary node in  $\mathcal V$ assign $\boldsymbol\psi\gets\boldsymbol 0$ for uniqueness;

  \For{$m = 1;\ m \leq 3;\ m = m + 1$}{

        apply loading under PBC with  $\boldsymbol\psi_{+}^{m}-\boldsymbol\psi_{-}^{m}\gets[1/2]\boldsymbol e_m \times [ \boldsymbol x_+ - \boldsymbol x_- ]$;

        solve the equation $\text{curl}_{\boldsymbol{x}}\,\boldsymbol{H}(\boldsymbol\psi^{m})=\boldsymbol{0}$ over the unit cell for $\boldsymbol\psi^{m}$;

        compute $\boldsymbol{B}(\boldsymbol\psi^{m})\gets\text{curl}_{\boldsymbol{x}}\,\boldsymbol\psi^{m}$;

        compute $\boldsymbol{H}(\boldsymbol\psi^{m})\gets\boldsymbol \mu^{-1} \cdot \boldsymbol{B}(\boldsymbol\psi^{m})$;

        compute $^\mathrm{M}\boldsymbol{H}^m\gets\dfrac{1}{|\mathcal{V}|}\,\int_{\mathcal{V}%
}\boldsymbol{H}(\boldsymbol\psi^{m})$;
        }
     \KwOut{%
        $[\boldsymbol{\beta}^\star]=
        \begin{bmatrix}
        ^\mathrm{M}{H}_1^1 & ^\mathrm{M}{H}_1^2 & ^\mathrm{M}{H}_1^3 \\
        ^\mathrm{M}{H}_2^1 & ^\mathrm{M}{H}_2^2 & ^\mathrm{M}{H}_2^3 \\
        ^\mathrm{M}{H}_3^1 & ^\mathrm{M}{H}_3^2 & ^\mathrm{M}{H}_3^3
        \end{bmatrix}\,.$
        }%
\label{algo:solution_vector}%
\end{algorithm}

\begin{algorithm}[ht!]
\label{A:2}
\SetAlgoLined
\caption{FE-based computational periodic homogenization for scalar potential formulation. Macroscopic loading is applied with PBC to result in $^\mathrm{M}\boldsymbol{H}^m\gets\boldsymbol e_m$. The notations $\psi_{+}^{m}=\psi^{m}(\boldsymbol{x}_{+})$ and $\psi_{-}^{m}=\psi^{m}(\boldsymbol{x}_{-})$ apply where $[ \boldsymbol x_+ - \boldsymbol x_- ]$ are position vectors of periodically located nodes on the surface $\partial \mathcal V$.}%
    \KwIn{%
         Periodically discretized FE model of the cell $\mathcal V$ with identified $[\boldsymbol{x}_+,\boldsymbol{x}_-]$;
         }%

         {\bf Initialization:} for an arbitrary node in  $\mathcal V$ assign $\psi\gets0$ for uniqueness;

  \For{$m = 1;\ m \leq 3;\ m = m + 1$}{

        apply loading under PBC with  $\psi_{+}^{m}-\psi_{-}^{m}\gets[ x_{+,m} -  x_{-,m} ]$;

solve the equation
 $\text{div}_{\boldsymbol{x}}\,\boldsymbol{B}(\psi^{m})=0$
  over the unit cell for
 $\psi^{m}$;

        compute $\boldsymbol{H}(\psi^{m})\gets-\boldsymbol{\nabla }_{\boldsymbol{x}}\,\psi^{m}$;

        compute $\boldsymbol{B}(\psi^{m})\gets\boldsymbol \mu \cdot \boldsymbol{H}(\psi^{m})$;

        compute $^\mathrm{M}\boldsymbol{B}^m\gets\dfrac{1}{|\mathcal{V}|}\,\int_{\mathcal{V}%
}\boldsymbol{B}(\psi^{m})$;
        }
     \KwOut{%
        $[\boldsymbol{\mu}^\star]=
        \begin{bmatrix}
        ^\mathrm{M}{B}_1^1 & ^\mathrm{M}{B}_1^2 & ^\mathrm{M}{B}_1^3 \\
        ^\mathrm{M}{B}_2^1 & ^\mathrm{M}{B}_2^2 & ^\mathrm{M}{B}_2^3 \\
        ^\mathrm{M}{B}_3^1 & ^\mathrm{M}{B}_3^2 & ^\mathrm{M}{B}_3^3
        \end{bmatrix}\,.$
        }%
\label{algo:solution_scalar}%
\end{algorithm}
\subsubsection{Magnetic Anisotropy Index}
In this part, we develop an index for quantification of the degree of anisotropy in the magnetic permeability of materials. To this end, we use a permeability ellipsoid (or ellipse in 2D) for the graphical representation of the second-order symmetric permeability tensor. Among other possible ellipsoids making the indicatrice surfaces geometrically representing second-order tensors, ours is \emph{the ellipsoid of the values of symmetric second-order tensor} \cite[p.\ 22]{Shuvalov1988}. For a second-order stress tensor, this corresponds to Lam\'{e}'s stress ellipsoid \cite[p.\ 215]{TimoshenkoGoodier1951}. The corresponding  meridional eccentricity of the ellipsoid, which is the eccentricity of the ellipse at the section of the longest and the shortest  radii, respectively denoted by $r_\mathrm{max}=\mu_\mathrm{max}$ and $r_\mathrm{min}=\mu_\mathrm{min}$, which gives $e=\sqrt{1-r_\mathrm{min}^2/r_\mathrm{max}^2}$ is then used as the anisotropy index with $0\leq e \leq 1$. Here, $\mu_\mathrm{max}$ and $\mu_\mathrm{min}$ are the largest and the smallest eigenvalues of the permeability tensor. For $e=0$, the permeability ellipsoid transforms onto a sphere which indicates isotropy in magnetic permeability. As $e$ increases, the directionality in the material's permeability also does so.
\section{Applications}
\subsection{Motivation: 2D Magnetostatics}
We are concerned only about two-dimensional problems for which the fields $\bs B$ and $\bs H$ have
merely in-plane components with $[\bs B]=[B_1, B_2]^\top$ and $[\bs H]=[H_1, H_2]^\top$. Considering two-phase composites, indexed with $i=1,2$, we let  $\boldsymbol{\mu}_i=\mu_i\,\boldsymbol{1}$ and $\boldsymbol{\beta}_i=\beta_i\,\boldsymbol{1}$ where $\mu_i=1/\beta_i$
with the assumption of isotropy at the microscale.
Here $\boldsymbol{1}$ is the second-order identity tensor.
 This allows representing the governing differential equations to be solved over the unit cell, see Algorithms\ 1 and 2,
 for both the vector and the scalar potential formulations, that is  $\text{curl}_{\boldsymbol{x}}\,%
\boldsymbol{H}=0$  and $\text{div}_{\boldsymbol{x}}\,%
\boldsymbol{B}=0$ respectively, in the following generic form in Cartesian components\footnote{Besides magnetostatics in scalar potential, the elliptic differential equation $\text{div}_{\boldsymbol{x}}\,%
\boldsymbol{\alpha}=0$, governs dielectrics, electrical conduction, and thermal conduction. Considering two-dimensional reduction, vector potential formulation can also be treated in the same formalism. This allows one to use any available  physics engine in the software to solve for magnetostatics by carefully mapping parameters.  In the current study, we exploit this condition and use \textsc{Abaqus} coupled
thermo-electrical analysis framework by omitting thermal fields. As opposed to \textsc{Abaqus} edge-based elements, which are generally used for modeling magnetostatics with vector potential, current use corresponds to a node-based approach. An important point to be considered is during post-processing. Although in scalar potential formulation, derivation of the magnetic field $\bs H=-\bs \nabla_{\bs x} \varrho$ which uses the exact form in electrical conductivity as  $\bs E=-\bs \nabla_{\bs x} \phi$ with $\bs E$ and $\phi$ denote the electrical field and scalar electric potential, respectively, which results in
$\bs H=\partial \varrho/\partial x_1 \bs e_1+\partial \varrho/\partial x_2 \bs e_2$ and $\bs E=\partial \phi/\partial x_1 \bs e_1+\partial \phi/\partial x_2 \bs e_2$. In magnetic vector  potential, however, $\bs B=\mathrm{curl}_{\bs x} \bs A$, which for the current condition where $\bs A=A_3 \bs e_3$ gives
$\bs B=\partial A_3/\partial x_2 \bs e_1+\partial A_3/\partial x_1 \bs e_2$.}
\begin{align}
\dfrac{\partial}{\partial x_1}
\left[\xi\dfrac{\partial \zeta}{\partial x_1}\right]+
\dfrac{\partial}{\partial x_2}
\left[\xi\dfrac{\partial \zeta}{\partial x_2}\right]=0\,.
\label{E:2D_generic}
\end{align}
Here, the vector potential formulation is recovered through replacements $\xi\leftarrow\beta$ and $\zeta\leftarrow A_3$, whereas for the scalar potential formulation, we use the substitutions $\xi\leftarrow\mu$ and $\zeta\leftarrow\varrho$.
\subsection{Considered Microstructures and Associated Model Generation}
In the current section, we consider regular and random  arrangements of nonoverlapping and overlapping monodisperse disks in a two-dimensional matrix. All generations are periodic. In three dimensions, these correspond to unidirectional circular cylindrical inclusions in a matrix material, which  extend indefinitely along the main axis of cylindrical inclusions. We take $\mu_\mathrm{i}=250\mu_0$ and $\mu_\mathrm{m}=\mu_0$ for the inclusion and the matrix, respectively. We also consider  phase-interchange, which corresponds to interchanged phases contemplating the same microstructure\cite{Torquato1991}.

\subsubsection{Regular Nonoverlapping Square Disk Arrangements}
\begin{figure*}[htb!]
% \vspace{5pt}
% trim=left bottom right top, clip
\centering
\frame{\includegraphics[height=0.125\textwidth,
trim=1317 294 1639 27, clip]{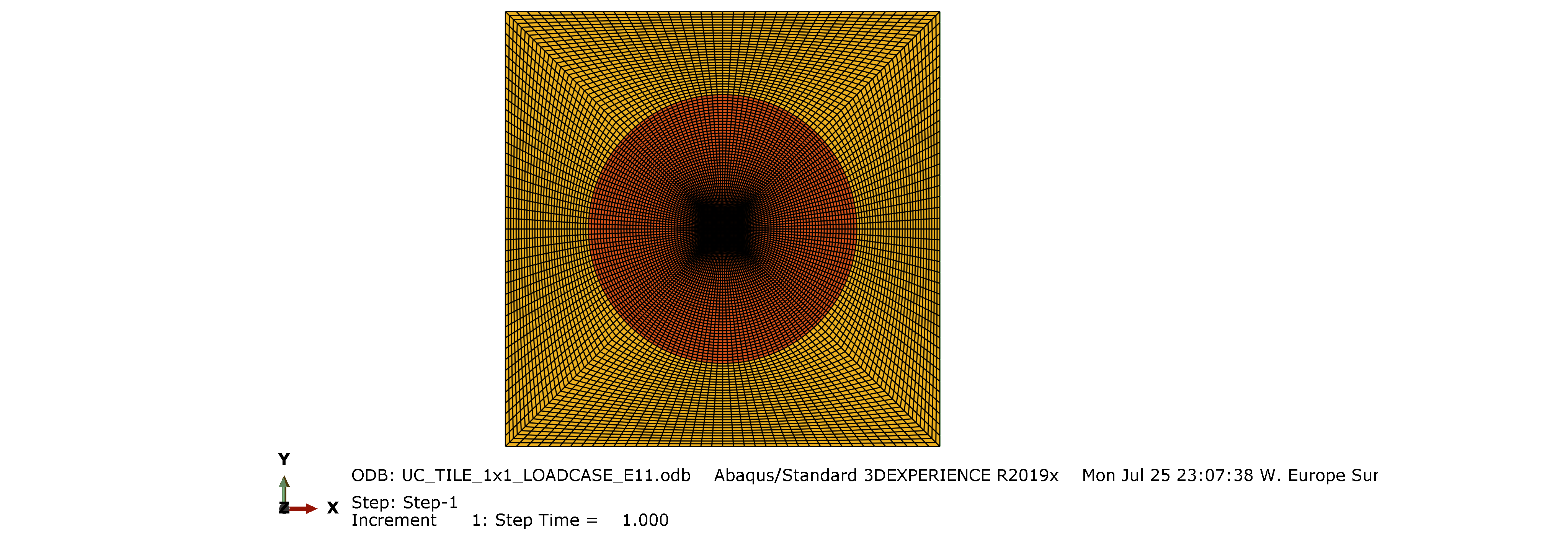}}
\frame{\includegraphics[height=0.125\textwidth,
trim=1238 324 1595 27, clip]{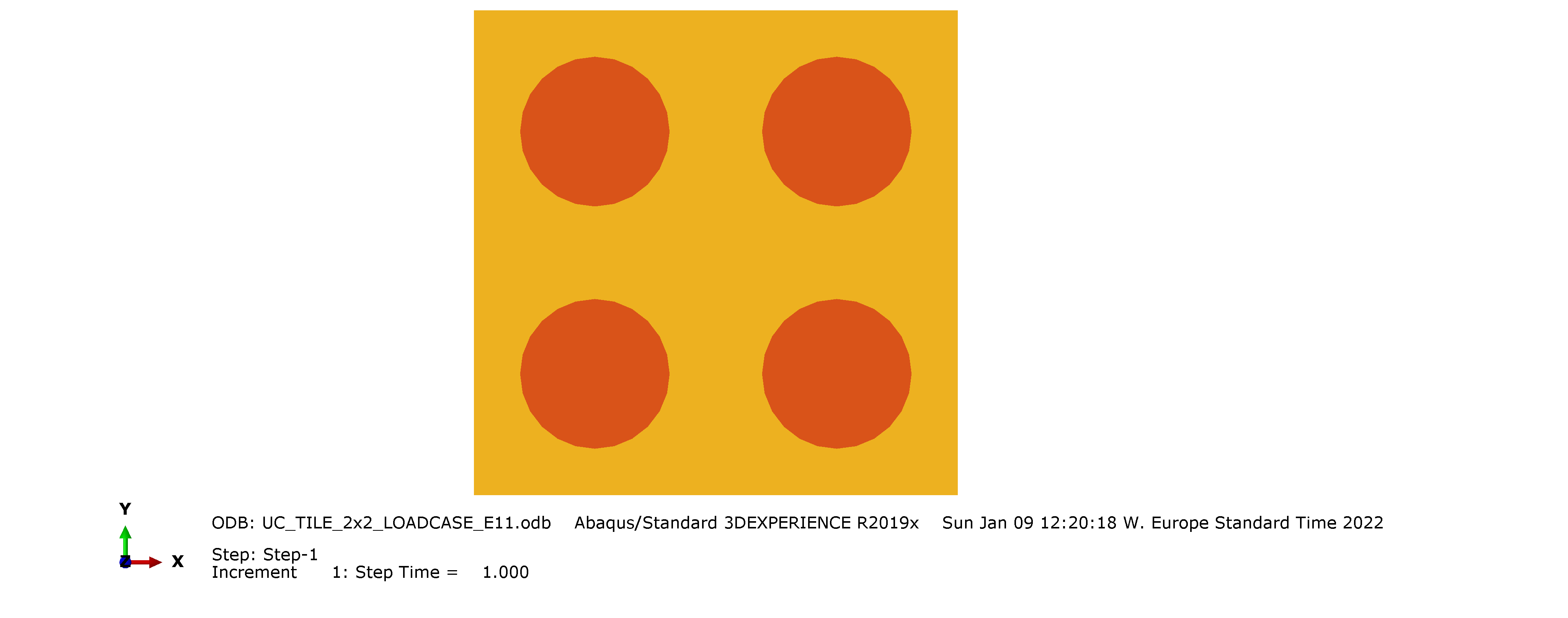}}
\frame{\includegraphics[height=0.125\textwidth,
trim=1238 324 1595 27, clip]{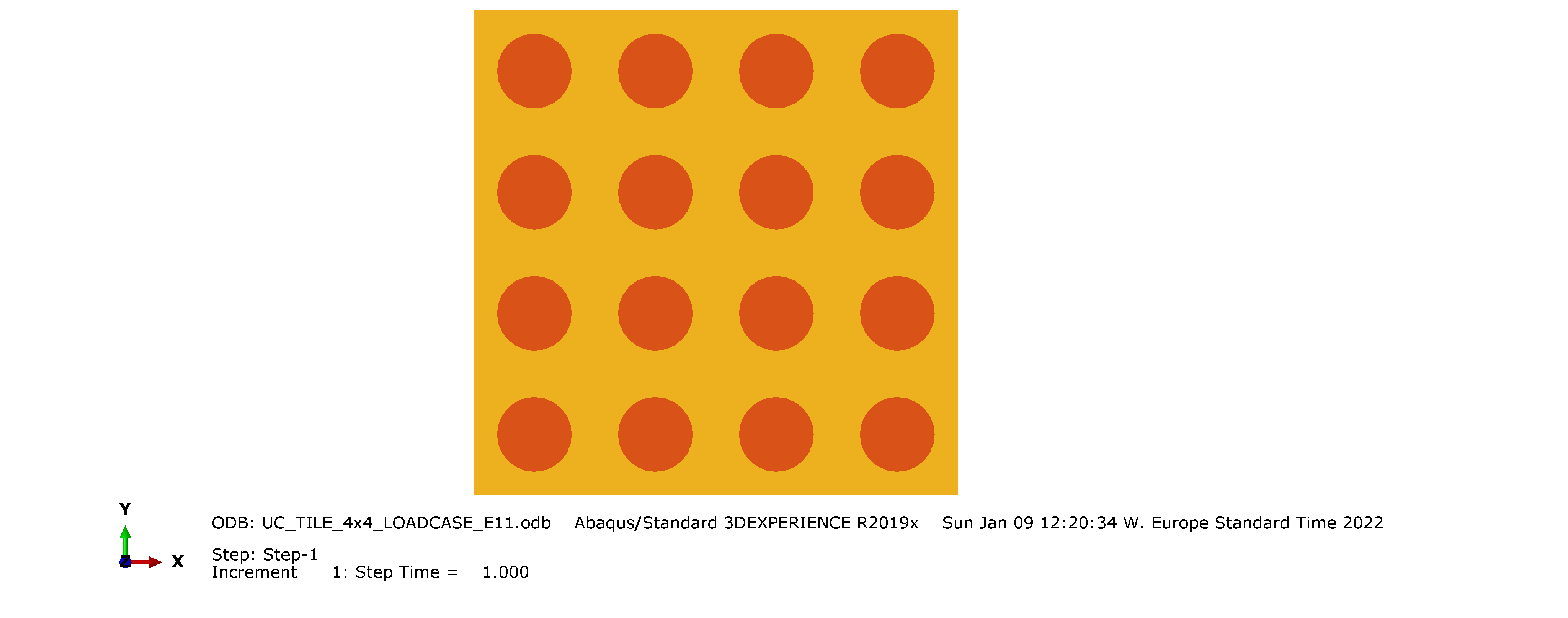}}
\frame{\includegraphics[height=0.125\textwidth,
trim=1238 324 1595 27, clip]{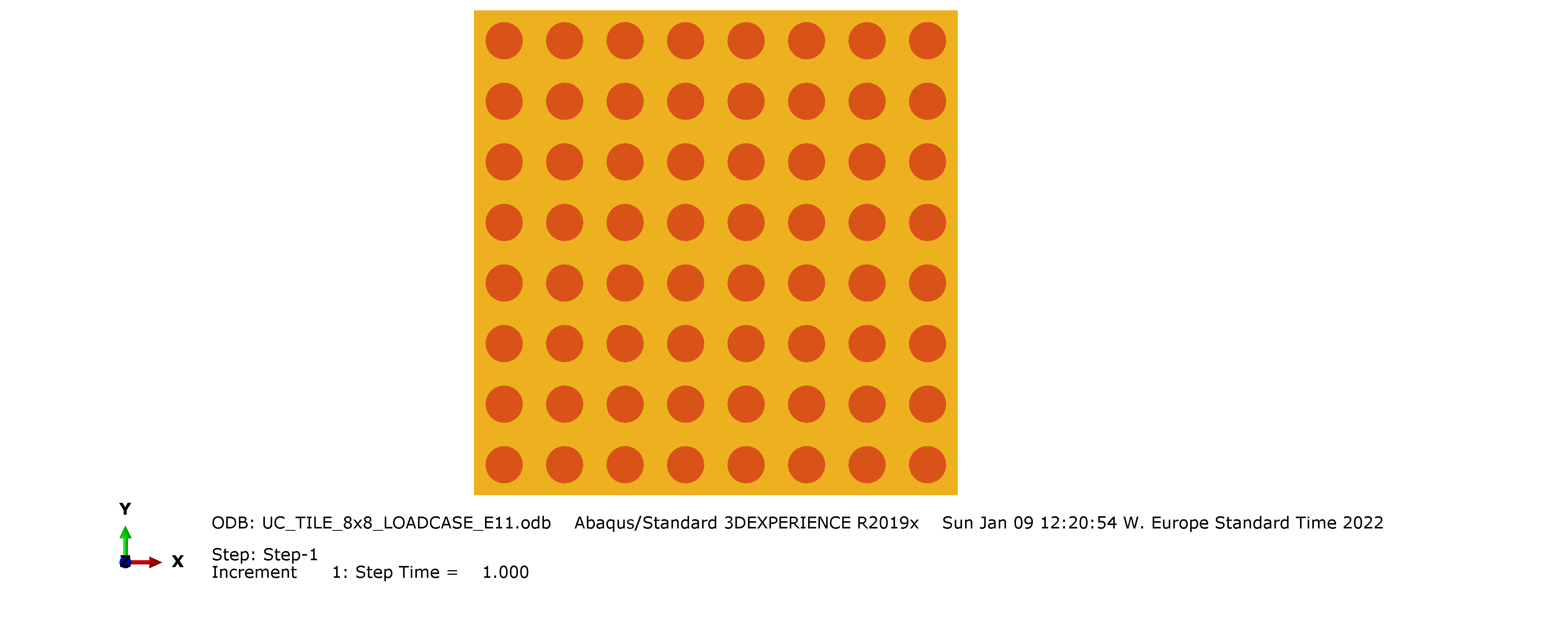}}
\frame{\includegraphics[height=0.125\textwidth,
trim=1238 324 1595 27, clip]{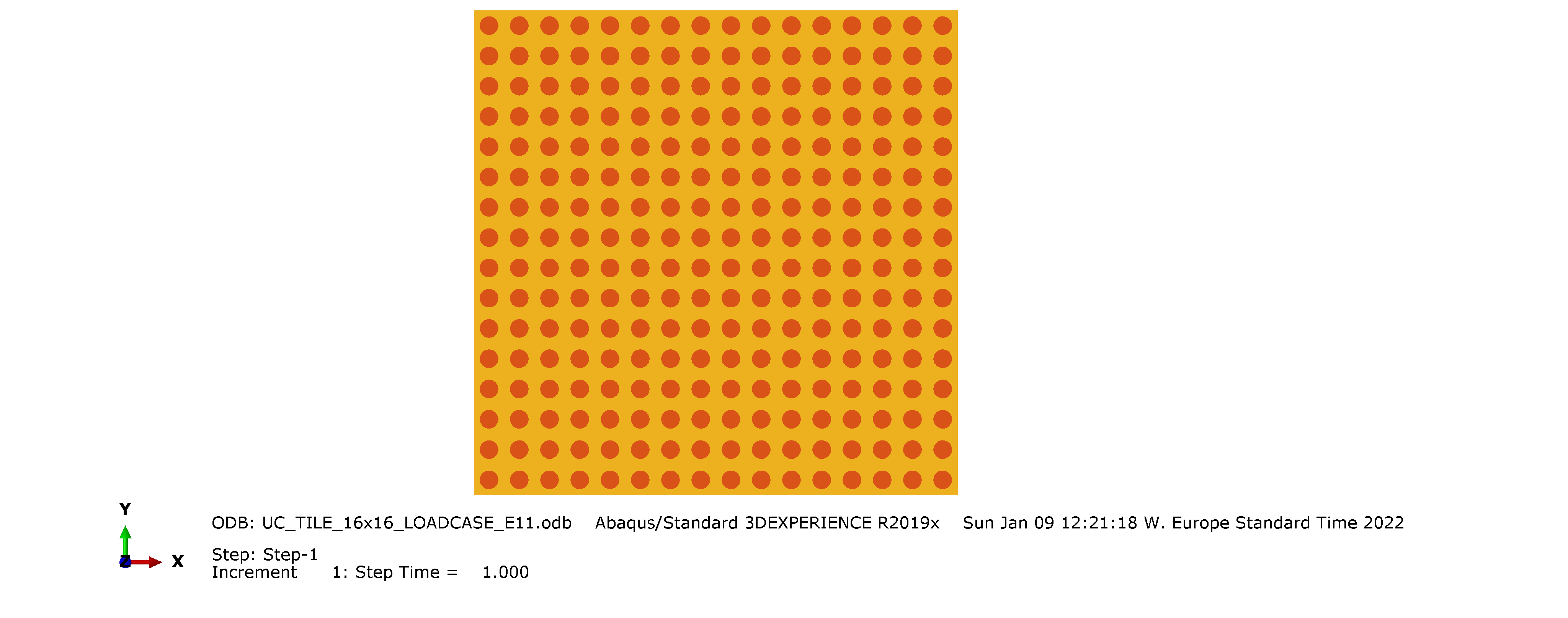}}
\frame{\includegraphics[height=0.125\textwidth,
trim=1238 324 1595 27, clip]{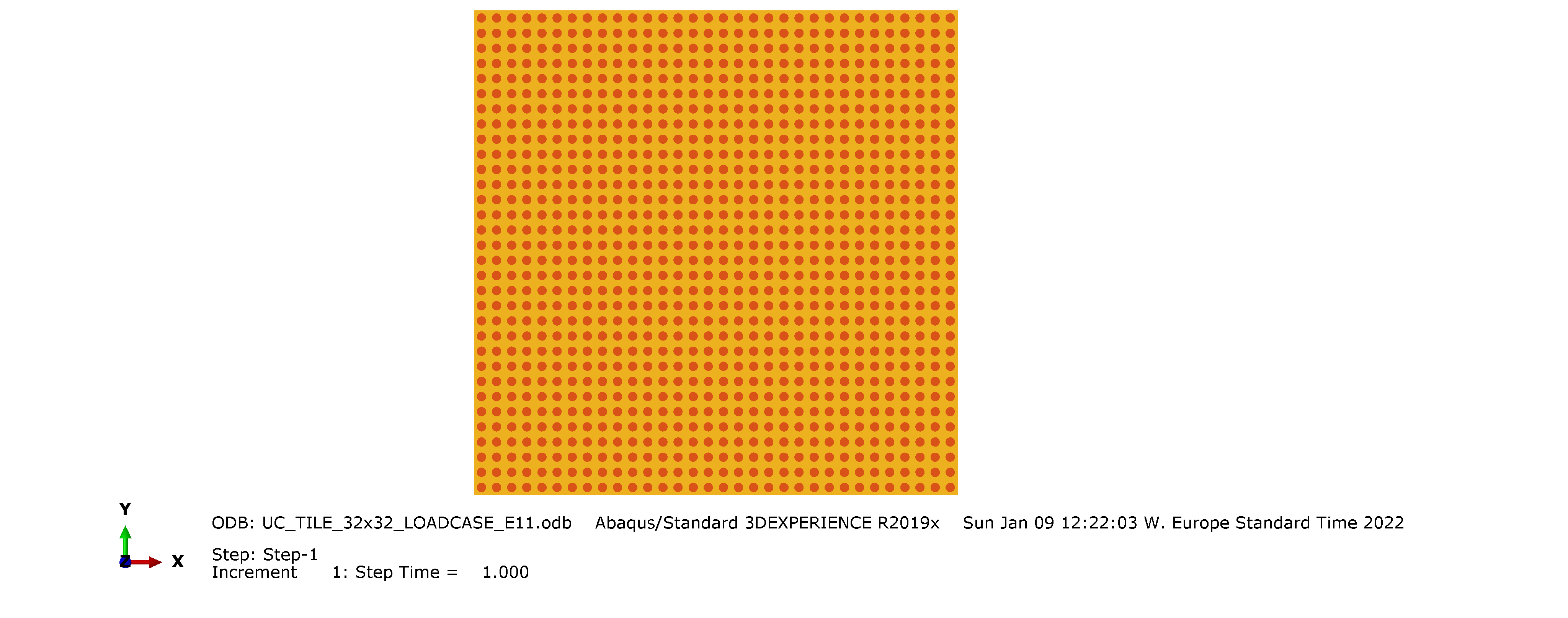}}\\
\vspace{2.5pt}
\frame{\includegraphics[height=0.125\textwidth,
trim=1317 294 1639 27, clip]{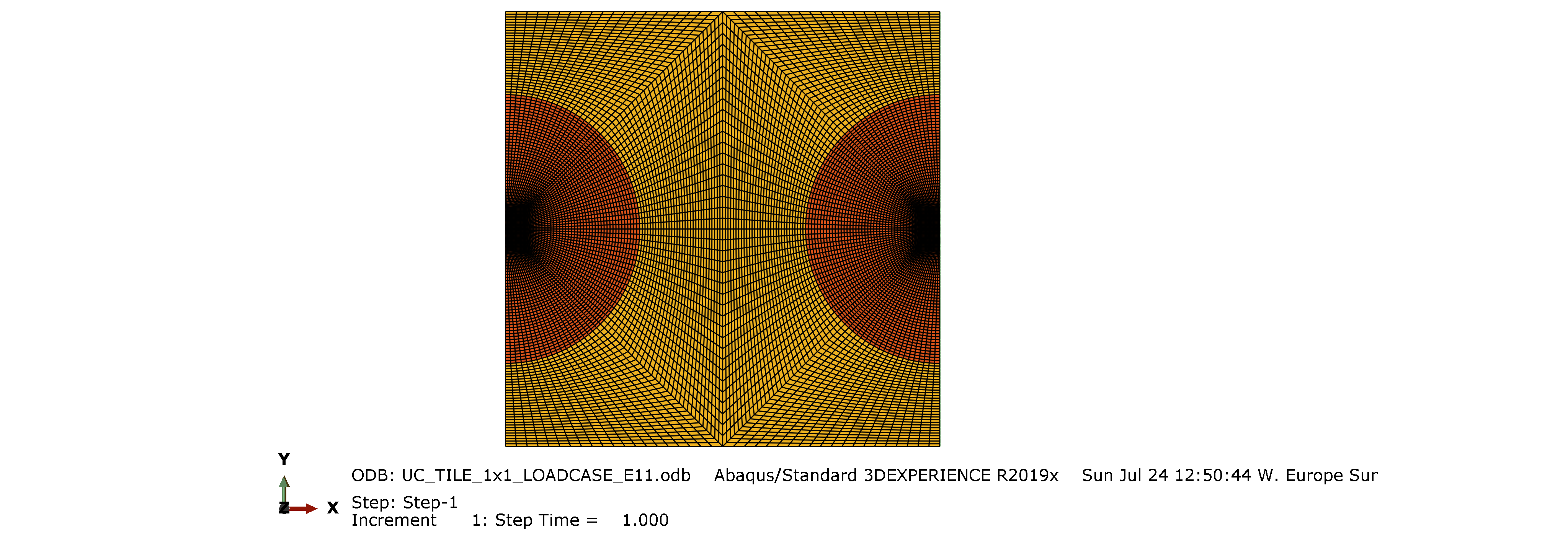}}
\frame{\includegraphics[height=0.125\textwidth,
trim=1238 324 1595 27, clip]{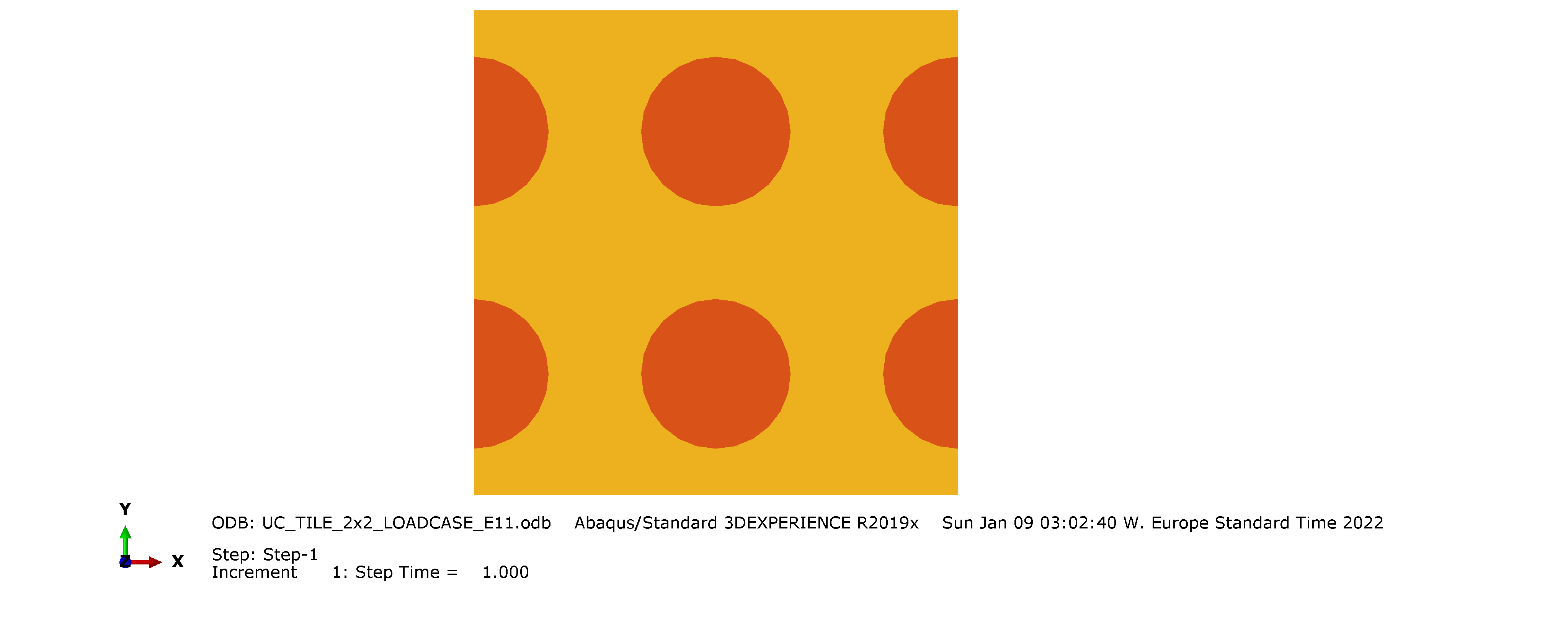}}
\frame{\includegraphics[height=0.125\textwidth,
trim=1238 324 1595 27, clip]{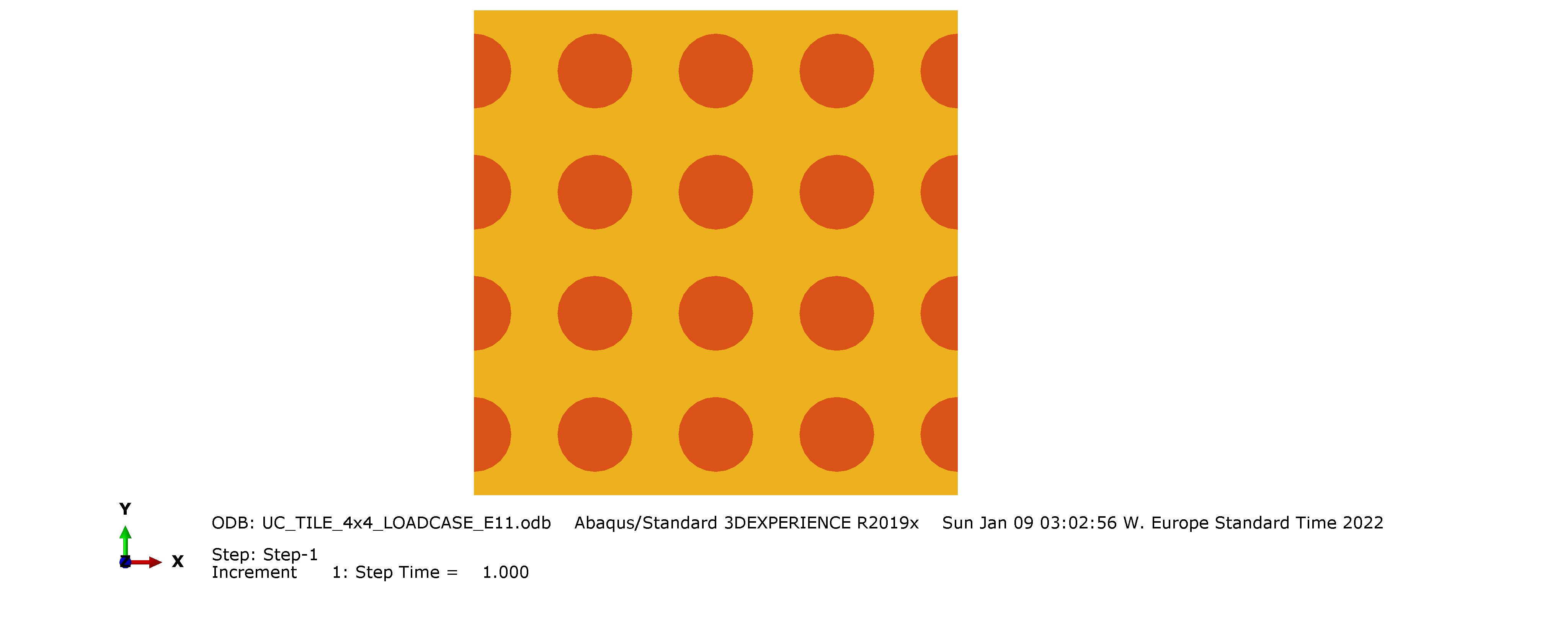}}
\frame{\includegraphics[height=0.125\textwidth,
trim=1238 324 1595 27, clip]{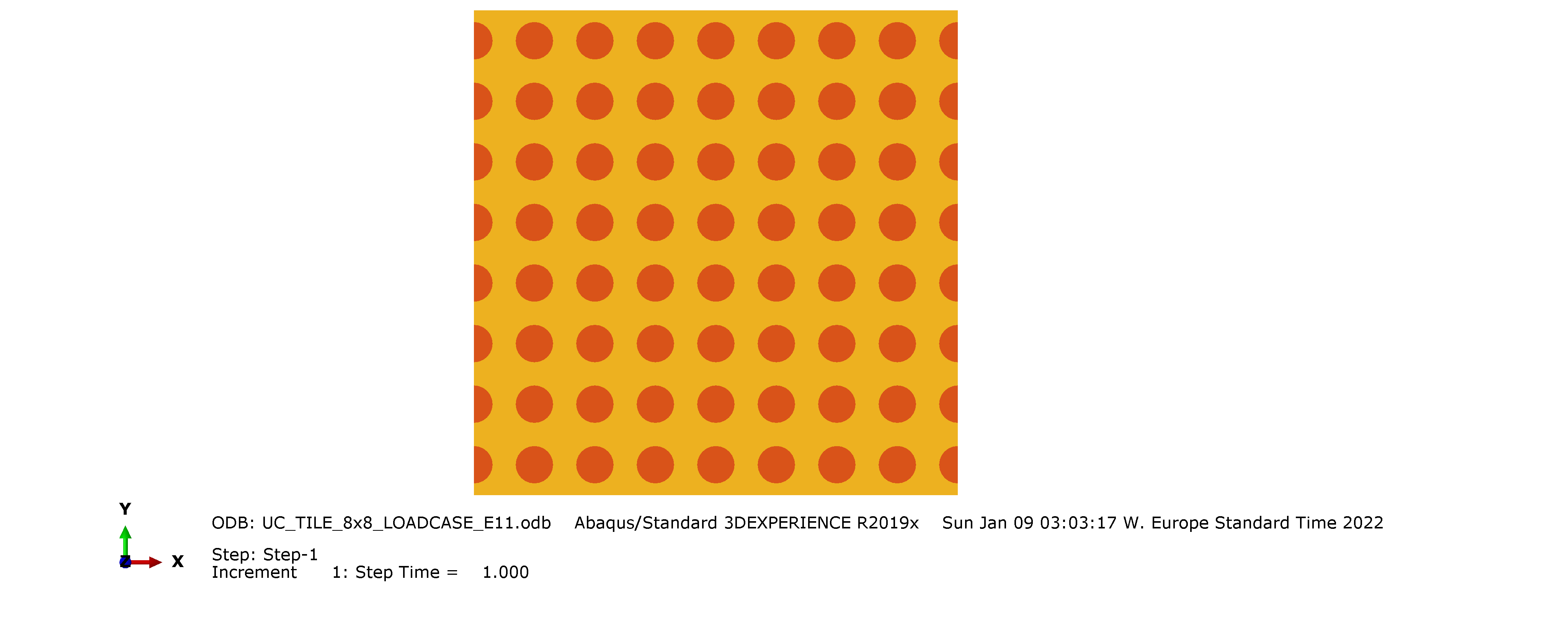}}
\frame{\includegraphics[height=0.125\textwidth,
trim=1238 324 1595 27, clip]{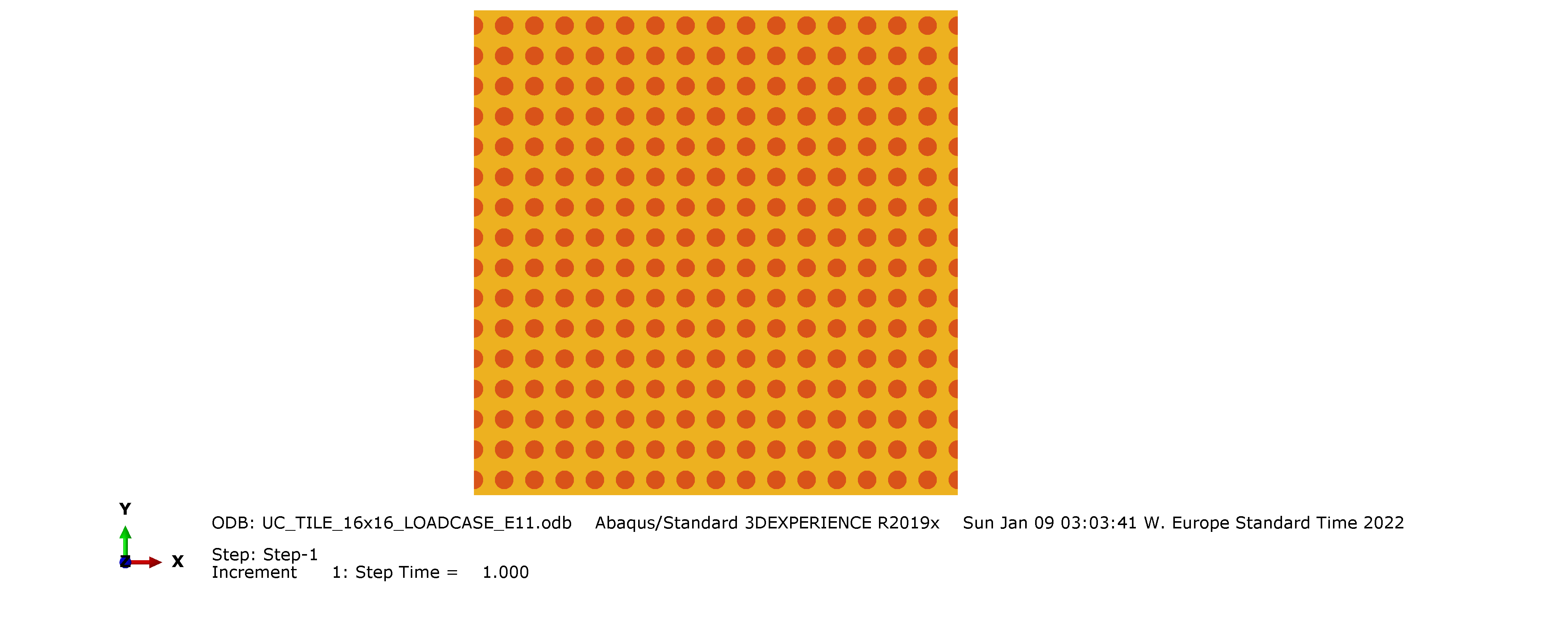}}
\frame{\includegraphics[height=0.125\textwidth,
trim=1238 324 1595 27, clip]{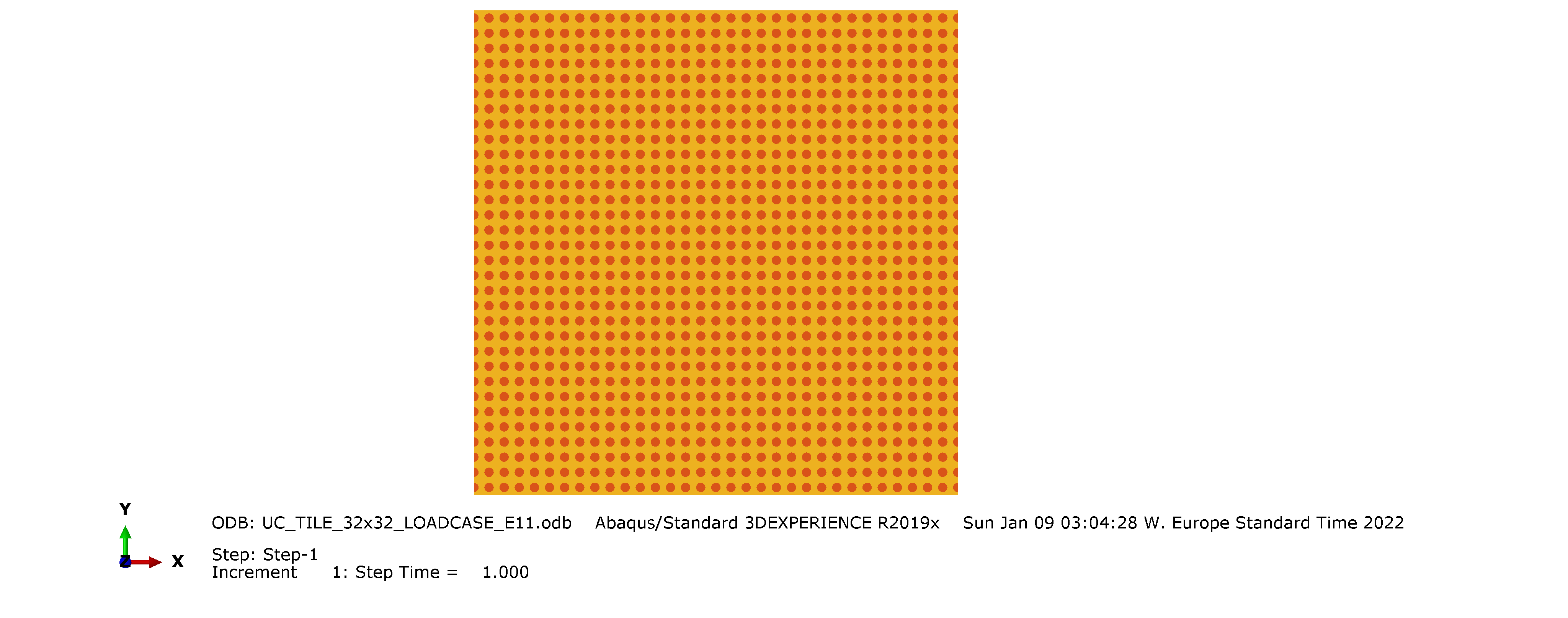}}\\
\vspace{2.5pt}
\frame{\includegraphics[height=0.125\textwidth,
trim=1317 294 1639 27, clip]{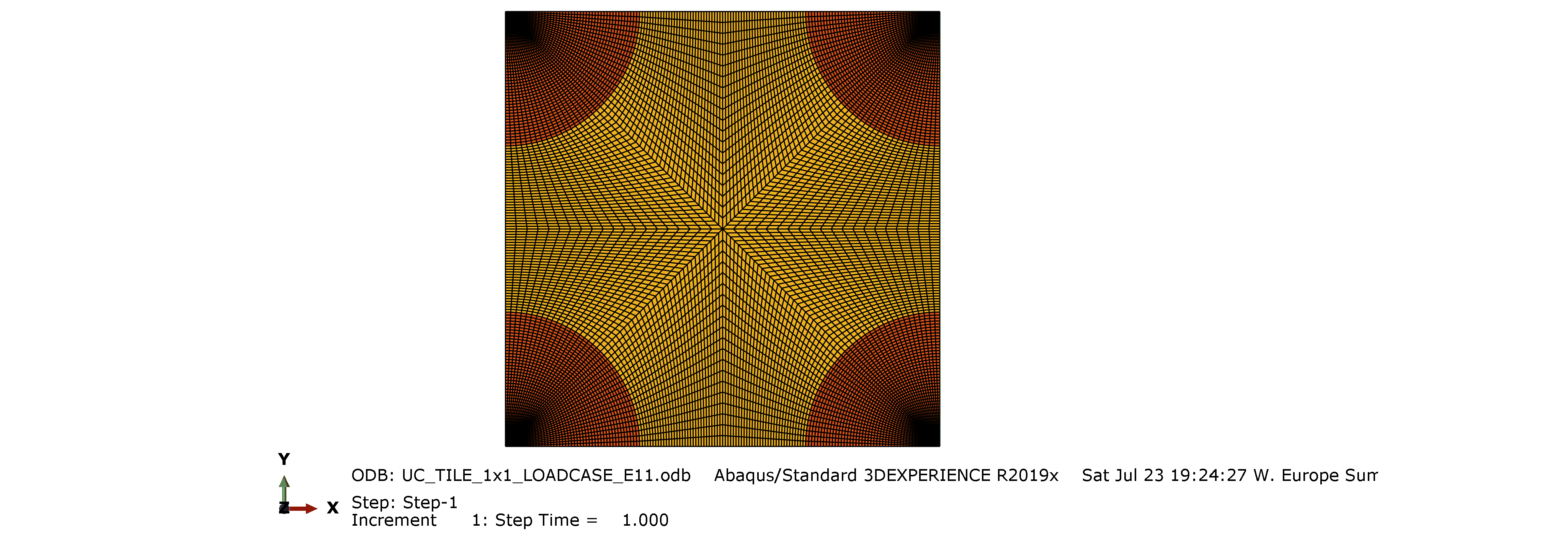}}
\frame{\includegraphics[height=0.125\textwidth,
trim=1238 324 1595 27, clip]{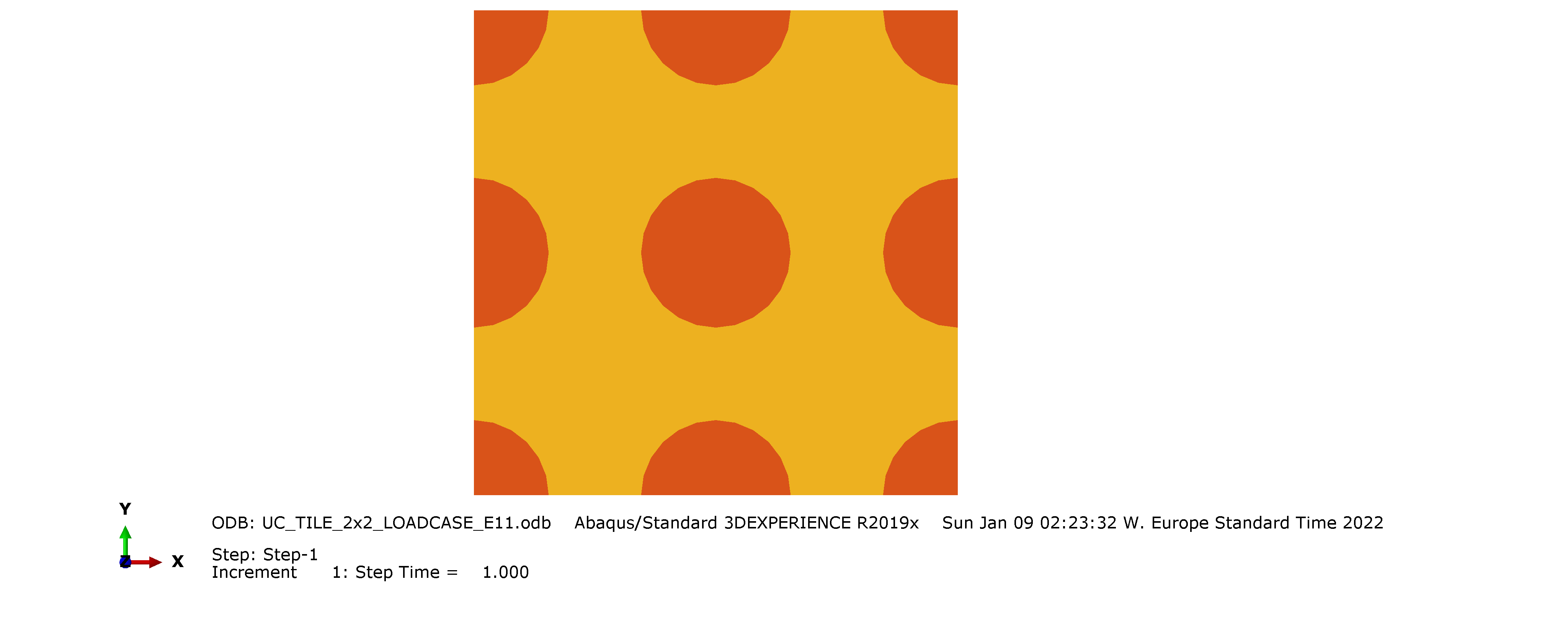}}
\frame{\includegraphics[height=0.125\textwidth,
trim=1238 324 1595 27, clip]{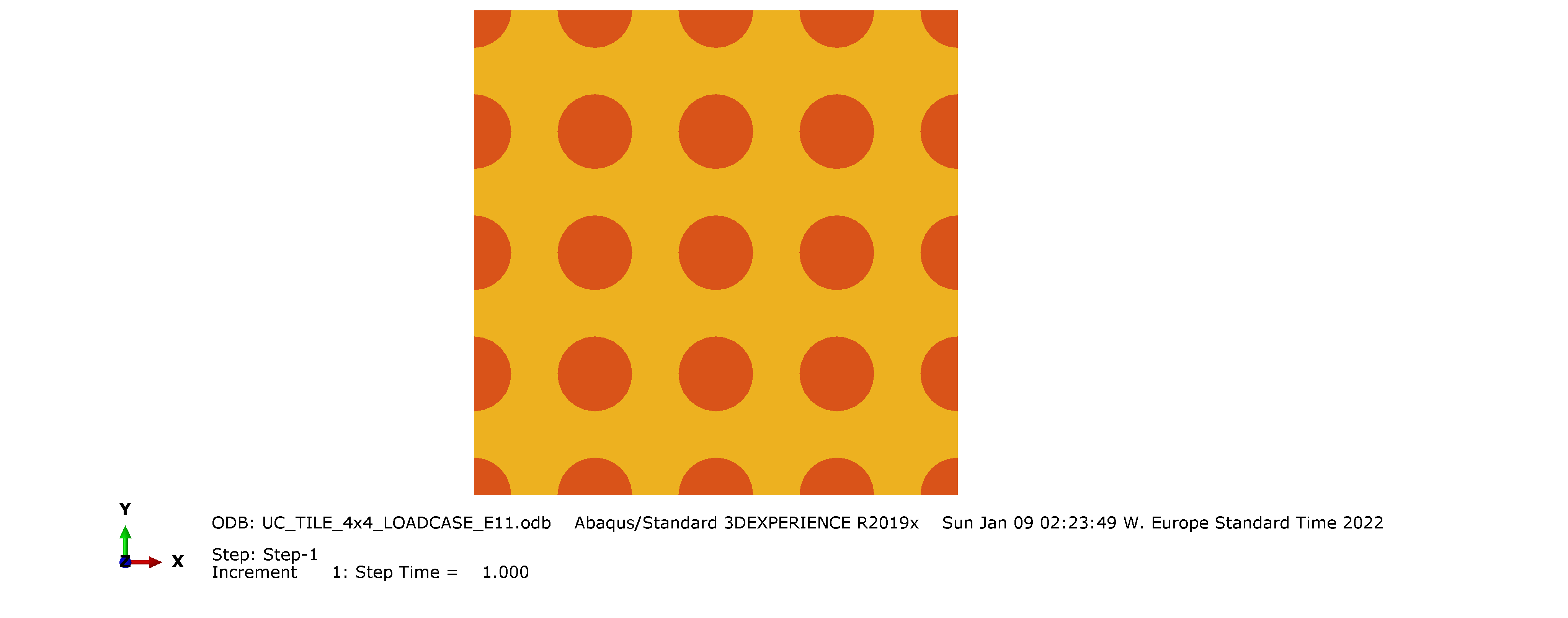}}
\frame{\includegraphics[height=0.125\textwidth,
trim=1238 324 1595 27, clip]{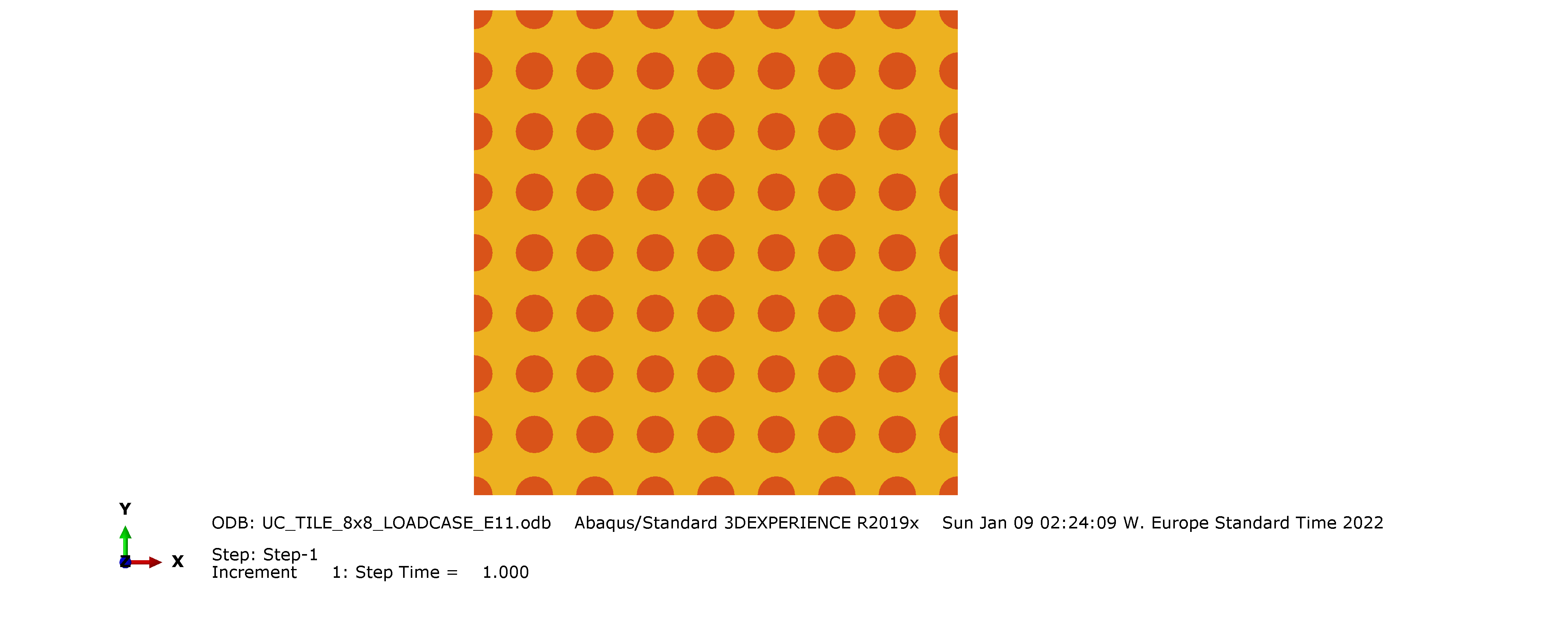}}
\frame{\includegraphics[height=0.125\textwidth,
trim=1238 324 1595 27, clip]{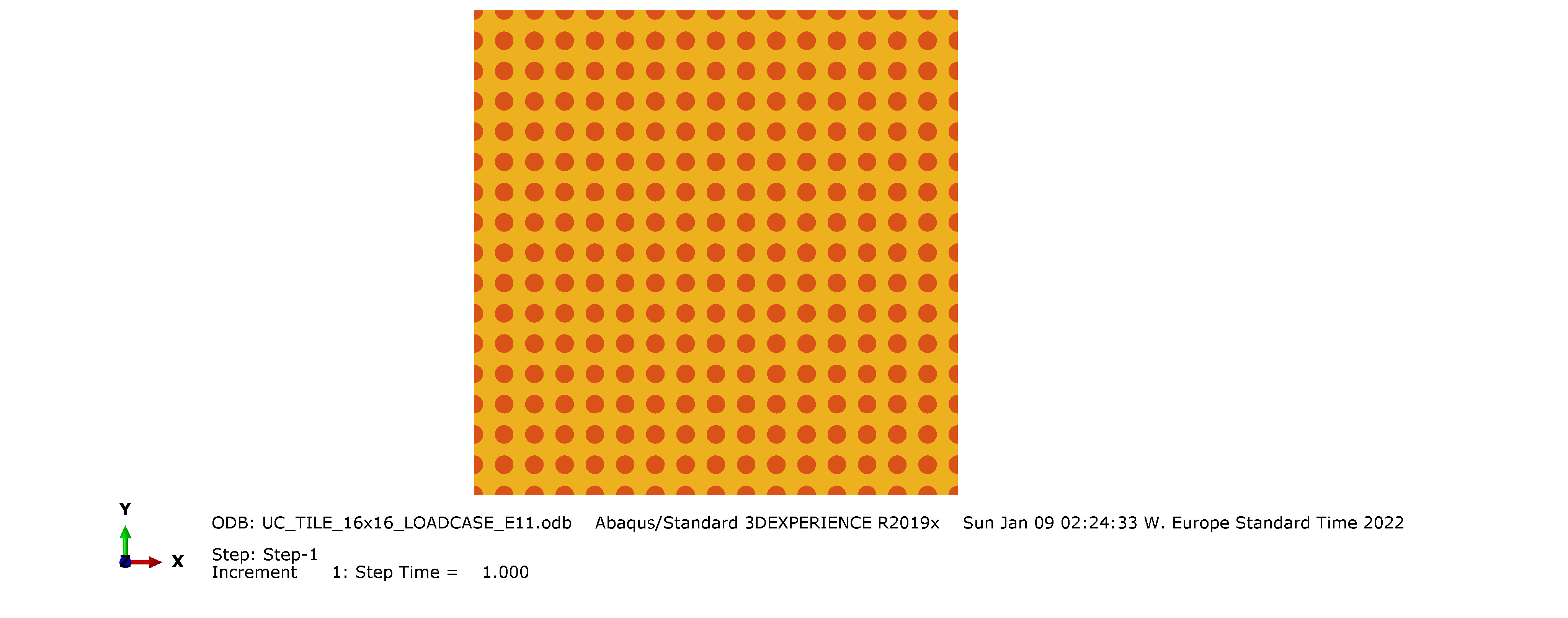}}
\frame{\includegraphics[height=0.125\textwidth,
trim=1238 324 1595 27, clip]{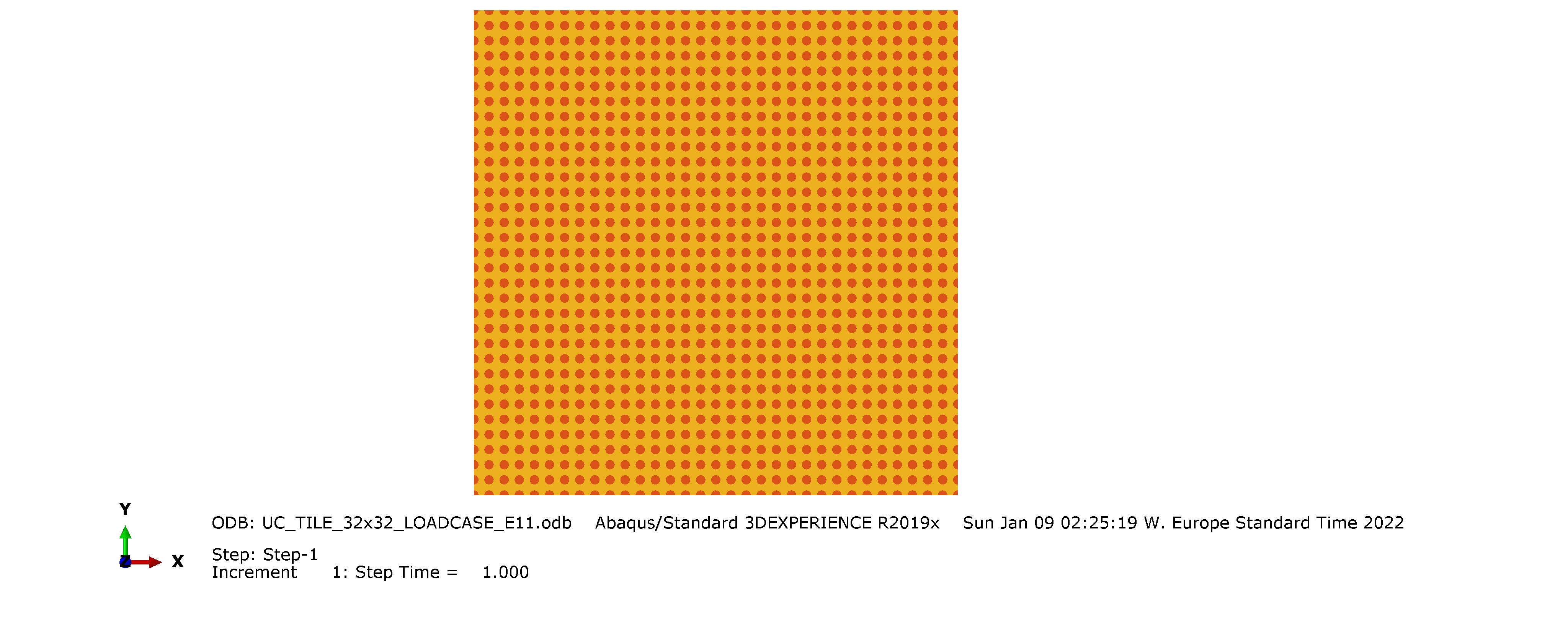}}
\caption{Three possible primitive unit cells for the periodic square disk arrangements, their discretizations into 12640 elements and $n \times n$ tiled configurations for $m=\{2,4,8,16,32\}$ for the inclusion volume fraction of $\phi_\mathrm{i}=0.30$.}
\label{F:2D_RVE_choices}
\end{figure*}

In periodic material systems, the choice for the primitive unit cell is not unique. Thus, we consider three possible primitive unit cells and their $n\times n$ spatial tilings for $n=\{2,4,8,16,32\}$,  henceforth  referred to as the volume elements (VEs), see Figure\ \ref{F:2D_RVE_choices}. As a first task, we focus on the effect of boundary conditions and the choice of the potential on which the formulation is based, in conjunction with the type and size of the used volume element on the effective properties. The type of the volume element controls its boundary property fluctuations, whereas its size controls the relative size of the emerging boundary layer. The models are subjected to periodic and uniform Dirichlet boundary conditions with both scalar and vector potential formulations of magnetostatics.
Here we fix the inclusion volume fraction at $\phi_\mathrm{i}=0.30$. A geometry-based approach, which  requires mesh seeds at the phase interface, is adopted to model phase boundaries accurately. The primitive unit cell domain is  discretized using 12640 node-based linear quadrilateral elements forming a structured mesh. To eliminate any mesh-related differences, the unit cells are meshed identically.

For the studied periodic square disk arrangements, a series expansion solution in terms of the disk volume fraction is given for the effective conductivity in \cite{Godin2013}. As a second task, we compare our numerical results with this analytical solution and consider the primitive unit cell with centered disk arrangement given in the first row of Figure\ \ref{F:2D_RVE_choices}. We created models with inclusion volume fractions of $\phi_\mathrm{i}=\{0.05,0.10,\ldots,0.55\}$. Each cell domain is finely discretized using over 10000 node-based linear quadrilateral elements forming a structured  mesh.
\subsubsection{Random Disk Arrangements}
In this part, we consider nonoverlapping and overlapping periodic disk arrangements which possess statistical ergodicity and uniformity. The microstructures are generated using a Monte Carlo method depending on a sequential adsorption process. The periodicity is enforced as follows: we consider a square region with size $L\times L$. We create a $3\times3$ continuous tiling of this region.
Letting $\mathcal{U}$ denote uniform probability distribution, we generate a disk with a fixed radius $r$, hence monodispersity, and random   central coordinates $(c_x,c_y)$ where $c_x \sim \mathcal{U}(0,L)$ and $c_y \sim \mathcal{U}(0,L)$. Taking the origin of  each square region as a reference for each disk coordinate,  we embed nine copies of this disk in our $3\times3$ square tiling. This process is repeated until the desired inclusion volume fraction is reached. Finally, we clip out the central square region, which possesses periodicity.

Generating a mesh with periodic nodal locations can be difficult for random microstructures. Although the analytical computation of inclusion volume fraction is trivial for nonoverlapping disks, for the overlapping ones, it is not.
Thus, to remedy both challenges, a voxel-based meshing strategy is followed in this part. However, this introduces a relatively higher error in the representation of volumes and surfaces compared to geometry-based meshing approaches.

\begin{figure}[htb!]
\centering
\begin{minipage}[]{0.75\textwidth}
\centering
% trim=left bottom right top, clip
    {\frame{\includegraphics[height=0.19\textwidth]{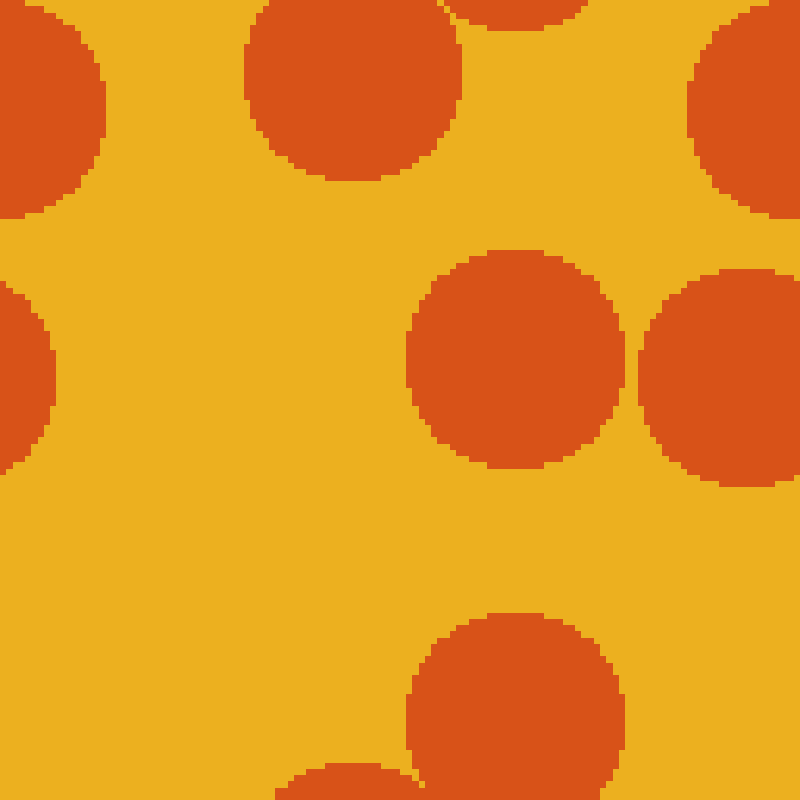}}}
    {\frame{\includegraphics[height=0.19\textwidth]{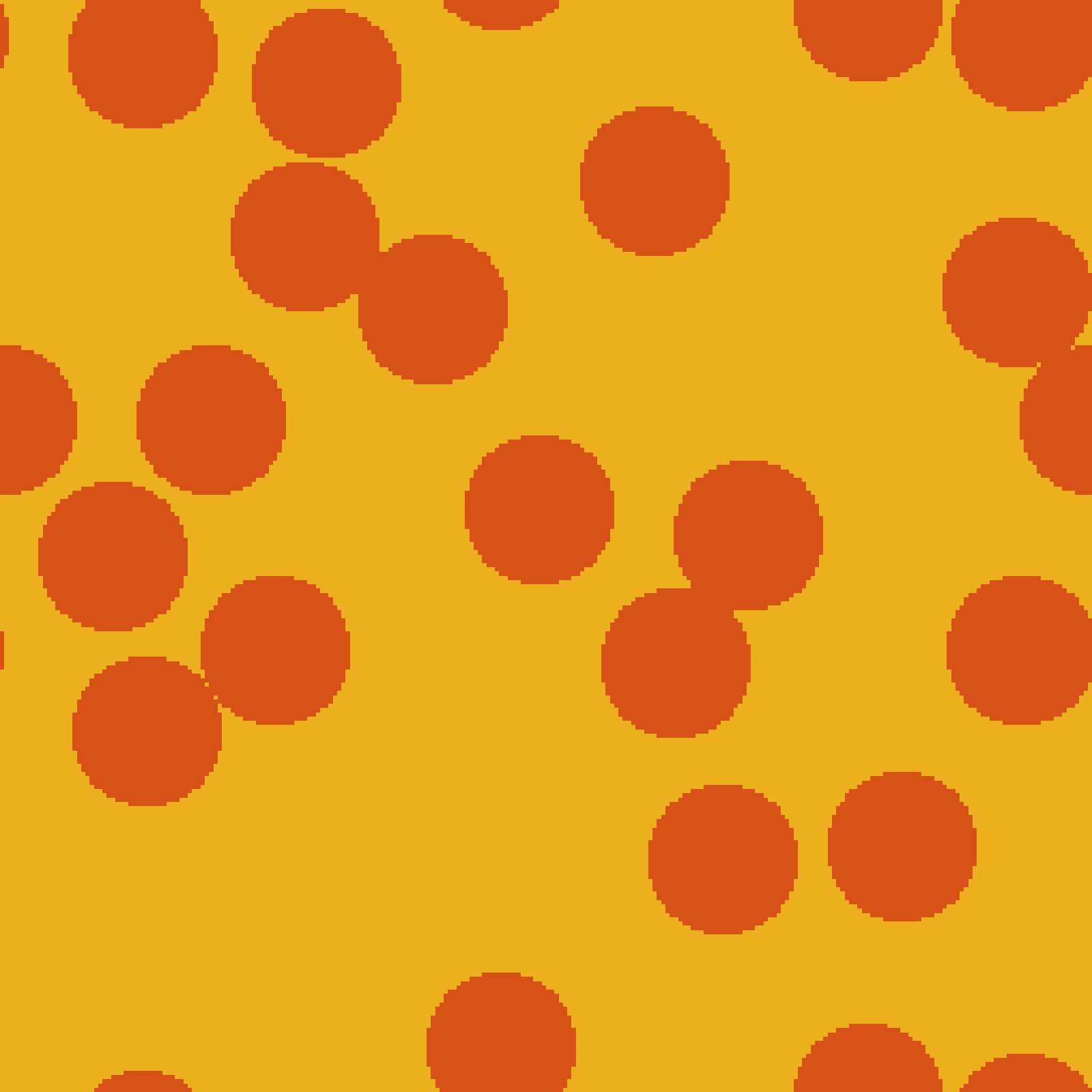}}}
    {\frame{\includegraphics[height=0.19\textwidth]{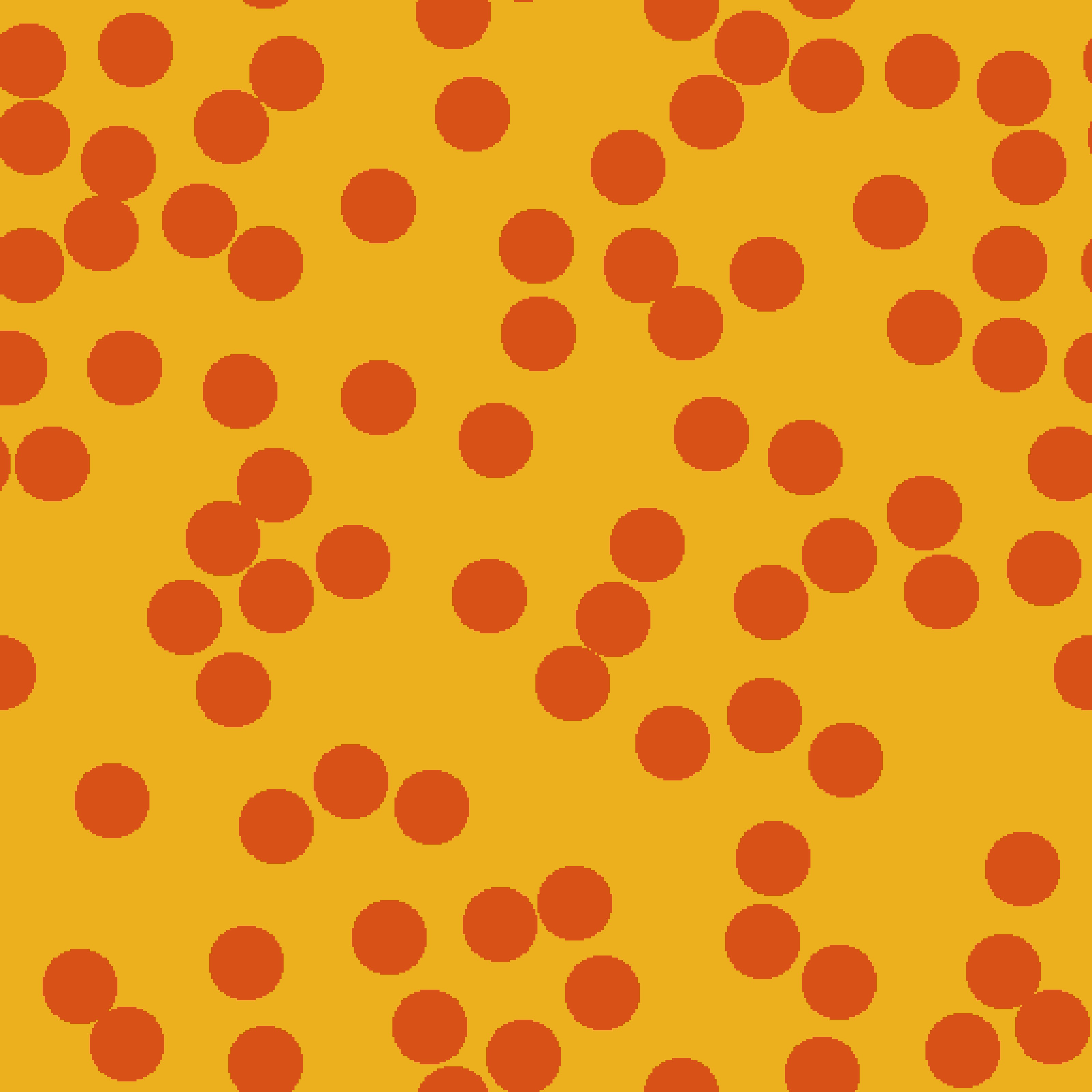}}}
    {\frame{\includegraphics[height=0.19\textwidth]{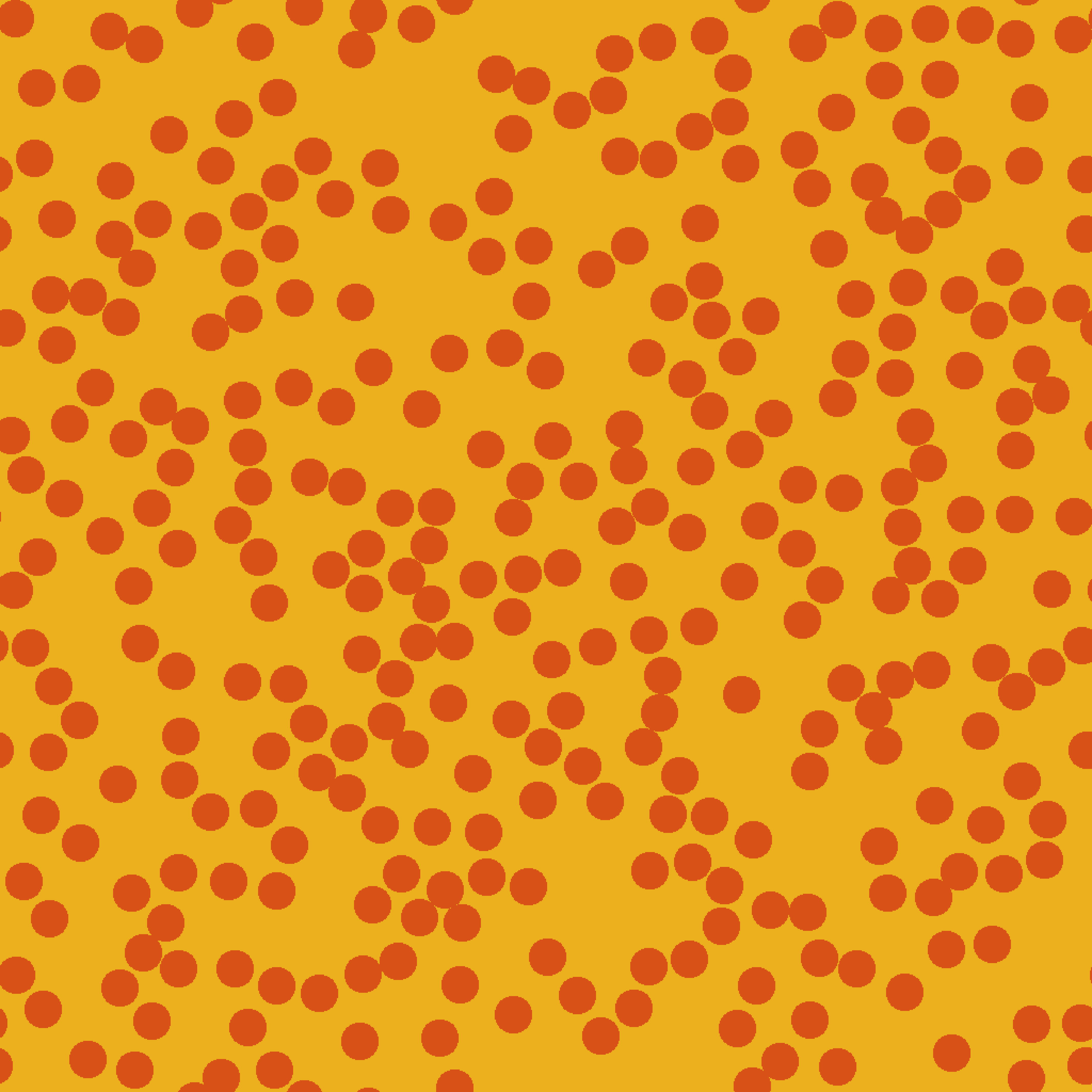}}}
    {\frame{\includegraphics[height=0.19\textwidth]{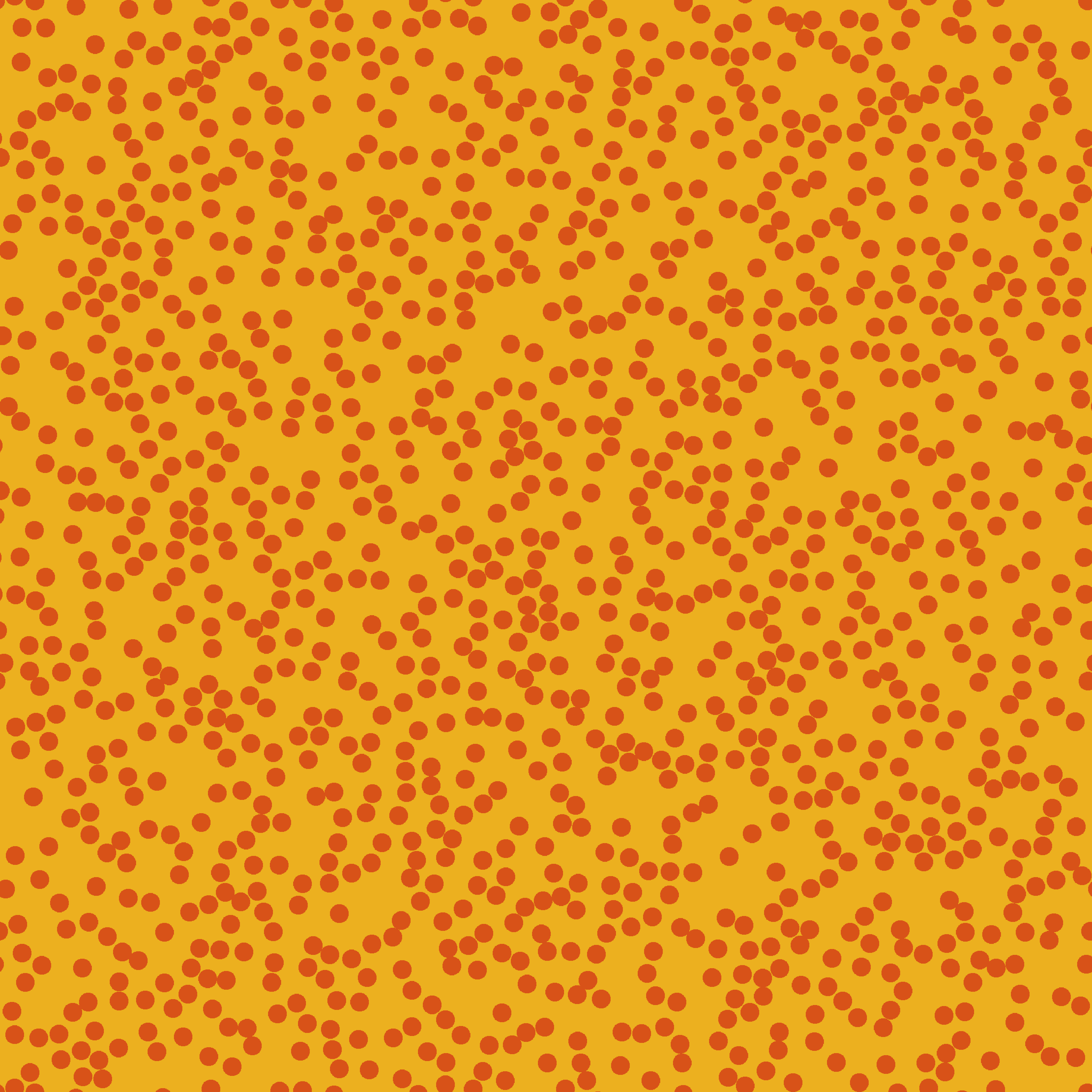}}}\vspace{2.5pt}
    {\frame{\includegraphics[height=0.19\textwidth]{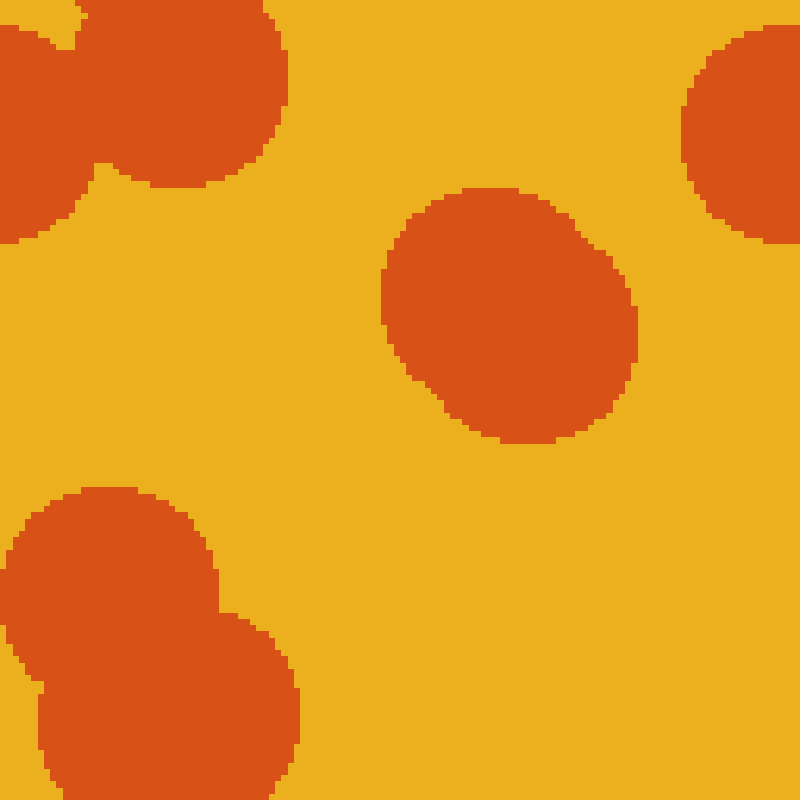}}}
    {\frame{\includegraphics[height=0.19\textwidth]{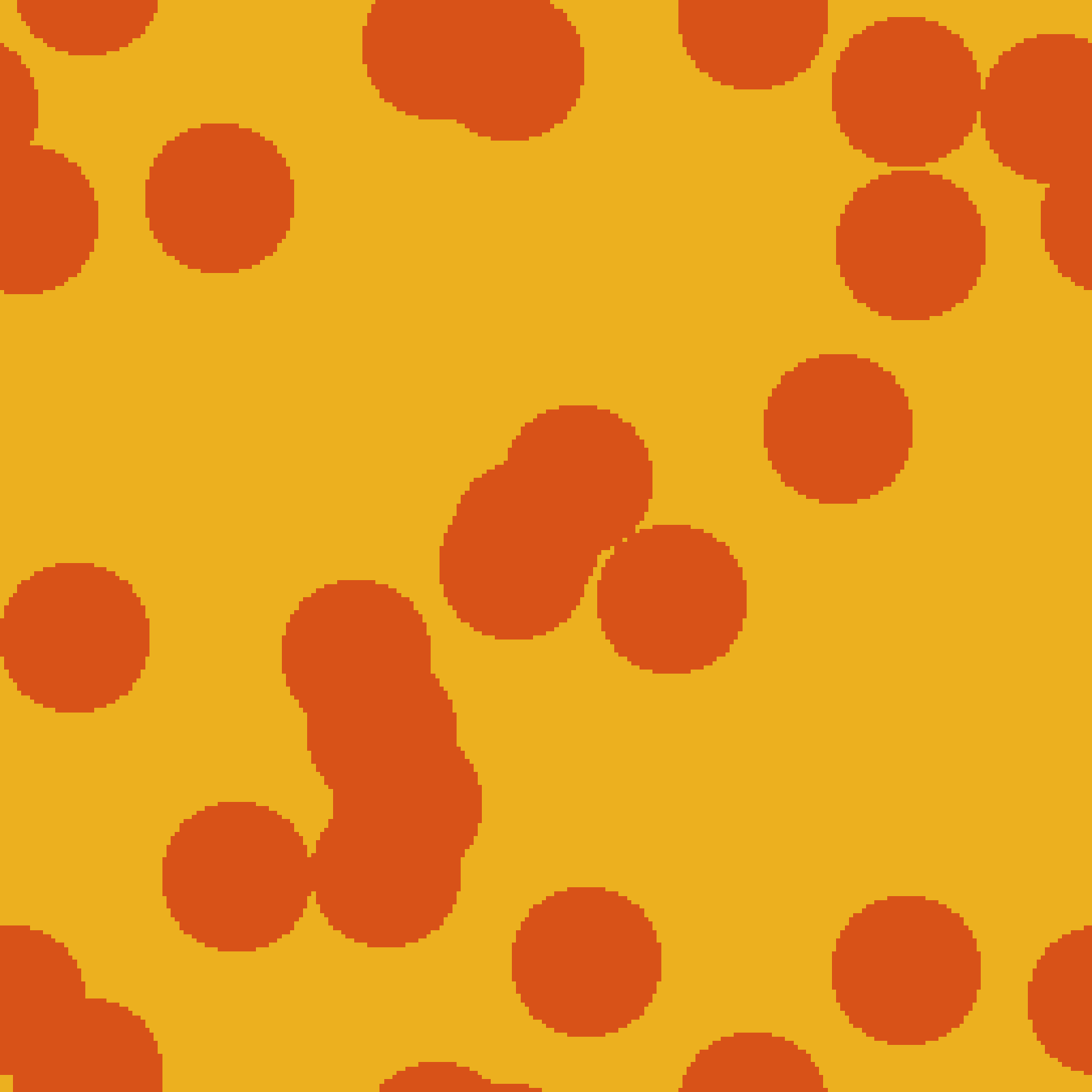}}}
    {\frame{\includegraphics[height=0.19\textwidth]{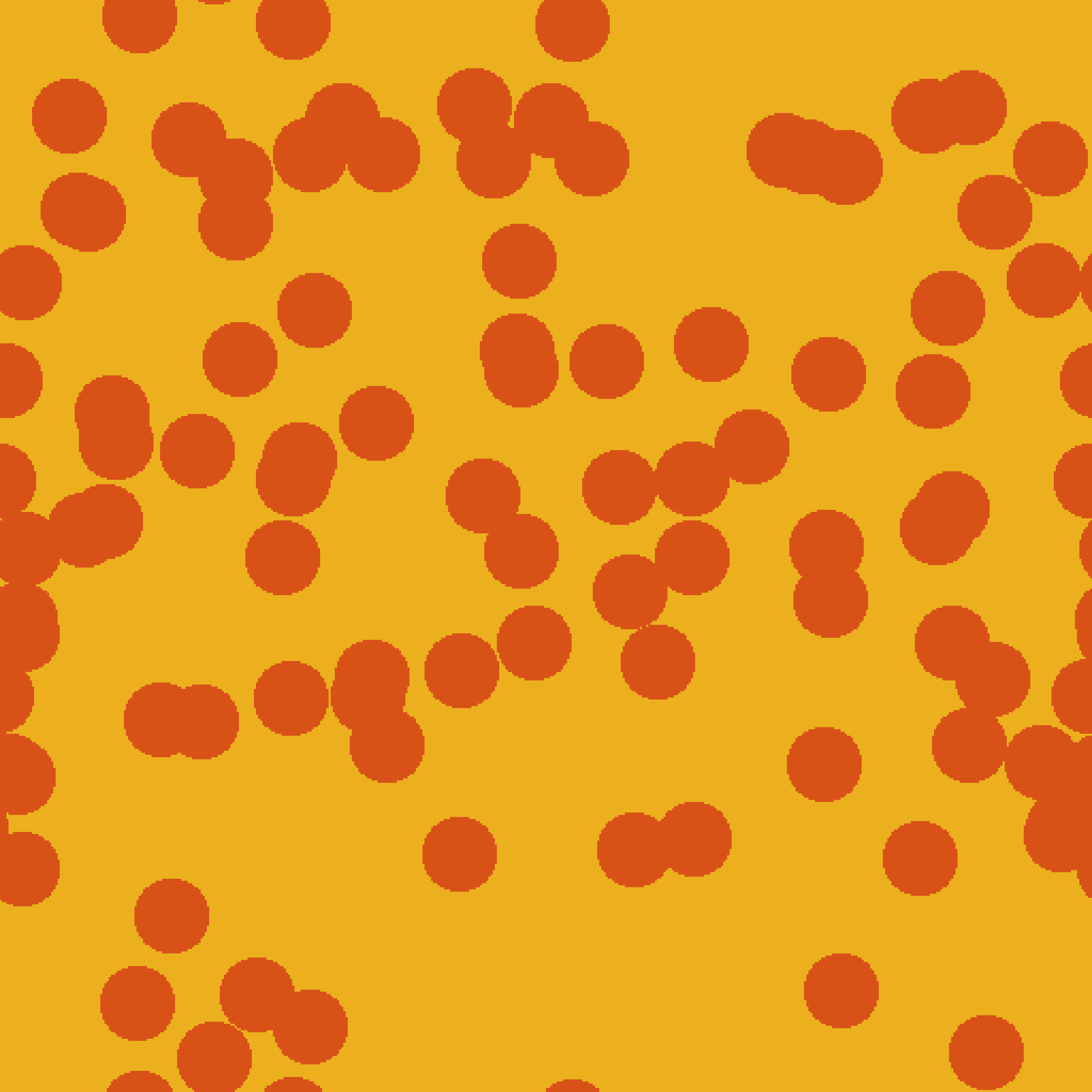}}}
    {\frame{\includegraphics[height=0.19\textwidth]{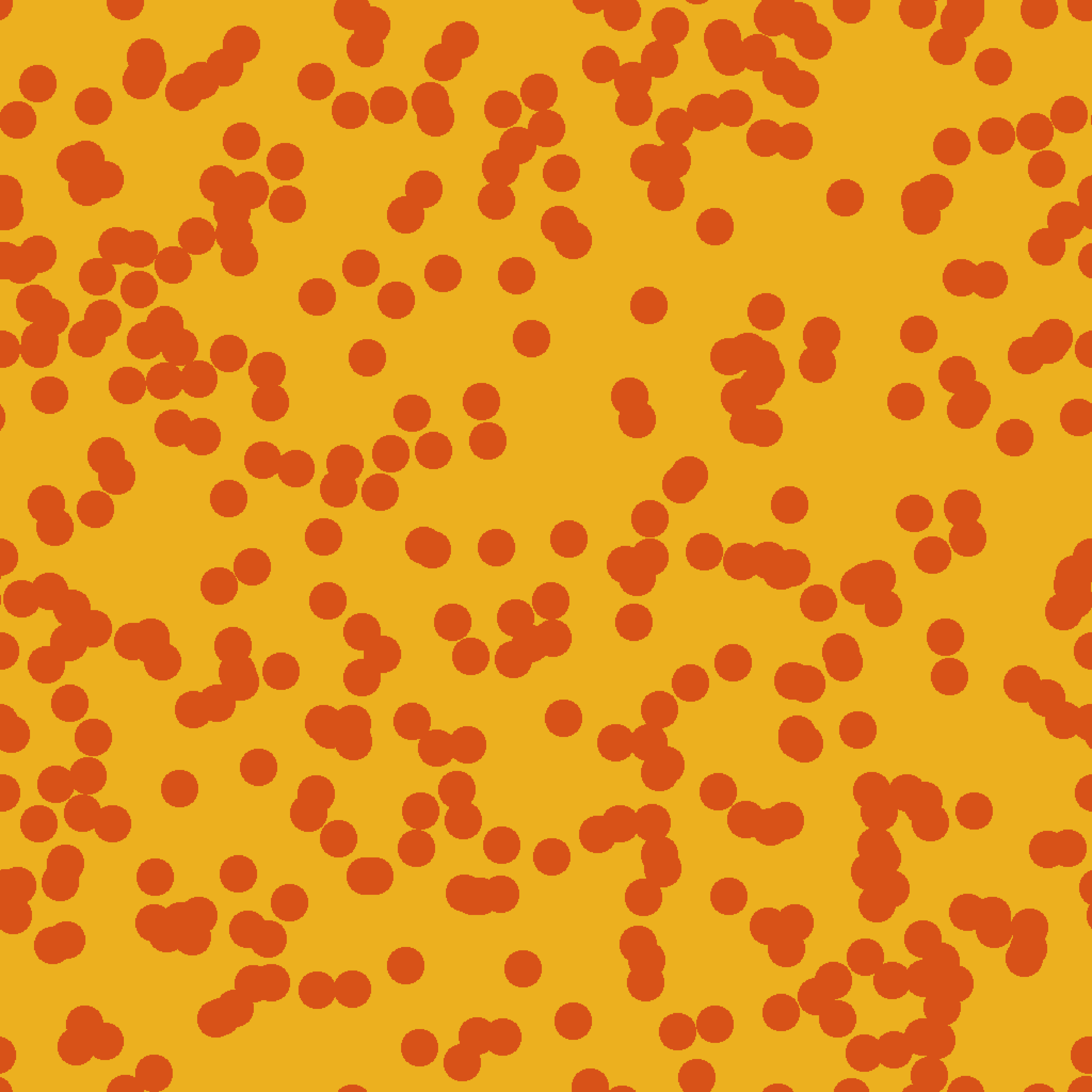}}}
    {\frame{\includegraphics[height=0.19\textwidth]{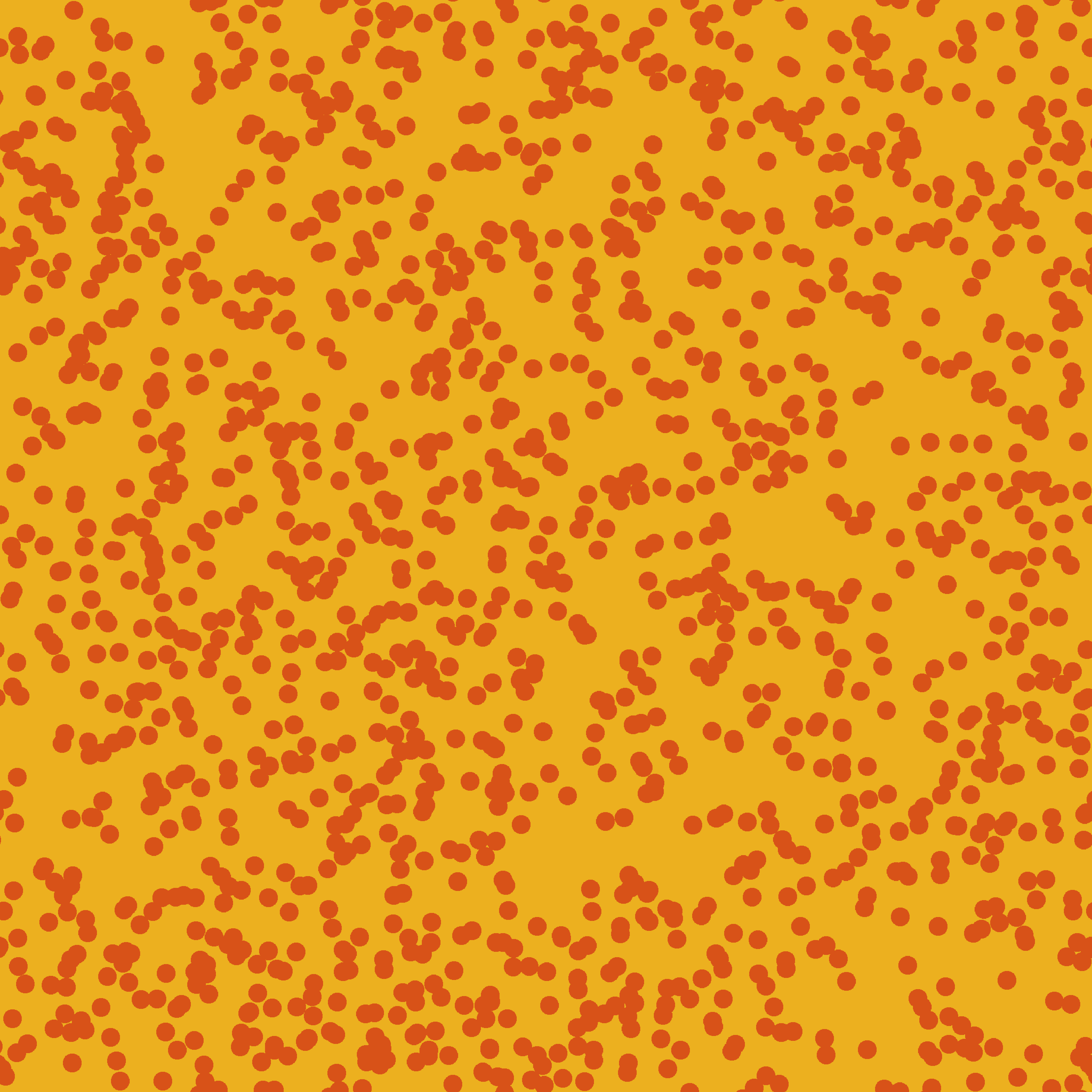}}}
    \end{minipage}
\caption{Example images of the voxel-based random microstructure realizations  for the inclusion volume fraction of $\phi_\mathrm{i}=0.30$ for nonoverlapping (top) overlapping (bottom) disks. The microstructures are generated using the indicated random sequential inhibition process under periodic boundary conditions. The VE sizes are 128, 256, 512, 1024, and 2048 pixels from left to right, respectively. For convenience, the image size is kept constant while demonstrating the VEs.}
\label{F:microstructures_nonoverlapping_overlapping_VE_size}
\end{figure}

According to Hill \cite{HILL1963}, a representative volume element (RVE) should yield the same  effective properties for the applied displacement or traction boundary conditions as far as they are macroscopically uniform.
For random composites whose constituents have
high-contrast physical properties, this is impractical as it results in substantial RVE sizes \cite{SAB1992,ShenBrinson2006}. In this part, we start with searching for the RVE size but adapting the definition, which defines it as the smallest material volume size that accurately represents the composite's average physical property \cite{DruganWillis1996}. For regular periodic composites, the primitive unit cell makes up the RVE. For random disk-matrix systems, this is not the case. Such systems require a  large set of unique microstructures of the same size and shape to be considered. What makes each microstructure unique is its underlying stochastic generation. These are also referred to as statistical volume elements (SVEs). The computation of effective properties requires an averaging over an ensemble of microstructures; an operation referred to as the ensemble averaging, see, e.g., \cite[p.85]{adamskalidindifullwood2013} and \cite{OstojaStarzewski2007}. The ergodic hypothesis asserts that averaging over all ensemble elements tends to volume averaging on any ensemble elements with increasing SVE size. In contrast, an equivalence is possible when the SVE size tends to infinity.
For random composites, the minimal RVE size is selected by monitoring the precision of the mean value of conducted realizations \cite{KanitForestJeulin20033647}.  This size is generally not unique, and it depends on the selected physical property, and the contrast of the constituent phase properties \cite{KanitForestJeulin20033647}.
Based on these concepts, we adopt a pragmatic methodology and consider sparsely populated ensembles with $N=15$ random microstructures, VEs (or SVEs); for each selected VE size, we compute mean $\upmu$, standard deviation $\upsigma$ and standard error of the mean  $\mathrm{SEM}=\upsigma/\sqrt{N}$ in the resultant effective properties while keeping the inclusion volume fraction constant with $\phi_\mathrm{i}=0.30$.
 Although not the case for overlapping disks, for nonoverlapping monodisperse disks, the volume fraction  increase with each added disk, and the associated relative error is known apriori as a function of the number of disks and the VE-size. This restricts the disk volume fractions that can be obtained, especially for small cell size-to-disk radius ratios. In our generation, we use a disk diameter of $D=34$ and a minimum VE size of $128$, which results in a $128\times128$ pixel-based discretization. This kept the associated absolute error in volume generation smaller than 0.01. In addition, with systematic doubling, we considered square VEs with $128\times128$, $256\times256$, $512\times512$, $1024\times1024$, and $2048\times2048$ discretizations. Exemplary periodic generations for nonoverlapping and overlapping disk arrangements are demonstrated in Figure\ \ref{F:microstructures_nonoverlapping_overlapping_VE_size}.

\begin{figure}[htb!]
\centering
\begin{minipage}[]{0.75\textwidth}
\centering
% trim=left bottom right top, clip
    {\frame{\includegraphics[height=0.19\textwidth]{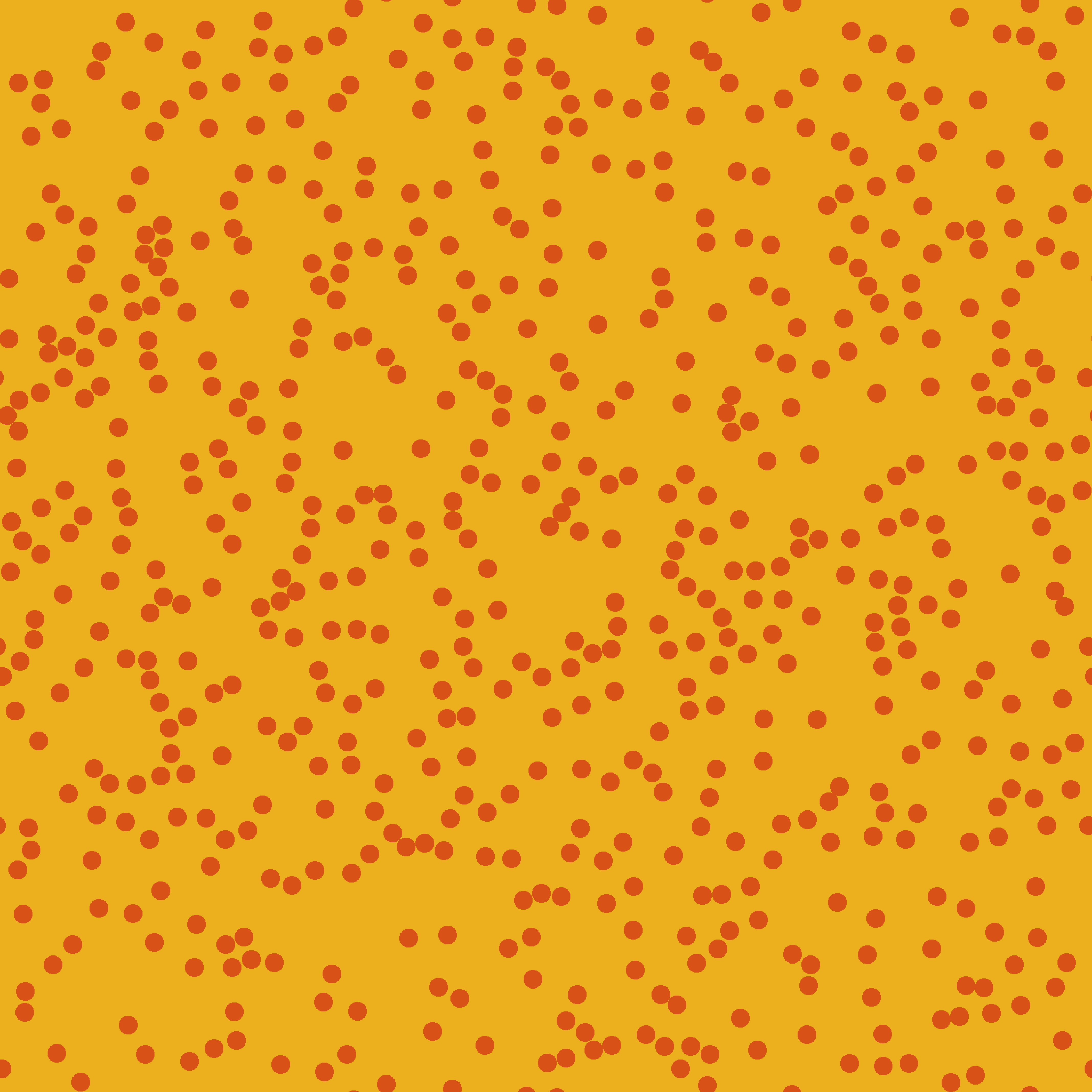}}}
    {\frame{\includegraphics[height=0.19\textwidth]{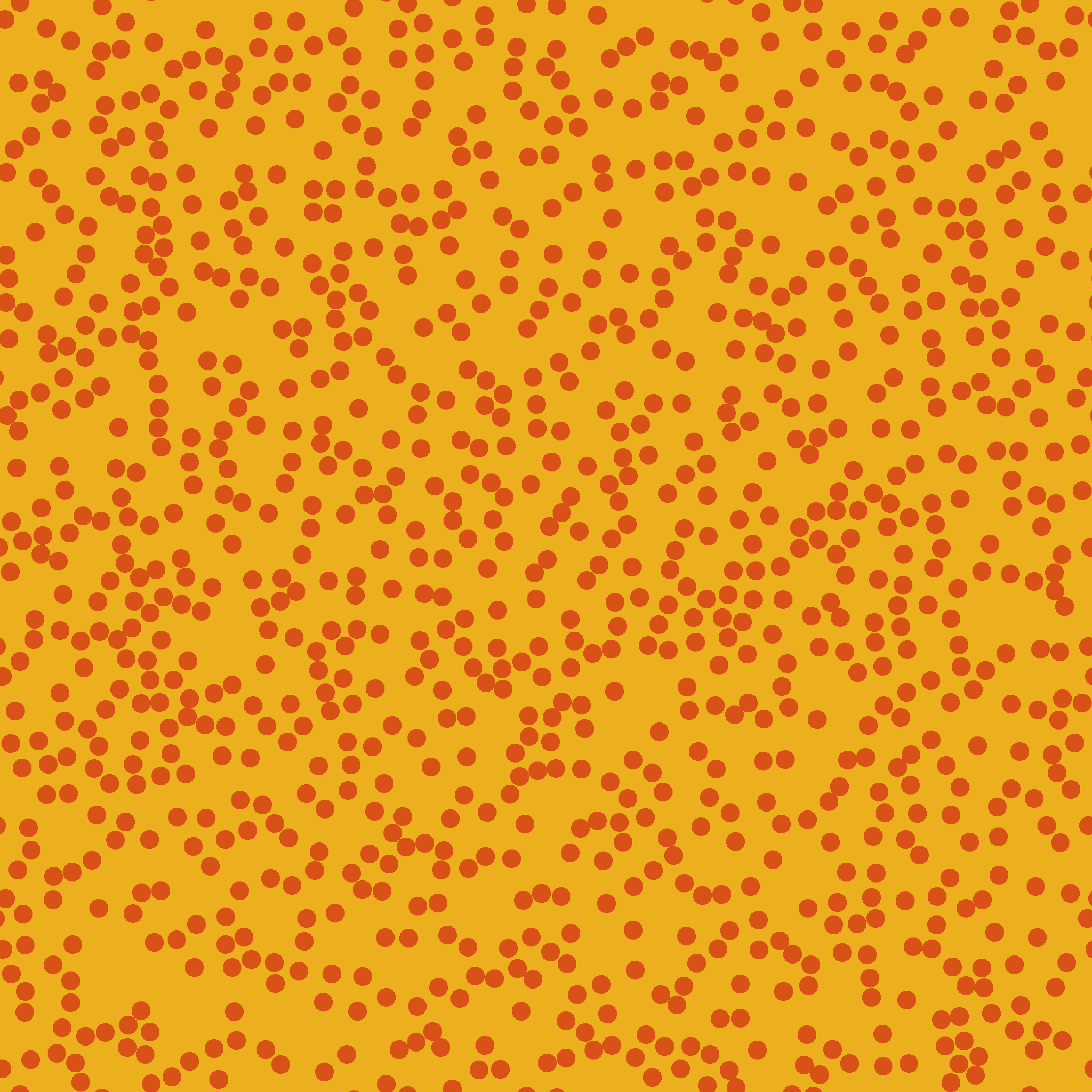}}}
    {\frame{\includegraphics[height=0.19\textwidth]{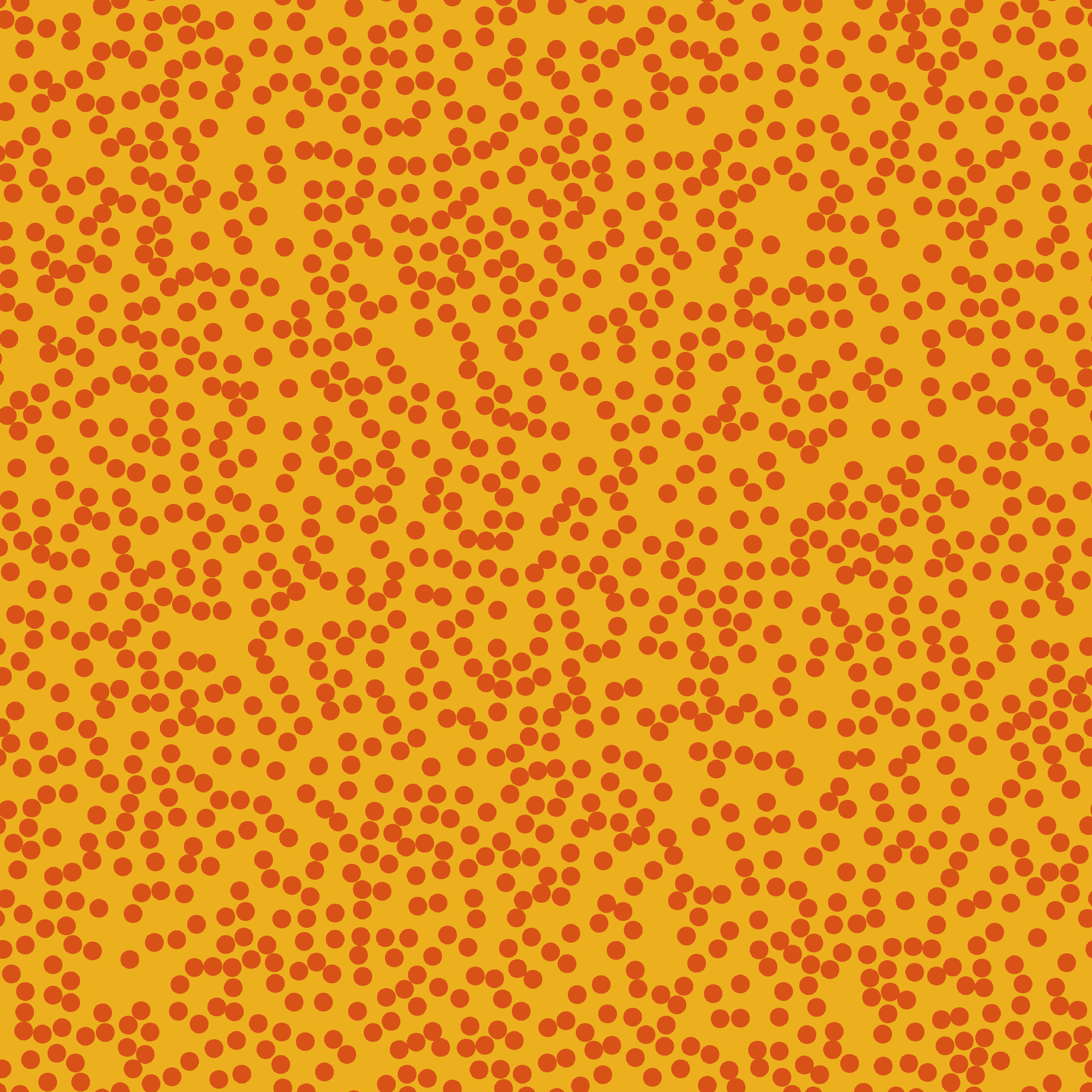}}}
    {\frame{\includegraphics[height=0.19\textwidth]{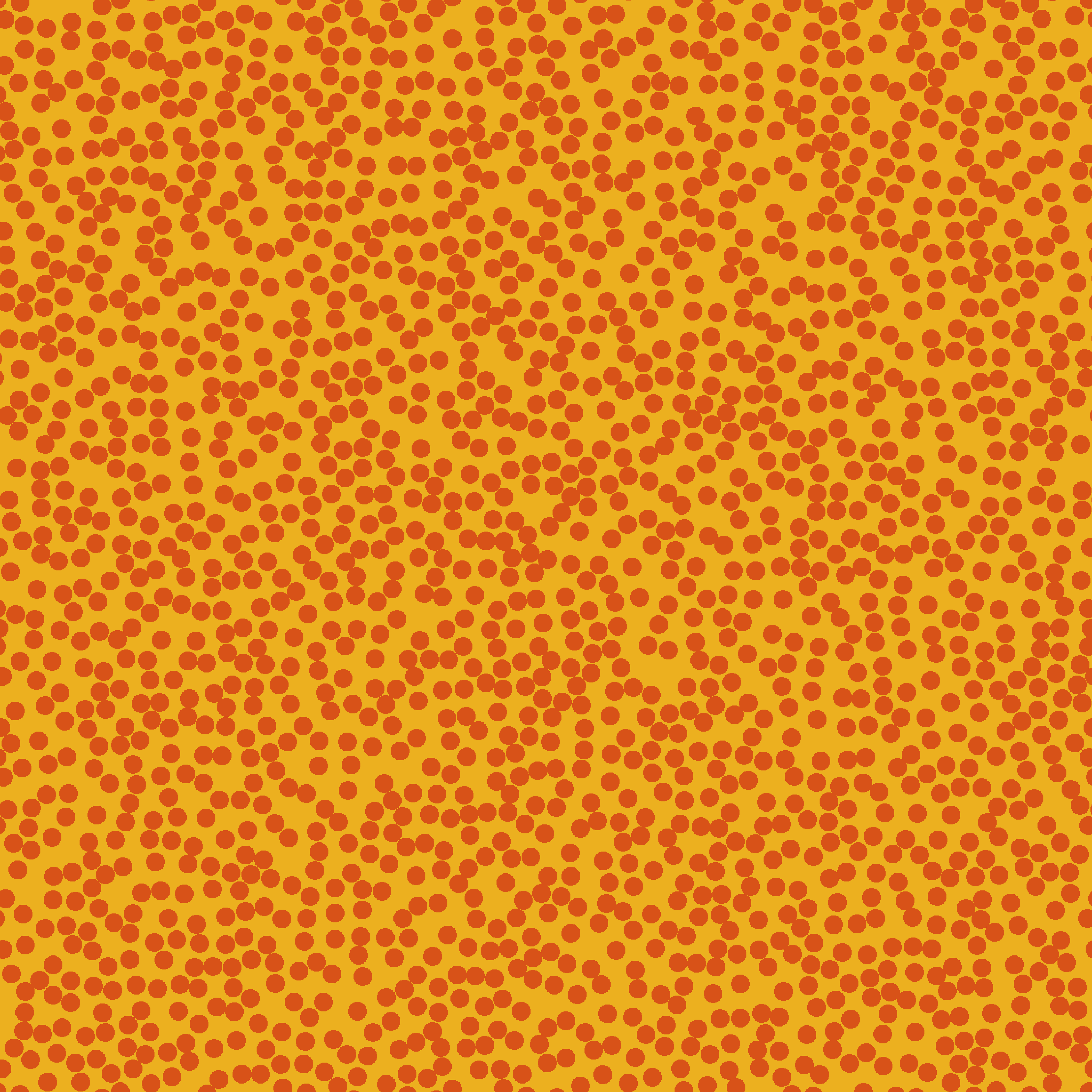}}}
    {\frame{\includegraphics[height=0.19\textwidth]{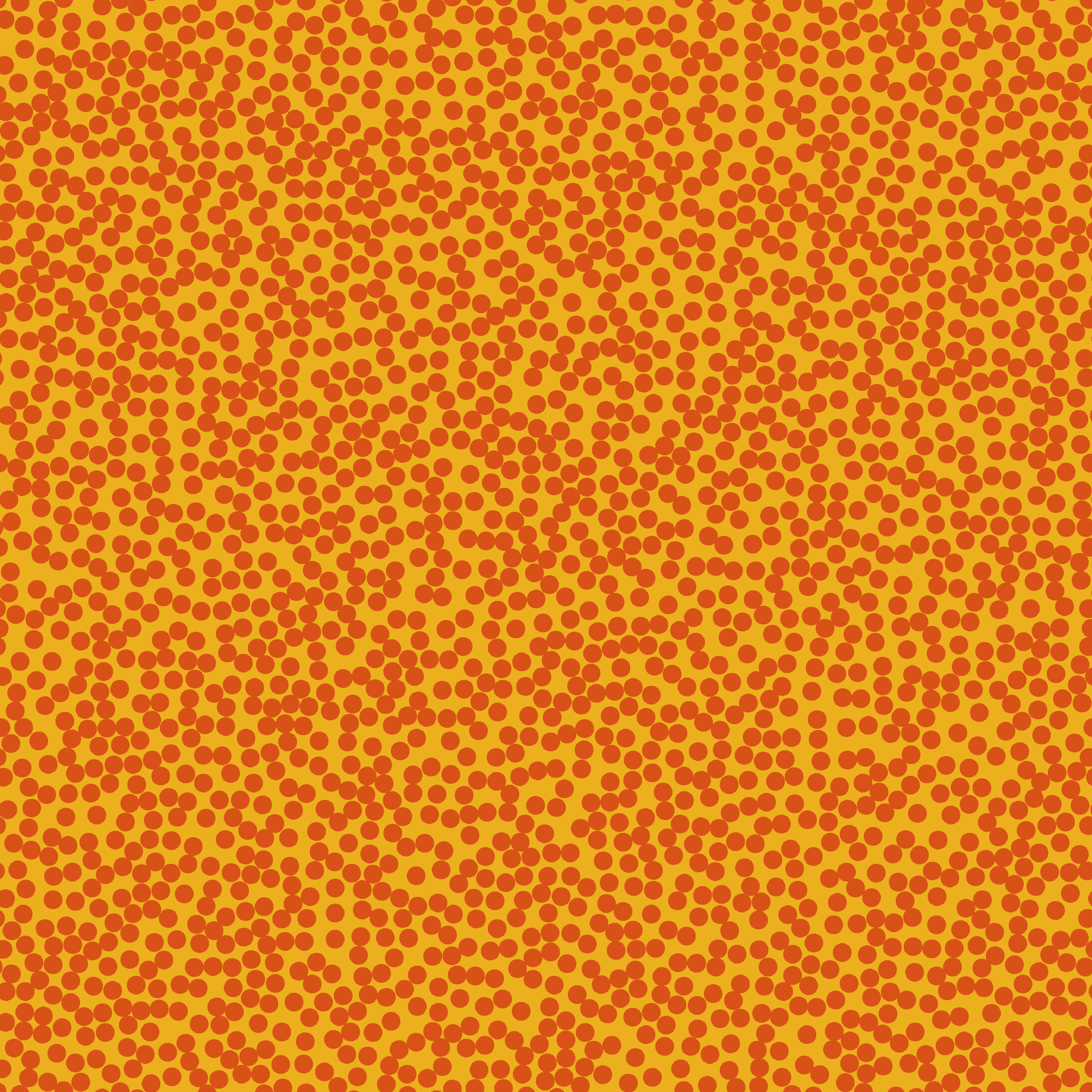}}}
    \end{minipage}
\caption{Example images of the voxel-based random microstructure realizations considering inclusion volume fractions of $\phi_\mathrm{i}=\{0.15, 0.25, \ldots, 0.55\}$, for nonoverlapping disks.
The VE size is 2048 pixels. The microstructures are generated using the indicated random sequential inhibition process under periodic boundary conditions.}
\label{F:microstructures_nonoverlapping_VOLFRAC}
\end{figure}

As a second task, we keep the VE size constant and vary the inclusion volume fraction to investigate the potential difference between overlapping and nonoverlapping disk system responses and between them and regular square disk arrangements. For nonoverlapping monodisperse disks, the random sequential adsorption process has the saturation limit at $\phi_\mathrm{s}\simeq0.55$ in two-dimensions \cite{Feder1980} whereas, for three-dimensions with spheres, it is  $\phi_\mathrm{s}\simeq0.38$ \cite{Cooper1988}. This is the limit of jamming beyond which no other random disks can be added to the system without overlapping. In Figure\ \ref{F:microstructures_nonoverlapping_VOLFRAC}, we demonstrate five randomly generated microstructures with nonoverlapping disk arrangements with volume fractions of $\phi_\mathrm{i}=\{0.15, 0.25, \ldots, 0.55\}$. For each volume fraction 15 generations were realized.

\begin{figure}[htb!]
\centering
\begin{minipage}[]{0.75\textwidth}
\centering
% trim=left bottom right top, clip
    {\frame{\includegraphics[height=0.19\textwidth]{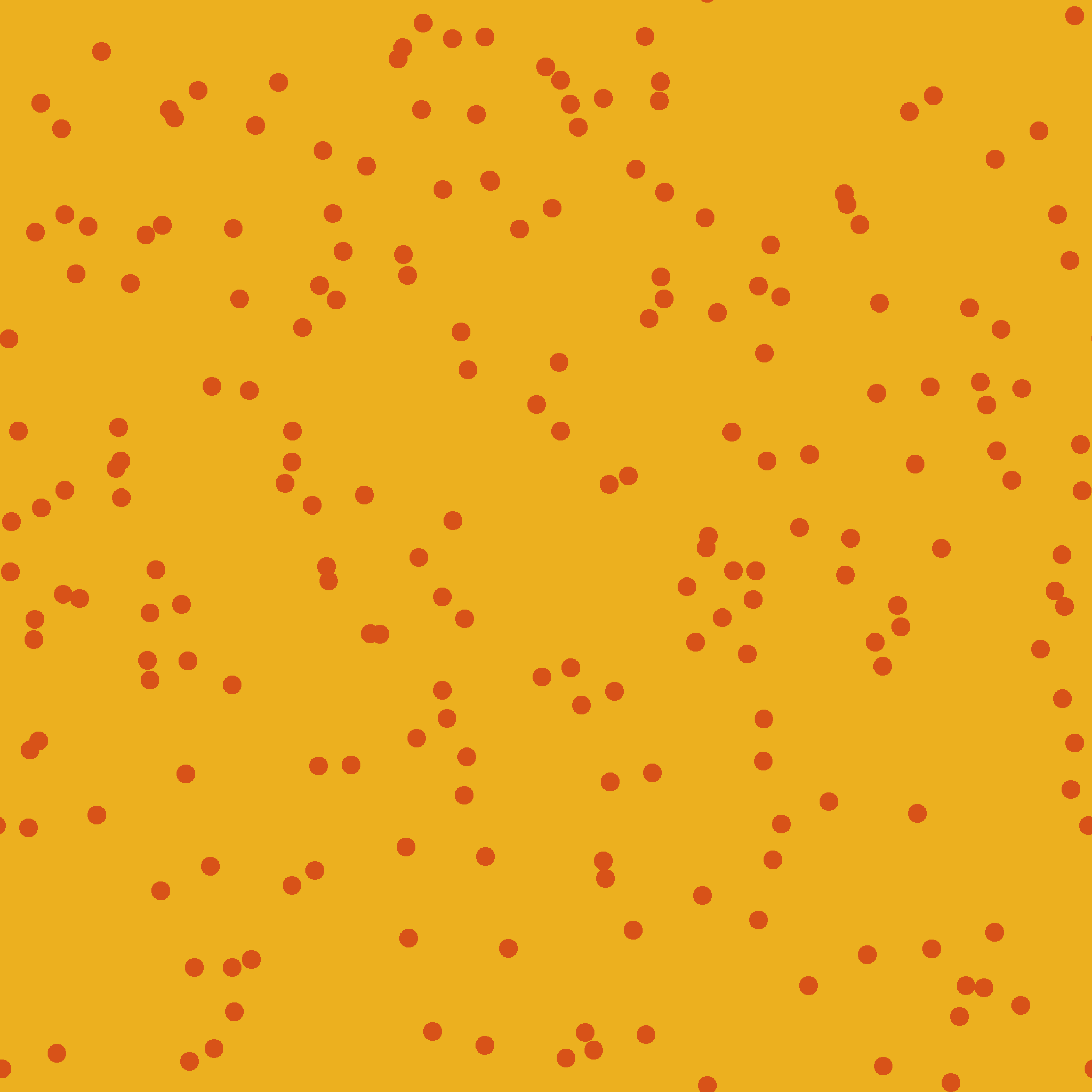}}}
    {\frame{\includegraphics[height=0.19\textwidth]{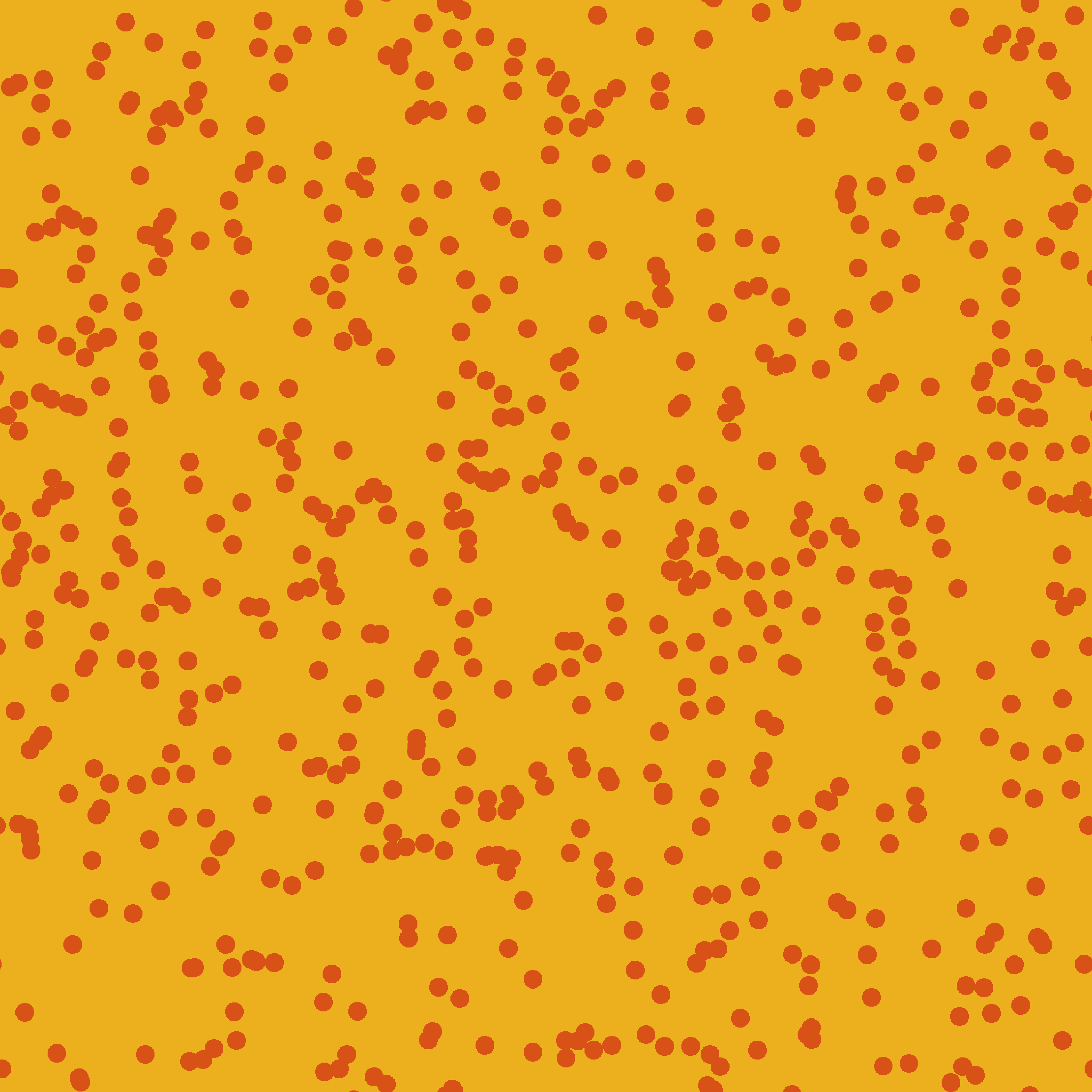}}}
    {\frame{\includegraphics[height=0.19\textwidth]{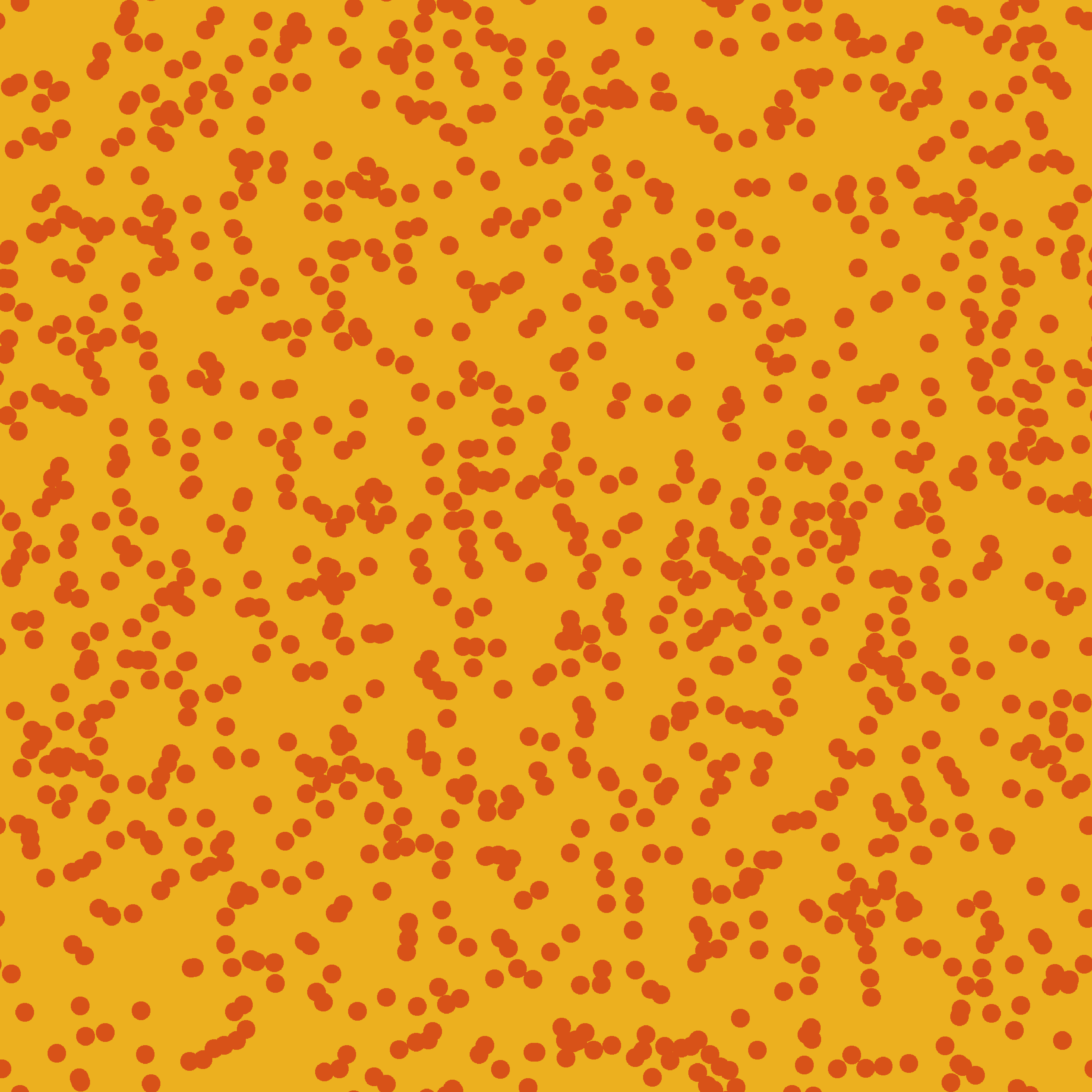}}}
    {\frame{\includegraphics[height=0.19\textwidth]{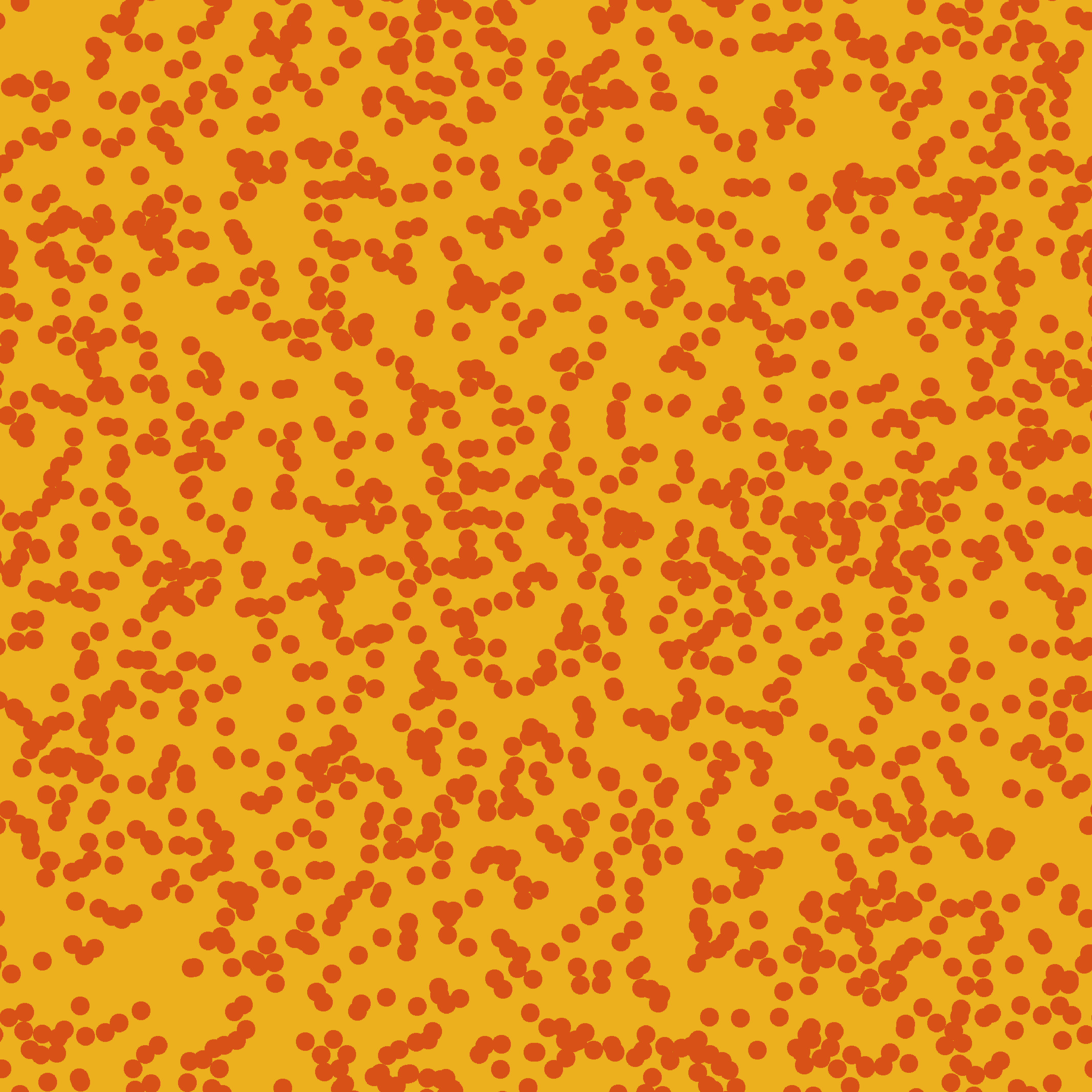}}}
    {\frame{\includegraphics[height=0.19\textwidth]{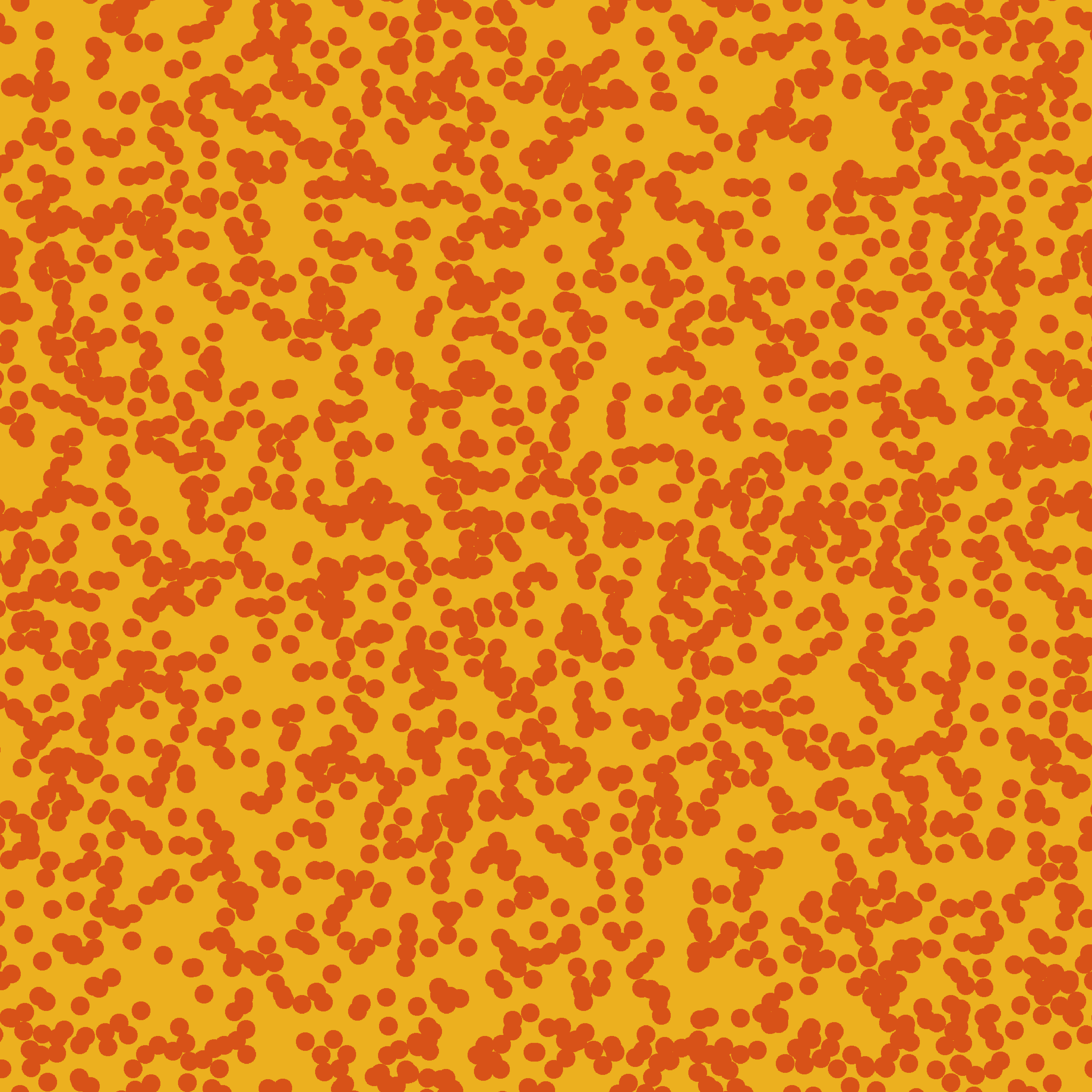}}}\vspace{2.5pt}
    {\frame{\includegraphics[height=0.19\textwidth]{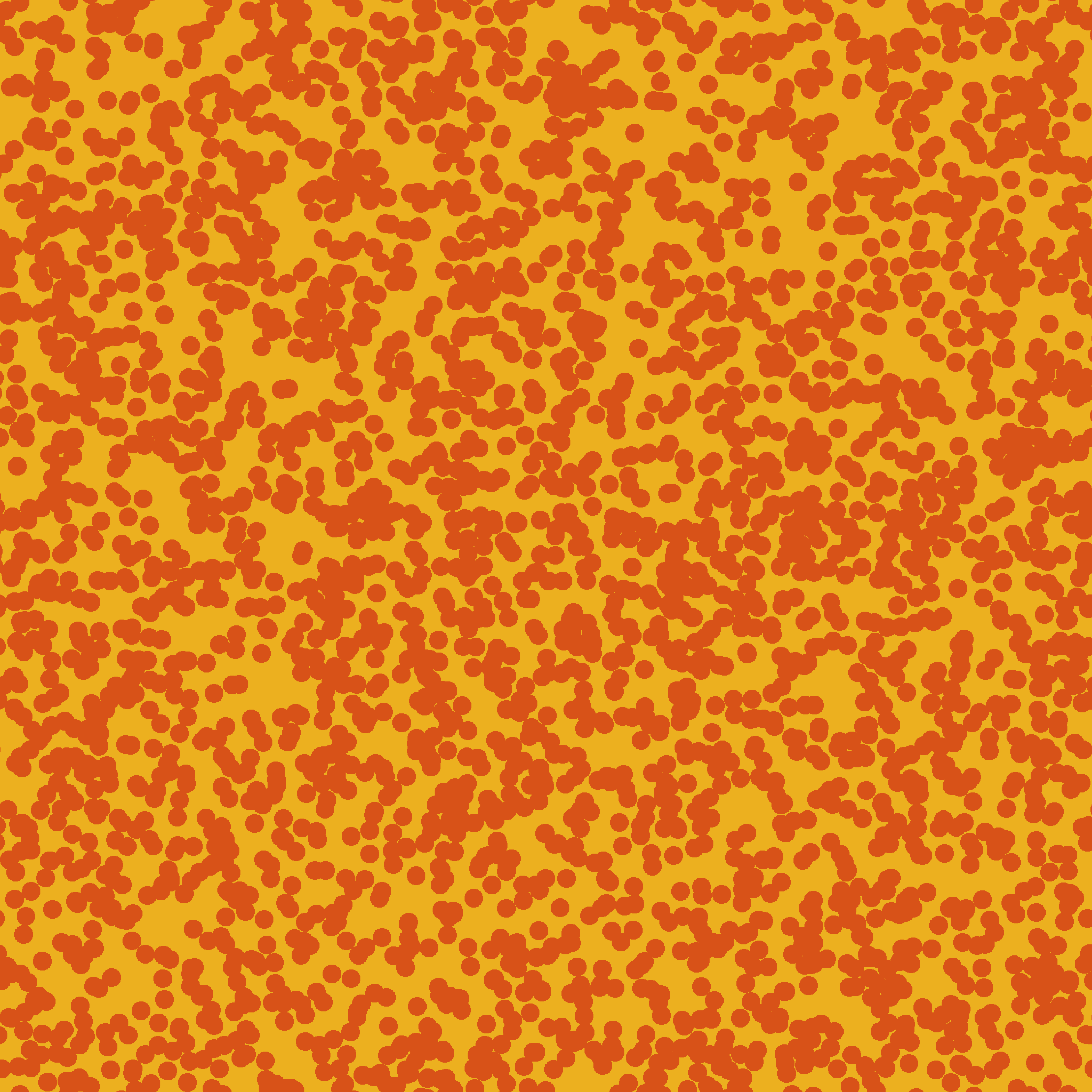}}}
    {\frame{\includegraphics[height=0.19\textwidth]{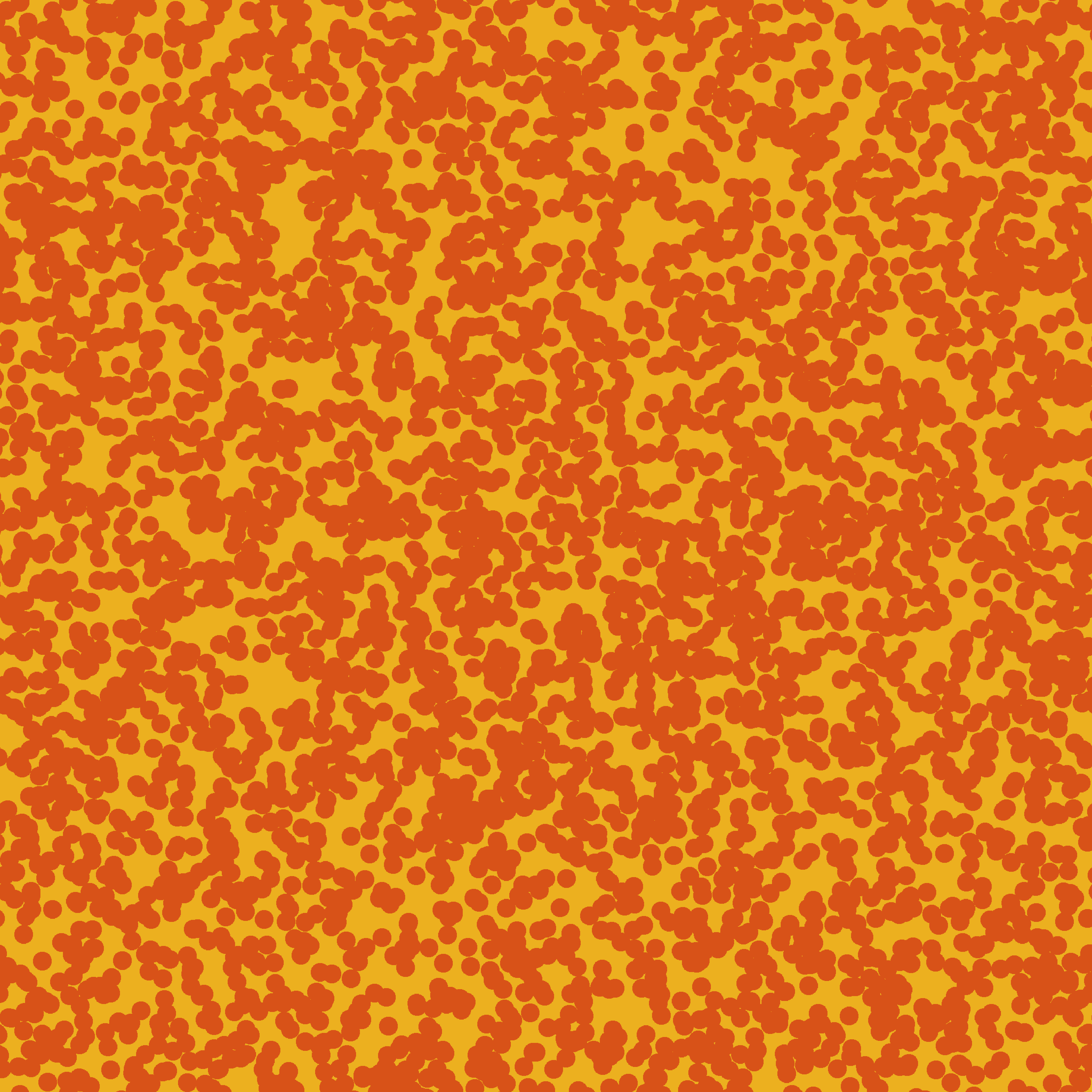}}}
    {\frame{\includegraphics[height=0.19\textwidth]{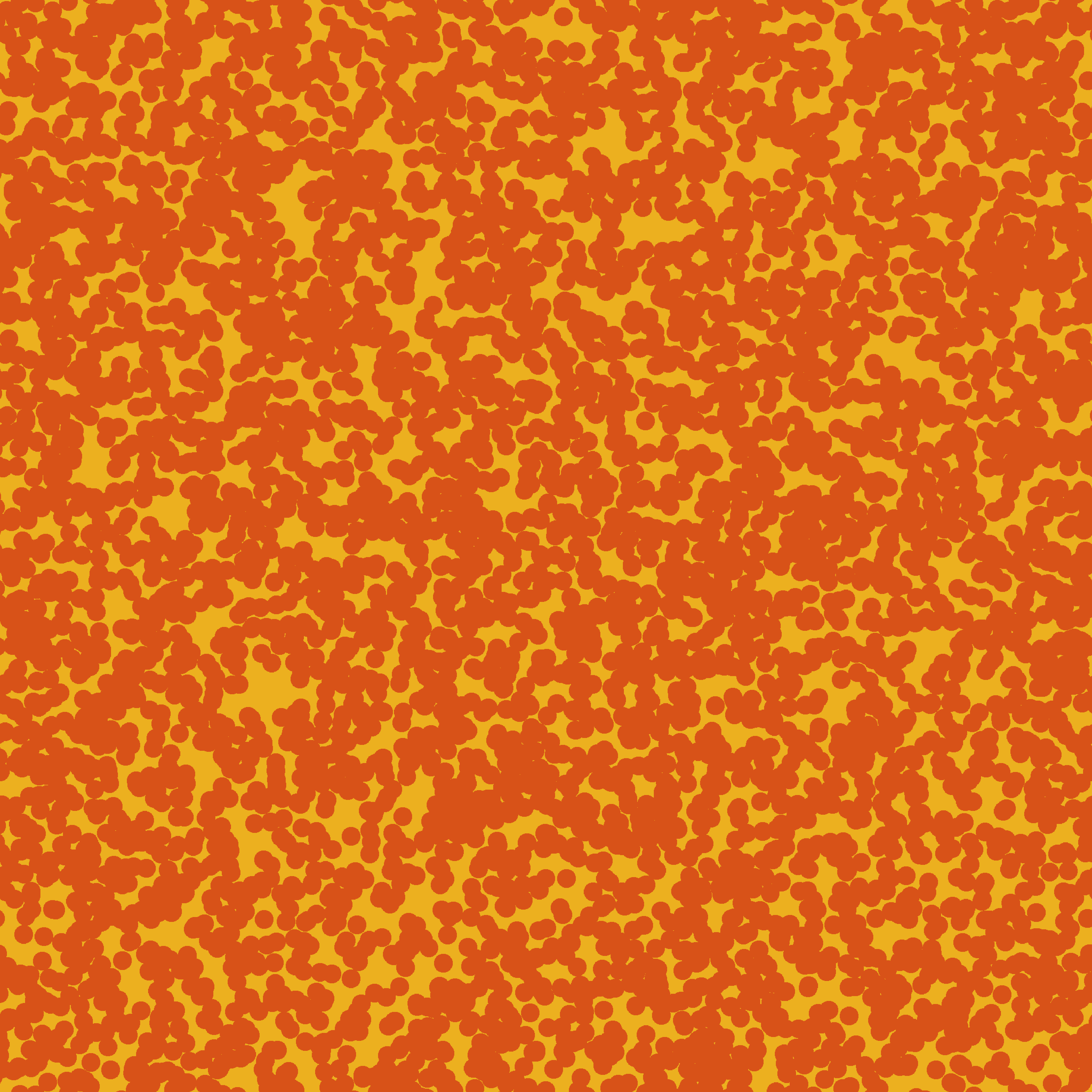}}}
    {\frame{\includegraphics[height=0.19\textwidth]{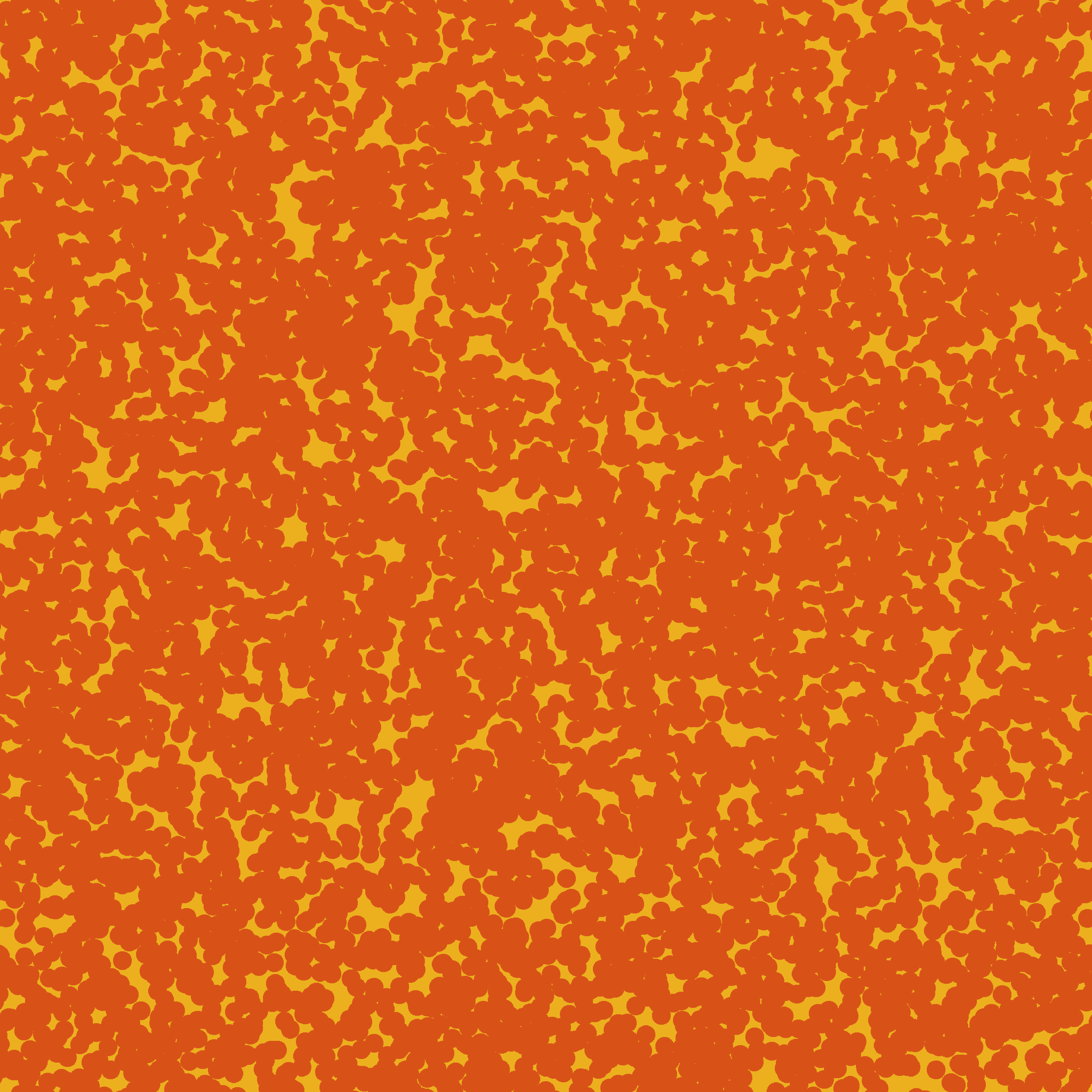}}}
    {\frame{\includegraphics[height=0.19\textwidth]{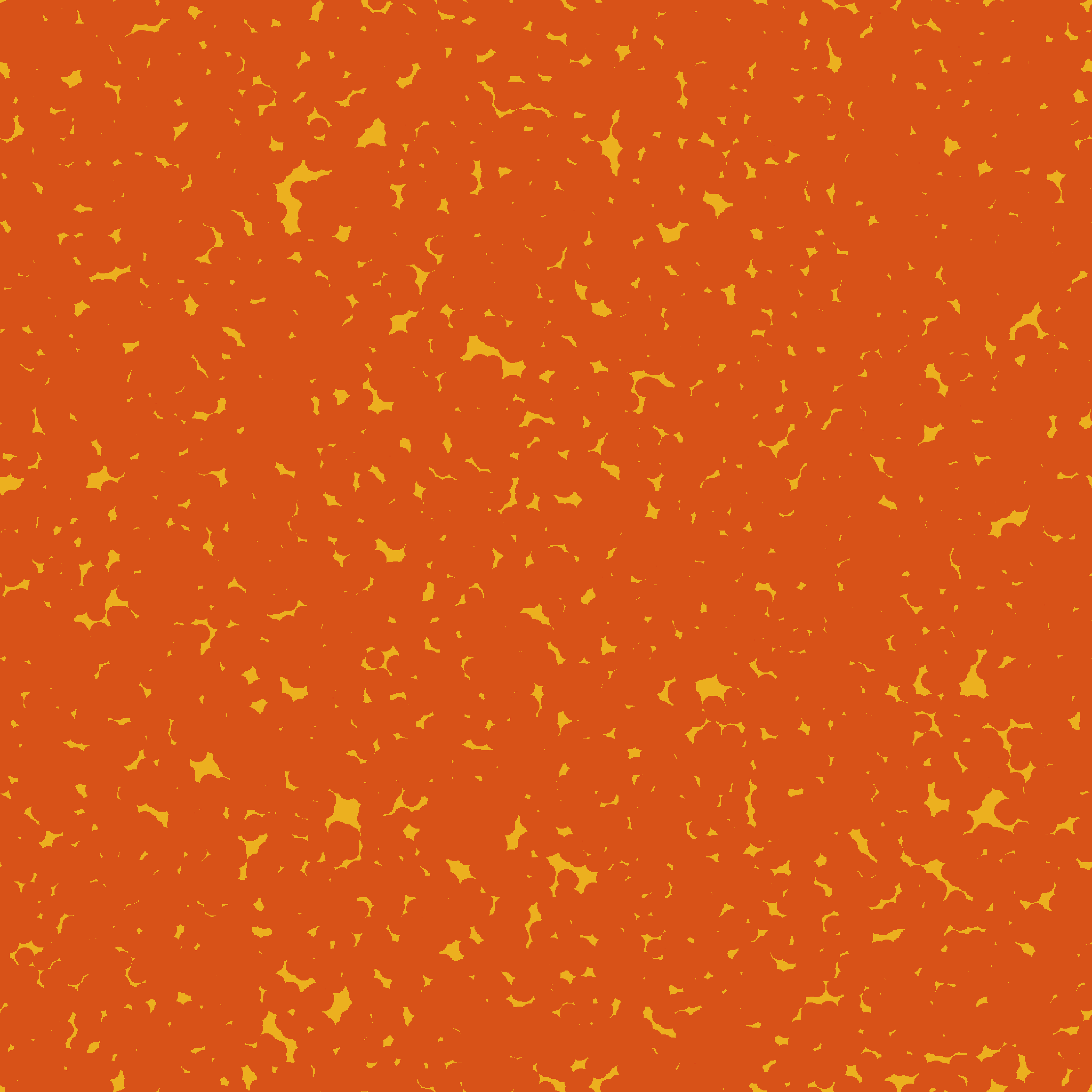}}}
    \end{minipage}
\caption{Example images of the voxel-based random microstructure realizations considering inclusion volume fractions of $\phi_\mathrm{i}=\{0.05, 0.10, \ldots, 0.95\}$, for overlapping disks. The VE size is 2048 pixels. The microstructures are generated using the indicated random sequential inhibition process under periodic boundary conditions.}
\label{F:microstructures_overlapping_VOLFRAC}
\end{figure}

As opposed to the case for nonoverlapping disks case, for overlapping disks, the attainable inclusion volume fraction is not limited. In Figure\ \ref{F:microstructures_overlapping_VOLFRAC}, we demonstrate five randomly generated microstructures with nonoverlapping disk arrangements with volume fractions $\phi_\mathrm{i}=\{0.15, 0.25, \ldots, 0.95\}$. Like before, 15 generations were realized for each volume fraction. Verification of the Monte Carlo generations of the overlapping and nonoverlapping microstructures with two-point probability functions is given in \ref{S:twopointprobabilityfunction}.

\subsection{Results}
\subsubsection{Regular Nonoverlapping Square Disk Arrangements}
\begin{figure*}[htb!]
% \vspace{5pt}
\centering
\subfigure[]{\begin{tikzpicture}
    \node[inner sep=0pt] (B11) at (0,0){\includegraphics[height=0.37\textwidth]{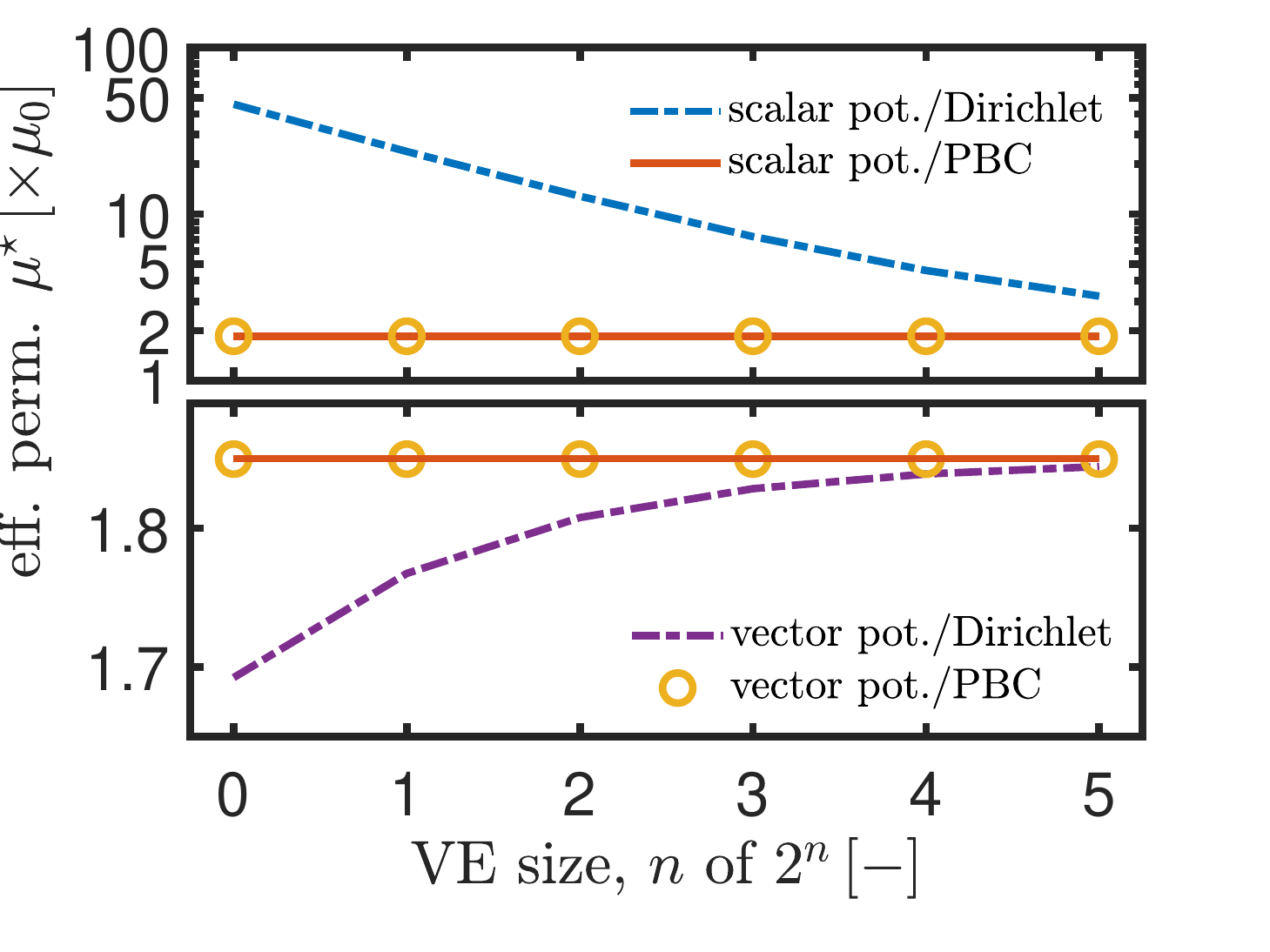}};
    \node (A) at (0,3.0) {\scriptsize $\mu_\mathrm{i}/\mu_\mathrm{m}=250/1$};
    \end{tikzpicture}}
\subfigure[]{\begin{tikzpicture}
    \node[inner sep=0pt] (B11) at (0,0){\includegraphics[height=0.37\textwidth]{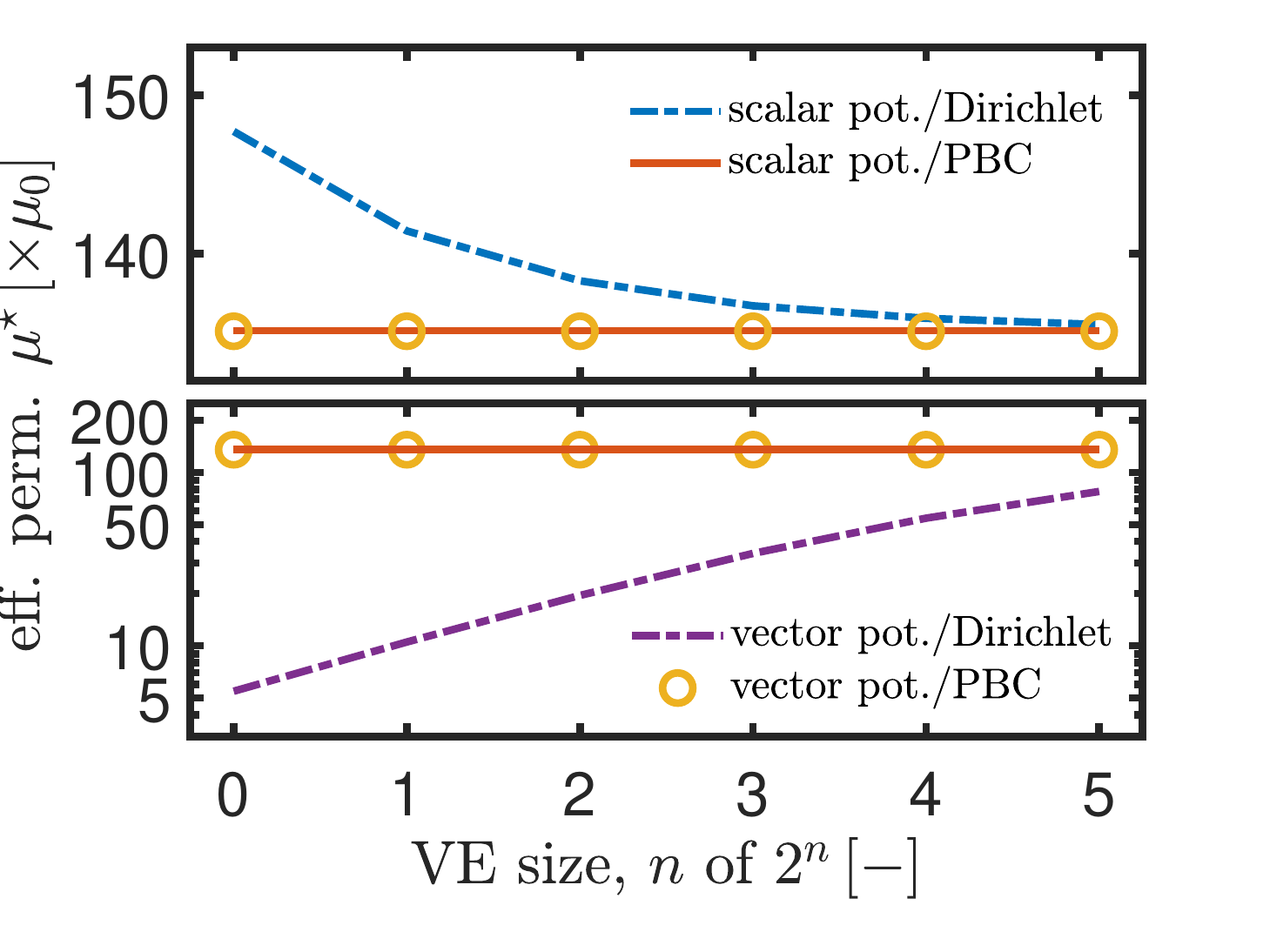}};
    \node (A) at (0,3.0) {\scriptsize $\mu_\mathrm{i}/\mu_\mathrm{m}=1/250$};
    \end{tikzpicture}}
\caption{Effective permeability predictions for the $n \times  n$  tilings of the unit cell with the inclusion volume fraction of $\phi_\mathrm{i}=0.30$ given in row 3 of Figure\ \ref{F:2D_RVE_choices} for $n = \{2, 4, 8, 16, 32\}$  for (a) $\mu_\mathrm{i}/\mu_\mathrm{e}=250/1$ and (b) $\mu_\mathrm{i}/\mu_\mathrm{e}=1/250$. The results for vector and scalar potential formulations with uniform Dirichlet and periodic boundary conditions are depicted.
The upper (Voigt) and lower (Reuss) bounds correspond to  $\mu_\mathrm{V}^\star\simeq75.70\mu_0$ and $\mu_\mathrm{R}^\star\simeq1.426\mu_0$, for the phase contrast  $\mu_\mathrm{i}/\mu_\mathrm{m}=250/1$, respectively, and  $\mu_\mathrm{V}^\star\simeq175.3\mu_0$ and $\mu_\mathrm{R}^\star\simeq3.303\mu_0$, for the phase-interchanged case. The effective permeability predictions with vector and scalar potential formulations with periodic boundary conditions give $1.850\mu_0$ and $135.1\mu_0$, considering the cases without and with phase-interchange, respectively.}
\label{F:2D_circular_cylindrical_results}
\end{figure*}

On account of the involved phase fluctuations at the cell boundaries, the $n\times n$ tilings of the unit cell given in row 3 of Figure\ \ref{F:2D_RVE_choices} are selected for computation of the effective permeability plots demonstrated in Figure\ \ref{F:2D_circular_cylindrical_results}  for $n=\{2,4,8,16,32\}$.
The vector and scalar potential formulations, once used in conjunction with  periodic boundary conditions, give results in agreement with each other with the effective permeability predictions reading  $1.850\mu_0$ and
$135.1\mu_0$, considering the cases without and with phase-interchange, respectively. Moreover, these results show size and unit cell type invariance. That is, the predicted effective permeability change neither with the tiling parameter $n$ nor the change of the unit cell type.  As can be calculated, the phase-interchange relation \cite{Keller1964} for a two-phase composite formulated by the equation  $\mu^\star(\mu_1,\mu_2)\,\mu^\star(\mu_2,\mu_1)=\mu_1\mu_2$ is satisfied within a numerical tolerance. Keller derives this relation for a medium consisting of a square lattice of monodisperse disks \cite{Keller1964}.

The size invariance, however, is not the case for the results with uniform Dirichlet boundary conditions. A monotonic convergence towards the prediction of periodic boundary conditions is observed. The direction of the convergence agrees with our remark on the effective permeability prediction of vector and scalar potential formulations for vanishing corrector functions. Without and with phase-interchange, the apparent permeability predictions for the VE with $1\times1$ tiling are $45.69\mu_0$ and $147.7\mu_0$, respectively, using scalar potential formulation. The VE with $64\times64$ tilings improves to give $3.220\mu_0$ and $135.5\mu_0$, respectively. Using vector potential formulation, without and with phase-interchange, the apparent permeability predictions for the VE with $1\times1$ tiling are $1.692\mu_0$ and $5.471\mu_0$, respectively. The VE with $64\times64$ tilings improves to give $1.844\mu_0$ and $77.63\mu_0$, respectively. For the phase contrast  $\mu_\mathrm{i}/\mu_\mathrm{m}=250/1$, Dirichlet boundary conditions give better results for vector potential formulation. In comparison, for $\mu_\mathrm{i}/\mu_\mathrm{m}=1/250$, scalar potential formulation predictions are the ones closer to the effective properties. This demonstrates the difficulty in locating the RVE size using uniform Dirichlet boundary conditions.
In light of these computations, it can be asserted that the effective permeability predictions with uniform Dirichlet boundary conditions constitute an upper bound to periodic boundary condition predictions when used with the scalar potential formulation. Otherwise, they comprise a lower bound. The computational results  are further bounded by the upper (Voigt) and lower (Reuss) bounds which, considering the current material and geometrical properties, are  $\mu_\mathrm{V}^\star\simeq75.70\mu_0$ and $\mu_\mathrm{R}^\star\simeq1.426\mu_0$, respectively. With phase-interchange giving  $\mu_\mathrm{i}/\mu_\mathrm{m}=250/1$ these read $\mu_\mathrm{V}^\star\simeq175.3\mu_0$ and $\mu_\mathrm{R}^\star\simeq3.303\mu_0$.

\begin{figure*}[htb!]
% \vspace{5pt}
% trim=left bottom right top, clip
\centering
{\frame{\includegraphics[height=0.15\textwidth,
trim=489 233 489 233, clip]{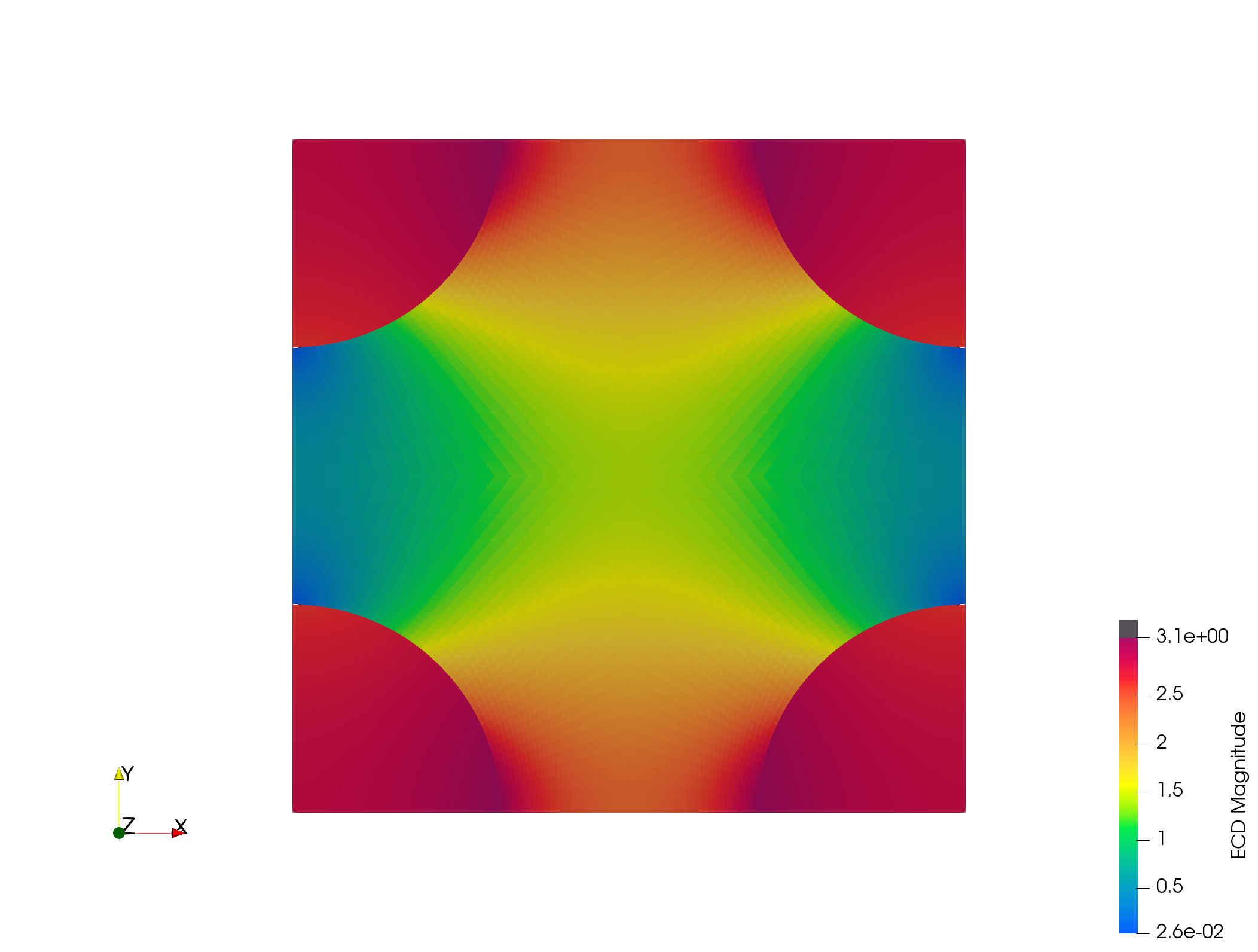}}}
{\frame{\includegraphics[height=0.15\textwidth,
trim=489 233 489 233, clip]{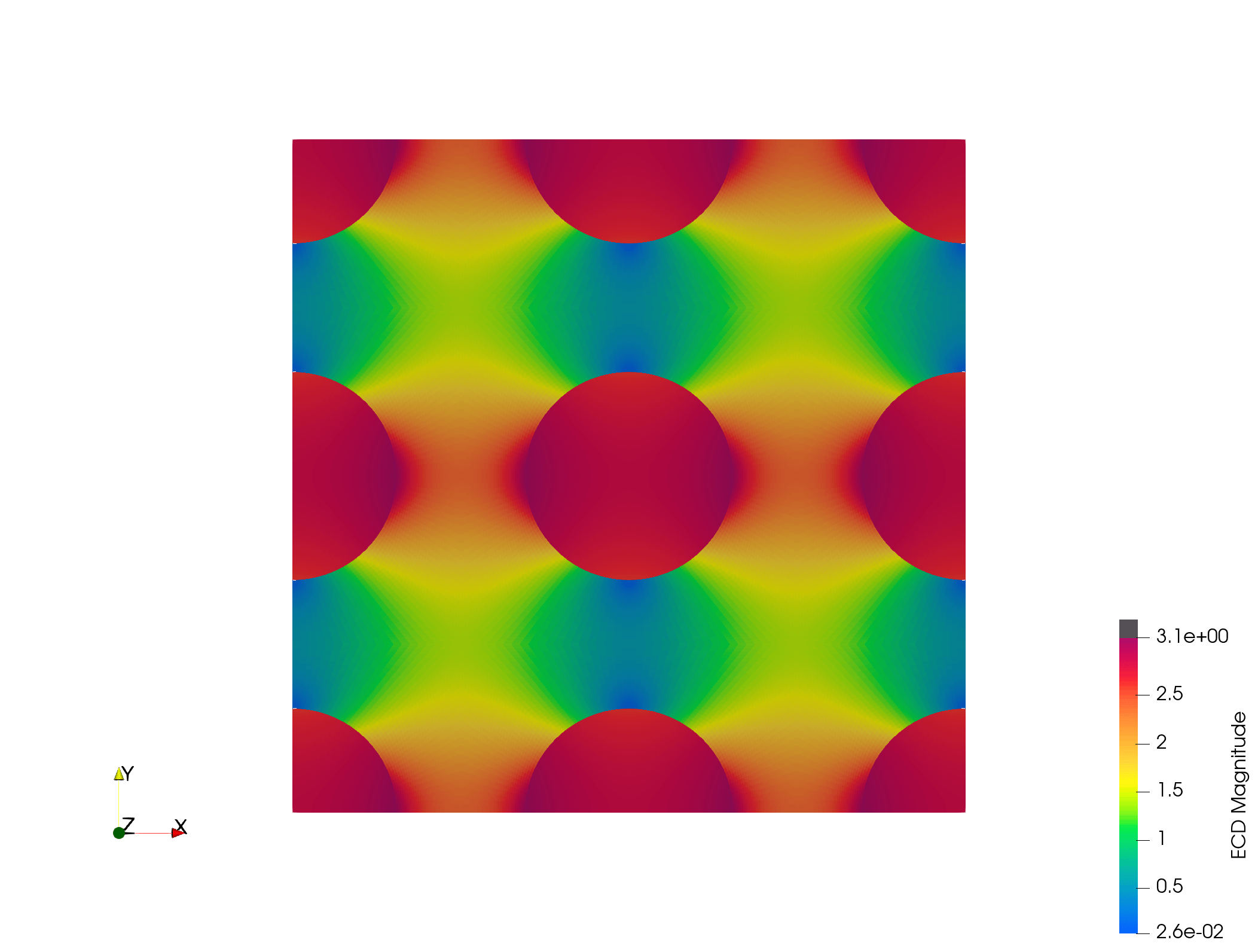}}}
{\frame{\includegraphics[height=0.15\textwidth,
trim=489 233 489 233, clip]{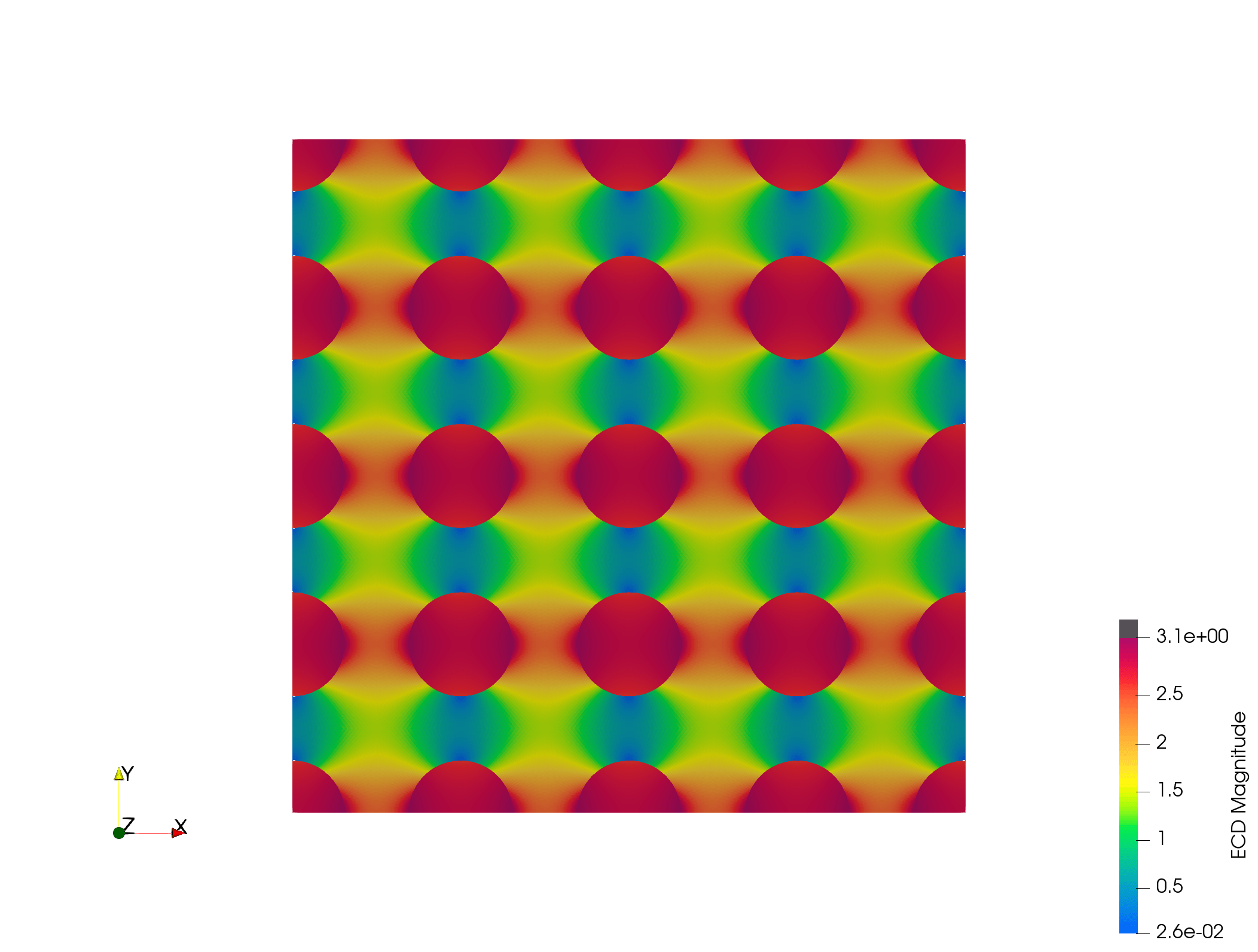}}}
{\frame{\includegraphics[height=0.15\textwidth,
trim=489 233 489 233, clip]{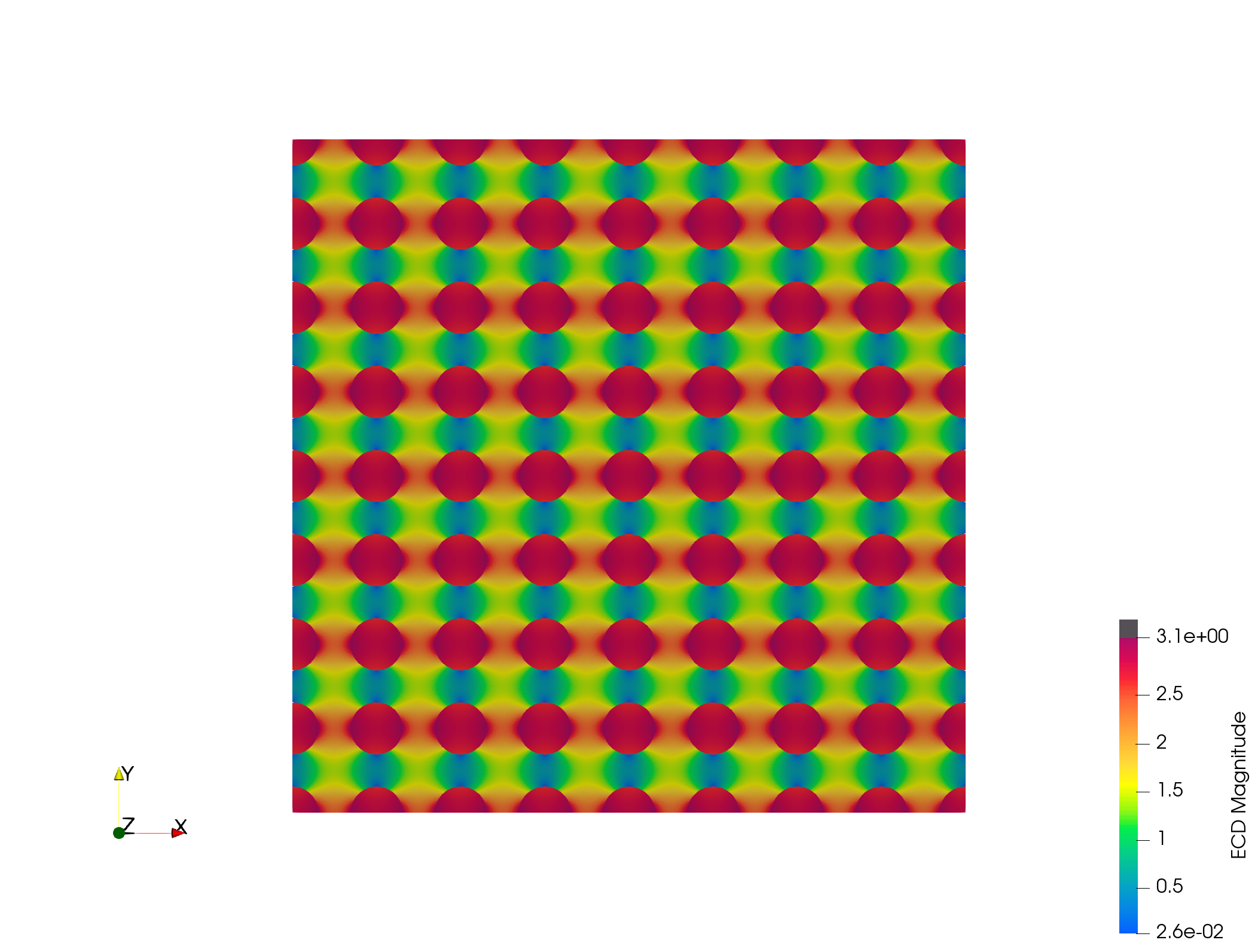}}}\\\vspace{2.5pt}
{\frame{\includegraphics[height=0.15\textwidth,
trim=489 233 489 233, clip]{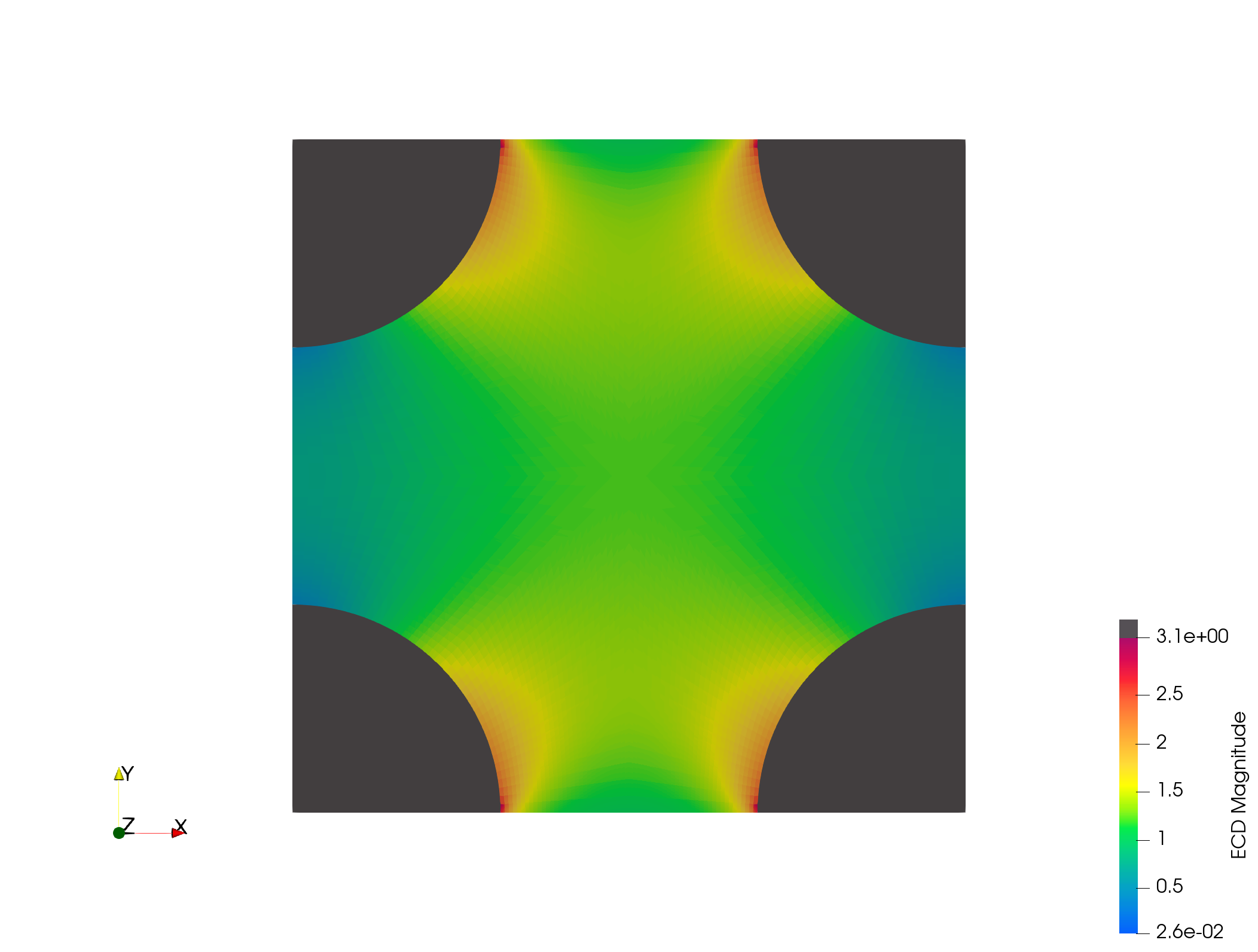}}}
{\frame{\includegraphics[height=0.15\textwidth,
trim=489 233 489 233, clip]{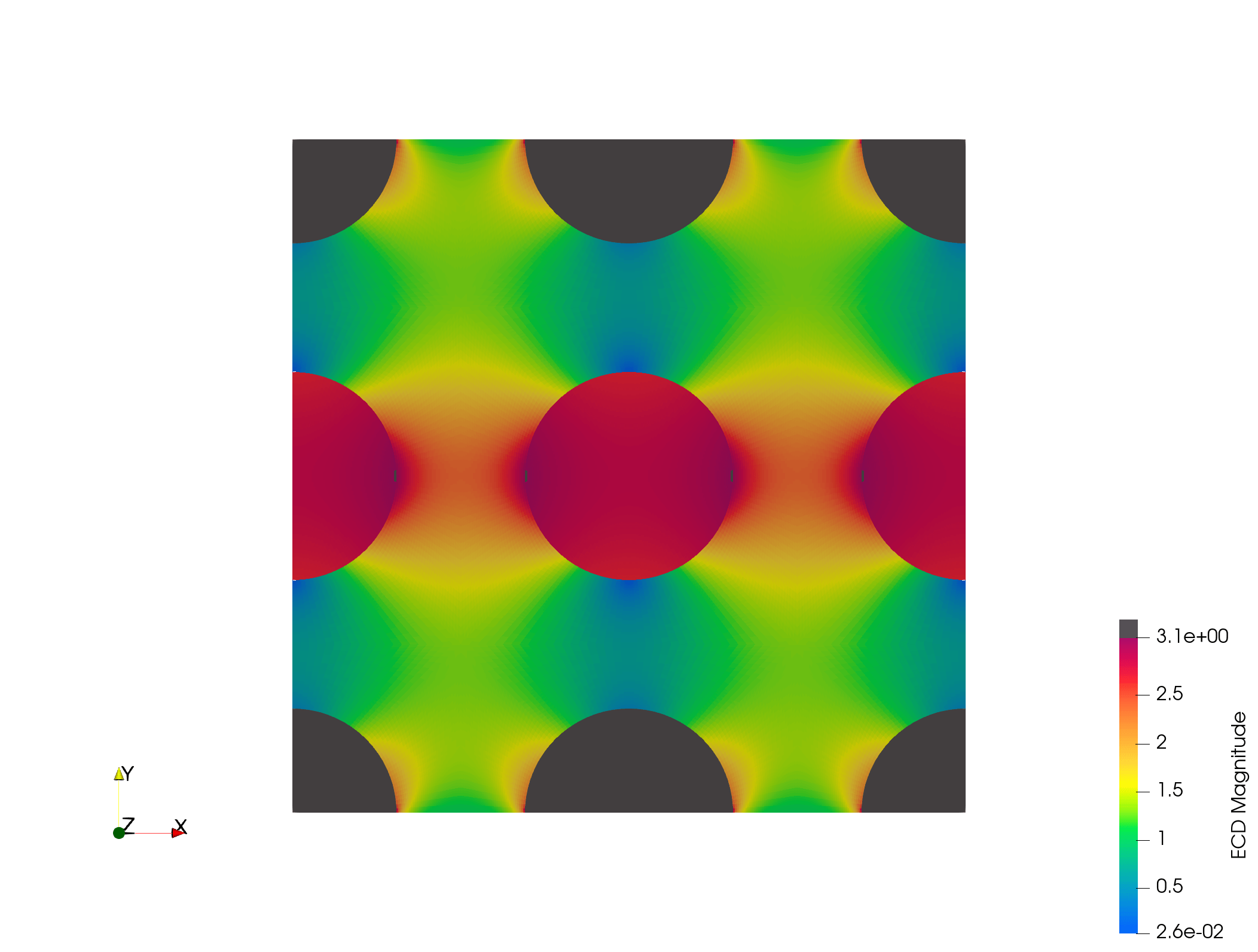}}}
{\frame{\includegraphics[height=0.15\textwidth,
trim=489 233 489 233, clip]{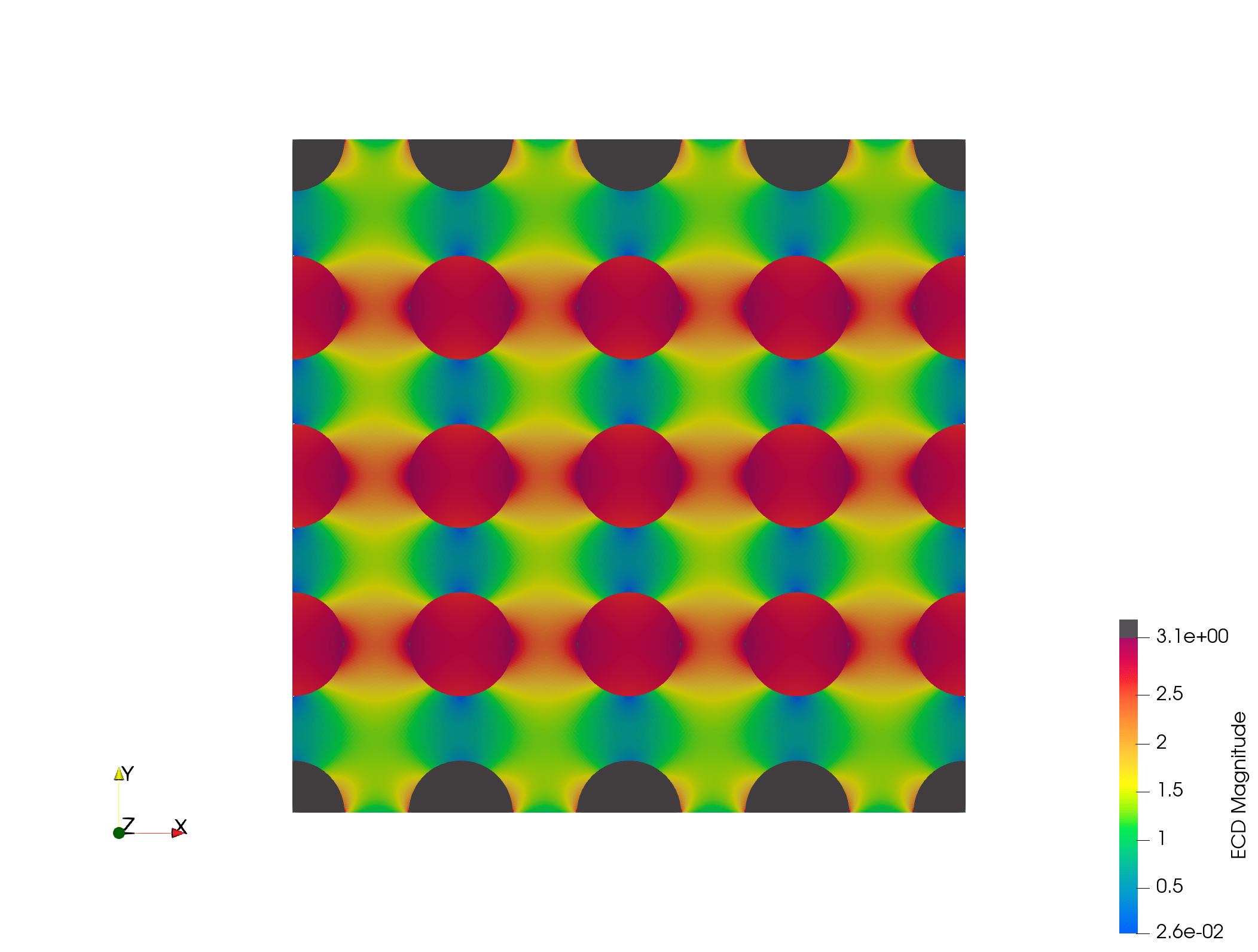}}}
{\frame{\includegraphics[height=0.15\textwidth,
trim=489 233 489 233, clip]{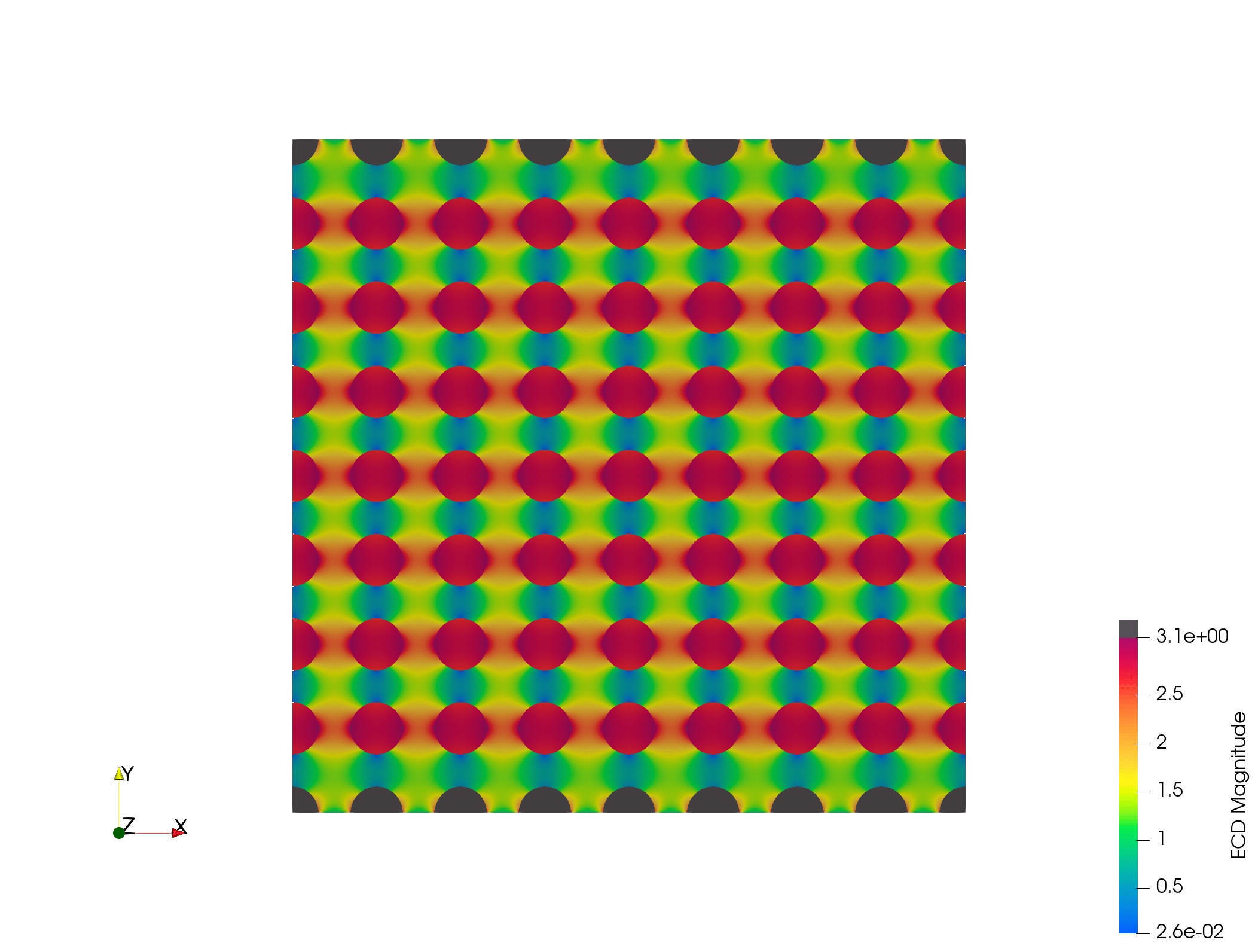}}}
    min$\,\,$\frame{\includegraphics[width=0.49\textwidth, trim=0 15 0 15, clip]{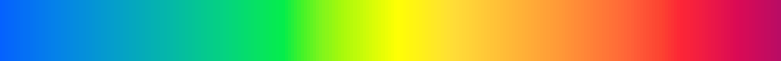}}$\,\,$max
\caption{Magnetic flux vector norm contour plots for $1\times1$, $2\times2$, $4\times4$, and $8\times8$  primitive unit cell tilings with the inclusion volume fraction of $\phi_\mathrm{i}=0.30$ for (top) periodic and (bottom) uniform Dirichlet boundary conditions with scalar potential formulation for the horizontally applied macroscopic magnetic field with $\bs H=[1, 0]^\top$. The results correspond to $\mu_\mathrm{i}/\mu_\mathrm{e}=250/1$.
For demonstration purposes, the intervals [min, max] of the contour plots are taken as $[0.026,3.094]$, which corresponds to the interval of the solution with periodic boundary conditions. The uniform Dirichlet boundary conditions result in a magnetic flux distribution with a maximum of $643.3$ corresponding to the dark grey region at the boundary inclusions. These results are normalized with respect to $\mu_0$. An effective permeability of $1.850\mu_0$ is predicted with vector and scalar potential formulations with periodic boundary conditions.}
\label{F:2D_circular_cylindrical_periodic_vs_dirichlet}
\end{figure*}

It is worth noting that uniform Dirichlet boundary conditions suppress fluctuation fields only at the unit cell boundaries but not at the interior domain. Figure\ \ref{F:2D_circular_cylindrical_periodic_vs_dirichlet} shows the influence of boundary conditions on magnetic flux distribution for periodic volume elements at which the high permeability phase occupies the boundary causing high property fluctuations at this region. When subjected to periodic boundary conditions, despite phase  property fluctuations at the edges, the solution remains periodic within the primitive unit cell. For uniform Dirichlet boundary conditions, however, this results in the formation of a boundary layer with unusually high magnetic flux concentration at the high permeability phase. This boundary layer is diminished fast towards the interior regions of the volume element such that the field distributions for the periodic boundary conditions are recovered. This implies that the accuracy can be improved by reducing the relative size of the boundary layer, hence by increasing the VE size, as shown above.

Having shown the invariance properties of the solutions with periodic boundary conditions in conjunction with scalar and vector potential formulations, we continue the rest of this section by investigating the inclusion volume fraction effect on the effective magnetic permeability of the composite. Here, we use the primitive unit cell given in row one of Figure\ \ref{F:2D_RVE_choices}.  We vary the inclusion volume fraction within the  interval of $\phi_\mathrm{i}\in[0,0.55]$.

\begin{figure*}[htb!]
% \vspace{5pt}
\centering
\subfigure[]{\begin{tikzpicture}
    \node[inner sep=0pt] (B11) at (0,0)
    {\includegraphics[height=0.37\textwidth]{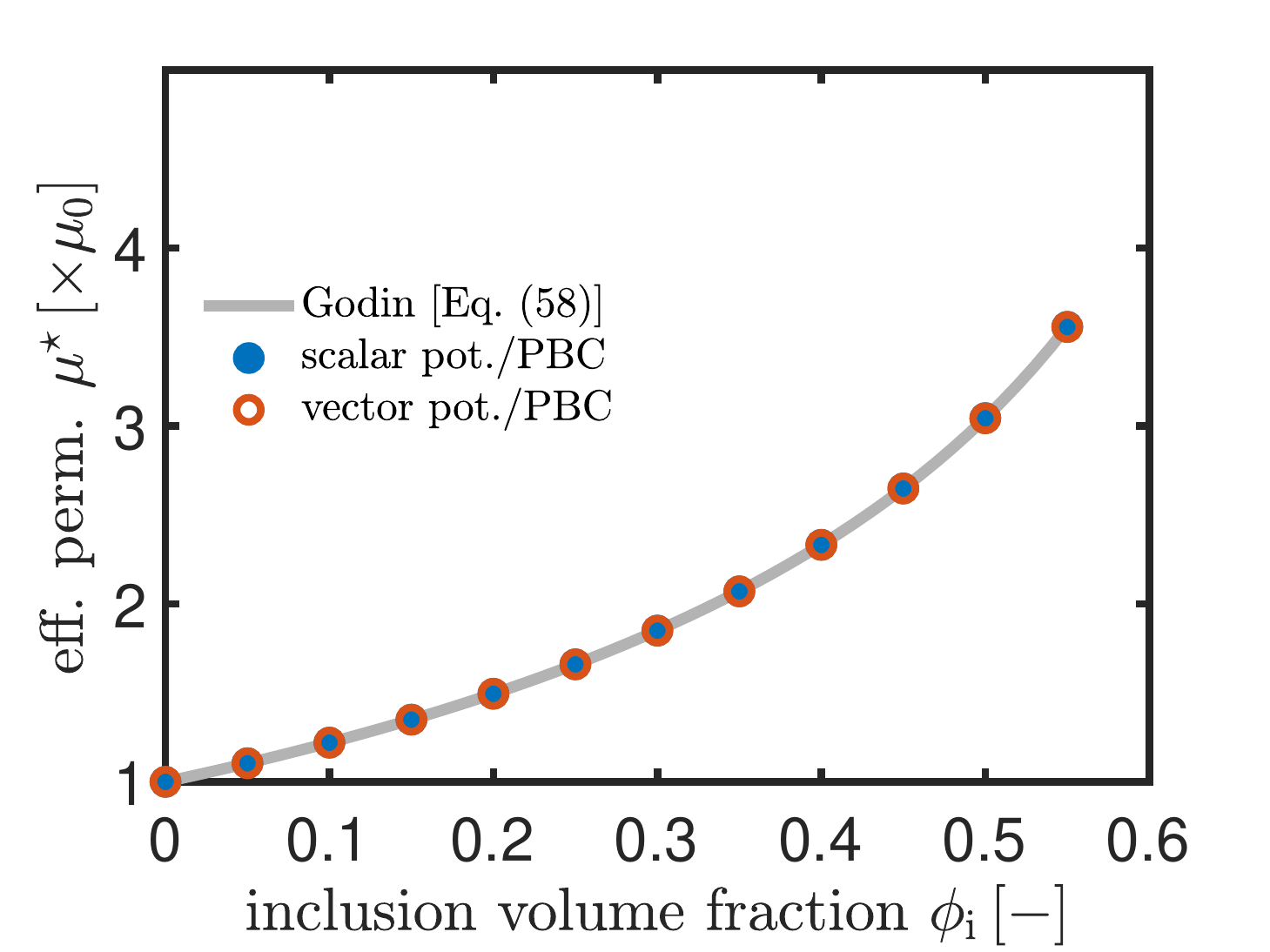}};
    \node[inner sep=0pt] (C1) at (-4.05+1.6,2.05)
    {{\includegraphics[height=0.0515\textwidth,
trim=1238 324 1595 27, clip]{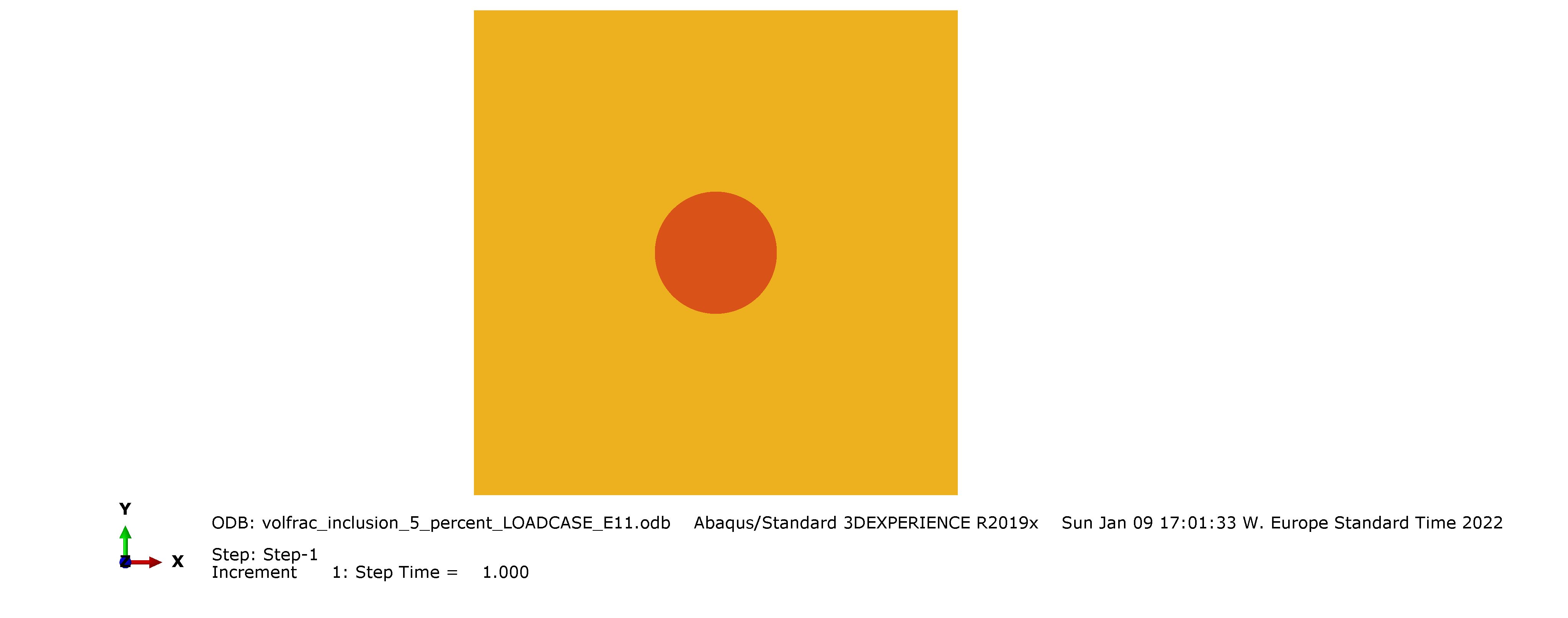}}};
    \node[inner sep=0pt] (C2) at (-4.05+1.6+1*1.04,2.05)
    {{\includegraphics[height=0.0515\textwidth,
trim=1238 324 1595 27, clip]{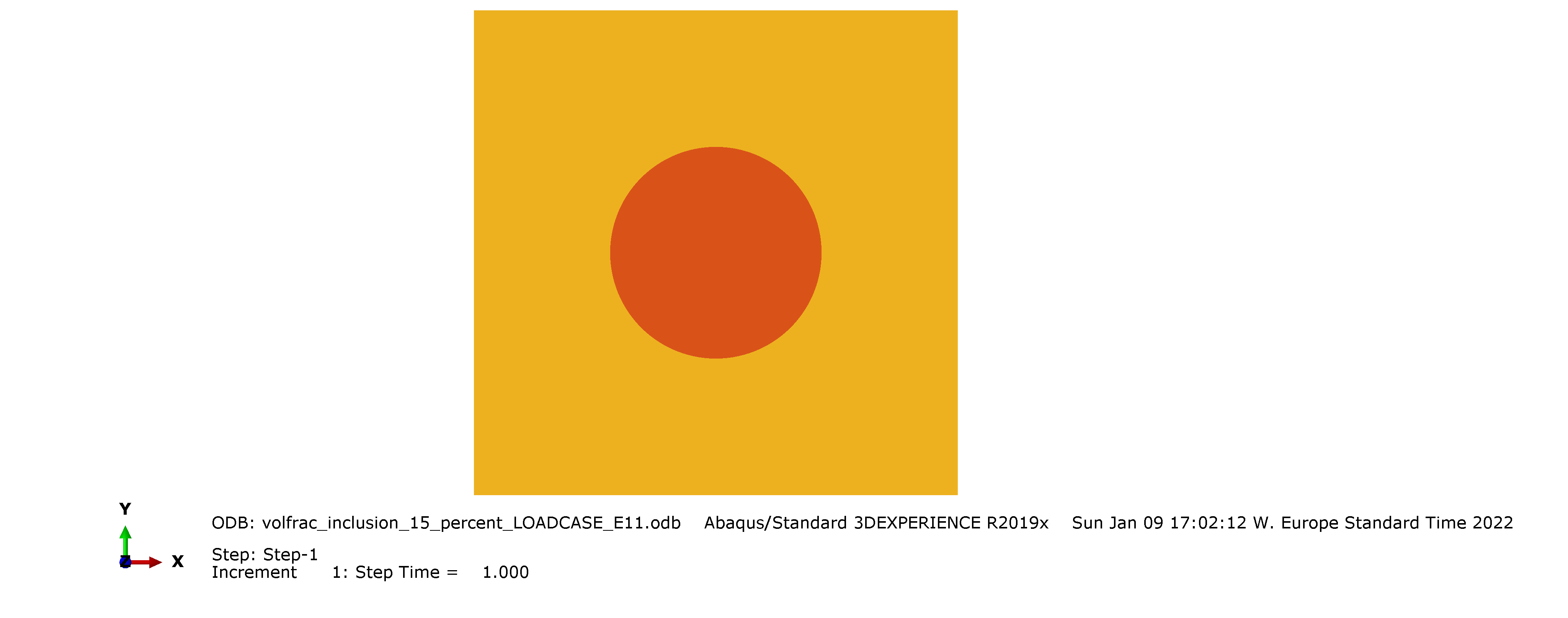}}};
    \node[inner sep=0pt] (C3) at (-4.05+1.6+2*1.04,2.05)
    {{\includegraphics[height=0.0515\textwidth,
trim=1238 324 1595 27, clip]{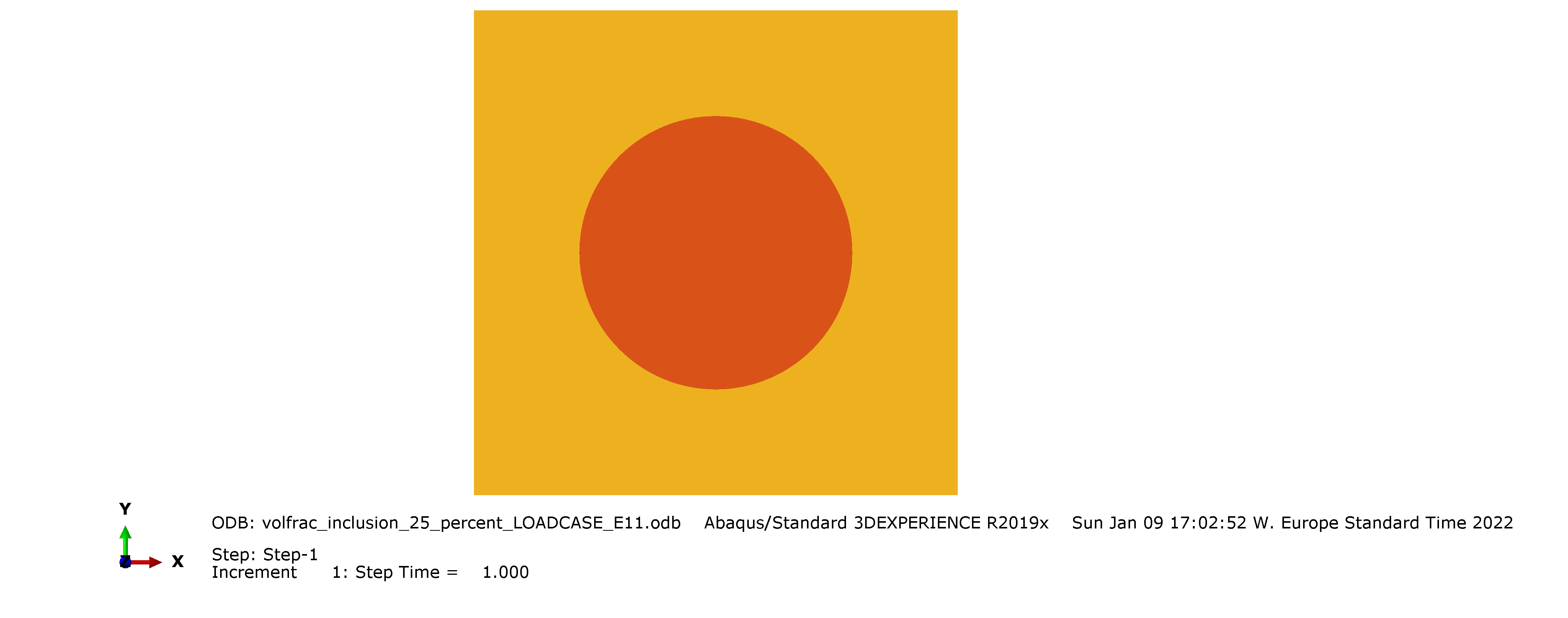}}};
    \node[inner sep=0pt] (C4) at (-4.05+1.6+3*1.04,2.05)
    {{\includegraphics[height=0.0515\textwidth,
trim=1238 324 1595 27, clip]{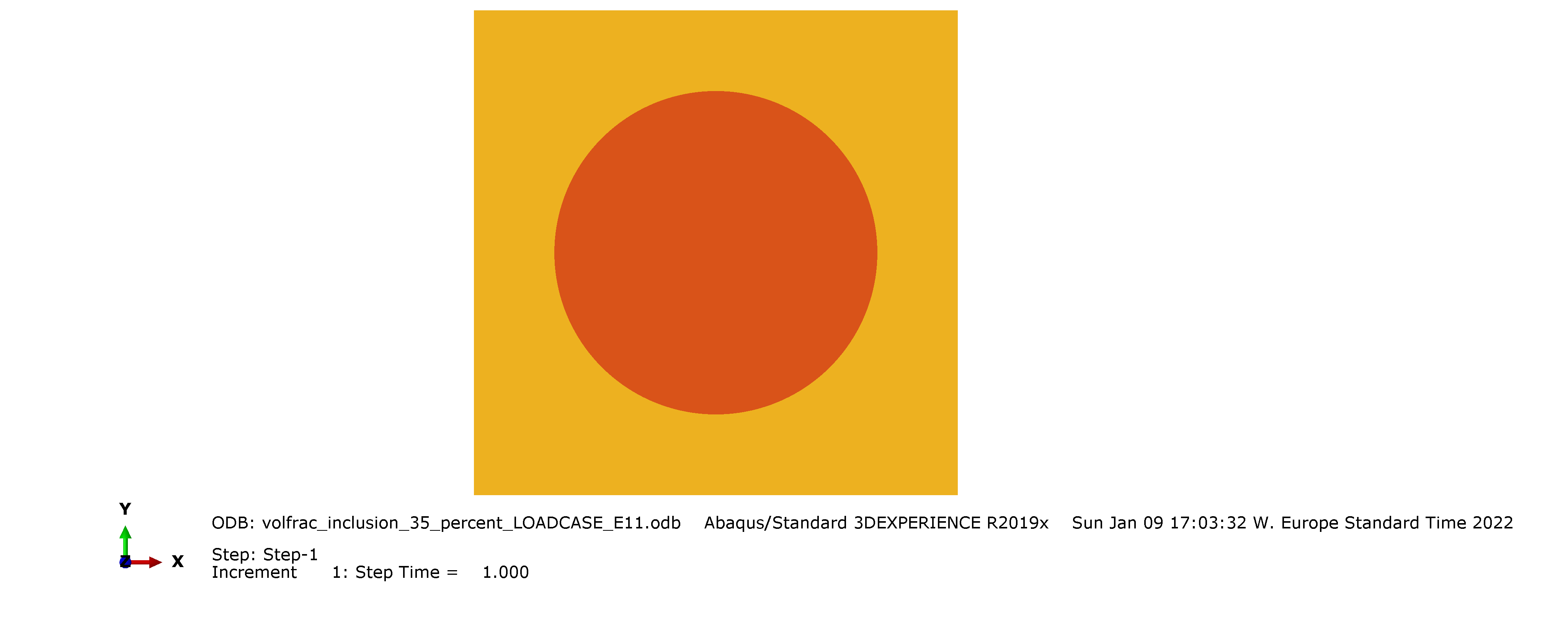}}};
    \node[inner sep=0pt] (C5) at (-4.05+1.6+4*1.04,2.05)
    {{\includegraphics[height=0.0515\textwidth,
trim=1238 324 1595 27, clip]{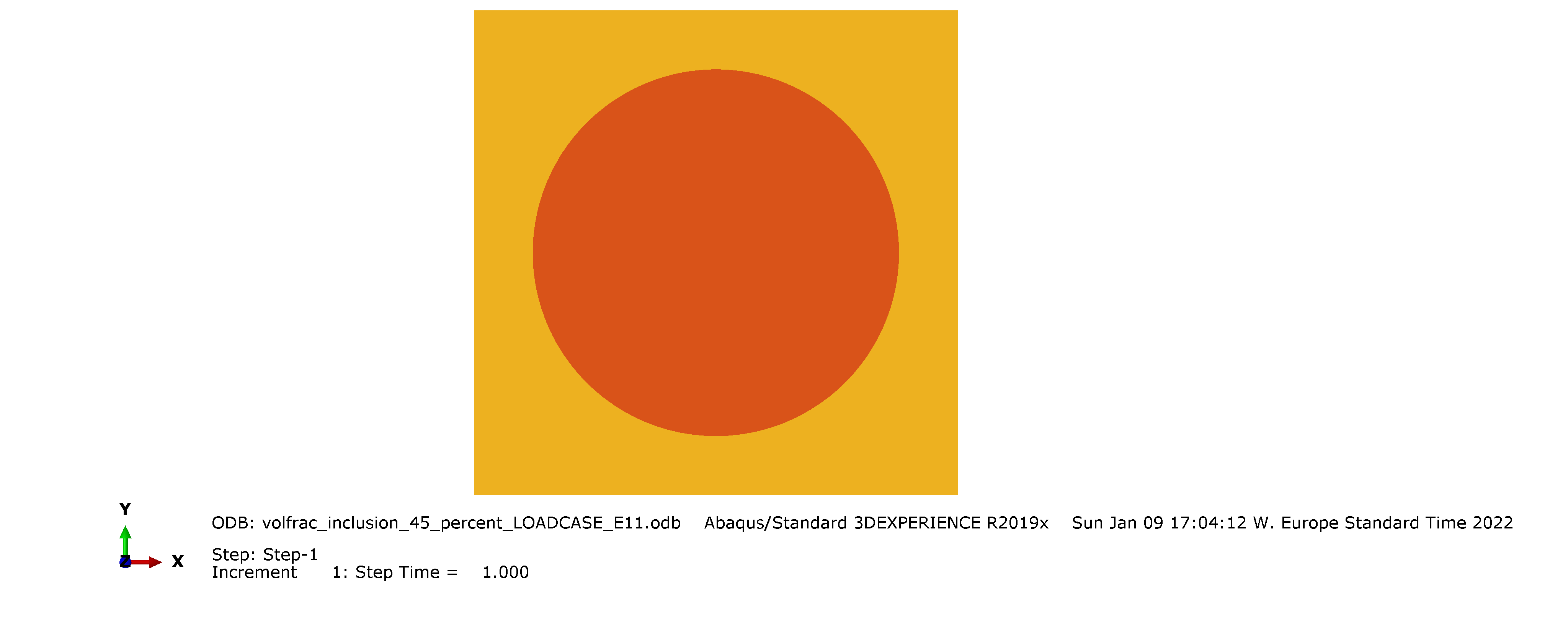}}};
    \node[inner sep=0pt] (C5) at (-4.05+1.6+5*1.04,2.05)
    {{\includegraphics[height=0.0515\textwidth,
trim=1238 324 1595 27, clip]{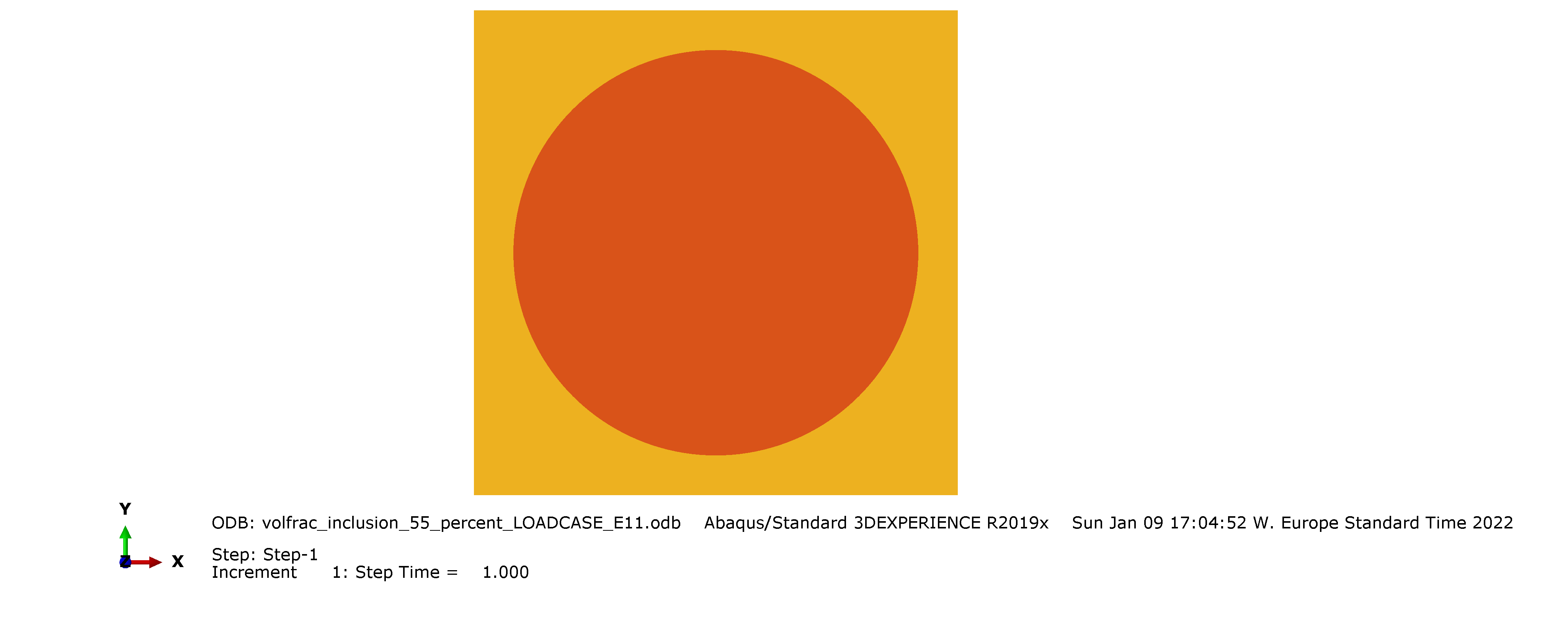}}};
    \node (A) at (0.0,3.0) {\scriptsize $\mu_\mathrm{i}/\mu_\mathrm{m}=250/1$};
\end{tikzpicture}}
\subfigure[]{\begin{tikzpicture}
    \node[inner sep=0pt] (B11) at (0,0)
    {\includegraphics[height=0.37\textwidth]{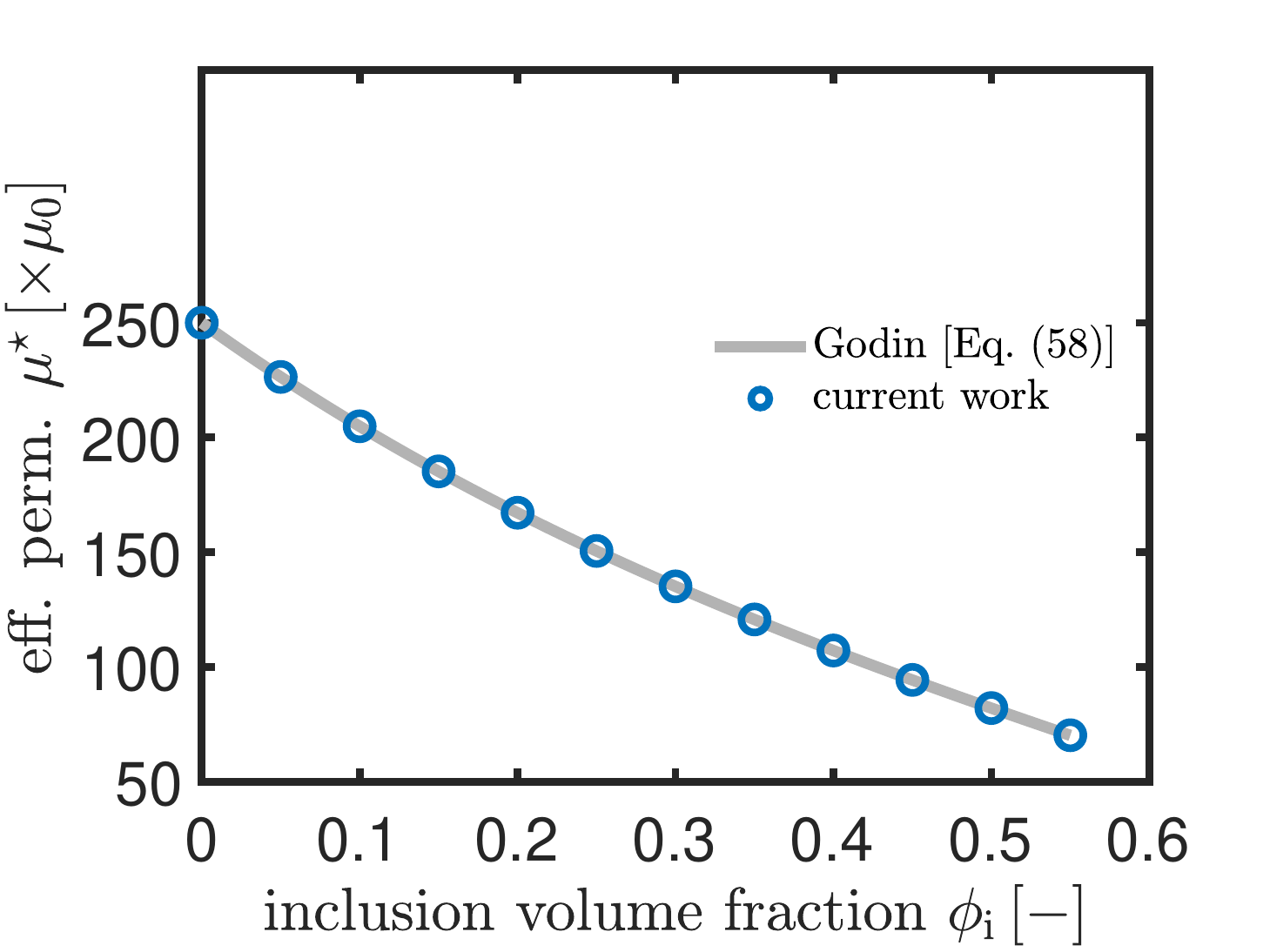}};
    \node[inner sep=0pt] (C1) at (-3.84+1.6,2.05)
    {{\includegraphics[height=0.0515\textwidth,
trim=1238 324 1595 27, clip]{volfrac_inclusion_5_NOT_MESHED}}};
    \node[inner sep=0pt] (C2) at (-3.84+1.6+1*1.0,2.05)
    {{\includegraphics[height=0.0515\textwidth,
trim=1238 324 1595 27, clip]{volfrac_inclusion_15_NOT_MESHED}}};
    \node[inner sep=0pt] (C3) at (-3.84+1.6+2*1.0,2.05)
    {{\includegraphics[height=0.0515\textwidth,
trim=1238 324 1595 27, clip]{volfrac_inclusion_25_NOT_MESHED}}};
    \node[inner sep=0pt] (C4) at (-3.84+1.6+3*1.0,2.05)
    {{\includegraphics[height=0.0515\textwidth,
trim=1238 324 1595 27, clip]{volfrac_inclusion_35_NOT_MESHED}}};
    \node[inner sep=0pt] (C5) at (-3.84+1.6+4*1.0,2.05)
    {{\includegraphics[height=0.0515\textwidth,
trim=1238 324 1595 27, clip]{volfrac_inclusion_45_NOT_MESHED}}};
    \node[inner sep=0pt] (C5) at (-3.84+1.6+5*1.0,2.05)
    {{\includegraphics[height=0.0515\textwidth,
trim=1238 324 1595 27, clip]{volfrac_inclusion_55_NOT_MESHED}}};
    \node (A) at (0.0,3.0) {\scriptsize $\mu_\mathrm{i}/\mu_\mathrm{m}=1/250$};
\end{tikzpicture}}
\caption{Effective permeability predictions for the primitive unit cell given in row 1 of Figure\ \ref{F:2D_RVE_choices} as a function of inclusion volume fraction for (a) $\mu_\mathrm{i}/\mu_\mathrm{e}=250/1$ and (b) $\mu_\mathrm{i}/\mu_\mathrm{e}=1/250$. The numerical results correspond to scalar potential formulation with periodic boundary conditions. Since Keller's phase-interchange relation $\mu^\star(\mu_1,\mu_2)\,\mu^\star(\mu_2,\mu_1)=\mu_1\mu_2$ given in \cite{Keller1964} is satisfied
multiplying individual components of the plotted resultant effective relative permeability vectors for (a) and its phase-interchanged version (b) always yields $1 \times 250$.}
\label{F:2D_circular_cylindrical_volume_fraction}
\end{figure*}

A matrix with a square and periodic arrangement of  nonoverlapping disks possesses magnetostatic isotropy. For the analytical solution quantifying the influence of the included content on composite's effective isotropic  permeability in periodic square disk arrangements, we adopt the truncated reiteration of the series expansion solution of Godin \cite{Godin2013} given in \cite{Ren2018}
\begin{equation}
\mu^\star=\left[\dfrac{1+ \phi_\mathrm{i}\,\lambda(\phi_\mathrm{i})}{1-\phi_\mathrm{i}\,\lambda(\phi_\mathrm{i})}\right]\,\mu_{\mathrm{m}}
\label{E:analytical_solution}
\end{equation}
with
\begin{equation}
\begin{split}
\lambda(\phi_i)=c_0+c_4\phi^4_\mathrm{i}+c_8\phi_\mathrm{i}^8+c_{12}\phi^{12}_\mathrm{i}+O(\phi^{16}_\mathrm{i})\,,
\end{split}
\end{equation}
where the coefficients as functions of $\alpha=[\mu_\mathrm{i}-\mu_\mathrm{m}][\mu_\mathrm{i}+\mu_\mathrm{m}]$ read
\begin{align}
c_0&=\alpha\,,\\
c_4&=0.305827\alpha^3\,,\\
c_8&=\alpha^3[0.0935304\alpha^2+0.0133615]\,,\\
c_{12}&=\alpha^3[0.0286042\alpha^4+0.437236\alpha^2+0.000184643]\,.
\end{align}
As demonstrated in Figure\ \ref{F:2D_circular_cylindrical_volume_fraction} our computational results
with scalar and vector potential formulations and  periodic boundary conditions
show an excellent agreement with the analytical solution  given in Eq.\ \eqref{E:analytical_solution}.
As before, Keller's phase-interchange relation encapsulated in the expression $\mu^\star(\mu_1,\mu_2)\,\mu^\star(\mu_2,\mu_1)=\mu_1\mu_2$ is satisfied within a numerical tolerance \cite{Keller1964}.

\begin{figure*}[htb!]
% \vspace{5pt}
% trim=left bottom right top, clip
\centering
\subfigure[]{\frame{\begin{tikzpicture}
    \node[inner sep=0pt] (B11) at (0,0)
    {\includegraphics[height=0.15\textwidth,
trim=400 233 964 233, clip]{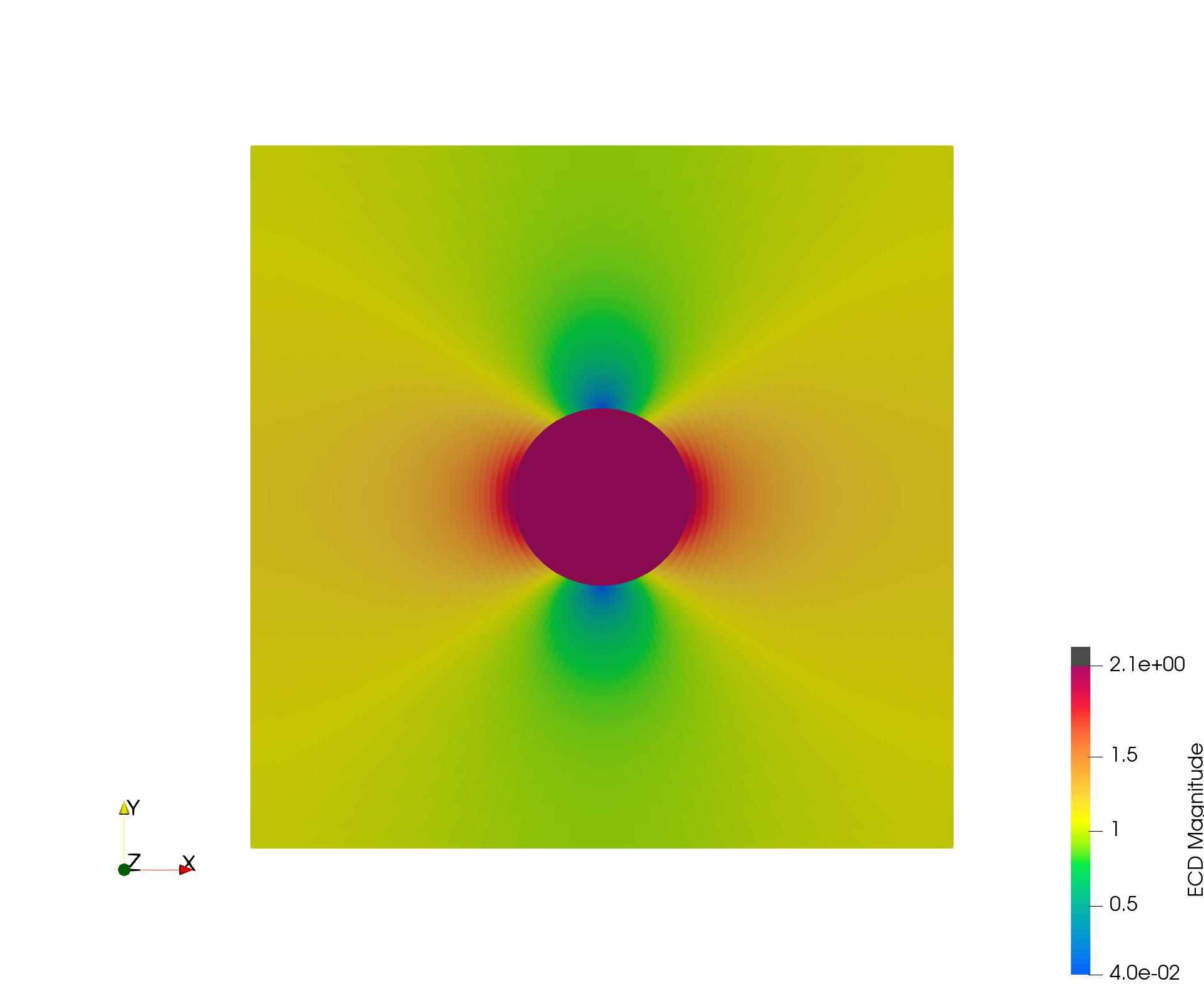}};
    \node[inner sep=0pt] (C1) at (1.24,0)
    {\includegraphics[height=0.15\textwidth,
trim=964 233 400 233, clip]{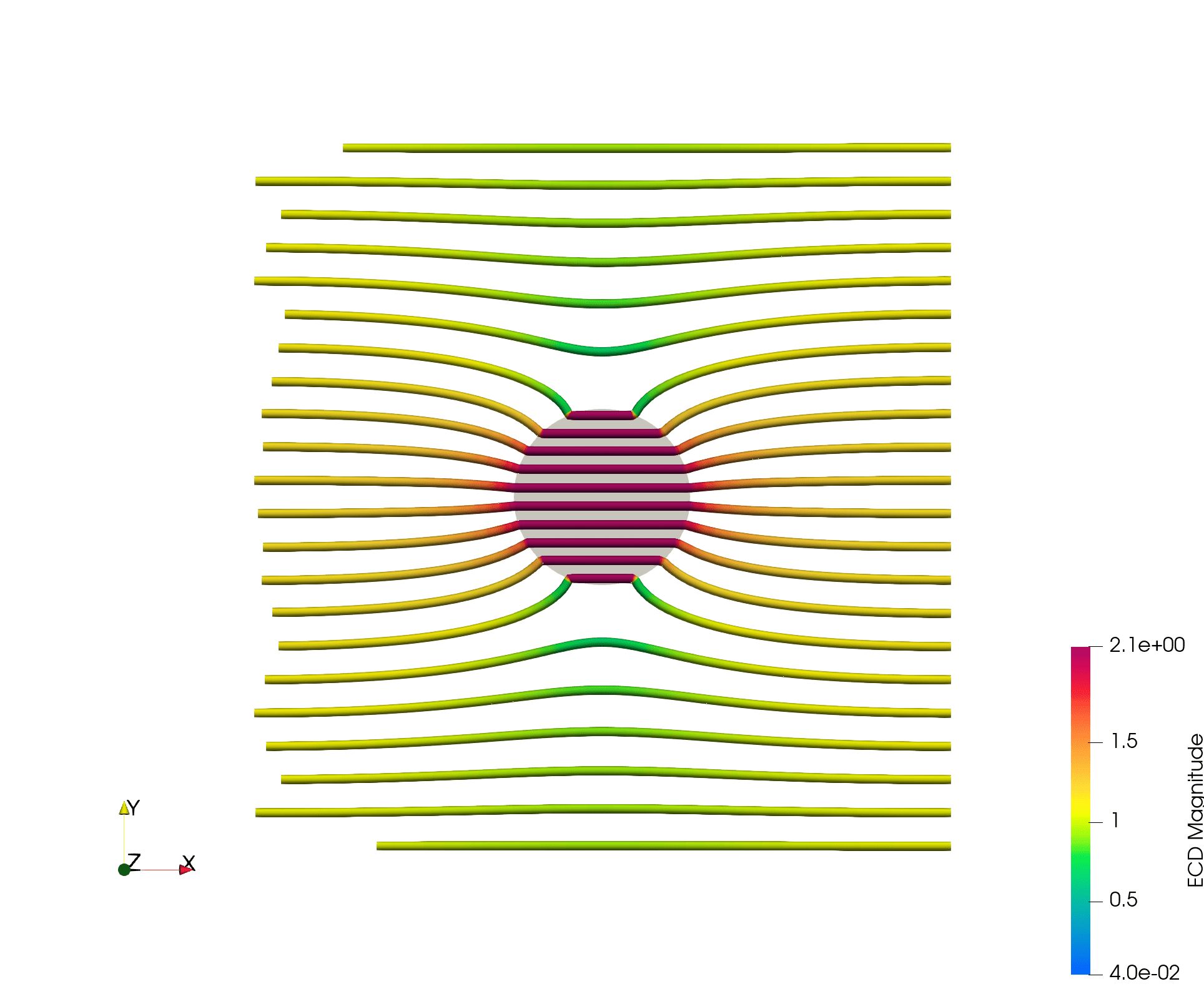}};
\end{tikzpicture}}}
\subfigure[]{\frame{\begin{tikzpicture}
    \node[inner sep=0pt] (B11) at (0,0)
    {\includegraphics[height=0.15\textwidth,
trim=400 233 964 233, clip]{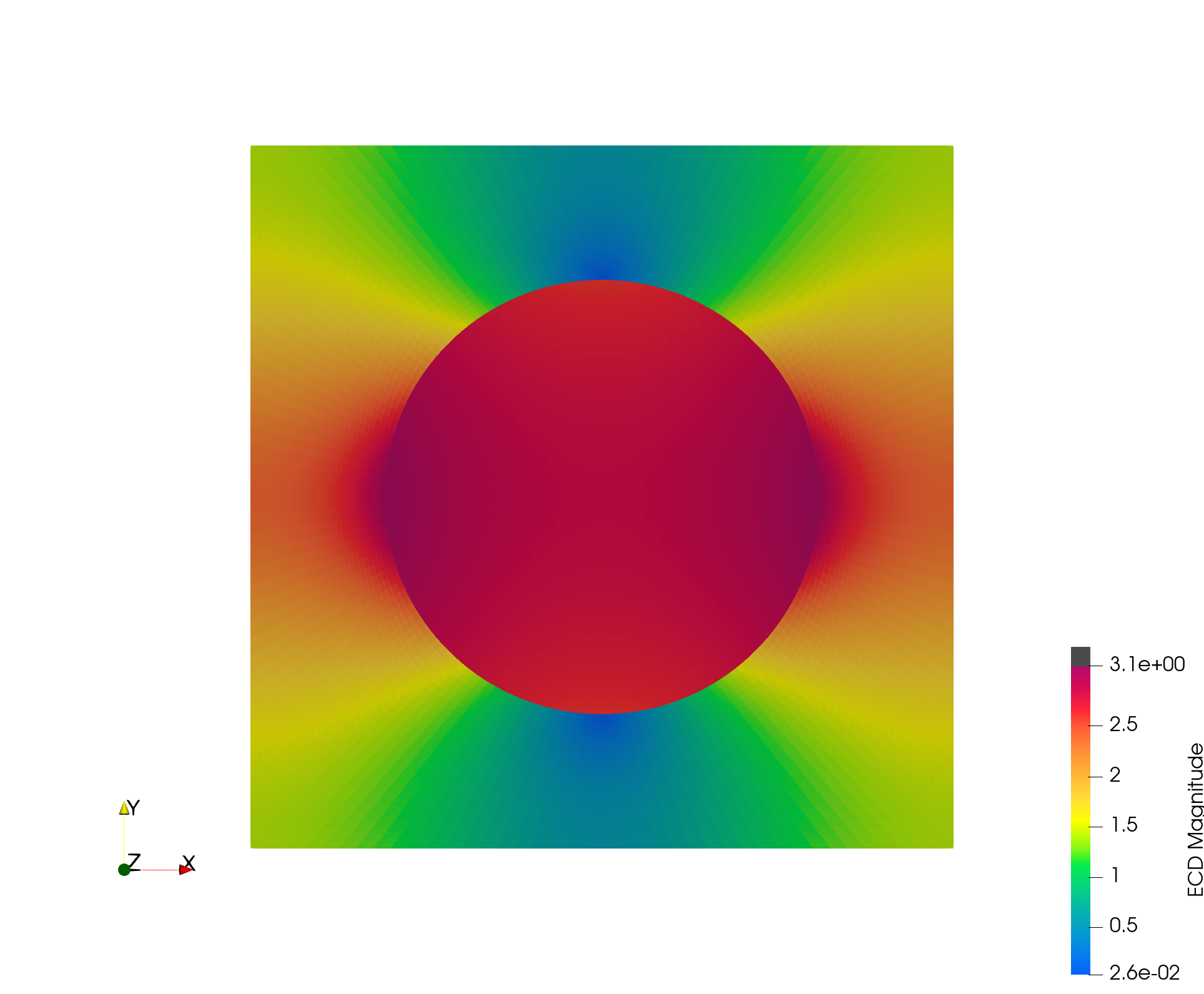}};
    \node[inner sep=0pt] (C1) at (1.24,0)
    {\includegraphics[height=0.15\textwidth,
trim=964 233 400 233, clip]{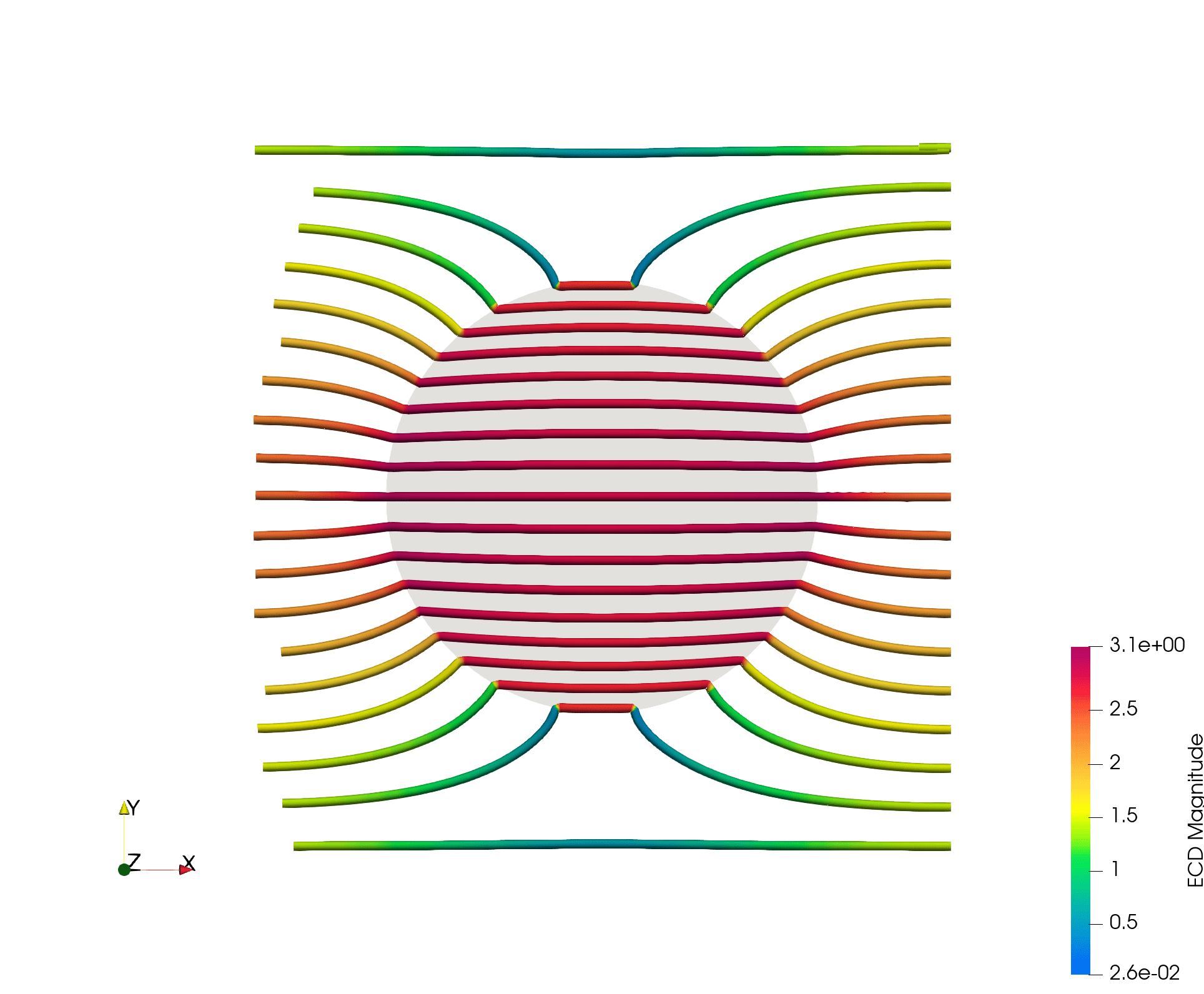}};
\end{tikzpicture}}}
\subfigure[]{\frame{\begin{tikzpicture}
    \node[inner sep=0pt] (B11) at (0,0)
    {\includegraphics[height=0.15\textwidth,
trim=400 233 964 233, clip]{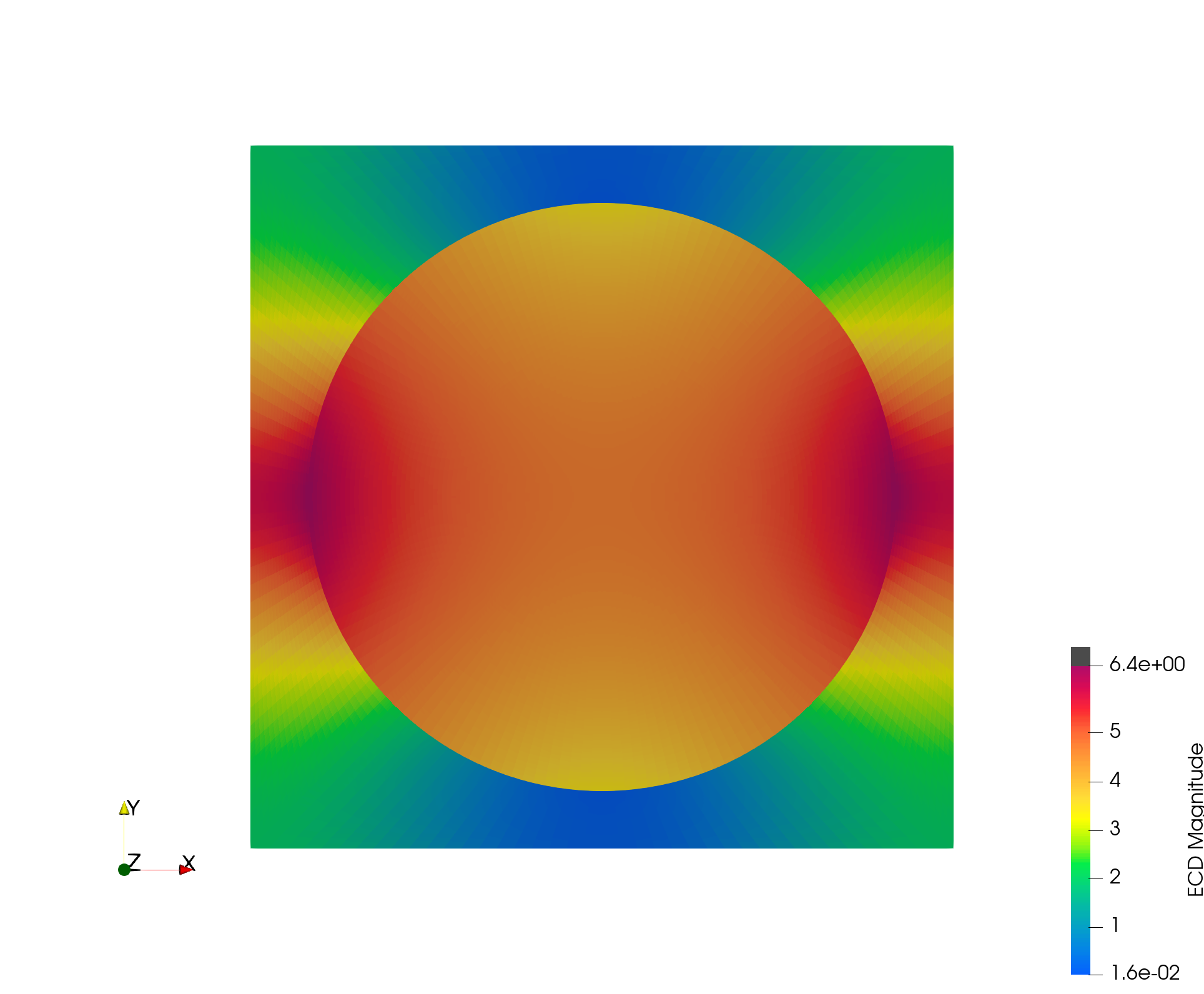}};
    \node[inner sep=0pt] (C1) at (1.24,0)
    {\includegraphics[height=0.15\textwidth,
trim=964 233 400 233, clip]{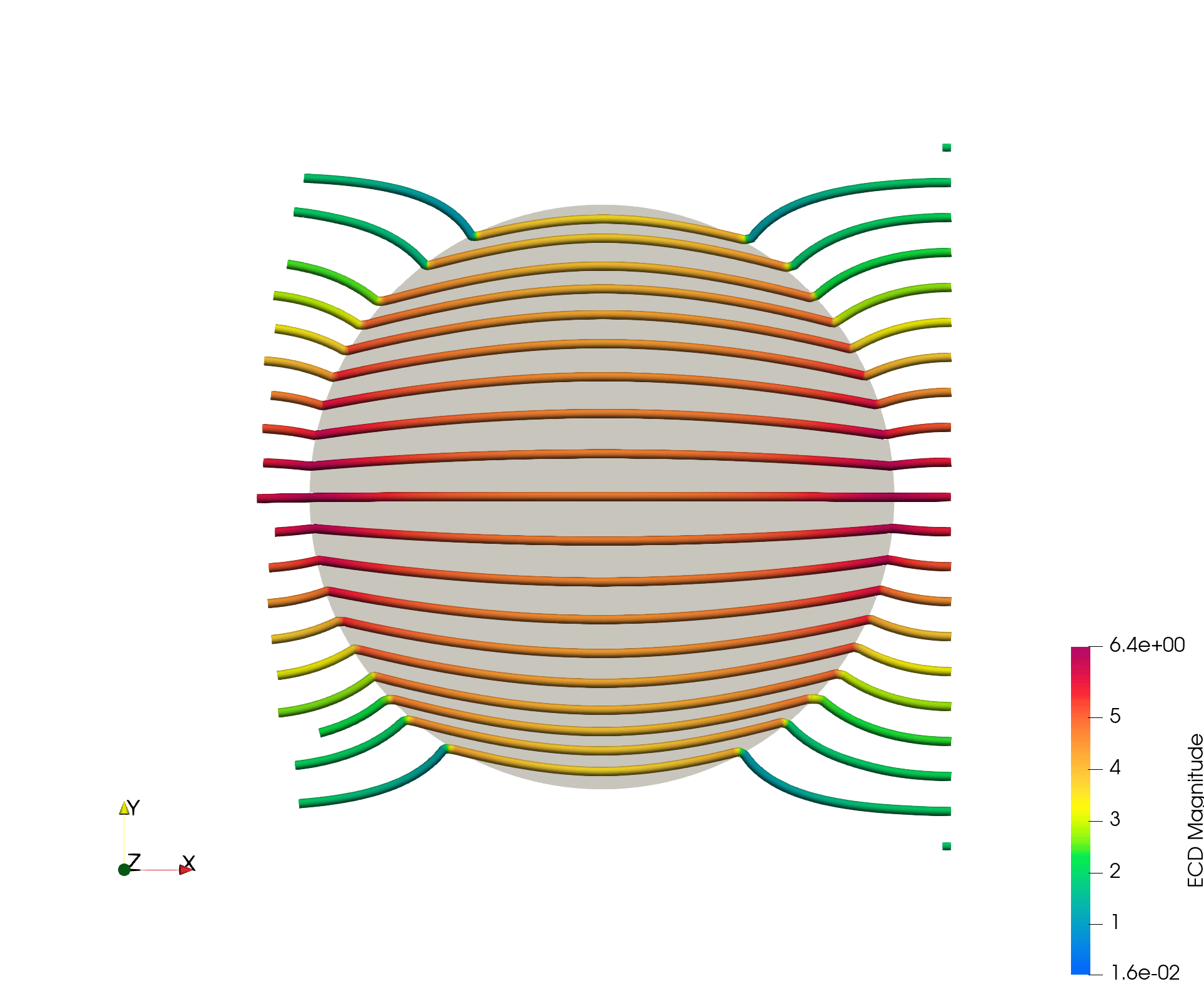}};
\end{tikzpicture}}}
\subfigure[]{\frame{\begin{tikzpicture}
    \node[inner sep=0pt] (B11) at (0,0)
    {\includegraphics[height=0.15\textwidth,
trim=400 233 964 233, clip]{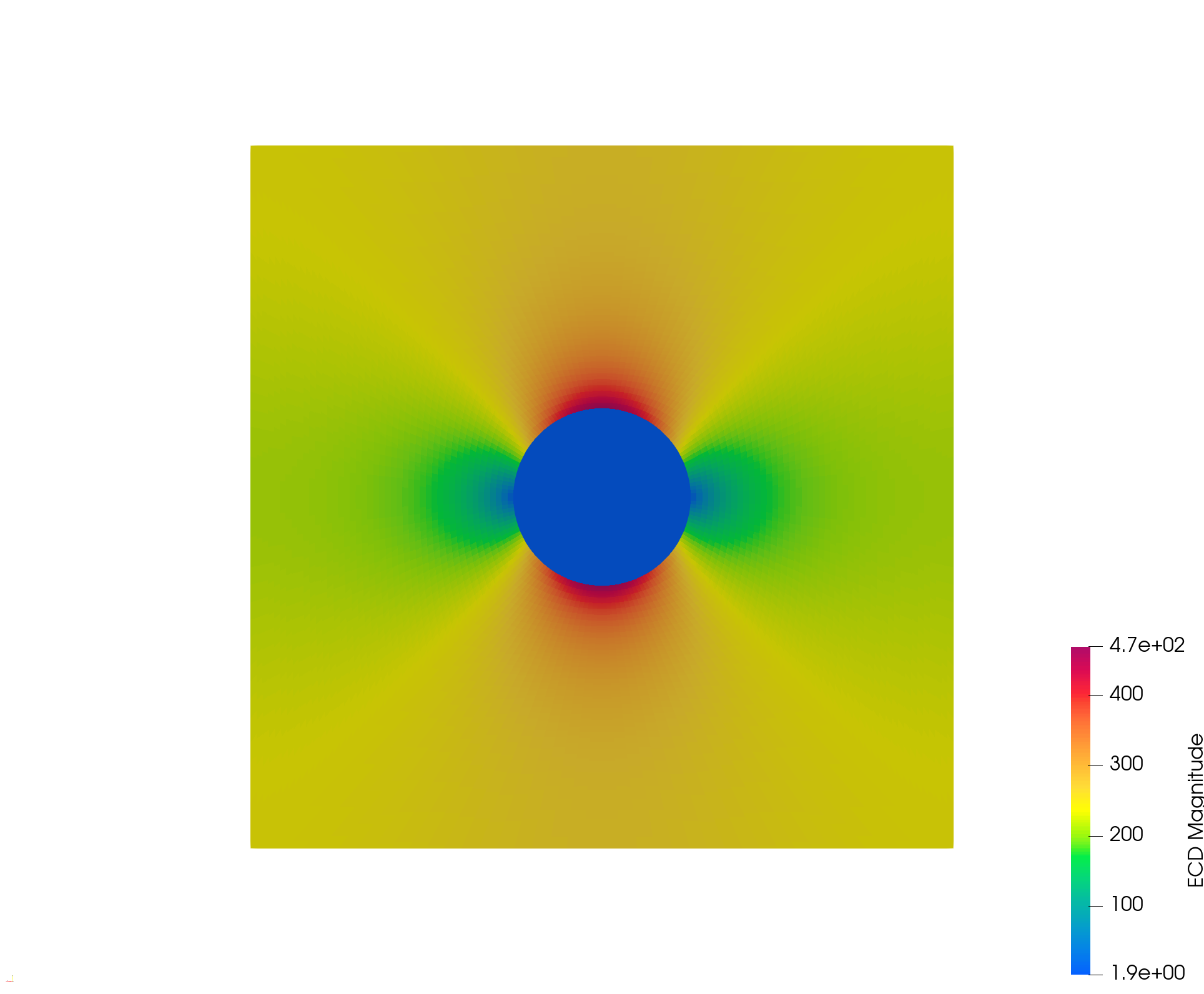}};
    \node[inner sep=0pt] (C1) at (1.24,0)
    {\includegraphics[height=0.15\textwidth,
trim=964 233 400 233, clip]{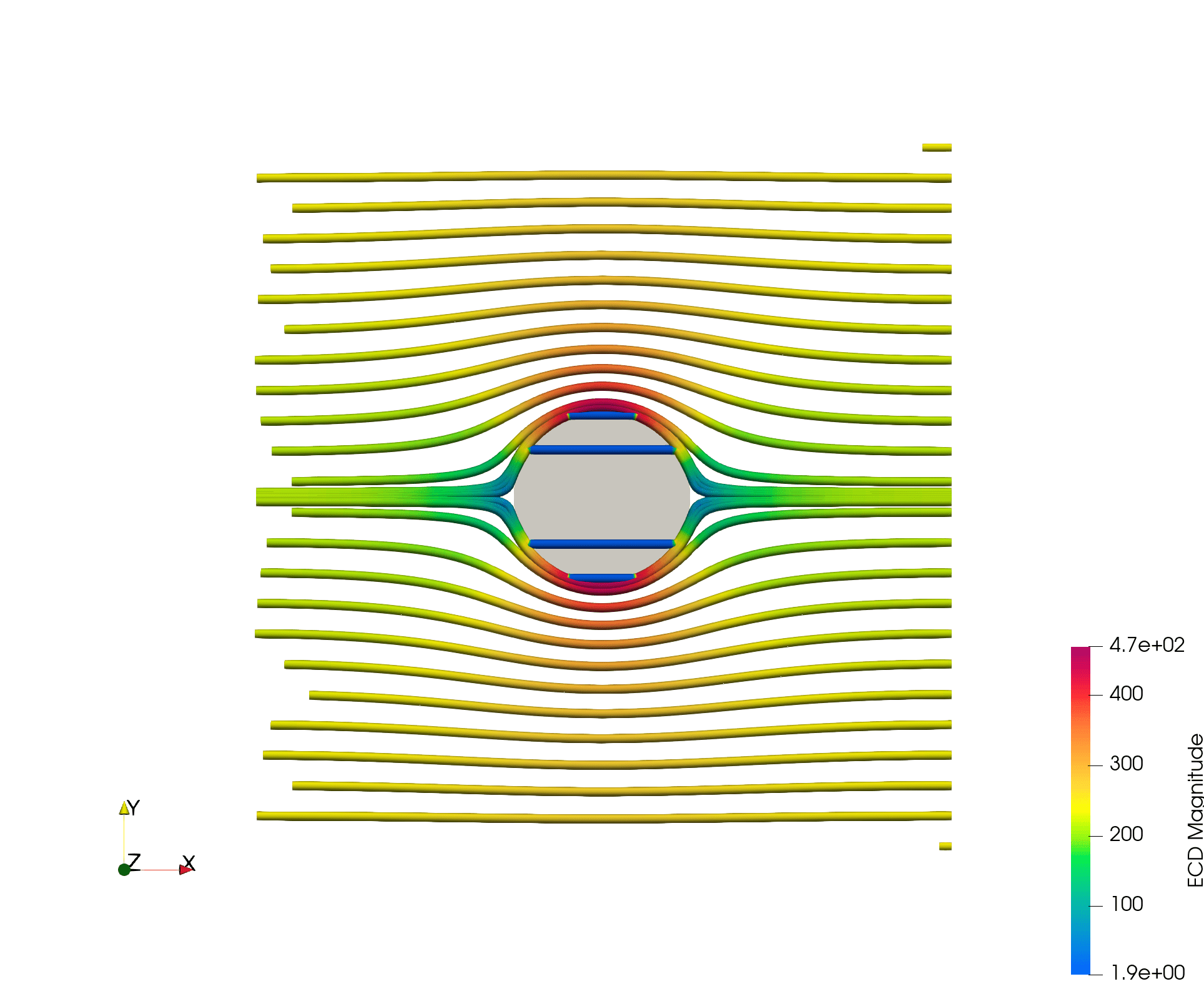}};
\end{tikzpicture}}}
\subfigure[]{\frame{\begin{tikzpicture}
    \node[inner sep=0pt] (B11) at (0,0)
    {\includegraphics[height=0.15\textwidth,
trim=400 233 964 233, clip]{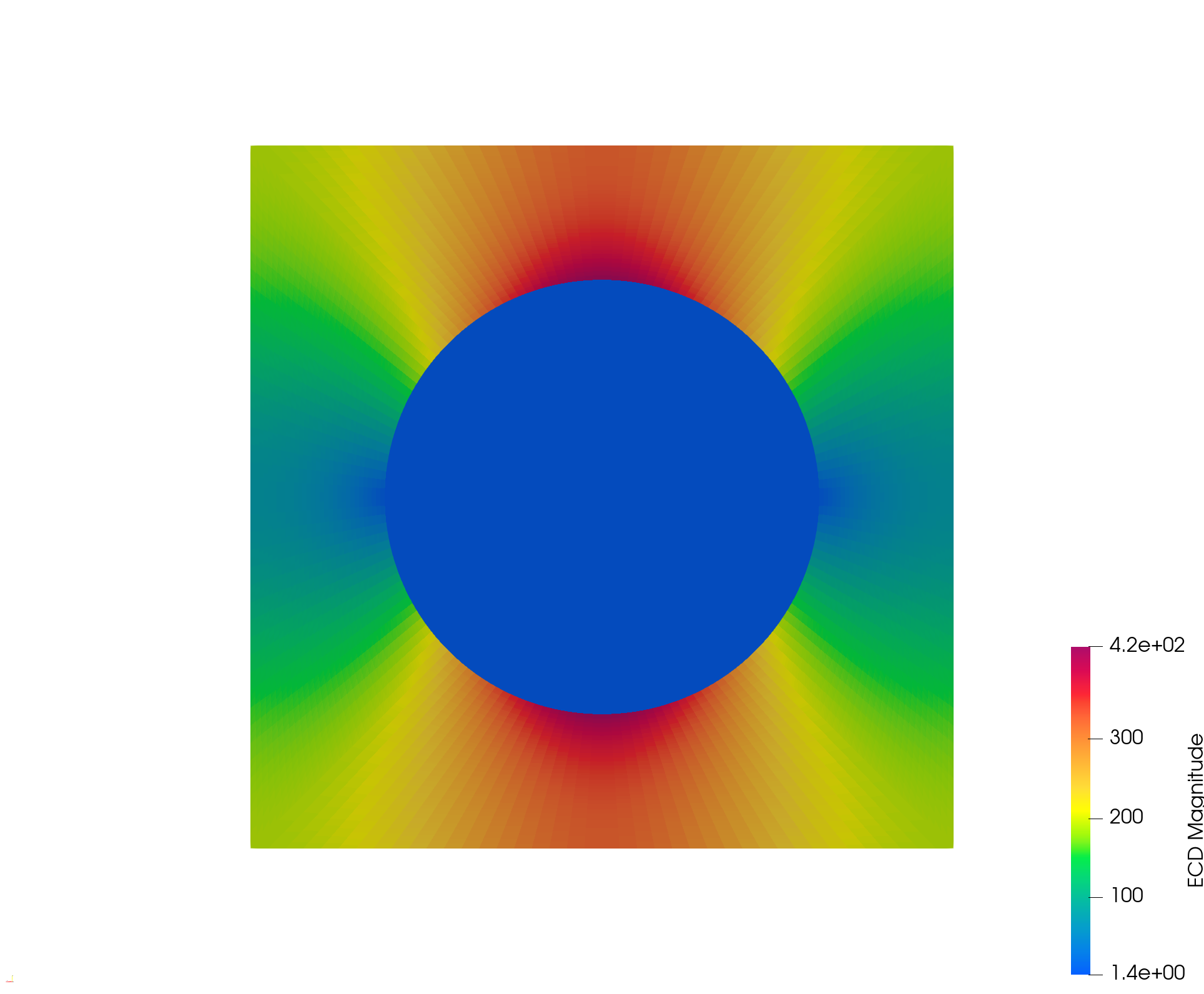}};
    \node[inner sep=0pt] (C1) at (1.24,0)
    {\includegraphics[height=0.15\textwidth,
trim=964 233 400 233, clip]{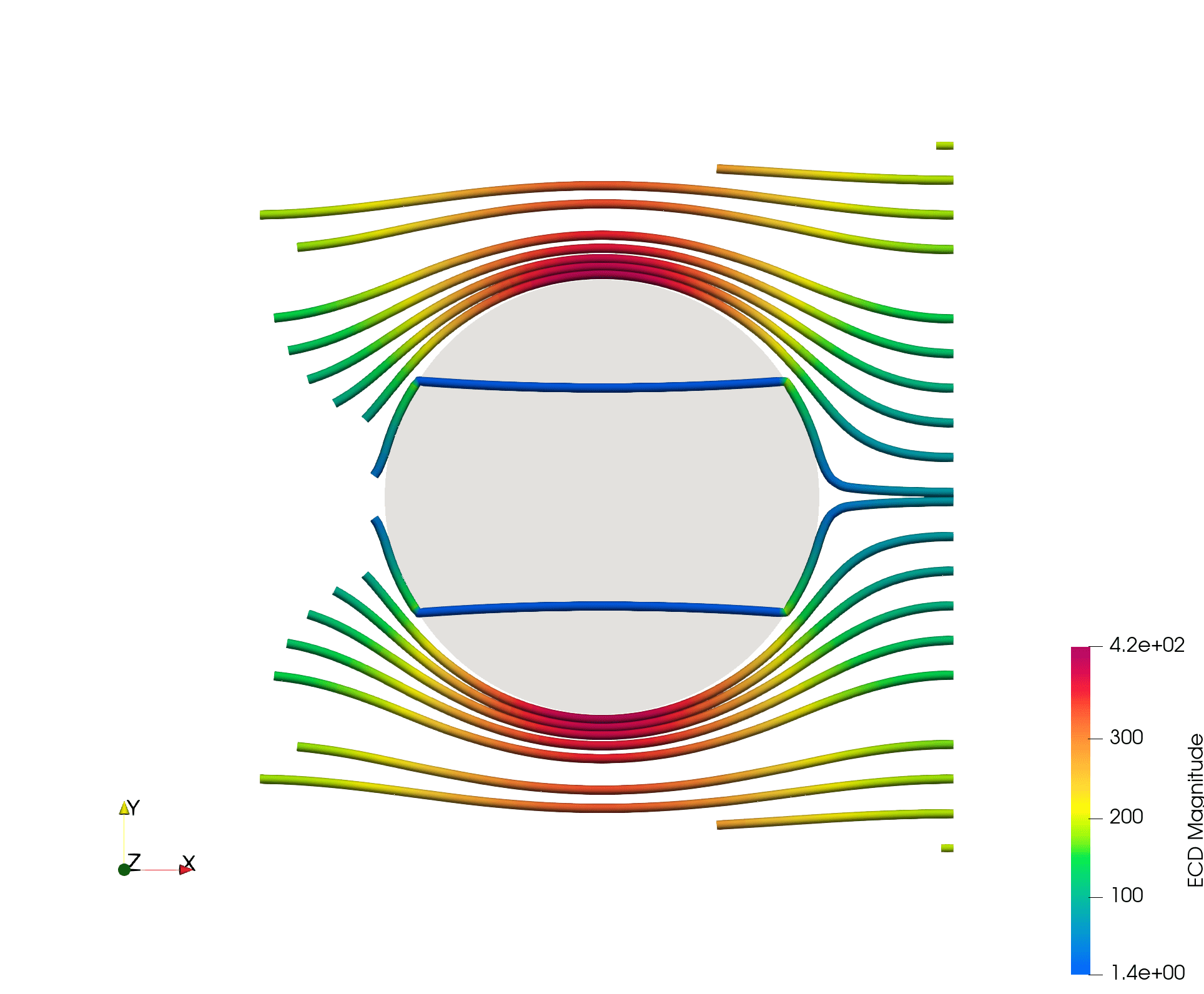}};
\end{tikzpicture}}}
\subfigure[]{\frame{\begin{tikzpicture}
    \node[inner sep=0pt] (B11) at (0,0)
    {\includegraphics[height=0.15\textwidth,
trim=400 233 964 233, clip]{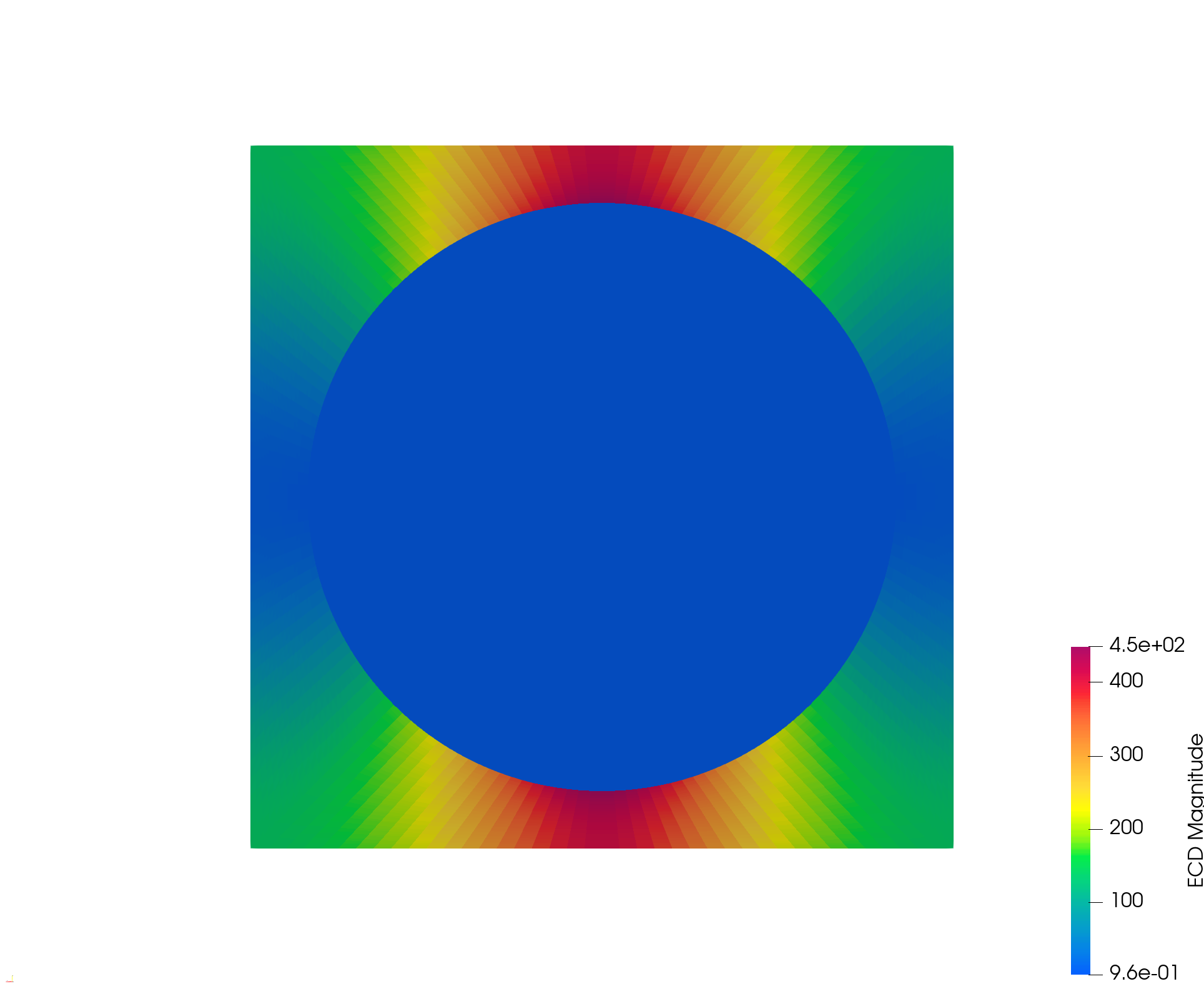}};
    \node[inner sep=0pt] (C1) at (1.24,0)
    {\includegraphics[height=0.15\textwidth,
trim=964 233 400 233, clip]{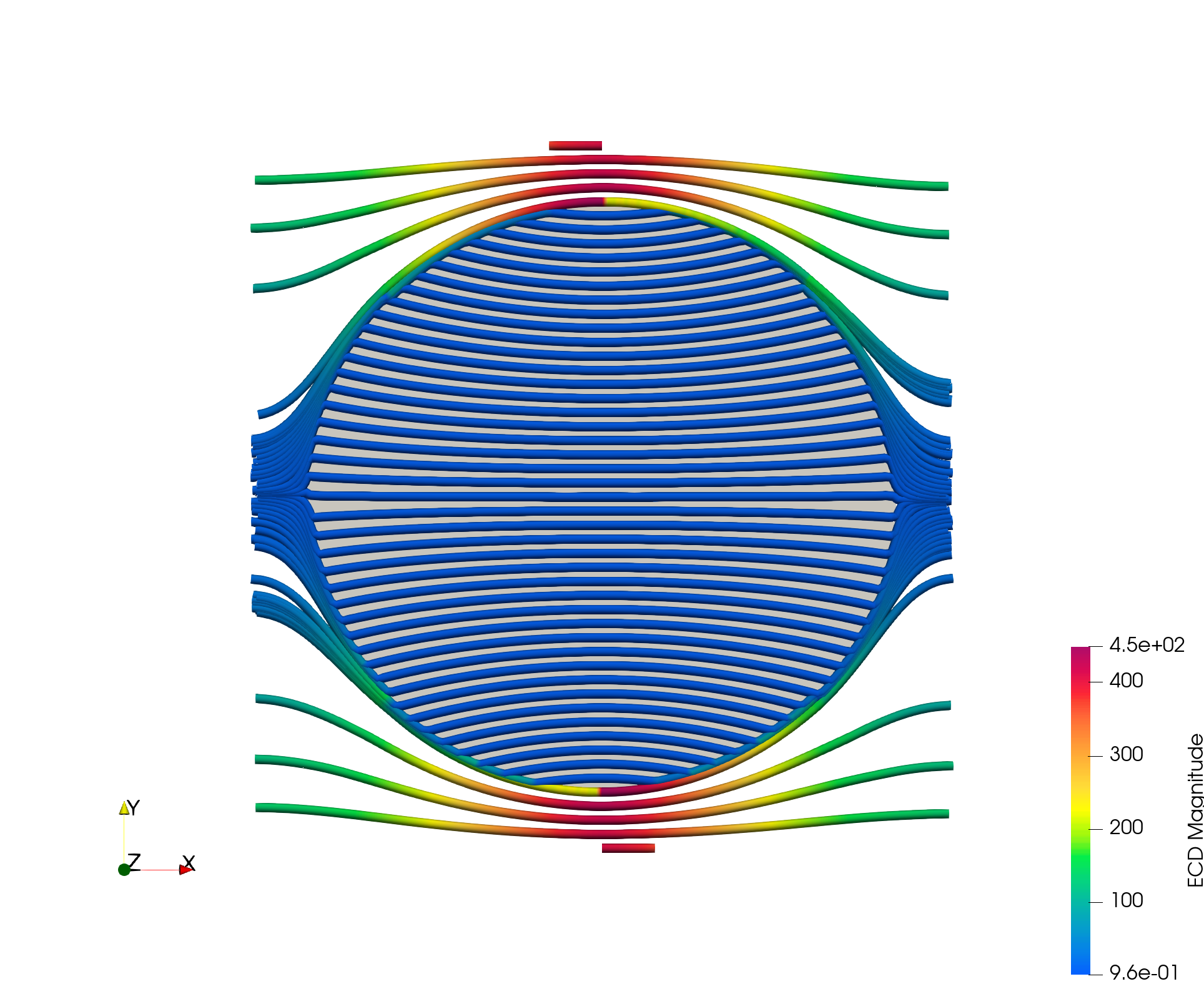}};
\end{tikzpicture}}}\\
    min$\,\,$\frame{\includegraphics[width=0.49\textwidth, trim=0 15 0 15, clip]{legend_2}}$\,\,$max
\caption{Magnetic flux vector norm contour plots and flux lines for the inclusion volume fractions of (a) and (d) $\phi_\mathrm{i}=0.05$, (b) and (e) $\phi_\mathrm{i}=0.30$ and (c) and (f) $\phi_\mathrm{i}=0.55$ for the horizontally applied macroscopic magnetic field with $\bs H=[1, 0]^\top$. The results correspond to $\mu_\mathrm{i}/\mu_\mathrm{e}=250/1$ for (a), (b), and (c) and to  $\mu_\mathrm{i}/\mu_\mathrm{e}=1/250$ for (d), (e) and (f).
Scalar potential formulation and periodic boundary conditions are used.
For demonstration purposes, the intervals [min, max] of each contour plot is individually determined to read (a)  $[0.040,2.101]$, (b)  $[0.026,3.094]$, (c) $[0.017,6.372]$, (d) $[1.893,469.1]$, (e)  $[1.408,416.8]$, (f)  $[0.958,448.4]$. These results are normalized with respect to $\mu_0$. For $\mu_\mathrm{i}/\mu_\mathrm{e}=250/1$ effective permeabilities of $1.104\mu_0$, $1.850\mu_0$ and $3.558\mu_0$ are predicted for inclusion volume fractions of $\phi_\mathrm{i}=0.05$, $0.30$, and $0.55$, respectively, with vector and scalar potential formulations with periodic boundary conditions. For $\mu_\mathrm{i}/\mu_\mathrm{e}=1/250$ these are $226.4\mu_0$, $135.1\mu_0$ and $70.30\mu_0$.}
\label{F:2D_circular_cylindrical_contour}
\end{figure*}

In understanding the influence of phase contrast and disk volume fraction on the effective magnetic properties of the composite, it is helpful to analyze the magnetic induction field intensities and lines.  Considering disk volume fractions of 0.05, 0.30, and 0.55,
Figure\ \ref{F:2D_circular_cylindrical_contour} demonstrates the results for the  contour plots for the norm of the magnetic field $\bs B$ as well as the associated flux lines for the horizontally applied macroscopic magnetic field with $\bs H=[1, 0]^\top$. Due to the microstructural symmetry, identical results are carried out for $\bs H=[0, 1]^\top$, however, with a $\pi/2$ in-plane rotation. As shown, the field lines are attracted toward the higher permeability material. For (a)-(c), the disconnected inclusions constitute the high permeability phase, whereas for (d)-(f), it is the continuous matrix. This explains the observed intensity difference in the flux fields.  For disk volume fraction of 0.30, the magnetic flux field in the neighborhood of circular disks is comparable to Stratton's electric field solution  for dilute high permittivity inclusion-low permittivity matrix systems \cite{Beran1968,Stratton1941}.

\subsubsection{Random Disk Arrangements}
\begin{figure}[htb!]
    \centering
% trim=left bottom right top, clip
    \subfigure[]{\begin{tikzpicture}
    \node[inner sep=0pt] (B11) at (0,0){\includegraphics[height=0.37\textwidth]{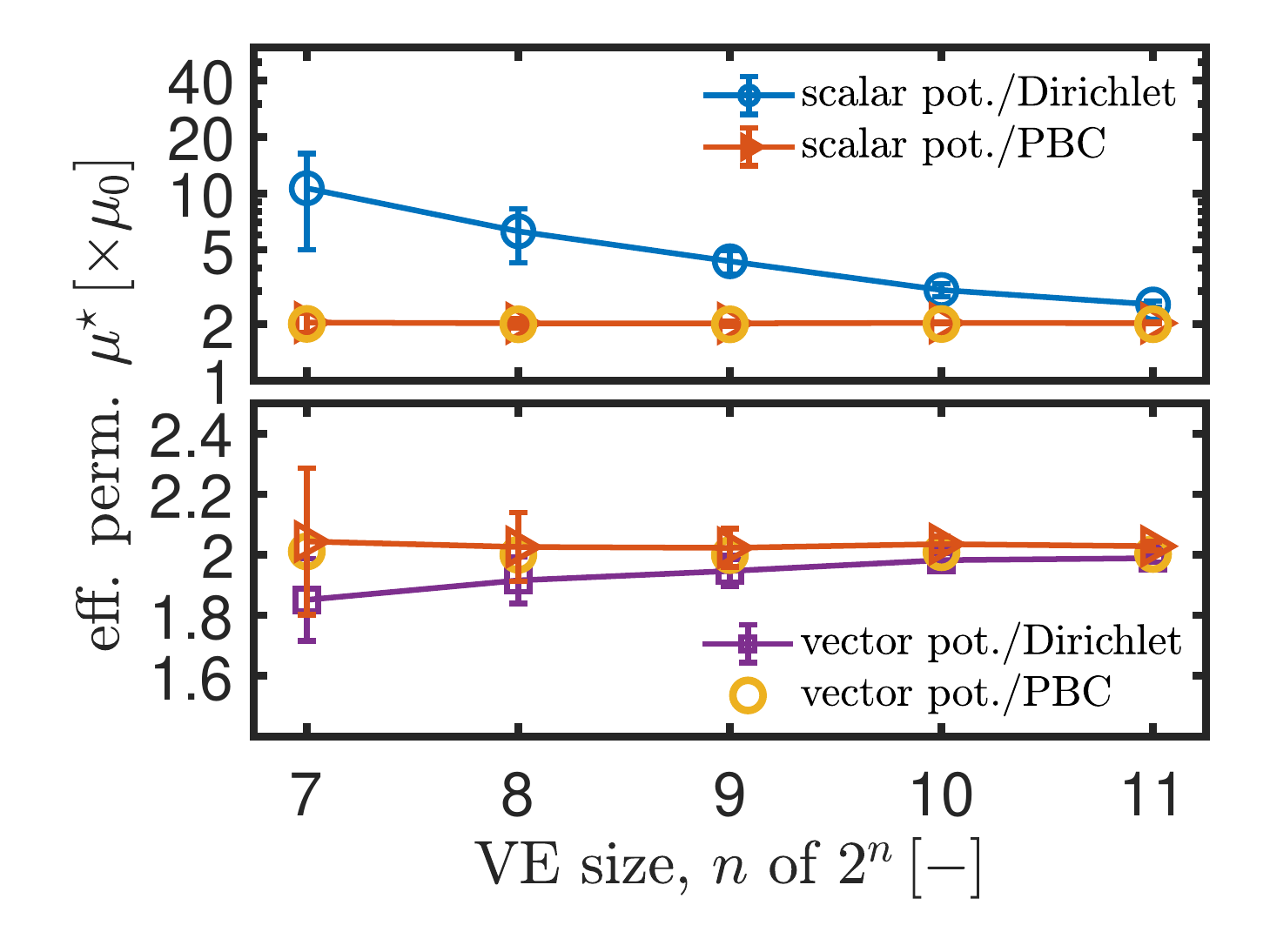}};
    \node (A) at (0.0,3.0) {\scriptsize nonoverlapping, $\mu_\mathrm{i}/\mu_\mathrm{m}=250/1$};
    \end{tikzpicture}}
    \subfigure[]{\begin{tikzpicture}
    \node[inner sep=0pt] (B11) at (0,0){\includegraphics[height=0.37\textwidth]{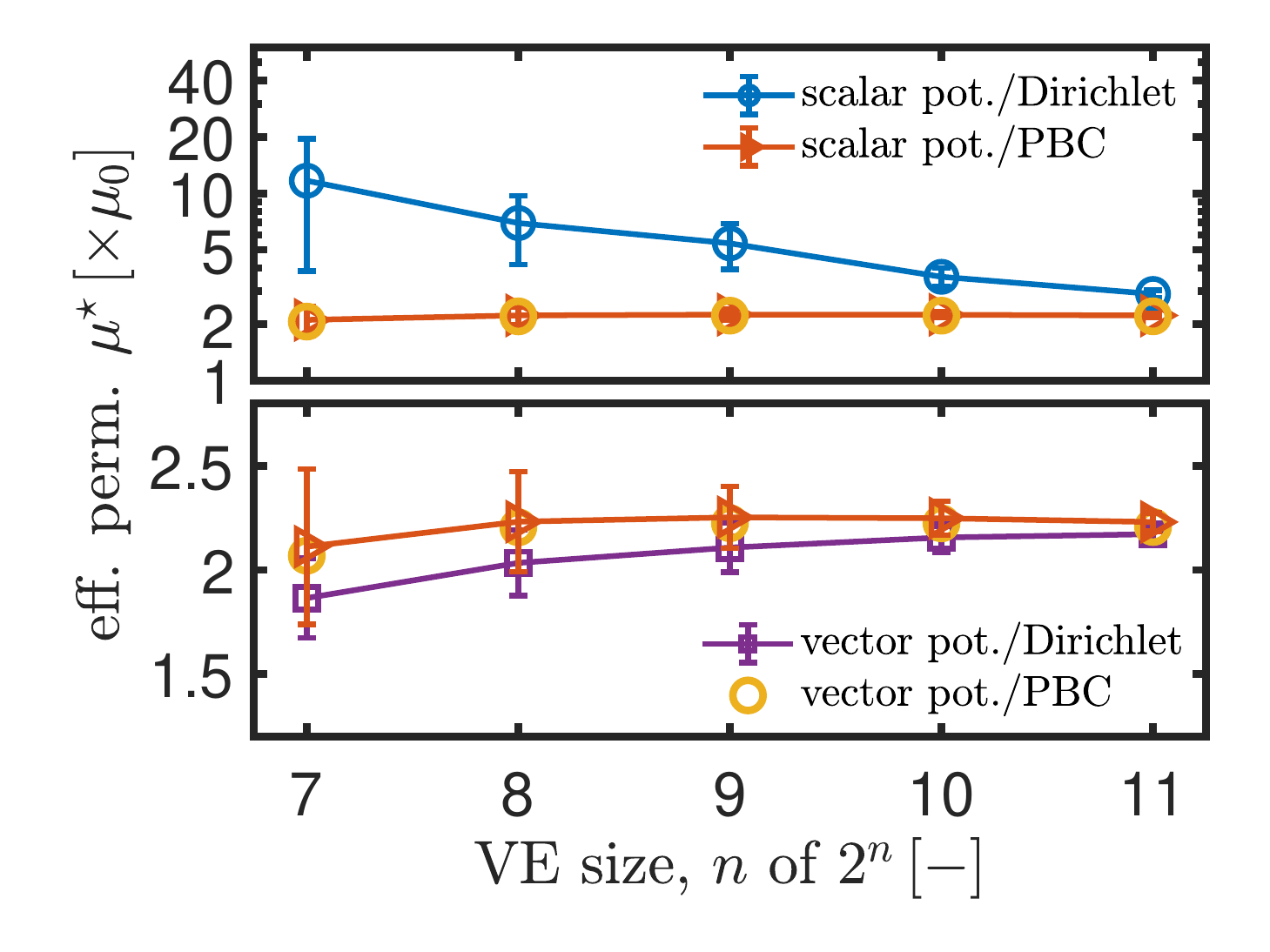}};
    \node (A) at (0.0,3.0) {\scriptsize overlapping, $\mu_\mathrm{i}/\mu_\mathrm{m}=250/1$};
    \end{tikzpicture}}
    \subfigure[]{\begin{tikzpicture}
    \node[inner sep=0pt] (B11) at (0,0){\includegraphics[height=0.37\textwidth]{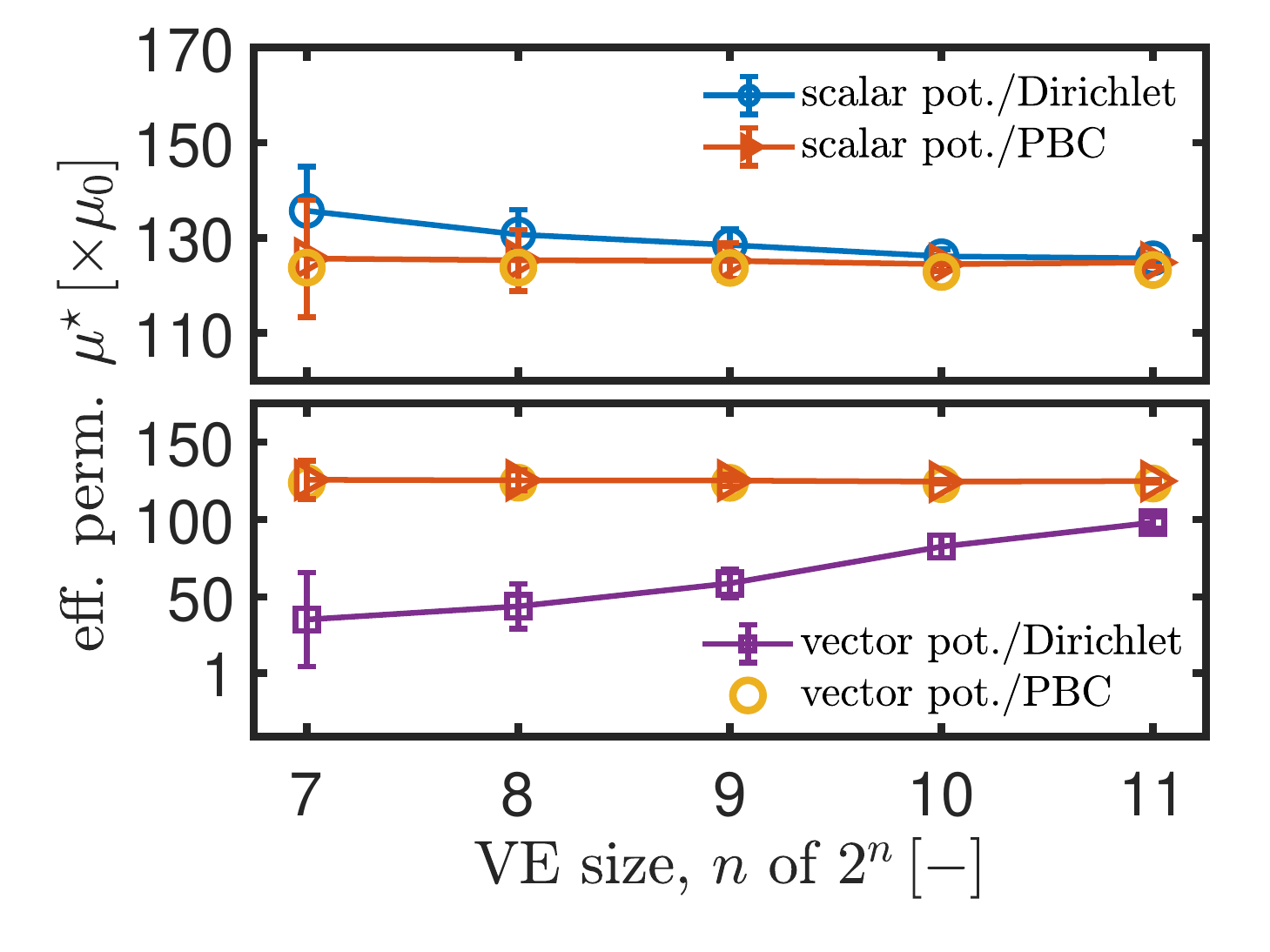}};
    \node (A) at (0.0,3.0) {\scriptsize nonoverlapping, $\mu_\mathrm{i}/\mu_\mathrm{m}=1/250$};
    \end{tikzpicture}}
    \subfigure[]{\begin{tikzpicture}
    \node[inner sep=0pt] (B11) at (0,0){\includegraphics[height=0.37\textwidth]{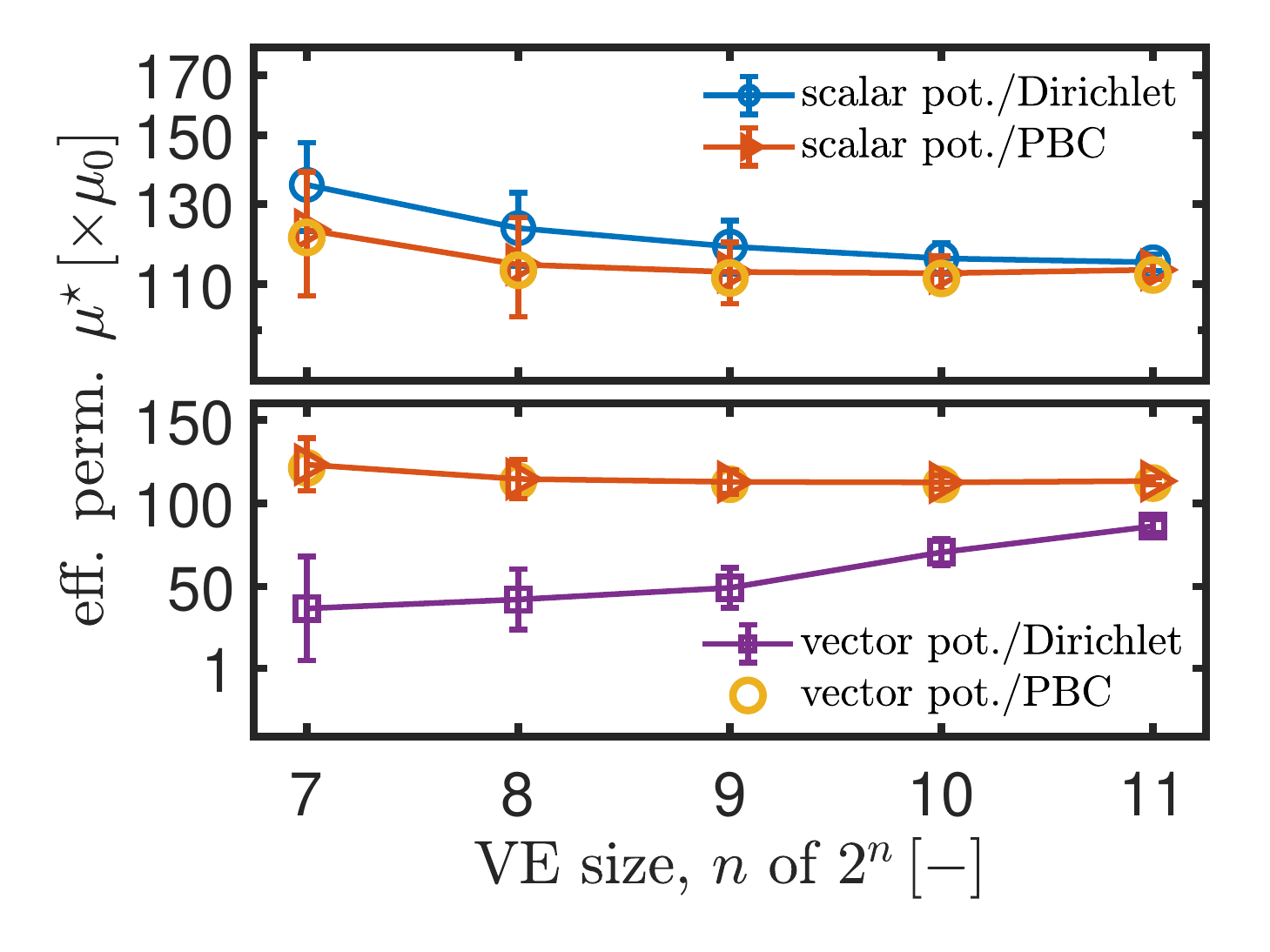}};
    \node (A) at (0.0,3.0) {\scriptsize overlapping, $\mu_\mathrm{i}/\mu_\mathrm{m}=1/250$};
    \end{tikzpicture}}
\caption{Effective permeability $\mu^\star$ predictions for random disk arrangements with  the inclusion volume fraction of $\phi_\mathrm{i}=0.30$ for various VE sizes  represented in terms of the powers $n$ of $2$. Thus, $n=7,8,\ldots,11$ corresponds to a VE size of $128,256,\ldots,2048$ pixels. The given predictions are for (a) and (b) $\mu_\mathrm{i}/\mu_\mathrm{e}=250/1$ and (c) and (d) $\mu_\mathrm{i}/\mu_\mathrm{e}=1/250$. In (a) and (c), nonoverlapping disk system results are given, whereas in (b) and (d), overlapping disks are considered. The results for vector and scalar potential formulations with uniform Dirichlet and periodic boundary conditions are depicted.
The upper (Voigt) and lower (Reuss) bounds correspond to  $\mu_\mathrm{V}^\star\simeq75.70\mu_0$ and $\mu_\mathrm{R}^\star\simeq1.426\mu_0$, for the phase contrast  $\mu_\mathrm{i}/\mu_\mathrm{m}=250/1$, respectively, and  $\mu_\mathrm{V}^\star\simeq175.3\mu_0$ and $\mu_\mathrm{R}^\star\simeq3.303\mu_0$, for the phase-interchanged case. Data are presented as mean value $\upmu$ (markers) and standard deviation $\upsigma$ from the analysis of 15 random realizations. For each volume fraction corresponding standard error of the mean $\mathrm{SEM}$  can be computed with $\mathrm{SEM}=\upsigma/\sqrt{15}$. Lines merge discrete computational results.}
\label{F:VE_size_on_mu}
\end{figure}
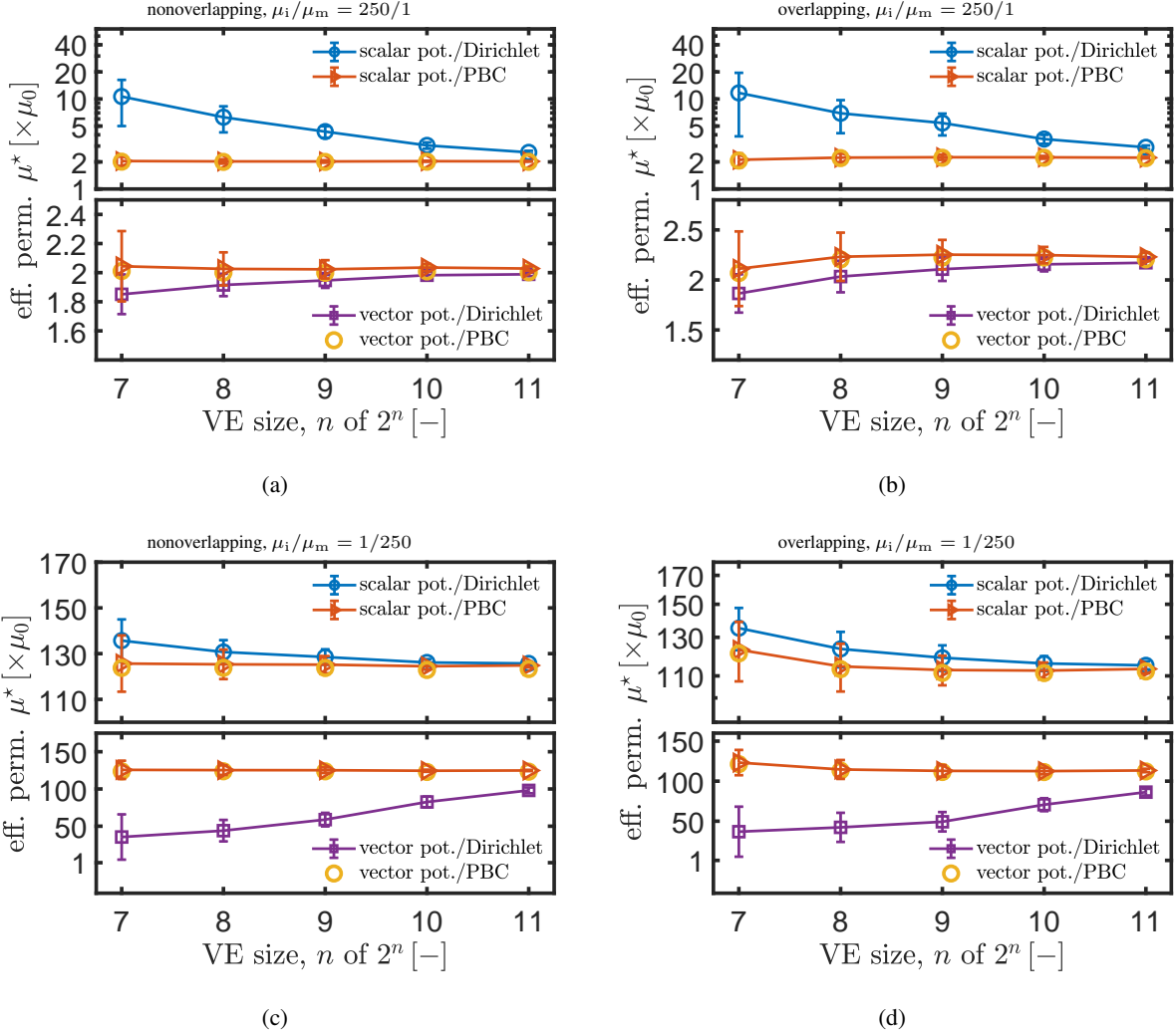

Figure\ \ref{F:VE_size_on_mu} demonstrates effective permeability $\mu^\star$ predictions as a function of the VE size considering both scalar and vector potential formulations as well as nonoverlapping and overlapping random disk arrangements with 0.30 inclusion volume fraction. Here the sizes of the pixelated VEs are represented in edge sizes in pixels for $\{128,256,\ldots,2048\}$, which are given as powers of two. Phase contrasts of $\mu_\mathrm{i}/\mu_\mathrm{e}=250/1$ and   $\mu_\mathrm{i}/\mu_\mathrm{e}=1/250$ with phase-interchange are considered. For the periodic boundary conditions, the observed size invariance concerning the selected VE size is not the case in contrast to the regular disk arrangements. This is due to the element of randomness.
Still, the relative difference between the periodic boundary condition-based predictions with vector and scalar potential formulations is below 1.3\% for cases without and with phase interchange. A convergence pattern towards the corresponding effective permeability is observed in the average predictions (ensemble average) for all vector and scalar potential formulations and boundary conditions with increasing VE size. While the mean values are converging, corresponding magnitudes of the standard deviations hence the standard errors of the mean, decrease. The convergence is strictly monotonic for Dirichlet and Neumann boundary conditions. The corresponding trends follow those indicated for size dependence  investigations for the periodic regular square disk arrangements. This time, random boundary property fluctuations are responsible for the observed trends. The fastest convergence is observed in the case of periodic boundary conditions.

As a result of these computations, the effective permeability of the nonoverlapping random disk arrangements for 0.30 disk volume fraction is computed as $[2.028\pm0.014]\mu_0$ and $[124.8\pm0.8]\mu_0$ for $\mu_\mathrm{i}/\mu_\mathrm{e}=250/1$ and $\mu_\mathrm{i}/\mu_\mathrm{e}=250/1$, respectively. These are respectively $[2.230\pm0.043]\mu_0$ and $[113.4\pm2.1]\mu_0$ for the overlapping disks case. Although there is relatively low disk interaction at low volume fractions, the effective permeability prediction for the overlapping disk systems is higher for $\mu_\mathrm{i}/\mu_\mathrm{e}=250/1$. As before, the computational results  are bounded by the upper (Voigt) and lower (Reuss) bounds which, considering the current material and geometrical properties, are  $\mu_\mathrm{V}^\star\simeq75.70\mu_0$ and $\mu_\mathrm{R}^\star\simeq1.426\mu_0$, respectively. With phase-interchange giving  $\mu_\mathrm{i}/\mu_\mathrm{m}=250/1$ these read $\mu_\mathrm{V}^\star\simeq175.3\mu_0$ and $\mu_\mathrm{R}^\star\simeq3.303\mu_0$.

\begin{figure}[htb!]
    \centering
% trim=left bottom right top, clip
    \subfigure[]{\begin{tikzpicture}
    \node[inner sep=0pt] (B11) at (0,0){\includegraphics[height=0.37\textwidth]{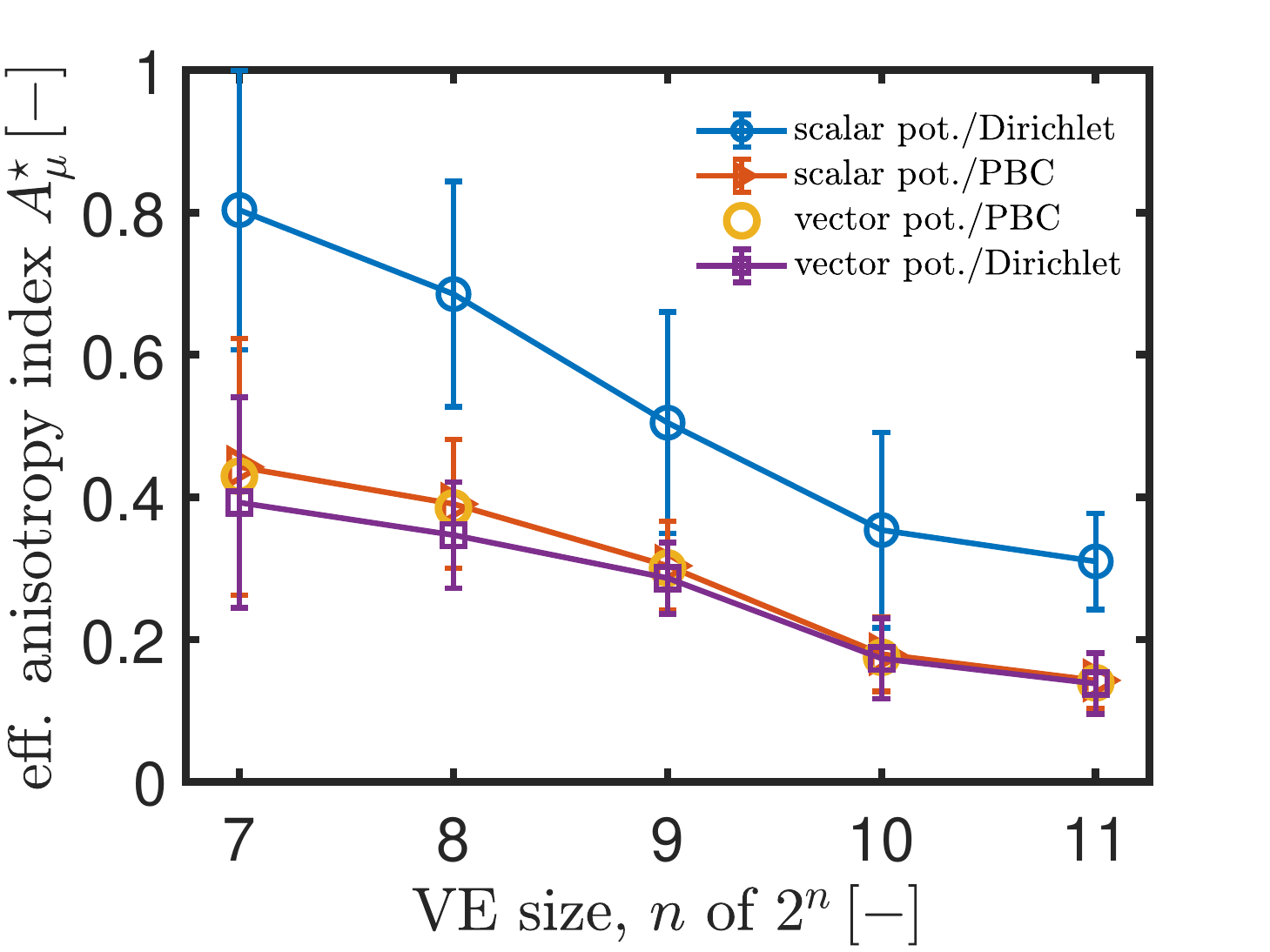}};
    \node (A) at (0.0,3.0) {\scriptsize nonoverlapping, $\mu_\mathrm{i}/\mu_\mathrm{m}=250/1$};
    \end{tikzpicture}}
    \subfigure[]{\begin{tikzpicture}
    \node[inner sep=0pt] (B11) at (0,0){\includegraphics[height=0.37\textwidth]{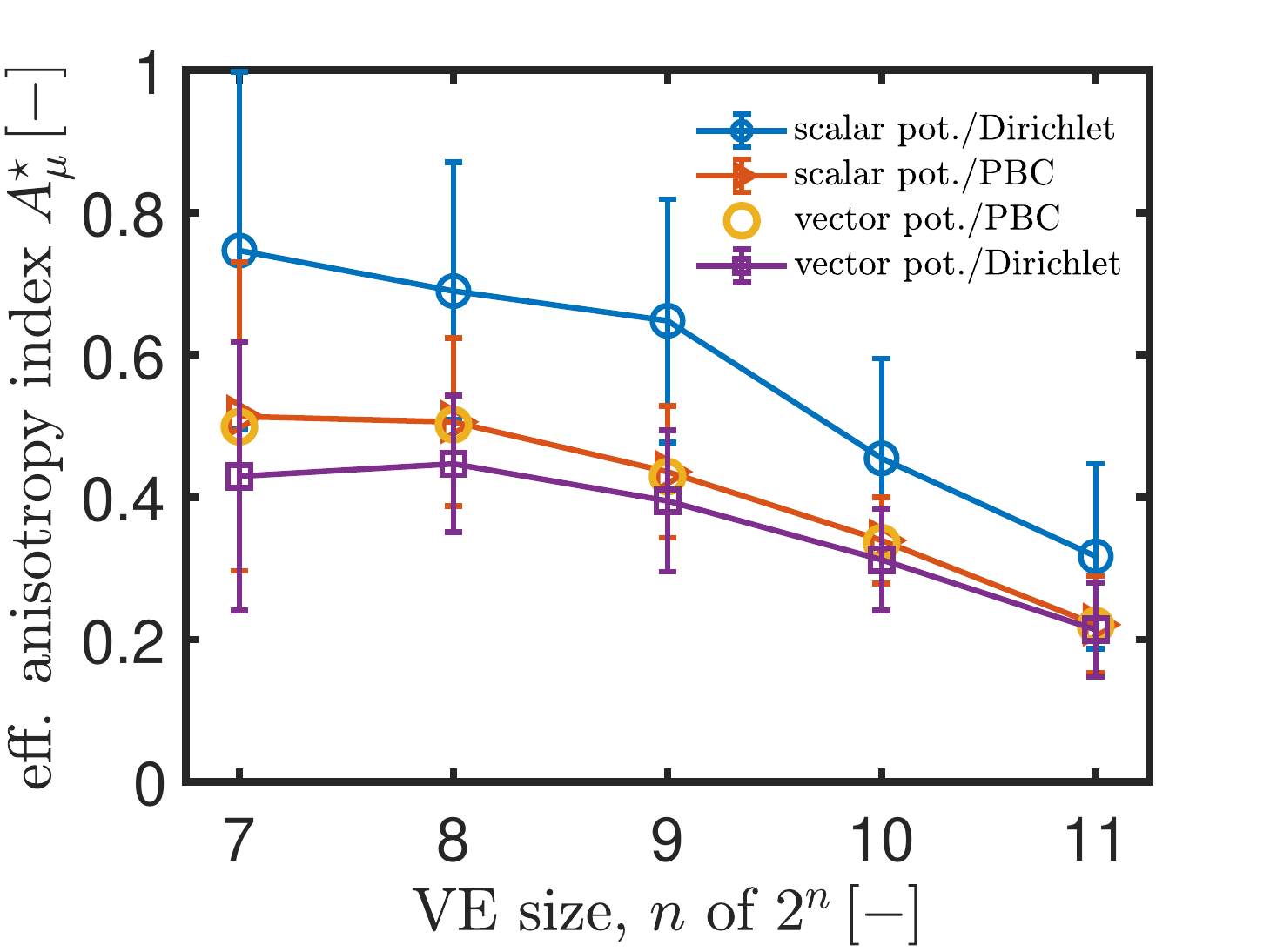}};
    \node (A) at (0.0,3.0) {\scriptsize overlapping, $\mu_\mathrm{i}/\mu_\mathrm{m}=250/1$};
    \end{tikzpicture}}
    \subfigure[]{\begin{tikzpicture}
    \node[inner sep=0pt] (B11) at (0,0){\includegraphics[height=0.37\textwidth]{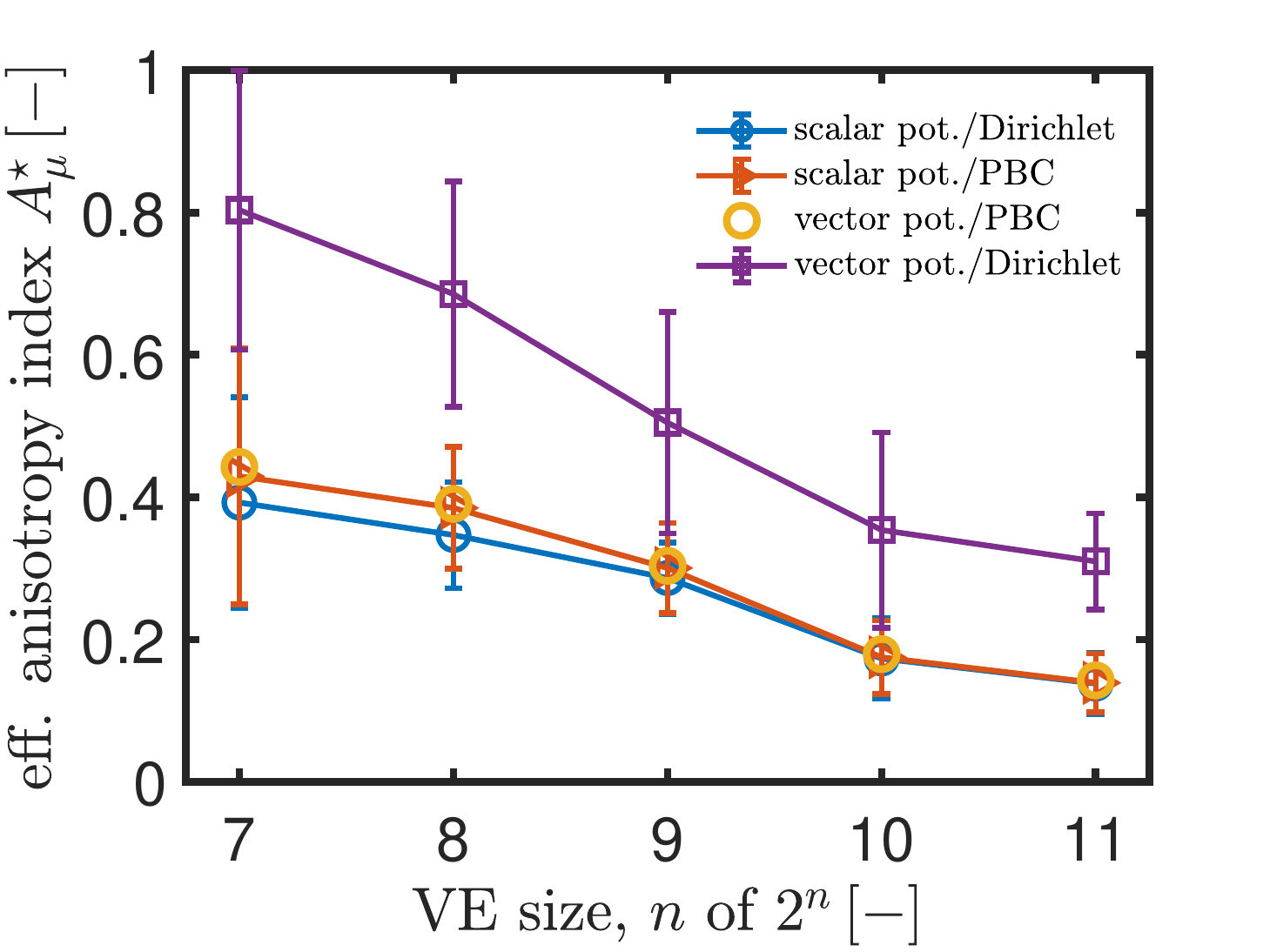}};
    \node (A) at (0.0,3.0) {\scriptsize nonoverlapping, $\mu_\mathrm{i}/\mu_\mathrm{m}=1/250$};
    \end{tikzpicture}}
    \subfigure[]{\begin{tikzpicture}
    \node[inner sep=0pt] (B11) at (0,0){\includegraphics[height=0.37\textwidth]{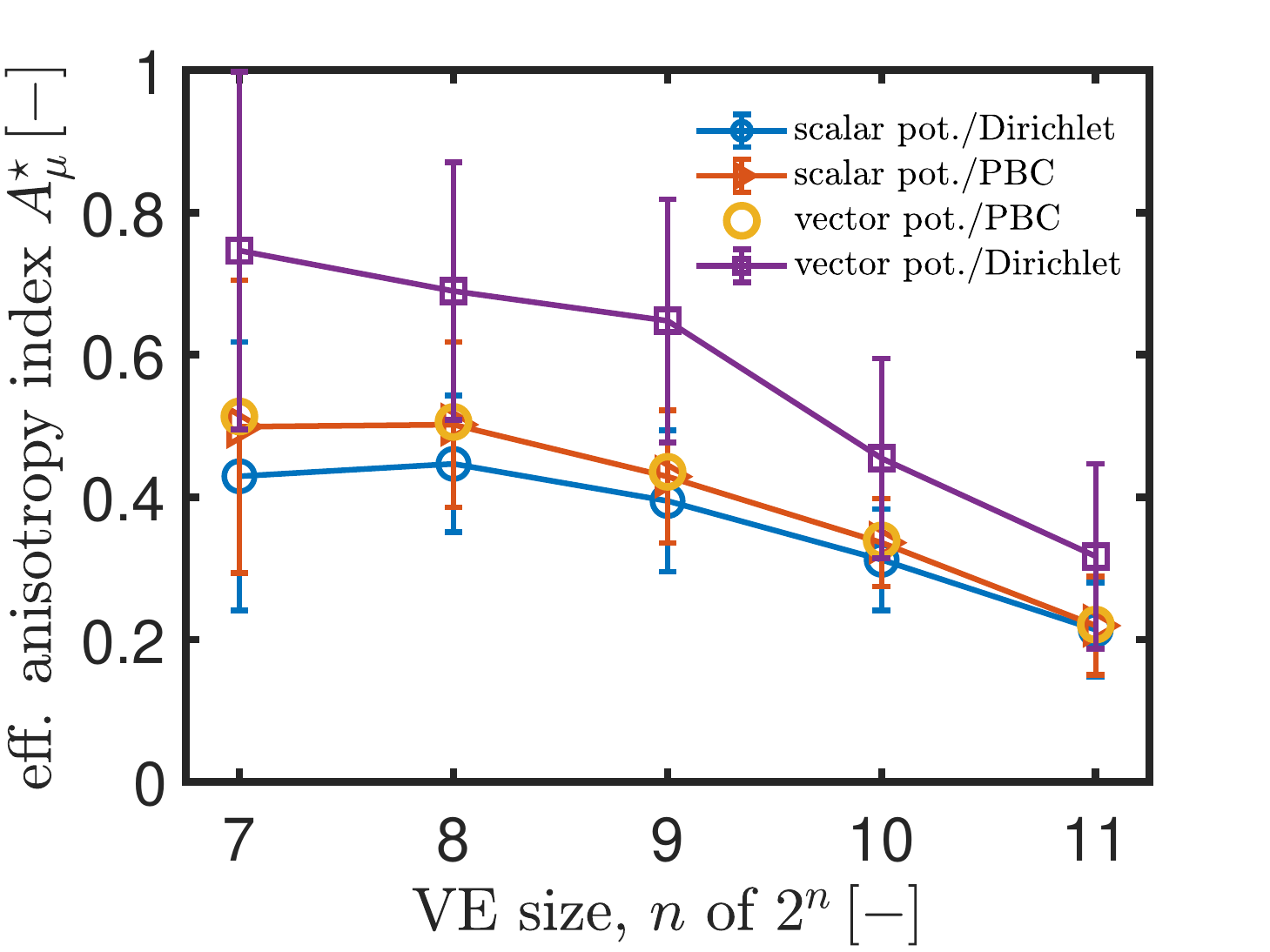}};
    \node (A) at (0.0,3.0) {\scriptsize overlapping, $\mu_\mathrm{i}/\mu_\mathrm{m}=1/250$};
    \end{tikzpicture}}
\caption{The effective anisotropy index $0\leq A_\mu^\star< 1$ predictions for the permeability tensor  for random disk arrangements with the  inclusion volume fraction of $\phi_\mathrm{i}=0.30$ for various VE sizes  represented in terms of the powers $n$ of $2$. Thus, $n=7,8,\ldots,11$ corresponds to a VE size of $128,256,\ldots,2048$ pixels. The given predictions are for (a) and (b) $\mu_\mathrm{i}/\mu_\mathrm{e}=250/1$ and (c) and (d) $\mu_\mathrm{i}/\mu_\mathrm{e}=1/250$. In (a) and (c), nonoverlapping disk system results are given, whereas in (b) and (d), overlapping disks are considered. The anisotropy index gives the eccentricity $e$ of the $2\times2$ matrix representation of the symmetric effective permeability tensor such that $A_\mu^\star=\sqrt{1-\mu^\star_2/\mu^\star_1}$ where $\mu^\star_i$ for $i=1,2$ are the ordered eigenvalues of the effective permeability tensor. Data are presented as mean value $\upmu$ (markers) and standard deviation $\upsigma$ from the analysis of 15 random realizations. For each volume fraction corresponding standard error of the mean $\mathrm{SEM}$  can be computed with $\mathrm{SEM}=\upsigma/\sqrt{15}$. Lines merge discrete computational results.}
\label{F:VE_size_on_Amu}
\end{figure}

For the VE size of 2048 pixels ($2^{11}$), the (symmetric) off-diagonal terms of the magnetic permeability tensor are generally two to three orders of magnitude smaller than the normal components, whereas, for smaller VE sizes, their relative magnitude may be more significant. However,  in-plane isotropy in the physical properties is anticipated for random disk systems. This should give a spherical two-dimensional tensor with only nonzero components being the main diagonals equal to each other. This is quantified using the indicated anisotropy index, which materializes the ellipse eccentricity as demonstrated in Figure\ \ref{F:VE_size_on_Amu}. To this end, the eigenvalues of the computed $2\times2$ permeability matrix are determined and ordered. It is shown that with increasing VE size, the mean anisotropy index gets closer to zero. This implies a reduction in the in-plane directional dependence of the effective magnetic response.

At the VE size of $2^{11}=2048$ pixels, the relative standard deviations in the computed effective permeabilities reduce below 2\%. Considering the 15 random realizations used in computations, this corresponds to a relative standard error of the sum below 0.5\%. The relative change of the mean effective permeability compared to the calculation for the VE size of 1024 pixels is below 1\%.
Moreover, this size provides considerably decreased directional dependence in the material's average in-plane magnetic permeability, with the corresponding anisotropy index falling below 0.2.
With these observations, even if for Dirichlet and Neumann boundary conditions, sufficient proximity to the effective solution is not met even at this size, RVE size is determined as 2048 pixels, and the remaining analysis is conducted using this size and periodic boundary conditions.
An RVE size, according to the definition of Hill, i.e., a size devoid of the Neumann and Dirichlet boundary condition effects as far as they are uniform, is to be much larger.\\

\noindent{\textbf{A Comparison to Effective Medium Approximations: }} Effective-medium approximations, which use simple microstructural information such as phase volume fraction and shape, prove helpful in revealing the qualitative trends in effective physical properties of the composites as a function of phase concentration. In the current study, we will consider three effective-medium approximations, namely, the Maxwell, the self-consistent,  and differential effective-medium approximations for which the $d-$dimensional generalizations are known   \cite{Torquato2002}. For convenience, we give these $d-$dimensional generalizations in  \ref{S:effective_medium} whereas, in the following, we present their two-dimensional specializations, i.e., for $d=2$. Starting with the two-dimensional form  of the Maxwell approximation given in Eq.\ \eqref{E:MaxwellGarnett} for the effective permeability  $\mu^\star_\mathrm{M}$ reads
\begin{equation}
\dfrac{\mu^\star_\mathrm{M}-\mu_1}
{\mu^\star_\mathrm{M}+\mu_1}=
\phi_2\dfrac{\mu_2-\mu_1}{\mu_2+\mu_1}\,.
\label{E:MaxwellGarnett_2D}
\end{equation}
Noting that $\phi_1=1-\phi_2$, the two-component self-consistent approximation \cite{Bruggeman1935,Landauer1978} for the effective permeability
$\mu^\star_\mathrm{SC}$ given in Eq.\ \eqref{E:selfconsistent} can be reiterated for two-dimensions as follows
\begin{equation}
\mu^\star_\mathrm{SC}=
\dfrac{\eta+\sqrt{\eta^2+4\mu_1\mu_2}}{2}\text{ where }\eta=[\mu_2-\mu_1][2\phi_2-1]\,.
\label{E:selfconsistent_2D}
\end{equation}
Finally, the two-dimensional differential effective-medium approximation for the effective permeability $\mu^\star_\mathrm{DEM}$ given in Eq.\ \eqref{E:differentialeffectivemedium} is given with the following implicit relation
\begin{equation}
\dfrac{\mu_2-\mu^\star_\mathrm{DEM}}{\mu_2-\mu_1}
\sqrt{\dfrac{\mu_1}{\mu^\star_\mathrm{DEM}}}=1-\phi_2\,,
\label{E:differentialeffectivemedium_2D}
\end{equation}
The $d-$dimensional generalizations of the effective-medium approximations given in Eqs.\ \eqref{E:MaxwellGarnett}, \eqref{E:selfconsistent} and \eqref{E:differentialeffectivemedium},  satisfy the indicated Voigt and Reuss  property  bounds as well as the limiting processes: $\mu^\star=\mu^\star_\mathrm{V}=\mu^\star_\mathrm{R}=\mu^\star_\mathrm{M}=\mu^\star_\mathrm{SC}=\mu^\star_\mathrm{DEM}\to \mu_1$ for $\phi_2\to0$ and $\mu^\star=\mu^\star_\mathrm{V}=\mu^\star_\mathrm{R}=\mu^\star_\mathrm{M}=\mu^\star_\mathrm{SC}=\mu^\star_\mathrm{DEM}\to \mu_2$ for $\phi_1\to0$.

\begin{figure}[htb!]
    \centering
% trim=left bottom right top, clip
\subfigure[]{\begin{tikzpicture}
    \node[inner sep=0pt] (B11) at (0,0){\includegraphics[height=0.37\textwidth]{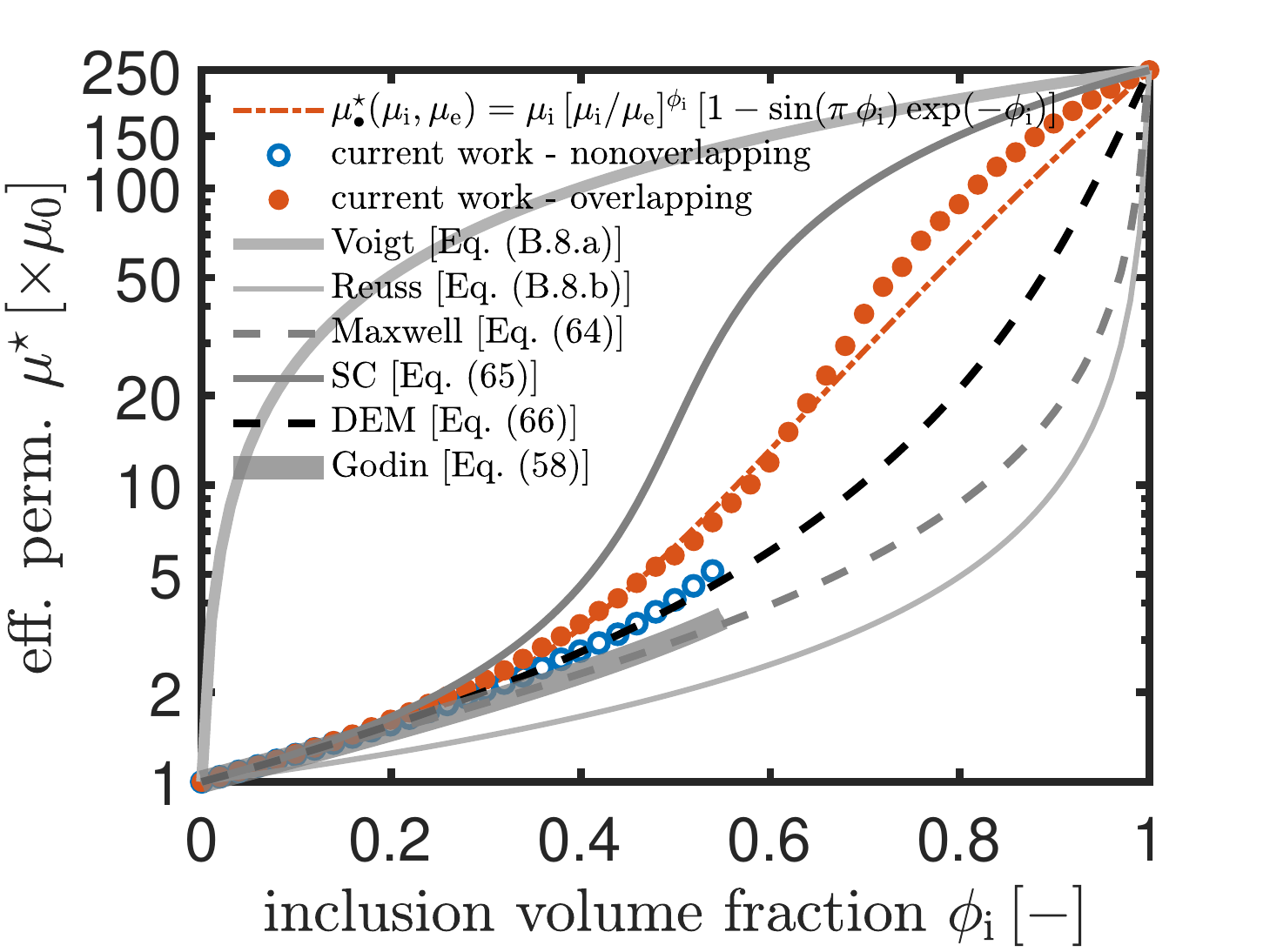}};
    \node (A) at (0.0,3.0) {\scriptsize $\mu_\mathrm{i}/\mu_\mathrm{m}=250/1$};
    \end{tikzpicture}}
\subfigure[]{\begin{tikzpicture}
    \node[inner sep=0pt] (B11) at (0,0){\includegraphics[height=0.37\textwidth]{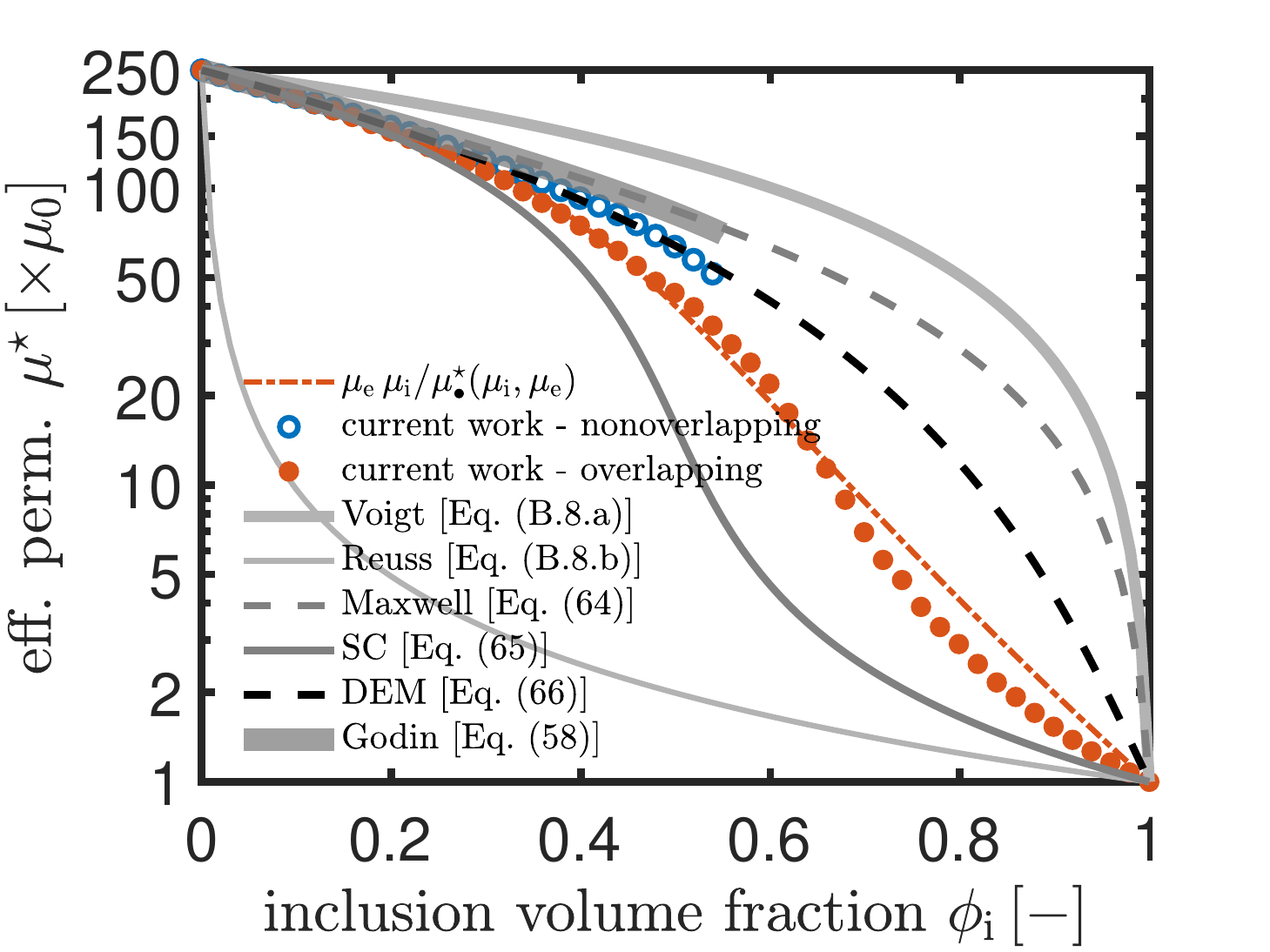}};
    \node (A) at (0.0,3.0) {\scriptsize $\mu_\mathrm{i}/\mu_\mathrm{m}=1/250$};
    \end{tikzpicture}}
\caption{Effective permeability predictions for the
considered nonoverlapping and overlapping random disk configurations and the selected RVE size of 2048 pixels as a function of the inclusion volume fraction for (a) $\mu_\mathrm{i}/\mu_\mathrm{e}=250/1$ and (b) $\mu_\mathrm{i}/\mu_\mathrm{e}=1/250$. The numerical results correspond to scalar potential formulation with periodic boundary conditions. The simple three-parameter expression  $\mu^\star_{\bullet}=\mu_\mathrm{e}\,[\mu_\mathrm{i}/\mu_\mathrm{e}]^{\phi_\mathrm{i}}\,[1-\sin(\pi\,\phi_\mathrm{i})\,\mathrm{exp}(-\phi_\mathrm{i})]$ given in Eq.\ \eqref{E:scalinglawthiswork} for $\phi_\mathrm{i}\geq\phi_\mathrm{e}$, shows a good agreement especially for volume fractions up to $\phi_\mathrm{i}=0.6$.  In agreement with Mendelson's generalization of Keller's relation, our  numerical predictions satisfy the phase-interchange relation, which gives $\mu^\star(\mu_1,\mu_2)\,\mu^\star(\mu_2,\mu_1)=\mu_1\mu_2$ given in \cite{Keller1964}. The abbreviations SC and DEM stand for self-consistent and differential effective-medium approximations.}
\label{F:curves}
\end{figure}

For the considered nonoverlapping and overlapping random disk configurations and the selected RVE size of 2048 pixels, Figure\  \ref{F:curves} depicts the effective permeability predictions as a function of the inclusion volume fraction without and with phase-interchange in a lin-log plot. As expected, increasing high permeability phase content monotonically increases the effective magnetic permeability.
The trend of the effective permeability predictions for the overlapping  disk systems in the current lin-log plot resembles an unsymmetric sigmoid function whose inflection point at which the corresponding highest rate of change occurs is located in the proximity of the aforementioned percolation threshold $\phi_{\mathrm{P}}\simeq0.68$. A similar trend is obtained by Helsing in the computation of effective electric properties of biphasic composites with randomly overlapping disks embedded in a homogeneous infinite medium  \cite{Helsing1998}. Considering the inclusion volume fractions up to 0.55, which is the range applicable to the nonoverlapping random disk microstructures, the effective permeability in overlapping disk systems is calculated to be up to 33\% higher\footnote{In the computation of relative differences, always the maximum of the two effective permeabilities is used in the denominator.}. The difference is even more drastic and reaches up to 56\% once the results for overlapping random disk systems and regular square disk systems are compared. This is explained via the inclusion interaction, which is minimal in regular systems since the inter-disk distances in the considered monodisperse system are maximized for a selected disk volume fraction. For the case of phase interchange, identical results are valid, whereas this time, the smallest effective permeability belongs to the overlapping disk systems.

For comparison purposes, the curves for the upper (Voigt) and the lower (Reuss) bounds as well as the two-dimensional Maxwell, self-consistent, and differential effective-medium approximations, respectively, given by Eqs.\ \eqref{E:MaxwellGarnett_2D}, \eqref{E:selfconsistent_2D} and \eqref{E:differentialeffectivemedium_2D} are also provided in Figure\ \ref{F:curves}. As anticipated, the Voigt and Reuss mixture rules bound the numerical solutions and the effective-medium approximations; nevertheless, they are far from being stringent.
As given in Ref.\ \cite{Torquato2002}, the Maxwell approximation for $\mu_2\geq\mu_1$ and $\mu_1\geq\mu_2$ coincide with the optimal lower and upper Hashin Shtrikman bounds, respectively. Our results are in complete agreement with this statement where for the phase contrast of  $\mu_\mathrm{i}/\mu_\mathrm{m}=250/1$, the Maxwell approximation constitutes the optimal lower bound. For the inverted system, it acts as the optimal upper bound to all other predictions. The Maxwell Garnet approximation agrees well with Godin's analytical solution for the periodic regular square disk arrangements with a maximum difference of 5\% for both phase contrasts. This is to show that the Maxwell effective-medium approximation  preserves validity beyond the dilute regime with $\phi_\mathrm{i}>0.05$. Nevertheless, this validity is limited to well-separated disk arrangements with the least interaction potential. Unlike other effective-medium approaches, the trend in self-consistent approximation shows sigmoid function  features similar to the curve of the nonoverlapping disks. Nevertheless, it is symmetric with an inflection point at $\phi_{\mathrm{i}}=0.5$.   Thus, its validity, for the considered random monodisperse disk systems for which the phase-interchange symmetry is not applicable, is only up to the dilute limit and for $\phi_{\mathrm{i}}>0.90$. The inflection point $\phi_{\mathrm{i}}=0.5$ corresponds to the built-in percolation threshold prediction for the inclusions in the self-consistent approximation. The Maxwell and the differential  effective-medium approximations consider that the phases remain connected until the other phase fills the space \cite{Torquato2002}.
Within the considered volume fraction range, the differential effective-medium approximation gives results with up to a 13\% difference from our computations for the nonoverlapping random disk systems. For overlapping random disk systems with  $\phi_1\geq\phi_2$, a correction $[1-\sin(\pi\,\phi_\mathrm{1})\,\mathrm{exp}(-\phi_\mathrm{1})]$ to the straight line
$[\mu_\mathrm{1}/\mu_\mathrm{2}]^{\phi_\mathrm{1}}$ in lin-log plot results in a simple three-parameter expression given in Eq.\ \eqref{E:scalinglawthiswork}  which gives a  remarkably better overall agreement with the computations than the considered effective medium approximations, especially for inclusion volume fractions up to $\phi_1=0.6$. Beyond $\phi_1=0.6$, the observed maximum relative error is around 30\%, around which occurs $\phi_1=0.8$. Moreover, Eq.\ \eqref{E:scalinglawthiswork} satisfies the indicated Voigt and Reuss  property  bounds as well as the limiting processes: $\mu^\star_{\bullet}\to \mu_1$ for $\phi_2\to0$ and $\mu^\star_{\bullet}\to \mu_2$ for $\phi_1\to0$.
\begin{equation}
\dfrac{\mu^\star_{\bullet}}{\mu_\mathrm{2}}=
\left[
\dfrac{\mu_\mathrm{1}}{\mu_\mathrm{2}}\right]^{\phi_\mathrm{1}}\,[1-\sin(\pi\,\phi_\mathrm{1})\,\mathrm{exp}(-\phi_\mathrm{1})]\,.
\label{E:scalinglawthiswork}
\end{equation}
\color{black}

Finally, Keller's phase-interchange relation  $\mu^\star(\mu_1,\mu_2)\,\mu^\star(\mu_2,\mu_1)=\mu_1\mu_2$ is satisfied for the considered random disk arrangements within a numerical tolerance as before in agreement with  Mendelson's generalization of Keller's relation \cite{Mendelson1975} to two-dimensional two-phase composites and considering the macroscopic isotropy in permeability of the two-dimensional random disk arrangements.

\begin{figure}[htb!]
    \centering
% trim=left bottom right top, clip
    \subfigure[]{\frame{\begin{tikzpicture}
    \node[inner sep=0pt] (B11) at (0,0)
    {\includegraphics[height=0.24\textwidth, trim=400 233 400 233, clip]{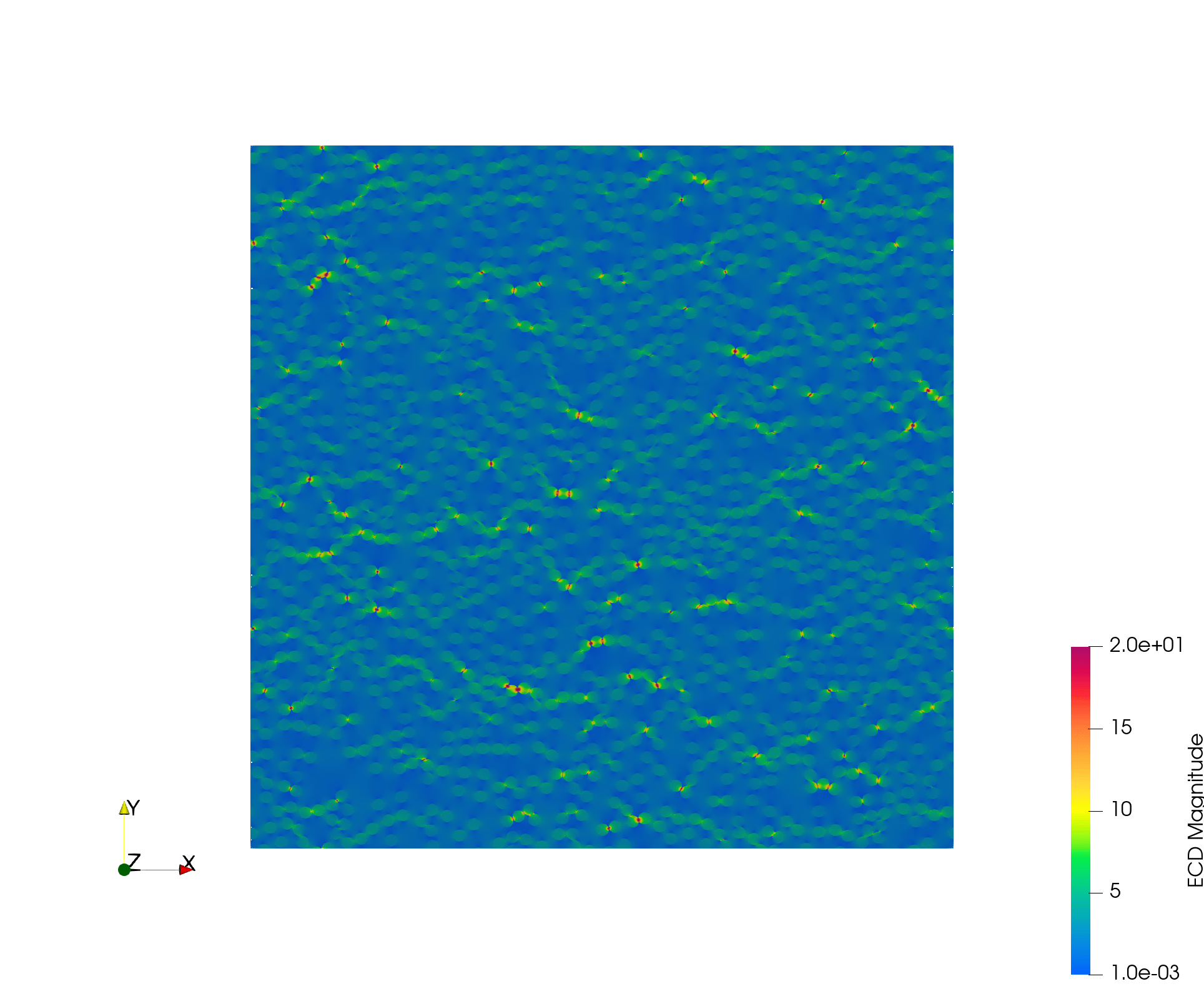}};
    \draw[thick, draw=white, solid, fill=white, fill opacity=0.25] (-0.50,-0.50) rectangle ++(1.0,1.0);
    \end{tikzpicture}}}
    \subfigure[]{\frame{\includegraphics[height=0.24\textwidth, trim=400 233 400 233, clip]{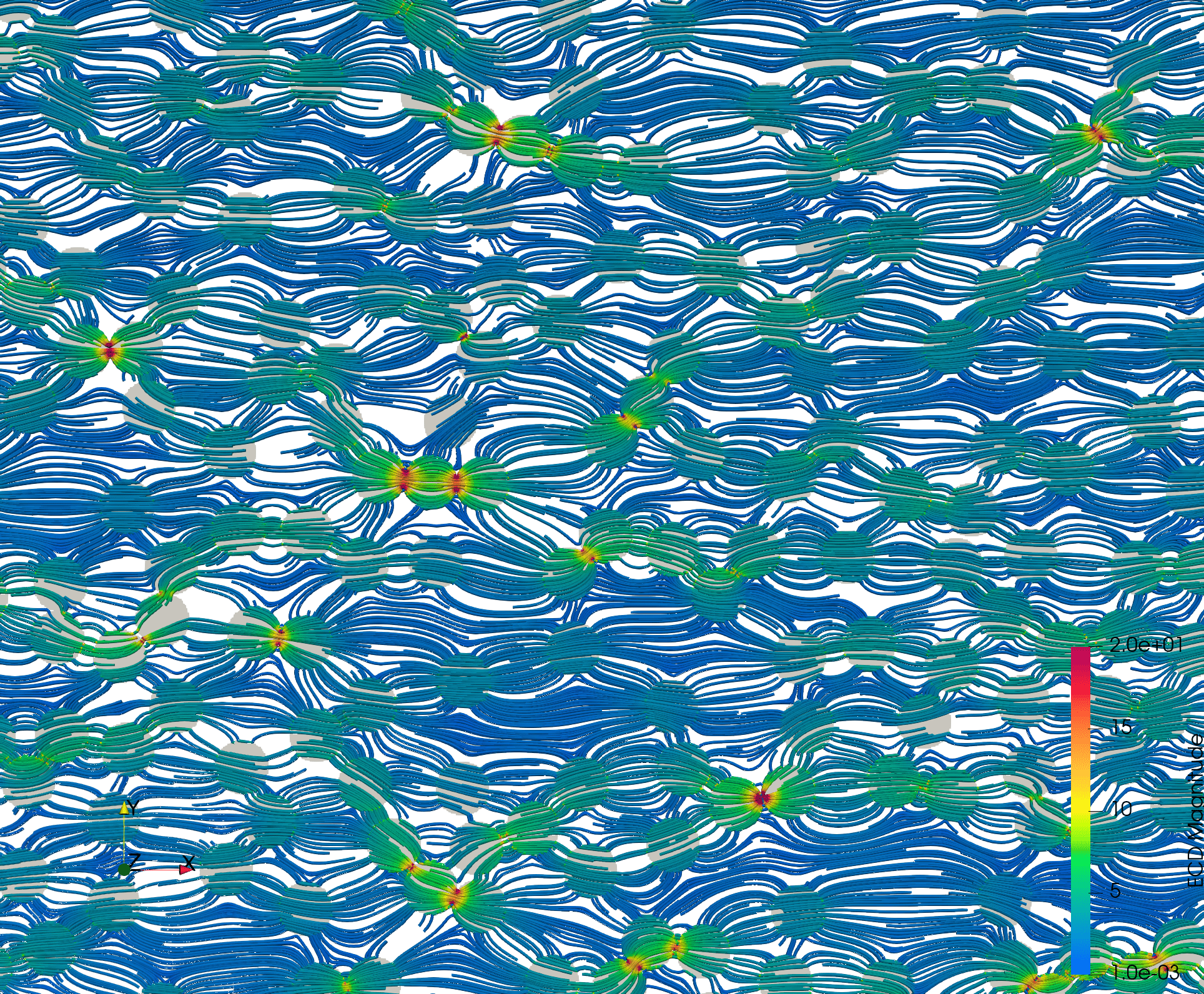}}}
    \subfigure[]{\frame{\begin{tikzpicture}
    \node[inner sep=0pt] (B22) at (0,0)
    {\includegraphics[height=0.24\textwidth, trim=400 233 400 233, clip]{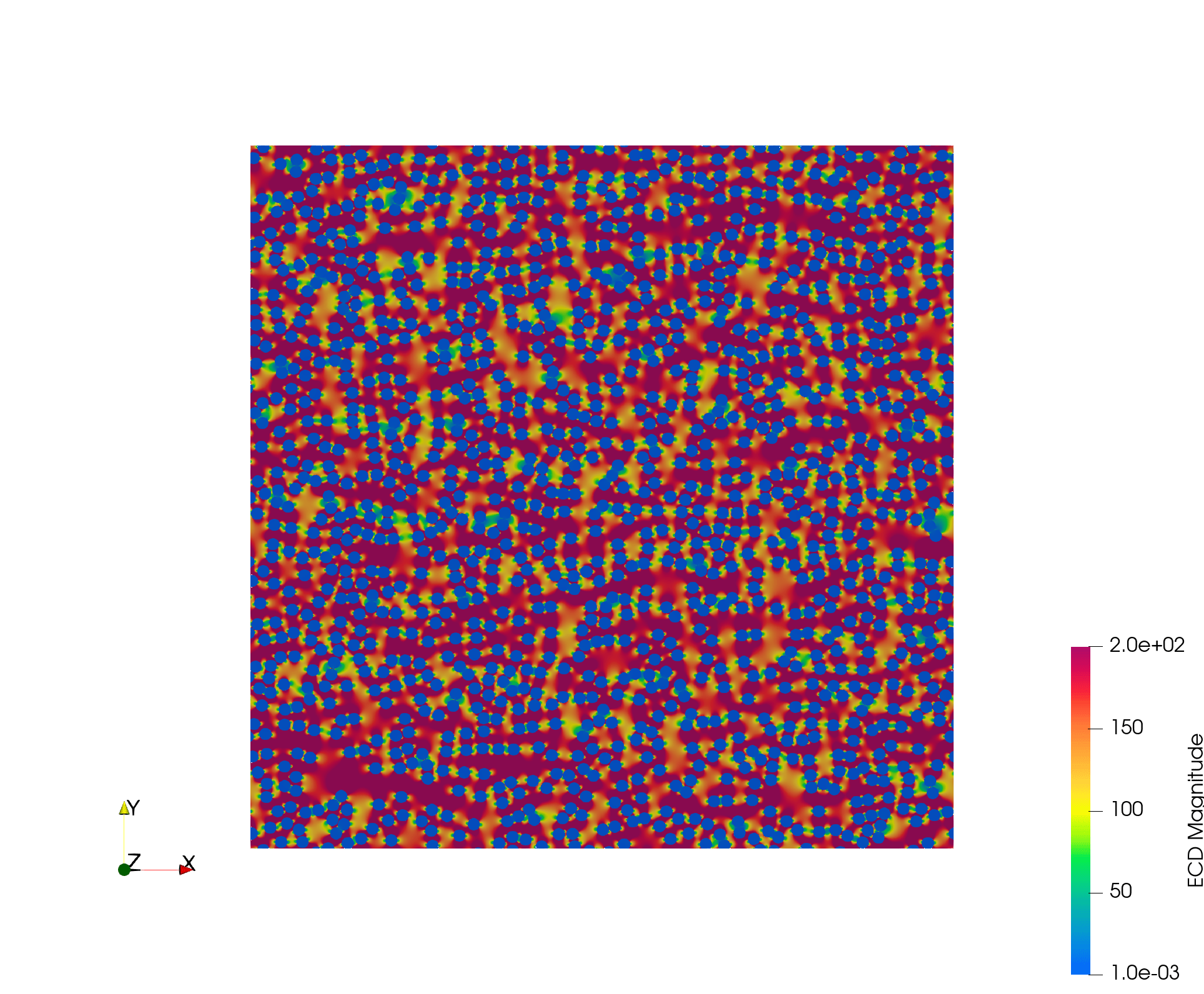}};
    \draw[thick, draw=white, solid, fill=white, fill opacity=0.25] (-0.50,-0.50) rectangle ++(1.0,1.0);
    \end{tikzpicture}}}
    \subfigure[]{\frame{\includegraphics[height=0.24\textwidth, trim=400 233 400 233, clip]{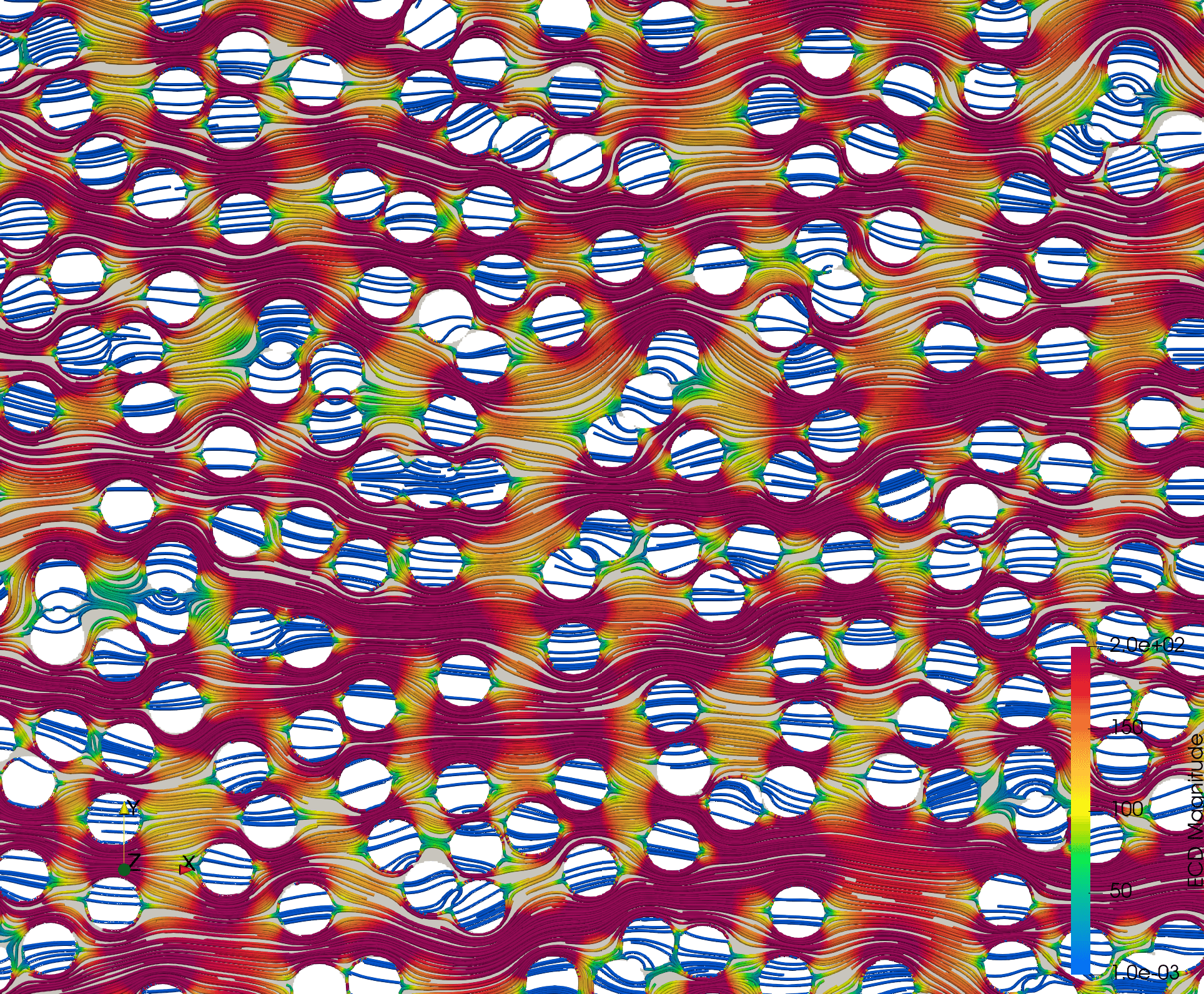}}}
    \subfigure[]{\frame{\begin{tikzpicture}
    \node[inner sep=0pt] (B11) at (0,0)
    {\includegraphics[height=0.24\textwidth, trim=400 233 400 233, clip]{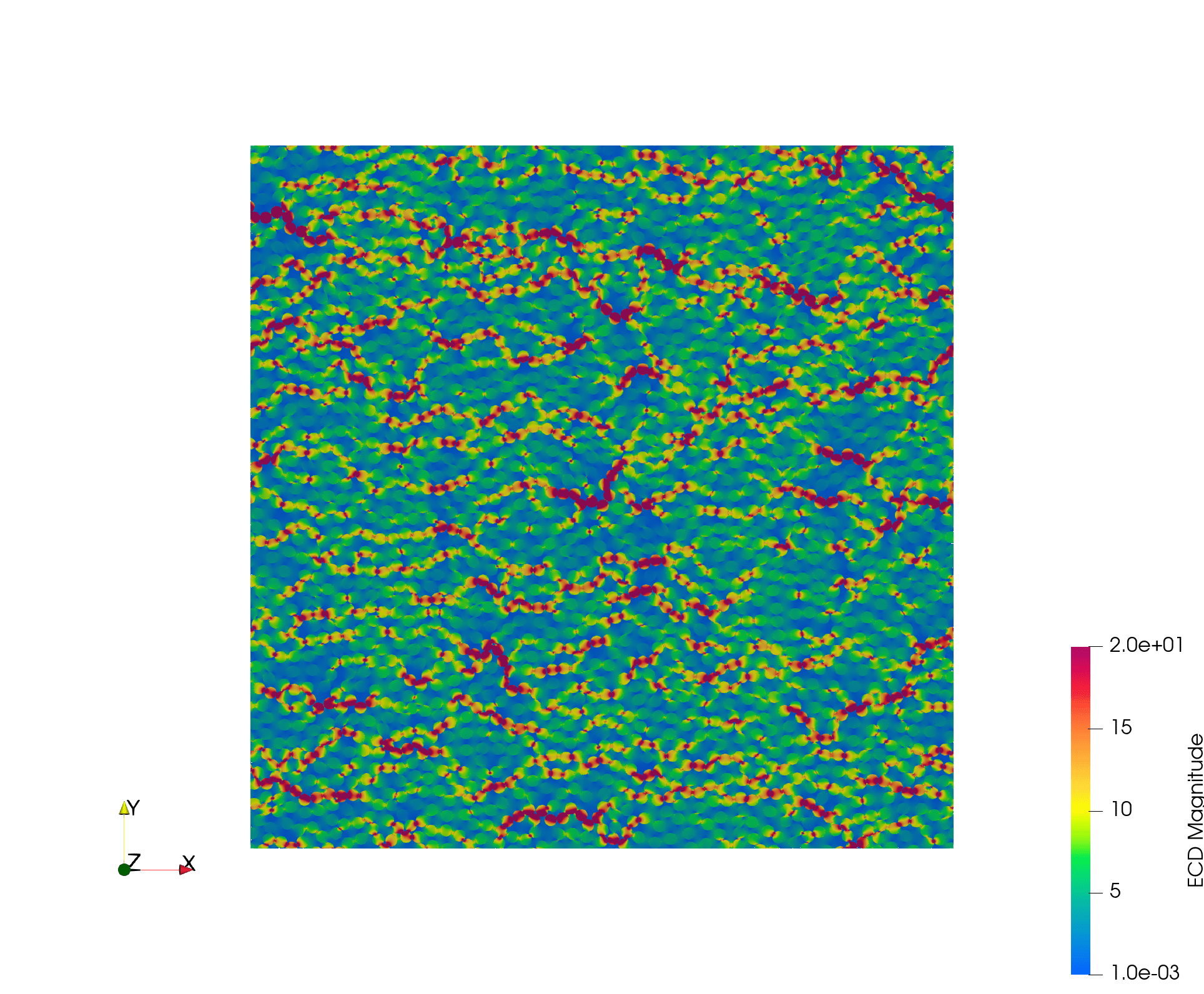}};
    \draw[thick, draw=white, solid, fill=white, fill opacity=0.25] (-0.50,-0.50) rectangle ++(1.0,1.0);
    \end{tikzpicture}}}
    \subfigure[]{\frame{\includegraphics[height=0.24\textwidth, trim=400 233 400 233, clip]{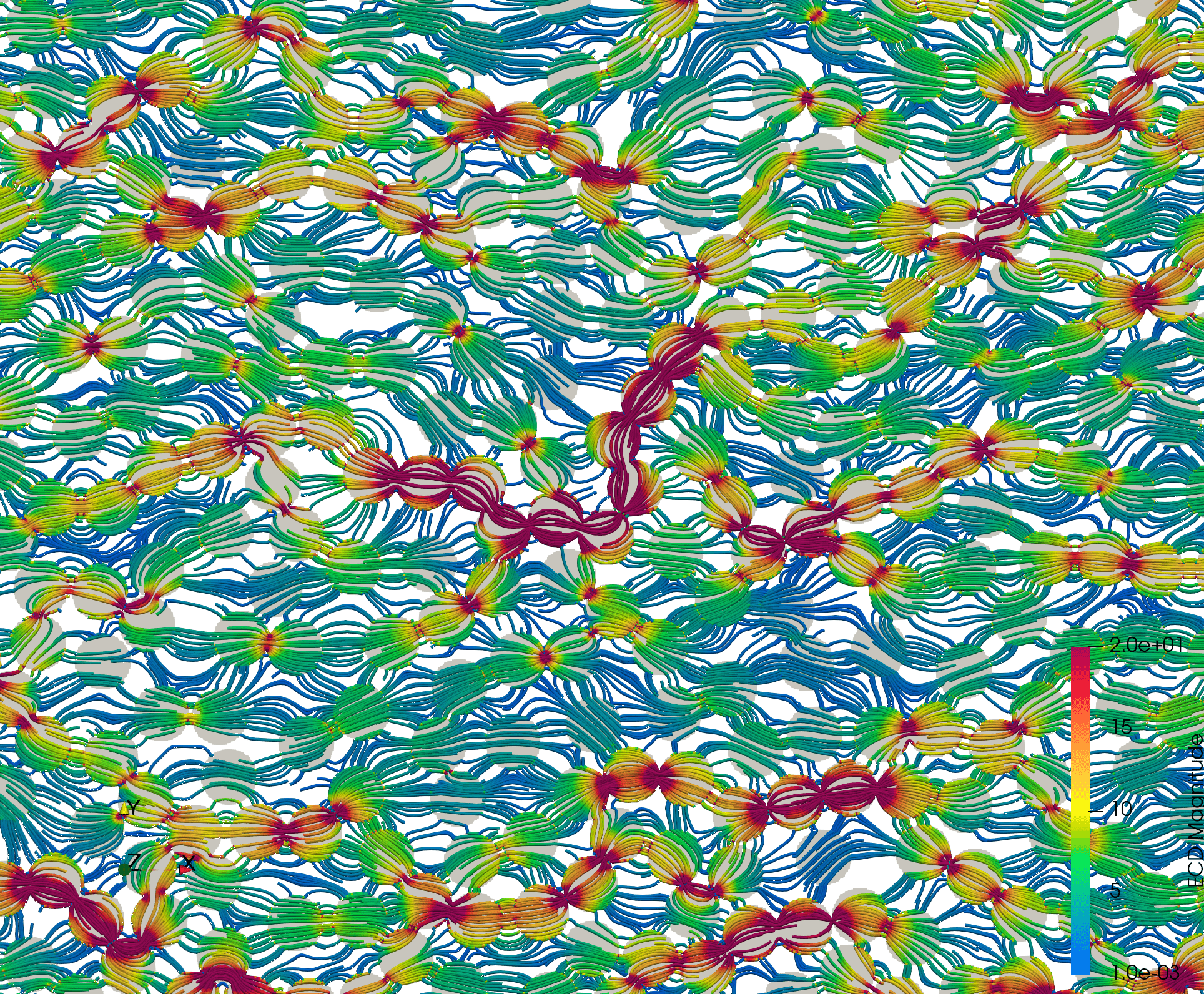}}}
    \subfigure[]{\frame{\begin{tikzpicture}
    \node[inner sep=0pt] (B22) at (0,0)
    {\includegraphics[height=0.24\textwidth, trim=400 233 400 233, clip]{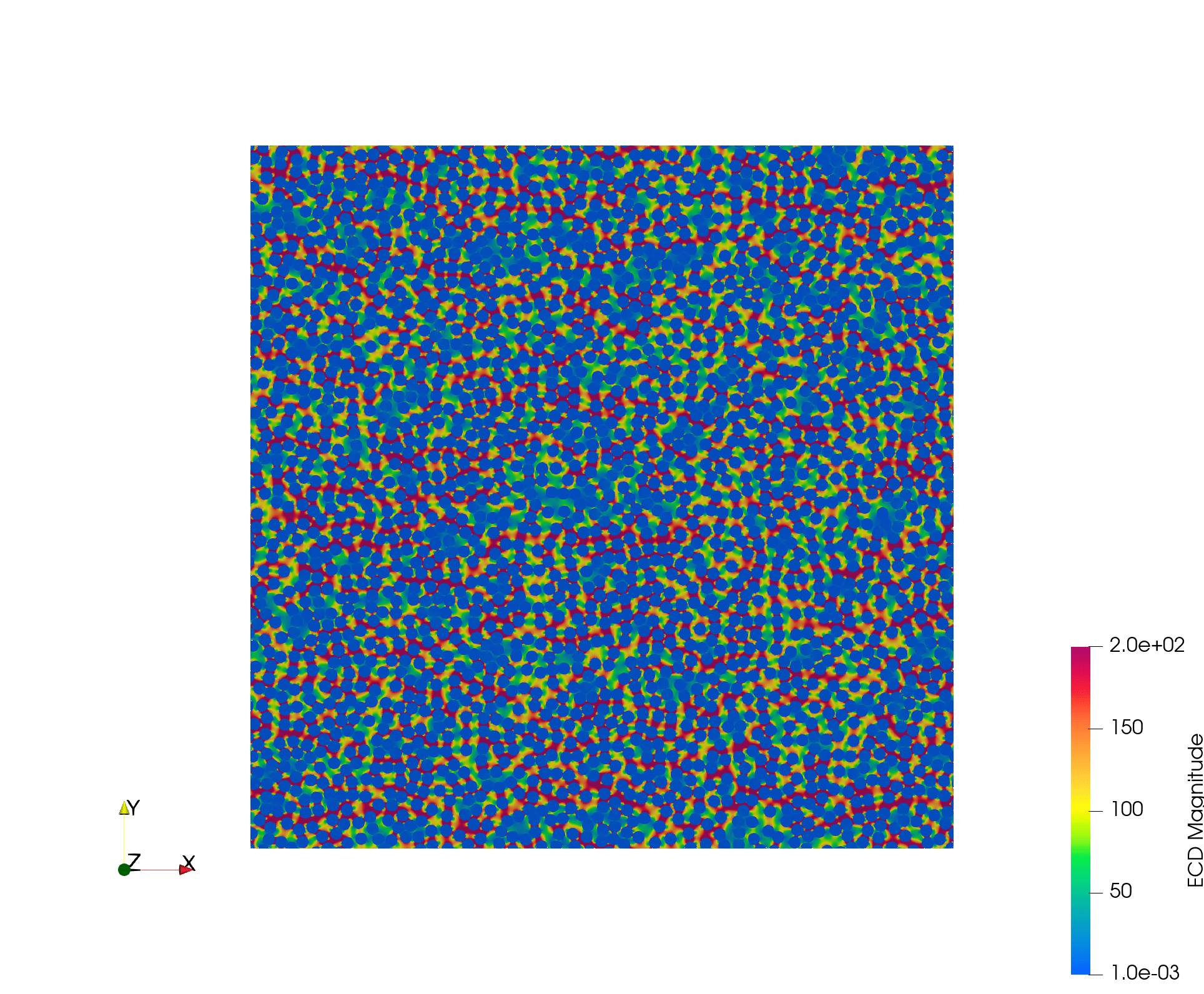}};
    \draw[thick, draw=white, solid, fill=white, fill opacity=0.25] (-0.50,-0.50) rectangle ++(1.0,1.0);
    \end{tikzpicture}}}
    \subfigure[]{\frame{\includegraphics[height=0.24\textwidth, trim=400 233 400 233, clip]{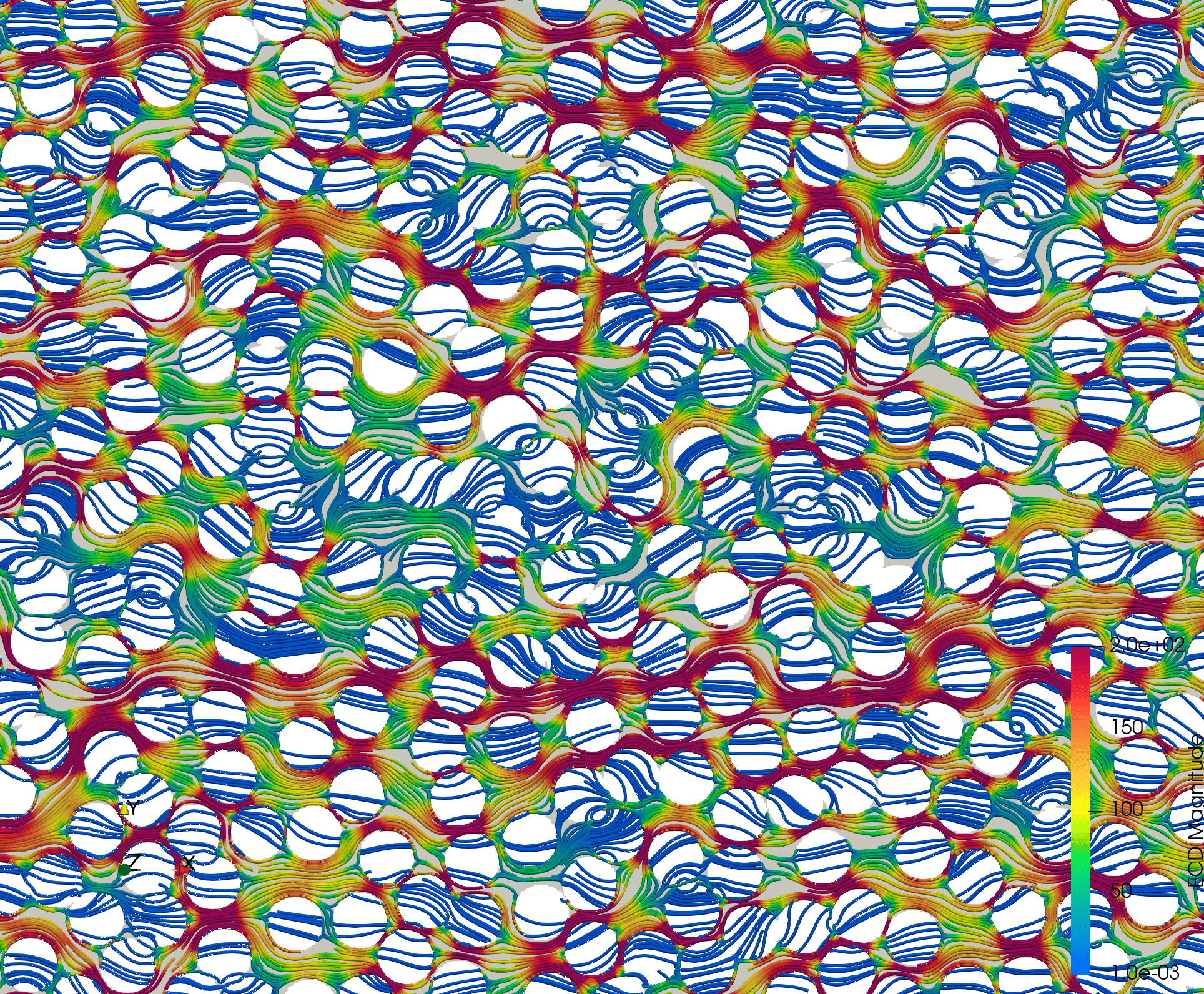}}}
    min$\,\,$\frame{\includegraphics[width=0.49\textwidth, trim=0 15 0 15, clip]{legend_2}}$\,\,$max
\caption{Magnetic flux field contours for the case of nonoverlapping disks  with inclusion volume fractions of (a), (b), (c) and (d)  $\phi_\mathrm{i}=0.30$, (e), (f), (g) and (h) $\phi_\mathrm{i}=0.55$ for the horizontally applied macroscopic magnetic field with $\bs H=[1, 0]^\top$. The results correspond to $\mu_\mathrm{i}/\mu_\mathrm{e}=250/1$ for (a), (b), (e), and (f) and to  $\mu_\mathrm{i}/\mu_\mathrm{e}=1/250$ for (c), (d), (g) and (h). In (a), (c), (e), and (g), the contour plots for the complete FE model are shown, whereas in (b), (d), (f), and (h), the magnetic flux lines are given for the central region marked with a white square in corresponding full images, with the background grey region denoting the high permeability phase. Scalar potential formulation and periodic boundary conditions are used. For demonstration purposes, the intervals [min, max] of each contour plot are individually determined to read (a) and (b)  $[0.001,291]$, (c) and (d) $[0.001,2487]$,
(e) and (f)  $[0.001,1125]$, (g) and (h) $[0.001,3672]$. These results are normalized with respect to $\mu_0$.
For $\mu_\mathrm{i}/\mu_\mathrm{e}=250/1$ effective permeabilities of $2.036\mu_0$ and $5.477\mu_0$ are predicted for inclusion volume fractions of $0.30$, and $0.55$, respectively. For $\mu_\mathrm{i}/\mu_\mathrm{e}=1/250$ these are $124.3\mu_0$ and $48.52\mu_0$.}
\label{F:Bfielddetails_nonoverlapping}
\end{figure}

\begin{figure}[htb!]
    \centering
% trim=left bottom right top, clip
    \subfigure[]{\frame{\begin{tikzpicture}
    \node[inner sep=0pt] (B11) at (0,0)
    {\includegraphics[height=0.24\textwidth, trim=400 233 400 233, clip]{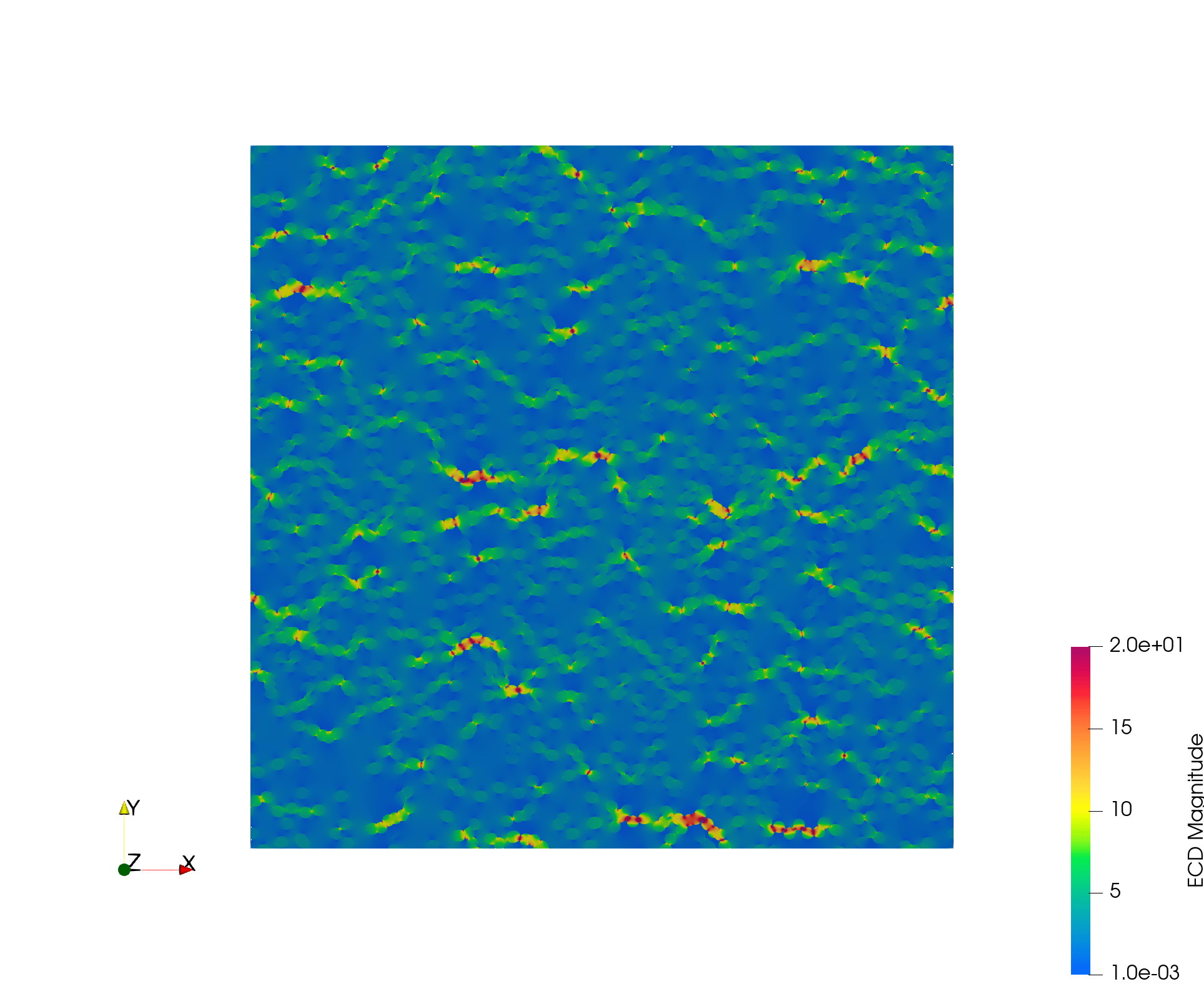}};
    \draw[thick, draw=white, solid, fill=white, fill opacity=0.25] (-0.50,-0.50) rectangle ++(1.0,1.0);
    \end{tikzpicture}}}
    \subfigure[]{\frame{\includegraphics[height=0.24\textwidth, trim=400 233 400 233, clip]{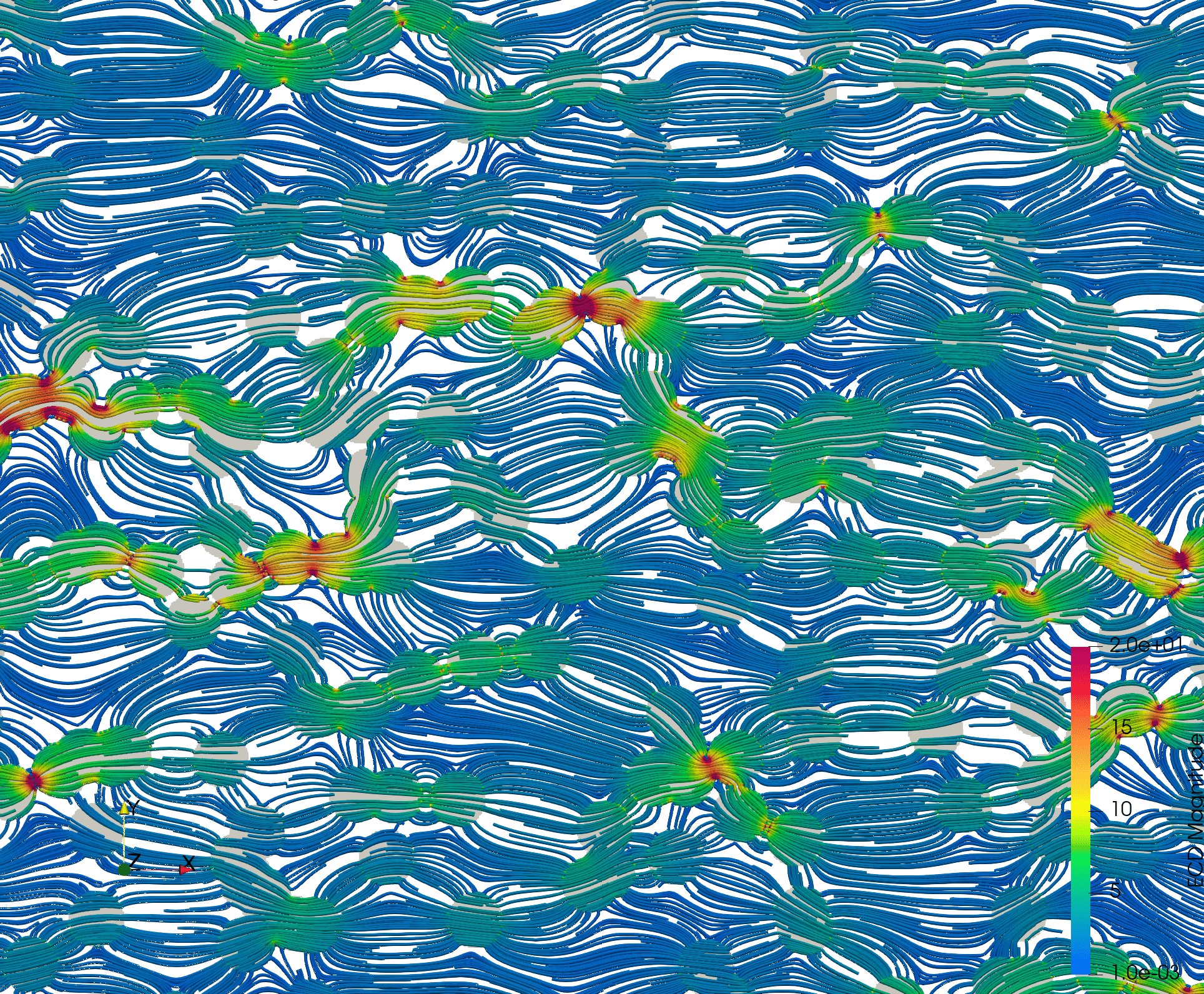}}}
    \subfigure[]{\frame{\begin{tikzpicture}
    \node[inner sep=0pt] (B22) at (0,0)
    {\includegraphics[height=0.24\textwidth, trim=400 233 400 233, clip]{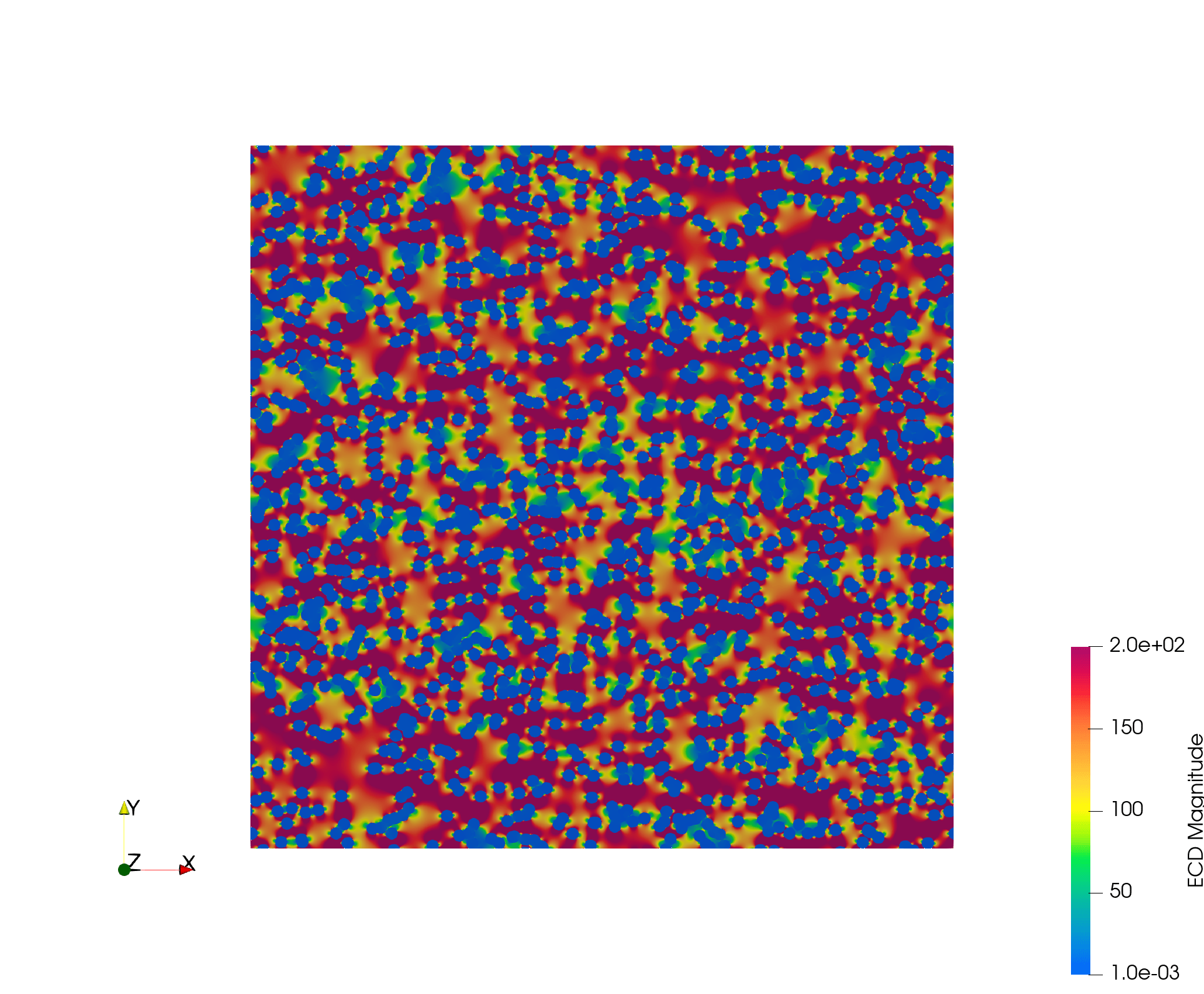}};
    \draw[thick, draw=white, solid, fill=white, fill opacity=0.25] (-0.50,-0.50) rectangle ++(1.0,1.0);
    \end{tikzpicture}}}
    \subfigure[]{\frame{\includegraphics[height=0.24\textwidth, trim=400 233 400 233, clip]{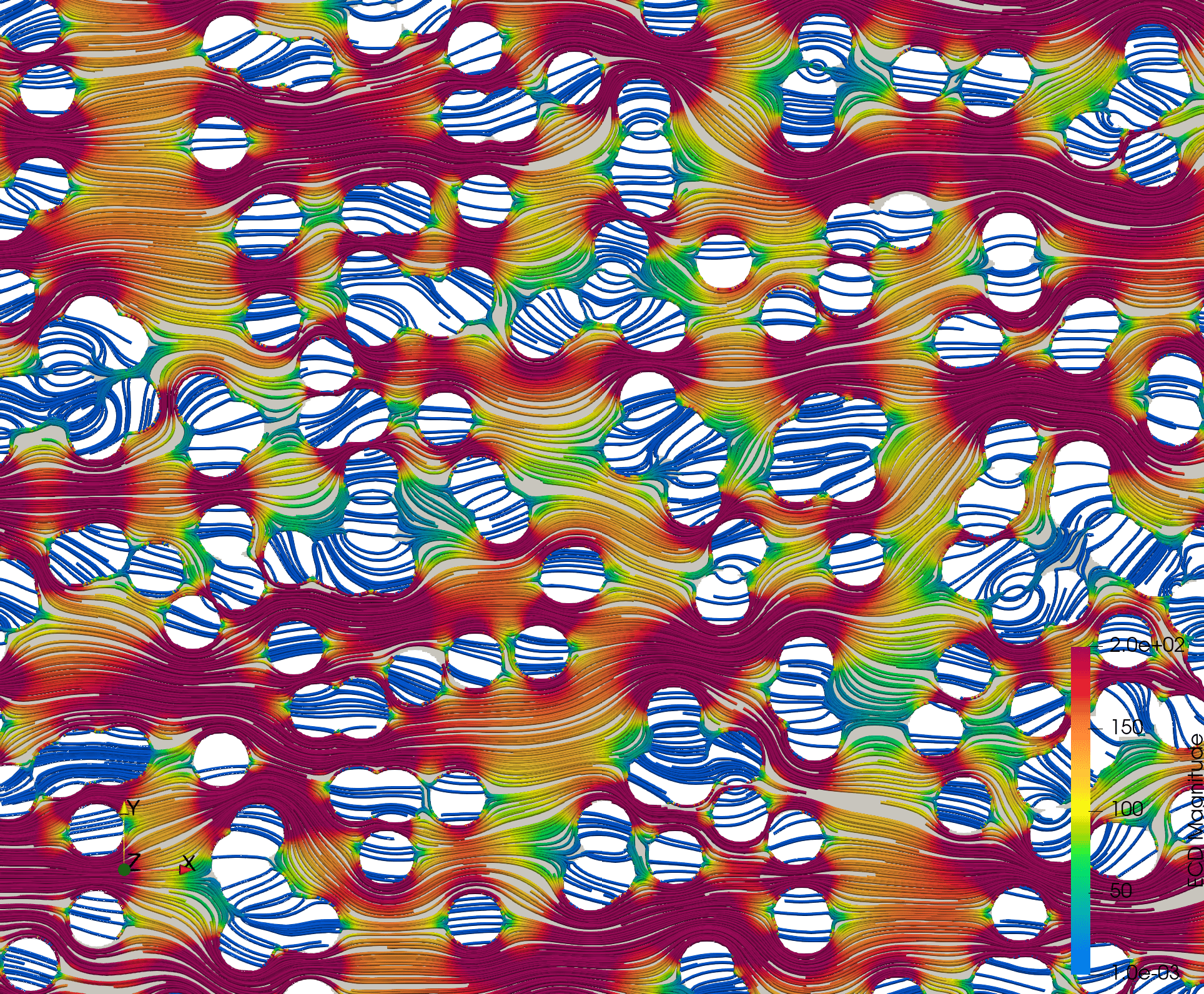}}}
    \subfigure[]{\frame{\begin{tikzpicture}
    \node[inner sep=0pt] (B11) at (0,0)
    {\includegraphics[height=0.24\textwidth, trim=400 233 400 233, clip]{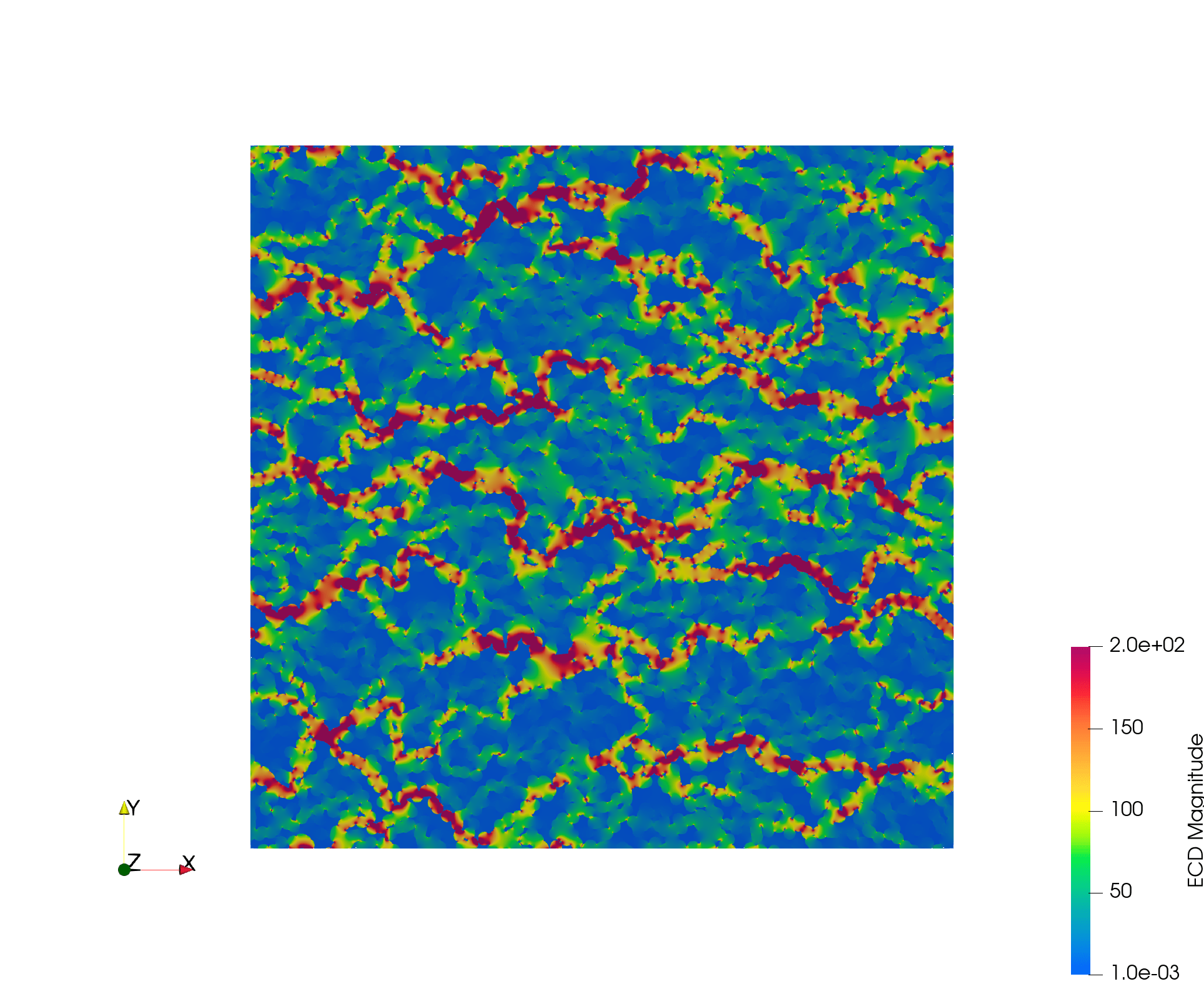}};
    \draw[thick, draw=white, solid, fill=white, fill opacity=0.25] (-0.50,-0.50) rectangle ++(1.0,1.0);
    \end{tikzpicture}}}
    \subfigure[]{\frame{\includegraphics[height=0.24\textwidth, trim=400 233 400 233, clip]{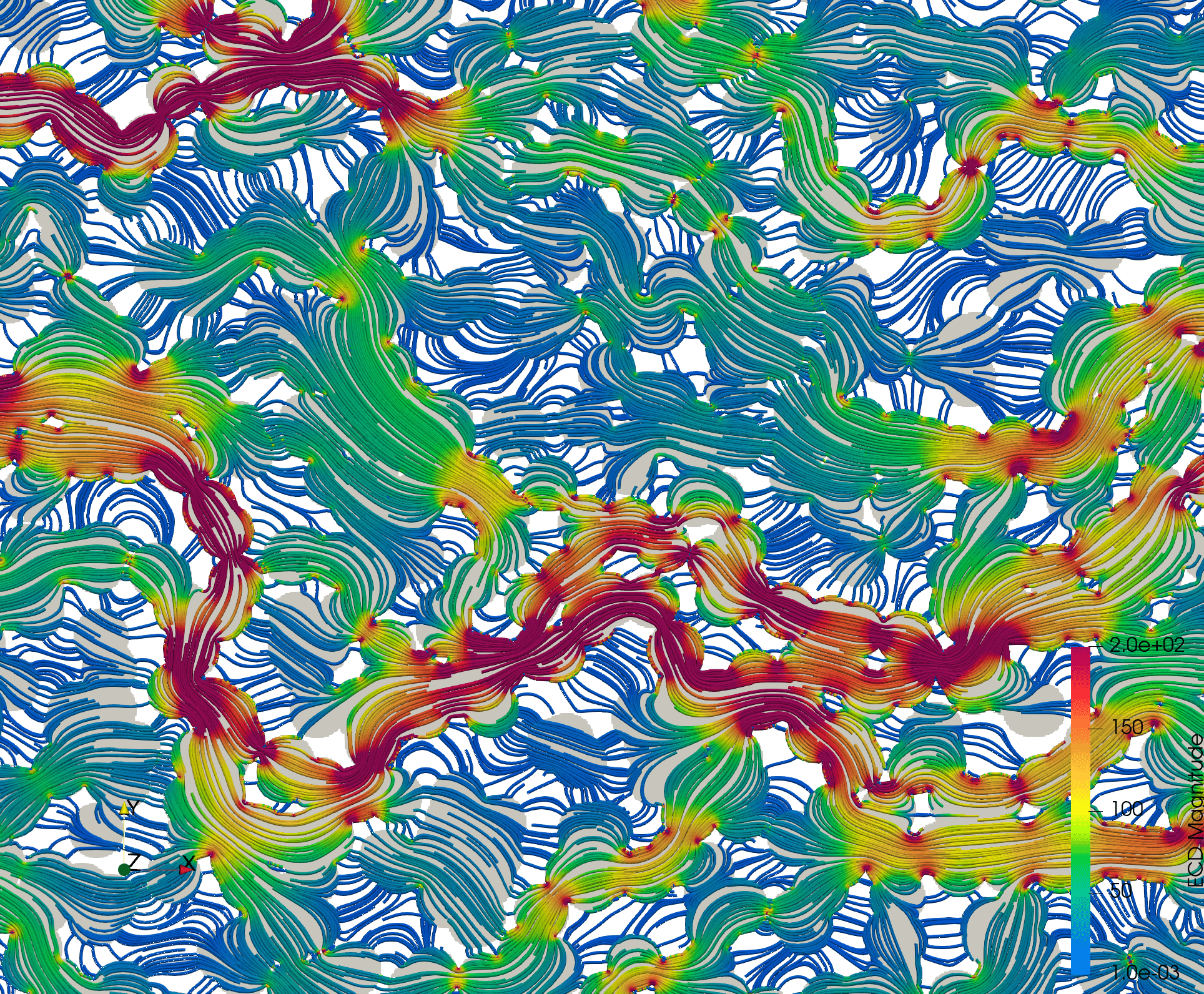}}}
    \subfigure[]{\frame{\begin{tikzpicture}
    \node[inner sep=0pt] (B22) at (0,0)
    {\includegraphics[height=0.24\textwidth, trim=400 233 400 233, clip]{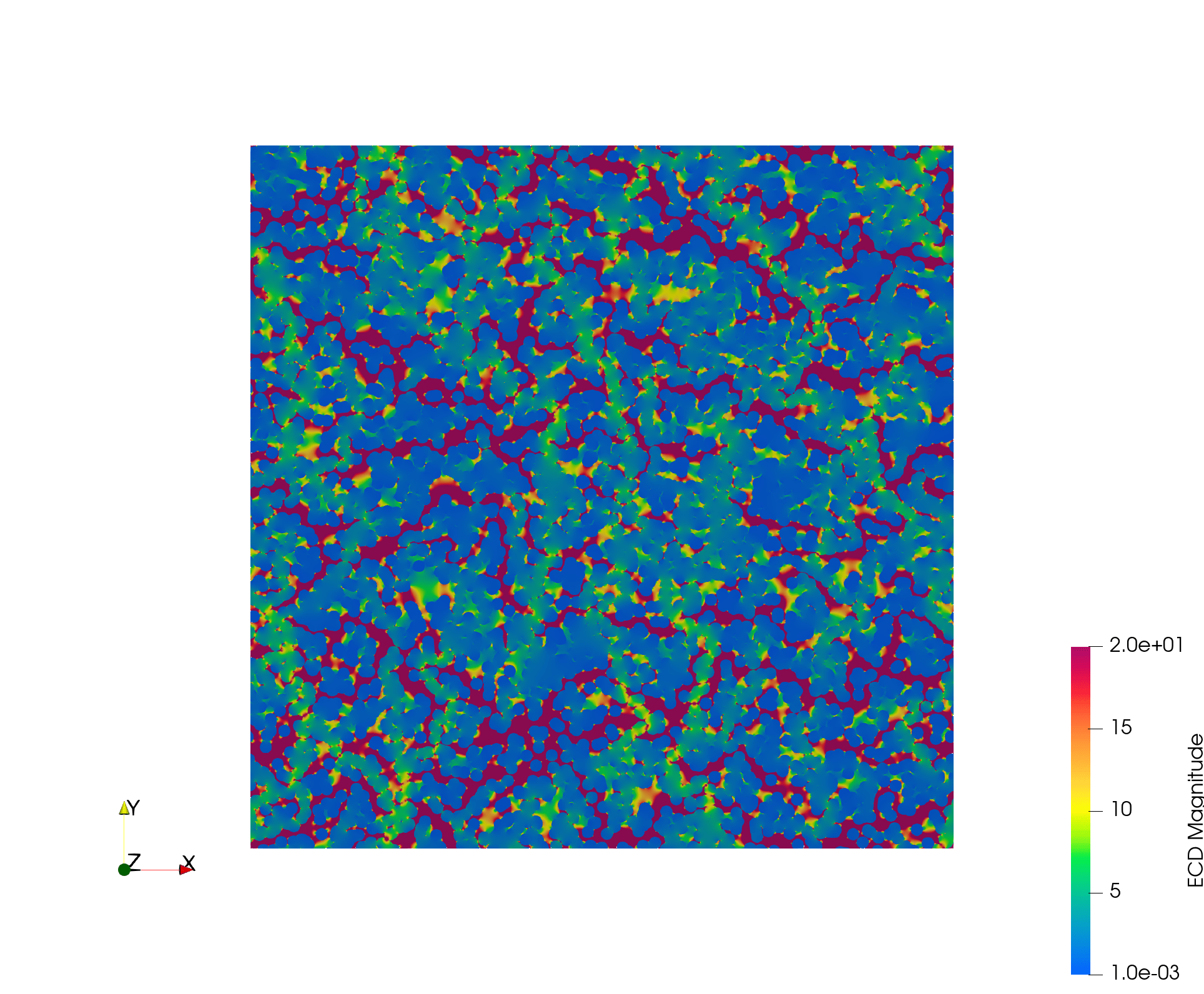}};
    \draw[thick, draw=white, solid, fill=white, fill opacity=0.25] (-0.50,-0.50) rectangle ++(1.0,1.0);
    \end{tikzpicture}}}
    \subfigure[]{\frame{\includegraphics[height=0.24\textwidth, trim=400 233 400 233, clip]{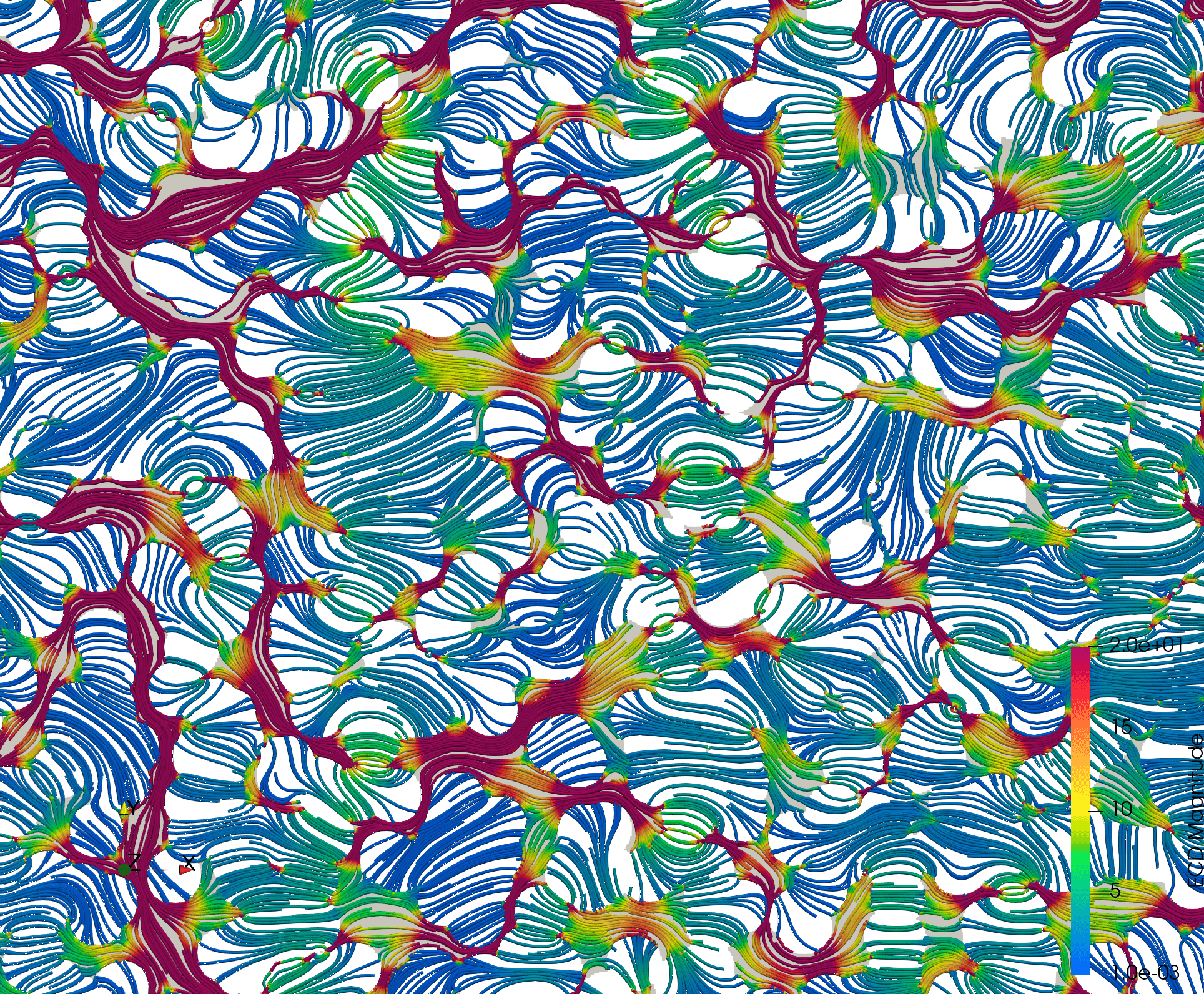}}}
    min$\,\,$\frame{\includegraphics[width=0.49\textwidth, trim=0 15 0 15, clip]{legend_2}}$\,\,$max
\caption{Magnetic flux field contours for the case of overlapping disks  with inclusion volume fractions of (a), (b), (c) and (d) $\phi_\mathrm{i}=0.30$, (e), (f), (g) and (h) $\phi_\mathrm{i}=0.70$ for the horizontally applied macroscopic magnetic field with $\bs H=[1, 0]^\top$. The results correspond to $\mu_\mathrm{i}/\mu_\mathrm{e}=250/1$ for (a), (b), (e), and (f) and to  $\mu_\mathrm{i}/\mu_\mathrm{e}=1/250$ for (c), (d), (g) and (h). In (a), (c), (e), and (g), the contour plots for the complete FE model are shown, whereas in (b), (d), (f), and (h), the magnetic flux lines are given for the central region marked with a white square in corresponding full images, with the background grey region denoting the high permeability phase. Scalar potential formulation and periodic boundary conditions are used. For demonstration purposes, the intervals [min, max] of each contour plot are individually determined to read (a) and (b)  $[0.001,454]$, (c) and (d) $[0.001,3575]$,
(e) and (f)  $[0.001,6084]$, (g) and (h) $[0.001,1864]$. These results are normalized with respect to $\mu_0$. For $\mu_\mathrm{i}/\mu_\mathrm{e}=250/1$ effective permeabilities of $2.211\mu_0$ and $60.64\mu_0$ are predicted for inclusion volume fractions of $0.30$, and $0.70$, respectively. For $\mu_\mathrm{i}/\mu_\mathrm{e}=1/250$ these are $114.4\mu_0$ and $4.281\mu_0$.}
\label{F:Bfielddetails_overlapping}
\end{figure}

Figures\ \ref{F:Bfielddetails_nonoverlapping} and \ref{F:Bfielddetails_overlapping} provide complementary information regarding the magnetic flux field and flux lines for several selected computations. In Figure\ \ref{F:Bfielddetails_nonoverlapping}, the distribution of magnetic flux densities is given for a horizontally applied unitary macroscopic field of $^\mathrm{M}\bs H=[1,0]^\top$ for nonoverlapping disk system. In the figure, disk volume fractions of 0.30 and 0.55 are considered with phase  inversion. As indicated previously, 0.55 is the saturation volume fraction of the microstructure generation of nonoverlapping disk systems using a sequential adsorption process. It is observed that, like in the case of a regular disk system, the magnetic flux lines are attracted toward the high permeability phase. In the nonoverlapping disk system, the disk interaction increases with increasing high permeability disks. For the phase-interchanged system, at 0.30 volume fraction, the matrix phase percolation effectively hosts the magnetic flux lines.

As opposed to nonoverlapping disk arrangements, overlapping disks allow the formation of long-range connections among inclusions. This is referred to as phase percolation; see, e.g., Fig.\ \ref{F:microstructures_overlapping_VOLFRAC}. In two-dimensions, the percolation threshold is $\phi_\mathrm{p}\simeq0.68$ \cite{QuintanillaTorquatoZiff2000} whereas for three-dimensions and with spheres it is $\phi_\mathrm{p}\simeq0.29$ \cite{RintoulTorquato1997}. For further information, the reader is referred to \cite[p.67, p.123]{Torquato2002} and the references therein. In Figure\ \ref{F:Bfielddetails_overlapping}, the distribution of magnetic flux densities is given for a horizontally applied unitary macroscopic field of $^\mathrm{M}\bs H=[1,0]^\top$ for an overlapping disk system. In the figure, disk volume fractions of 0.30 and 0.70 are considered with phase contrasting, where 0.70 is selected as a volume fraction slightly above the percolation threshold in two dimensions. It is shown that at 0.30 of phase volume fraction, which is below the percolation threshold, the behavior is similar to the nonoverlapping disk system, such that long-range high-intensity flux lines are not developed.
Nevertheless, the overlapping provides the formation of higher hot spot regions. With increasing volume fraction to 0.70, the continuous pathways starting from the left face and reaching the right face of the volume element are developed.  These are the percolating phase paths at which long-range connectivity is developed. The magnetic flux lines tend towards these percolating high permeability phase paths.

\section{Conclusion}
Inspired by the displacement method of Lukassen et al. \cite{LUKKASSEN1995519}, we developed an asymptotic homogenization method with a finite element method in the computation of the initial effective magnetic permeability of composites. We considered vector and scalar potential formulations with periodic and uniform Dirichlet boundary conditions assuming linear magnetostatics. We showed that the vector/scalar potential formulation yields the upper (Voigt)/lower (Reuss) bound in the computed effective magnetic permeability for completely suppressed perturbation fields. Using the concept of the eccentricity of an ellipse, we proposed an anisotropy index to quantify the degree of directionality in the material's permeability. We applied the developed framework to various monodisperse disk arrangements in two dimensions. We computationally demonstrated that while periodic boundary conditions with both vector and scalar potential formulations provide unit cell size-invariance for regular periodic material systems, the predictions with the uniform Dirichlet boundary conditions are size-dependent. With increasing size, the results converge to the effective behavior where the vector potential formulation results merge from lower magnitudes, and scalar potential formulation results converge from upper. For random material systems, convergence to effective properties is much faster for periodic boundary conditions.

Our results show that even for the selected limited class of material systems constituting monodisperse disk configurations, neither the analytical bounds nor the selected popular effective-medium approximations are universally applicable.
Although the effective-medium approximations reveal qualitative trends in the physical properties of random systems, they fail to provide accurate quantitative predictions. This is because they only use volume fractions and inclusion shapes as descriptors. Thus, their applicability is generally limited to conditions with low contrast of properties and low inclusion volume fractions.
This constitutes an obstacle in front of their generalization to random composites. On the other hand, the proposed computational asymptotic homogenization with finite element method remedies this limitation by providing an accurate account of the underlying phase geometry, which relates to the shape and size of constituent phases, and topology, which describes constituent connectivity. Using our numerical computations, we proposed a simple three-parameter approximation for the effective permeability of overlapping random disk systems. This shows a remarkable predictive capability for inclusion volume fractions up to 0.60.
\section*{Acknowledgments}
This work was supported by the STW project 15472 of the STW Smart Industry
2016 program. This support is gratefully acknowledged.
\appendix
\section{Nondimensionalization in Magnetostatics}
\label{s:nondimensionalization}
In the following, we start with reiterating standard equations of
magnetostatics in a dimensionalized and their most general form. Let $%
\overline{\boldsymbol{H}}$, $\overline{\boldsymbol{B}}$ and $\overline{^%
\mathrm{M} \boldsymbol{x}}$ denote the magnetic field, magnetic induction
field and the position of the particle, respectively. The free current
density $\overline{\boldsymbol{J}}$ is linked to $\overline{\boldsymbol{H}}$
with
\begin{equation}
\mathrm{curl}_{\overline{^\mathrm{M}\boldsymbol{x}}} \overline{\boldsymbol{H}}
= \overline{\boldsymbol{J}}\,.  \label{E:freecurrent}
\end{equation}
The governing differential equation in magnetostatics is given by Gauss's
law for magnetism
\begin{equation}
\text{div}_{\overline{^\mathrm{M}\boldsymbol{x}}}\overline{\boldsymbol{B}}=%
\dfrac{\partial\overline{ B}_i}{\partial \overline{^\mathrm{M}x}_i}=0\, \quad%
\text{in }\mathcal{B}\,.  \label{E:dimequlibriumNEU1}
\end{equation}
Considering linear magnetostatics, we have the following constitutive form
which links $\overline{\boldsymbol{B}}$ to the magnetic field $\overline{%
\boldsymbol{H}}$
\begin{equation}
\overline{\boldsymbol{B}}=\overline{\boldsymbol{\mu}}\cdot\overline{%
\boldsymbol{H}} \quad\text{in }\mathcal{B}\,,
\label{E:dimconst}
\end{equation}
Here, $\overline{\boldsymbol{\mu}}$ is the symmetric second-order magnetic
permeability tensor is, $\overline{\mu}_{ij}=\overline{\mu}_{ji}$.\newline

\noindent \textbf{Magnetostatic Vector Potential} As a natural extension of
above theory, In vector potential formulation, one has
\begin{equation}
\overline{\boldsymbol{B}}=\mathrm{curl}_{\overline{^\mathrm{M}\boldsymbol{x}}}%
\overline{\boldsymbol{A}}\,,  \label{E:vecpotA}
\end{equation}
\noindent \textbf{Magnetostatic Scalar Potential} With the assumption of $%
\overline{\boldsymbol{J}}\to \boldsymbol{0}$ in view of Eq.\ %
\eqref{E:freecurrent}, one has $\mathrm{curl}_{\overline{^\mathrm{M}%
\boldsymbol{x}}} \overline{\boldsymbol{H}} = \boldsymbol{0}$. Then, there
exists a magnetic scalar potential $\overline{\varrho}$, for which
\begin{equation}
\overline{\boldsymbol{H}}=-\boldsymbol{\nabla}_{\overline{^\mathrm{M}%
\boldsymbol{x}}} \overline{\varrho}\,,  \label{E:scalarpotvarrho}
\end{equation}
In Tables\ \ref{T:nondim1} and \ref{T:nondim2}, we introduce
nondimensionalization of the independent and dependent variables. With this nondimensionalization of the coordinates, the differential operators read

\begin{equation}
\boldsymbol{\nabla}_{\overline{^\mathrm{M}\boldsymbol{x}}} =\dfrac{1}{{^\text{M%
}}x_{\mathrm{ref}}} \boldsymbol{\nabla}_{^\mathrm{M}\boldsymbol{x}}\,,\quad
\mathrm{div}_{\overline{^\mathrm{M}\boldsymbol{x}}} = \dfrac{1}{{^\mathrm{M}}x_{%
\mathrm{ref}}}\mathrm{div}_{^\mathrm{M}\boldsymbol{x}}\,, \quad\text{and}\quad
\mathrm{curl}_{\overline{^\mathrm{M}\boldsymbol{x}}} = \dfrac{1}{{^\mathrm{M}}x_{%
\mathrm{ref}}}\mathrm{curl}_{^\mathrm{M}\boldsymbol{x}}\,.
\label{E:dimconst2}
\end{equation}

\begin{table}[h!]
\caption{Nondimensionalization of the variables and parameters for vector
potential formulation.}
\label{T:nondim2}\centering
\begin{tabular}{llll}
\hline
dimensional form & reference size & nondimensional form &  \\ \hline
${\overline{^\mathrm{M}\boldsymbol{x}}}$ & $^\mathrm{M}x_{\mathrm{ref}}$ & ${^%
\mathrm{M}\boldsymbol{x}}={\overline{^\mathrm{M}\boldsymbol{x}}}/{^\mathrm{M}x_{%
\mathrm{ref}}}$ &  \\
$\overline{\boldsymbol{A}}$ & $A_{\mathrm{ref}}$ & $\boldsymbol{A}=\overline{%
\boldsymbol{A}}/A_{\mathrm{ref}}$ &  \\
$\overline{\boldsymbol{\mu}}$ & $\mu_{\mathrm{ref}}$ & $\boldsymbol{\mu}=%
\overline{\boldsymbol{\mu}}/\mu_{\mathrm{ref}}$ &  \\
$\overline{\boldsymbol{H}}$ & $A_{\mathrm{ref}}/[\mu_{\mathrm{ref}}{^\mathrm{M}%
}x_{\mathrm{ref}}]$ & $\boldsymbol{H}=\overline{\boldsymbol{H}}/[A_{\mathrm{%
ref}}/[\mu_{\mathrm{ref}}{^\mathrm{M}}x_{\mathrm{ref}}]]$ &  \\
$\overline{\boldsymbol{B}}$ & $A_{\mathrm{ref}}/{^\mathrm{M}}x_{\mathrm{ref}}$
& $\boldsymbol{B}=\overline{\boldsymbol{B}}/[A_{\mathrm{ref}}/{^\mathrm{M}}x_{%
\mathrm{ref}}]$ &  \\
$\overline{\boldsymbol{J}}$ & $\mu_{\mathrm{ref}}/A_{\mathrm{ref}}$ & $%
\boldsymbol{J}=\overline{\boldsymbol{J}}/[\mu_{\mathrm{ref}}/A_{\mathrm{ref}%
}]$ &  \\ \hline
\end{tabular}%
\end{table}

\begin{table}[h!]
\caption{Nondimensionalization of the variables and parameters for scalar
potential formulation.}
\label{T:nondim1}\centering
\begin{tabular}{llll}
\hline
dimensional form & reference size & nondimensional form &  \\ \hline
${\overline{^\mathrm{M}\boldsymbol{x}}}$ & $^\mathrm{M}x_{\mathrm{ref}}$ & ${^%
\mathrm{M}\boldsymbol{x}}={\overline{^\mathrm{M}\boldsymbol{x}}}/{^\mathrm{M}x_{%
\mathrm{ref}}}$ &  \\
$\overline{\varrho}$ & $\varrho_{\mathrm{ref}}$ & $\varrho=\overline{\varrho}%
/\varrho_{\mathrm{ref}}$ &  \\
$\overline{\boldsymbol{\mu}}$ & $\mu_{\mathrm{ref}}$ & $\boldsymbol{\mu}=%
\overline{\boldsymbol{\mu}}/\mu_{\mathrm{ref}}$ &  \\
$\overline{\boldsymbol{H}}$ & $\varrho_{\mathrm{ref}}/{^\mathrm{M}}x_{\mathrm{%
ref}}$ & $\boldsymbol{H}=\overline{\boldsymbol{H}}/[\varrho_{\mathrm{ref}}/{^%
\mathrm{M}}x_{\mathrm{ref}}]$ &  \\
$\overline{\boldsymbol{B}}$ & $\mu_{\mathrm{ref}}\varrho_{\mathrm{ref}}/{^%
\mathrm{M}}x_{\mathrm{ref}}$ & $\boldsymbol{B}=\overline{\boldsymbol{B}}/[\mu_{%
\mathrm{ref}}\varrho_{\mathrm{ref}}/{^\mathrm{M}}x_{\mathrm{ref}}]$ &  \\
\hline
\end{tabular}%
\end{table}
\section{Derivation of the Voigt and the Reuss Estimates}
\label{s:derivationvoigtreuss}
To find the Voigt and Reuss estimates, we impress  $\overline{\bs{H}}$ and $\overline{\bs{B}}$ fields, respectively, to create uniform $\bs{H}$ and $\bs{B}$ fields inside the RVE as follows
\begin{align}
\bs{H}(\bs{x})&=\overline{\bs{H}}(\bs{x})\,,\quad \bs{x}\in \mathcal{V}\quad\text{(Voigt)}\,,\\
\bs{B}(\bs{x})&=\overline{\bs{B}}(\bs{x})\,,\quad \bs{x}\in \mathcal{V}\quad\text{(Reuss)}\,.
\end{align}
Consequently, the following are trivially satisfied
\begin{align}
\left\langle \bs{H} \right \rangle &= \overline{\bs{H}}\,,\quad\text{(Voigt)}\,,\\
\left\langle \bs{B} \right \rangle &= \overline{\bs{B}}\,,\quad\text{(Reuss)}\,,
\end{align}
in which the notation $\left\langle \{\bullet\} \right \rangle=[1/|\mathcal{V}|]\,\int_{\mathcal{V}}
\{\bullet\}\mathrm{d}V$ is used. Considering linear magnetostatics, with  $\bs{B}=\bs{\mu}\cdot\bs{H}$, and, hence, $\bs{H}=\bs{\mu}^{-1}\cdot\bs{B}$, where $\bs{\mu}$ is the micro-permeability field and $\bs{\mu}^{-1}$ is its inverse, we obtain:
\begin{align}
\left\langle \bs{B} \right \rangle &=
\left\langle \bs{\mu}\cdot\bs{H} \right \rangle=
\left\langle \bs{\mu}\cdot\overline{\bs{H}} \right \rangle=
\left\langle \bs{\mu} \right \rangle \cdot\overline{\bs{H}}=
\left\langle \bs{\mu} \right \rangle \cdot \left\langle \bs{H} \right \rangle\,,\quad\text{(Voigt)}\,,\\
\left\langle \bs{H} \right \rangle &=
\left\langle \bs{\mu}^{-1}\cdot\bs{B} \right \rangle=
\left\langle \bs{\mu}^{-1}\cdot\overline{\bs{B}} \right \rangle=
\left\langle \bs{\mu}^{-1} \right \rangle \cdot\overline{\bs{B}}=
\left\langle \bs{\mu}^{-1} \right \rangle \cdot \left\langle \bs{B} \right \rangle\,,\quad \text{(Reuss)}\,.
\end{align}
The above derivations imply that the Voigt  and Reuss  estimates for the macro (effective) permeability, respectively denoted by $\bs{\mu}^\star_\mathrm{V}$ and $\bs{\mu}^\star_\mathrm{R}$, are given by the following averages, respectively
\begin{align}
\bs{\mu}^\star_\mathrm{V}=\left\langle \bs{\mu} \right \rangle\quad\text{and}\quad
\bs{\mu}^{\star}_\mathrm{R}=\left\langle \bs{\mu}^{-1} \right \rangle^{-1}\,.
\end{align}
For material systems with magnetically isotropic constituents for which the volume fraction of phases are referred to as $\phi_1$ and $\phi_2$ with $\phi_1+\phi_2=1$, these, respectively, correspond to the following mixing rules
\begin{align}
{\mu}^\star_\mathrm{V}=\phi_1 \mu_1 + \phi_2 \mu_2\quad\text{and}\quad
{\mu}^\star_\mathrm{R}=[\phi_1 {\mu_1}^{-1} + \phi_2 {\mu_2}^{-1}]^{-1}\,.
\end{align}

\section{Effective-Medium Theories: $d-$dimensional Generalizations}
\label{S:effective_medium}
In the following, using Ref.\ \cite{Torquato2002}, we give the $d-$dimensional generalizations of the Maxwell, self-consistent, and differential effective-medium approximations for two-component medium. The Maxwell (also referred to as Maxwell Garnett) approximation for the effective permeability  $\mu^\star_\mathrm{M}$ is given as
\begin{equation}
\dfrac{\mu^\star_\mathrm{M}-\mu_1}
{\mu^\star_\mathrm{M}+[d-1]\mu_1}=
\phi_2\dfrac{\mu_2-\mu_1}{\mu_2+[d-1]\mu_1}\,.
\label{E:MaxwellGarnett}
\end{equation}
The self-consistent approximation \cite{Bruggeman1935,Landauer1978} for the effective permeability
$\mu^\star_\mathrm{SC}$ reads \cite{Torquato2002}
\begin{equation}
\mu^\star_\mathrm{SC}=
\dfrac{\eta+\sqrt{\eta^2+4[d-1]\mu_1\mu_2}}{2[d-1]}
\text{ where }
\eta=\mu_1[d\phi_1-1]+\mu_2[d\phi_2-1]\,.
\label{E:selfconsistent}
\end{equation}
Finally, the differential effective-medium approximation or the effective permeability $\mu^\star_\mathrm{DEM}$ requires a solution of the following implicit relation
\begin{equation}
\dfrac{\mu_2-\mu^\star_\mathrm{DEM}}{\mu_2-\mu_1}
\left[\dfrac{\mu_1}{\mu^\star_\mathrm{DEM}}\right]^{[1/d]}=1-\phi_2\,.
\label{E:differentialeffectivemedium}
\end{equation}
\section{Stacked-Layer Microstructures}
This part investigates the effect of the constituent phase volume fractions on the stacked-layer composite's magnetic permeability. Only periodic boundary conditions with scalar and vector potential formulations are applied in the computations. The volume fractions of the band are studied from 0 to 1 with a volume fraction step size of   $0.05$, which required 21 model generations. The magnetic permeabilities of phases 1 and 2 are assumed to be  $\mu_1=250\,\mu_0$ and $\mu_2=\mu_0$, leading to a phase contrast of $\mu_1/\mu_2=250/1$.
\begin{figure*}[htb!]
% \vspace{5pt}
% trim=left bottom right top, clip
\centering
\begin{minipage}[]{0.75\textwidth}
\centering
{\frame{\includegraphics[height=0.089\textwidth,
trim=1238 324 1595 27, clip]{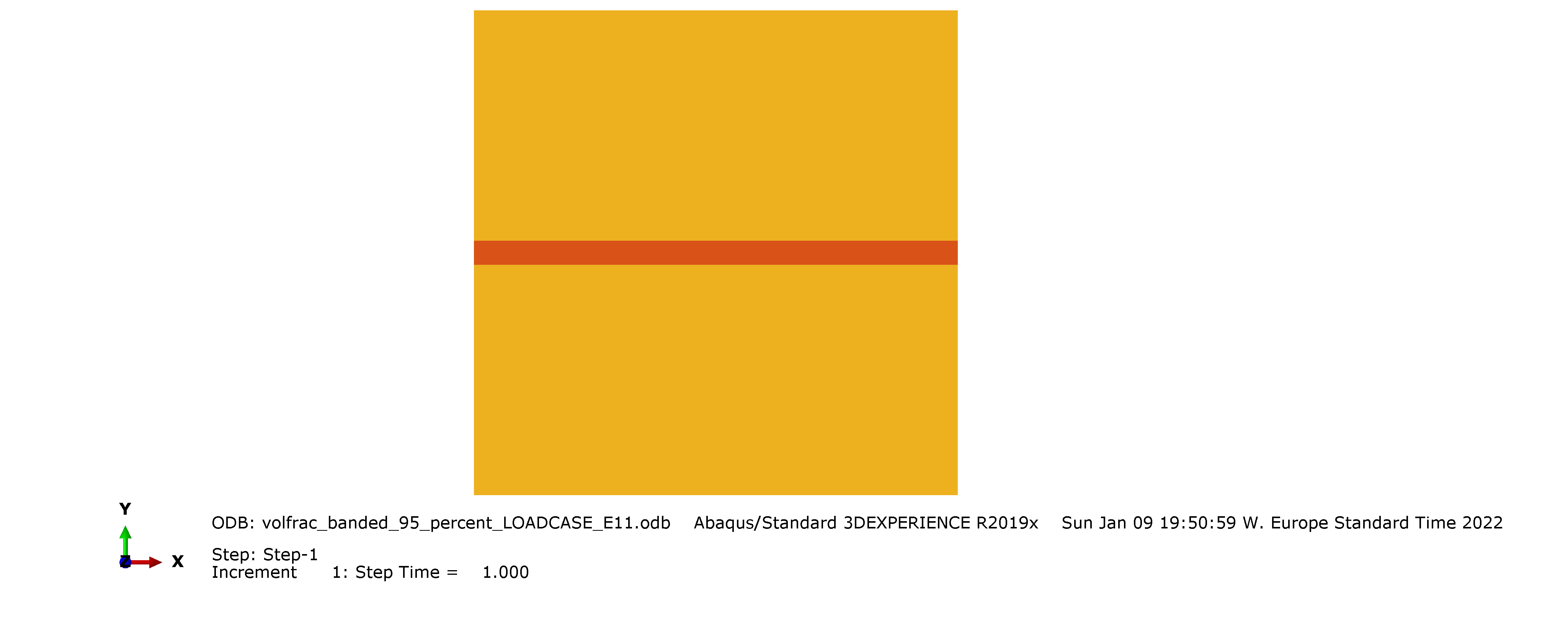}}}
{\frame{\includegraphics[height=0.089\textwidth,
trim=1238 324 1595 27, clip]{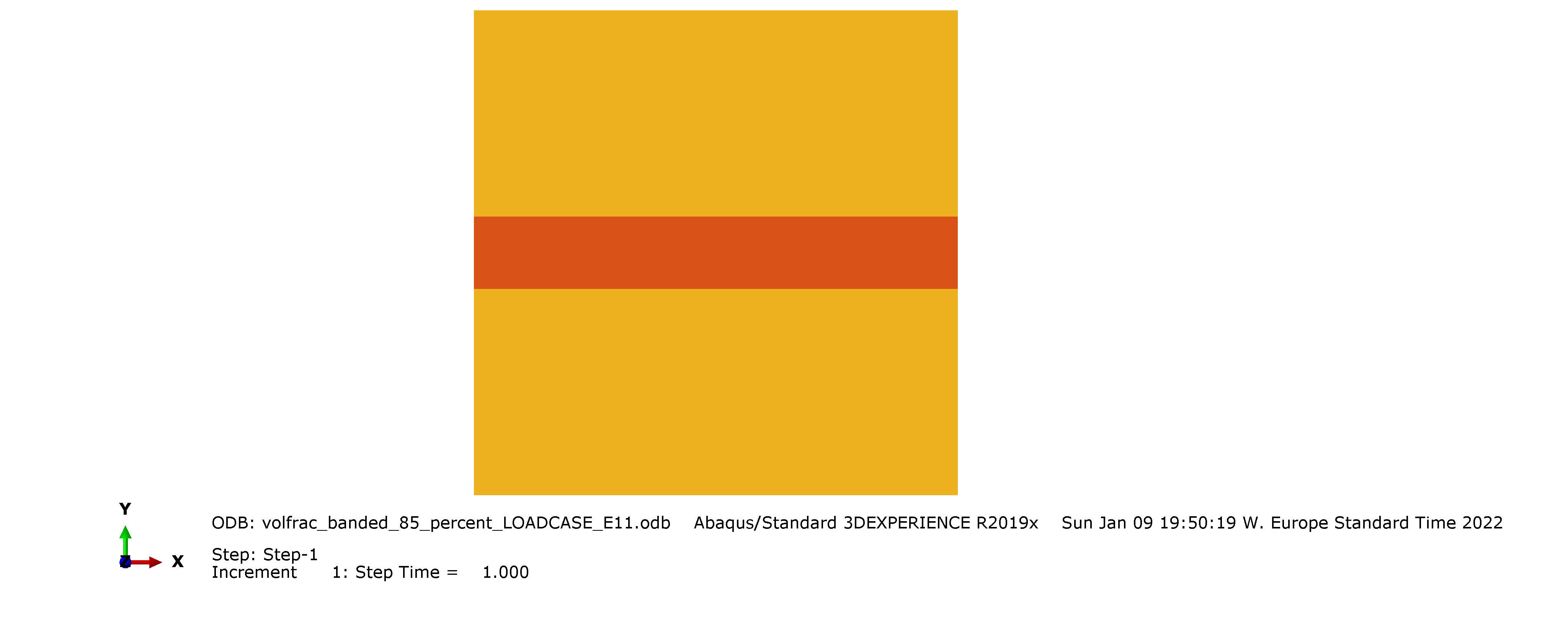}}}
{\frame{\includegraphics[height=0.089\textwidth,
trim=1238 324 1595 27, clip]{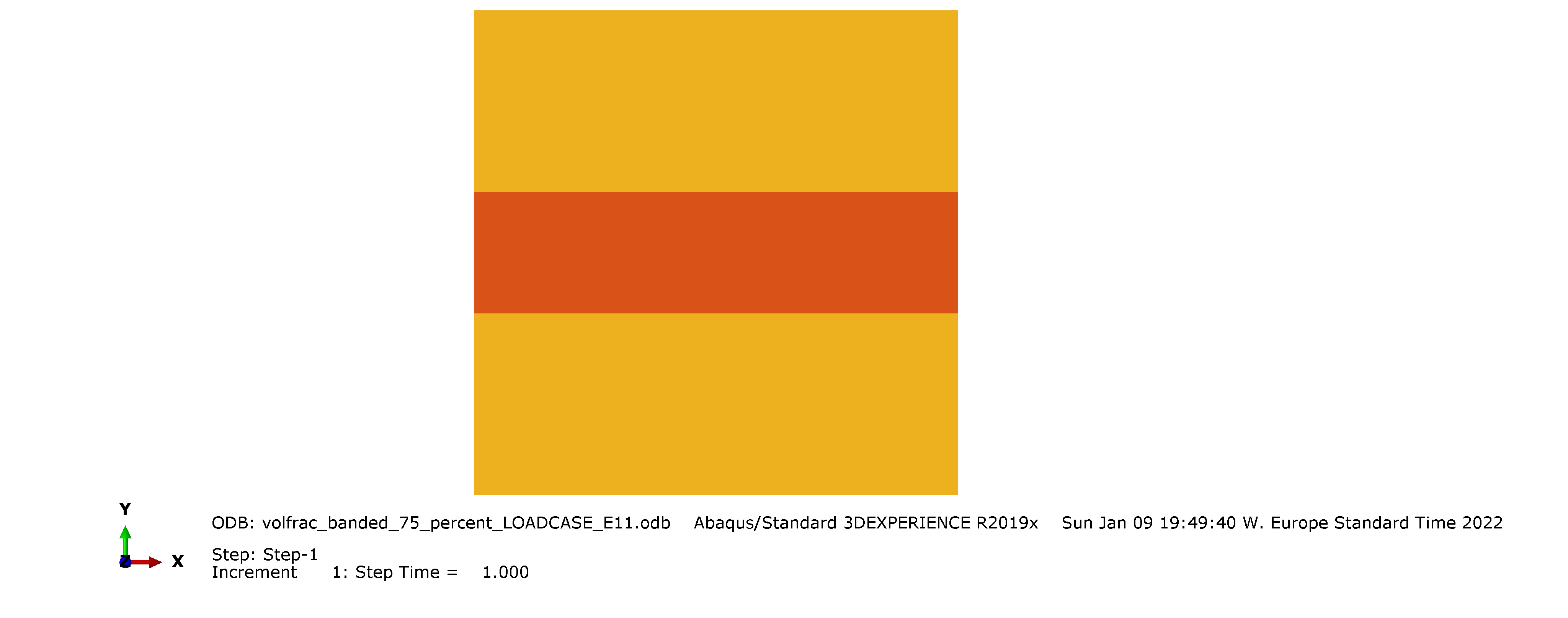}}}
{\frame{\includegraphics[height=0.089\textwidth,
trim=1238 324 1595 27, clip]{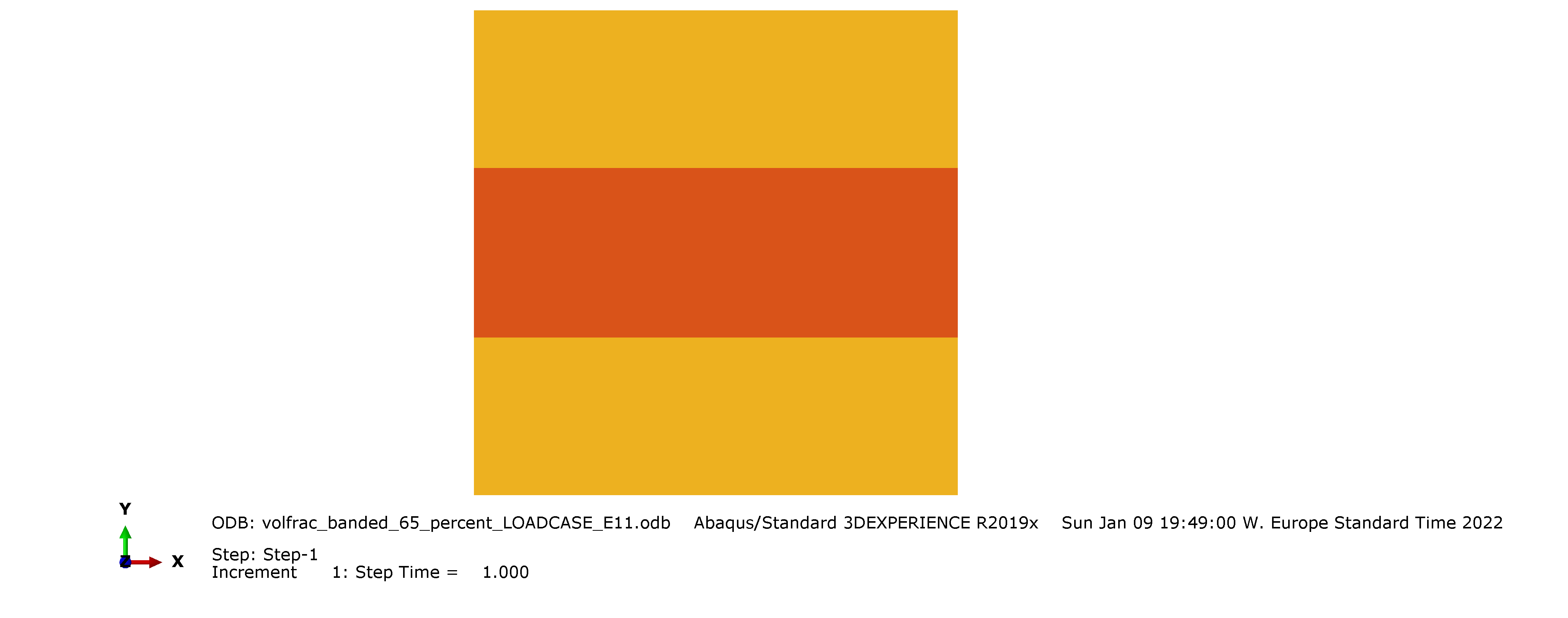}}}
{\frame{\includegraphics[height=0.089\textwidth,
trim=1238 324 1595 27, clip]{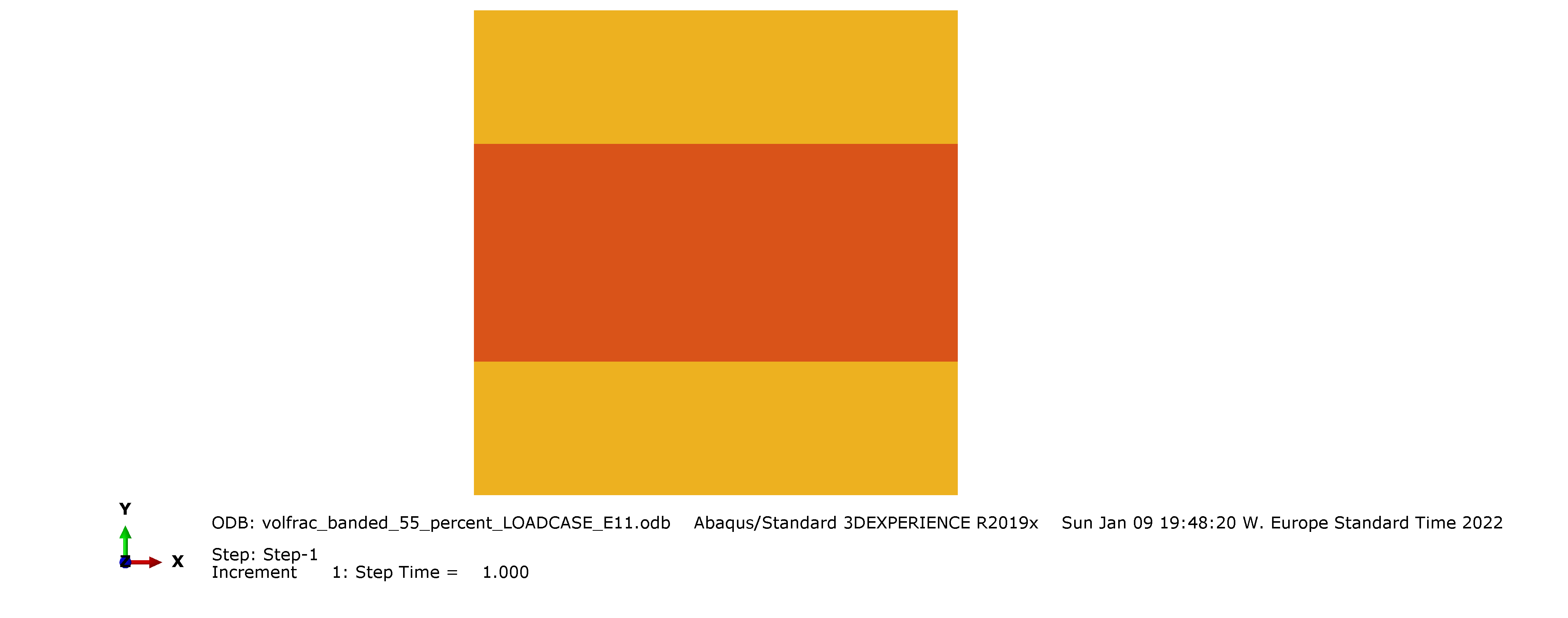}}}
{\frame{\includegraphics[height=0.089\textwidth,
trim=1238 324 1595 27, clip]{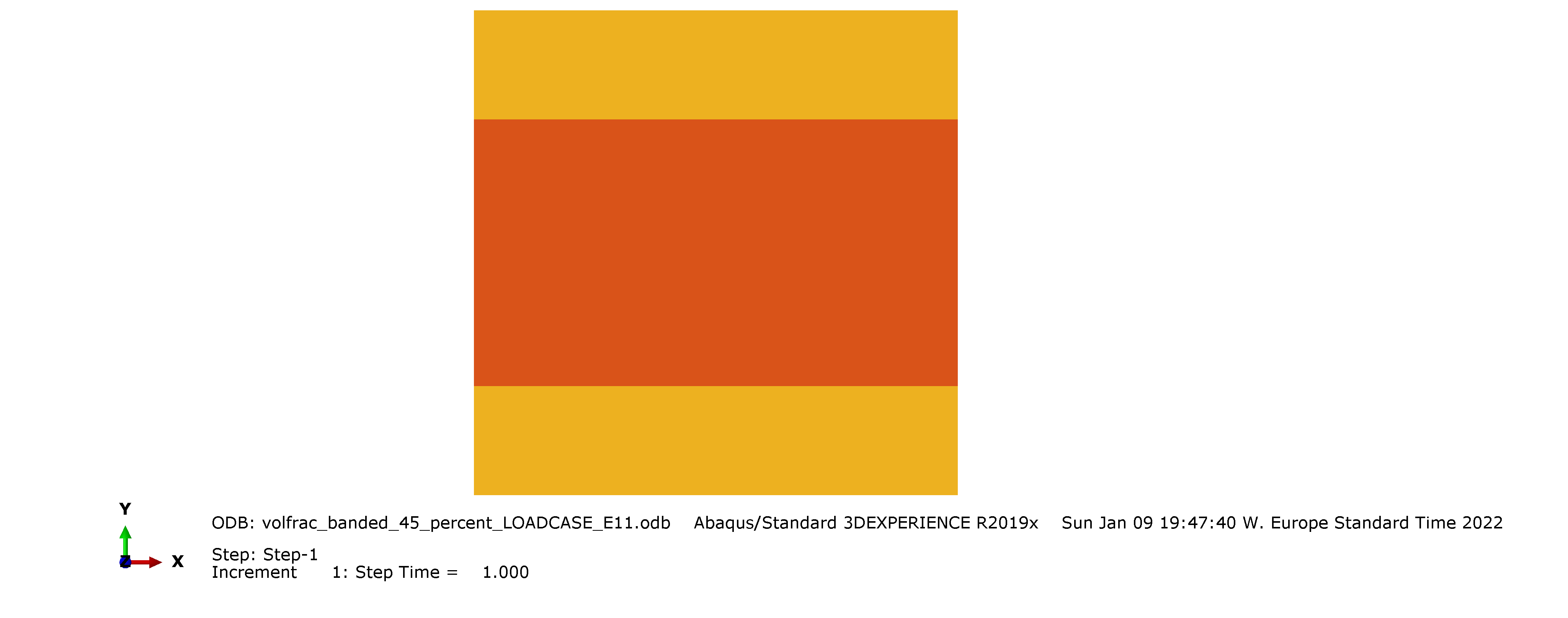}}}
{\frame{\includegraphics[height=0.089\textwidth,
trim=1238 324 1595 27, clip]{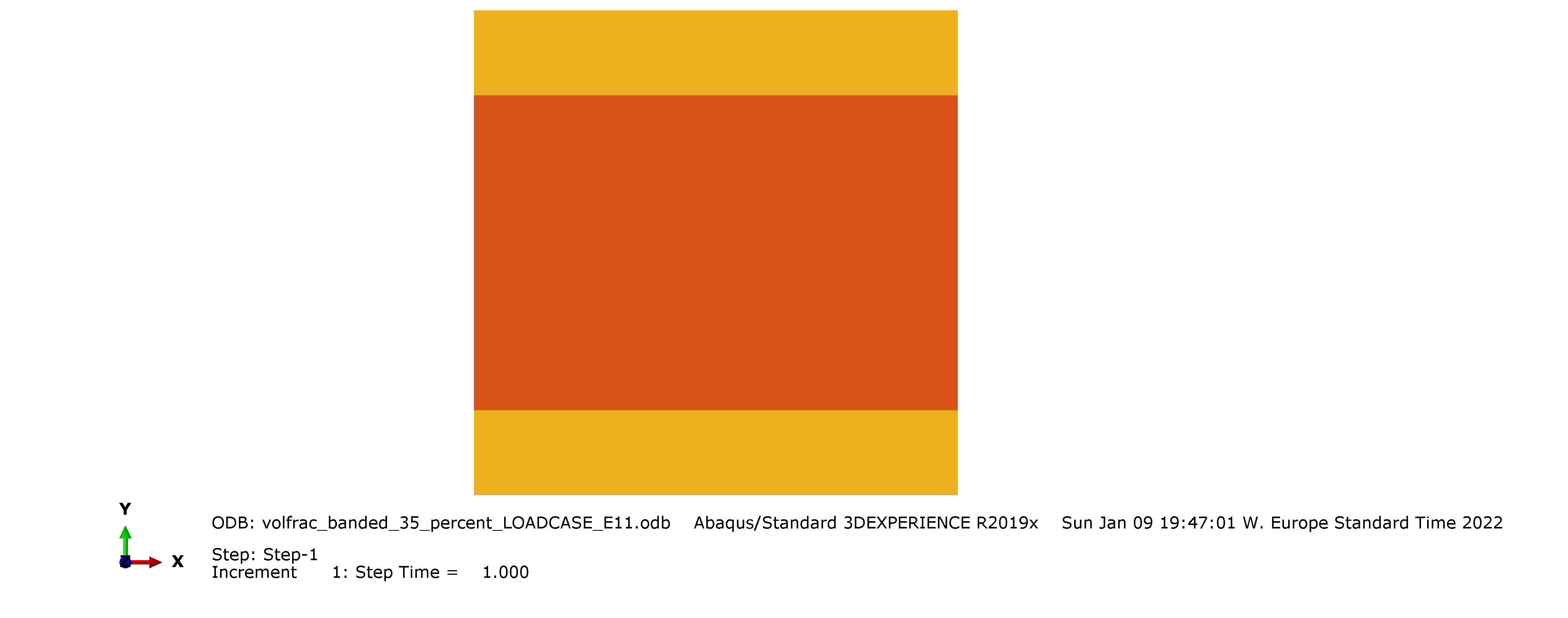}}}
{\frame{\includegraphics[height=0.089\textwidth,
trim=1238 324 1595 27, clip]{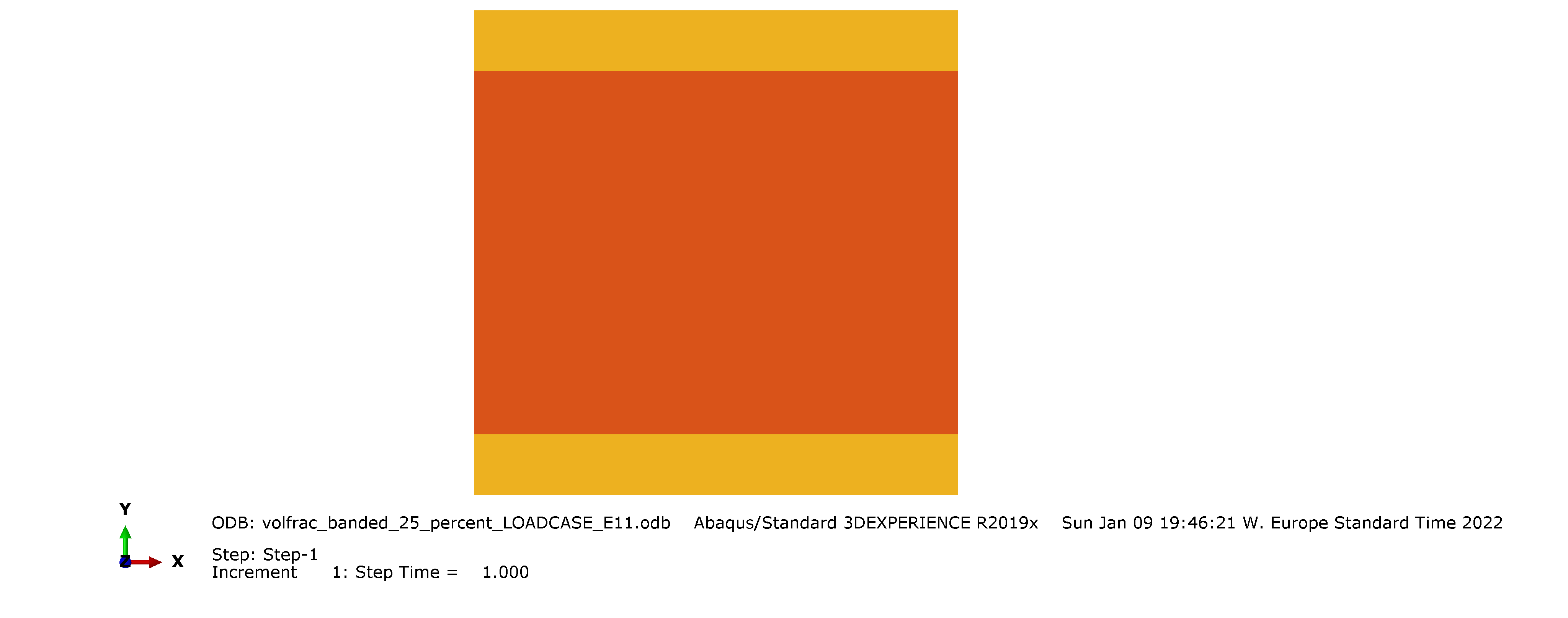}}}
{\frame{\includegraphics[height=0.089\textwidth,
trim=1238 324 1595 27, clip]{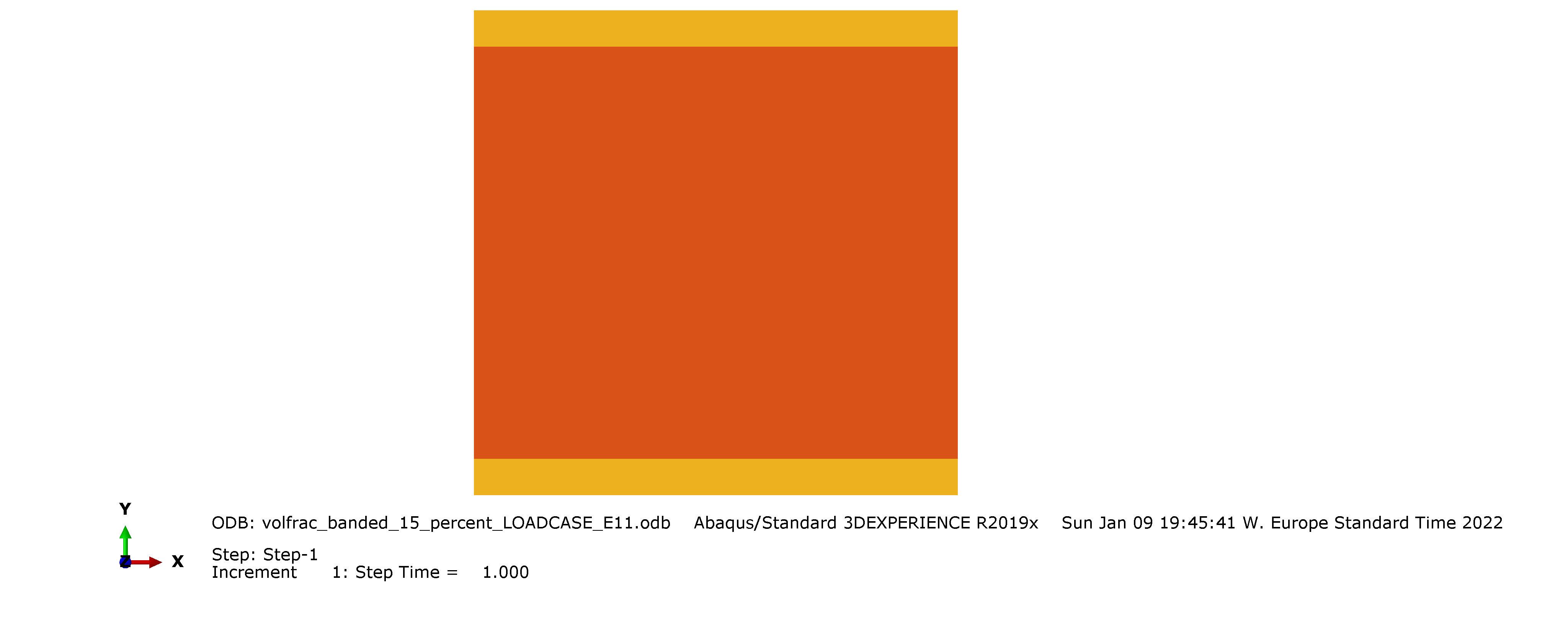}}}
{\frame{\includegraphics[height=0.089\textwidth,
trim=1238 324 1595 27, clip]{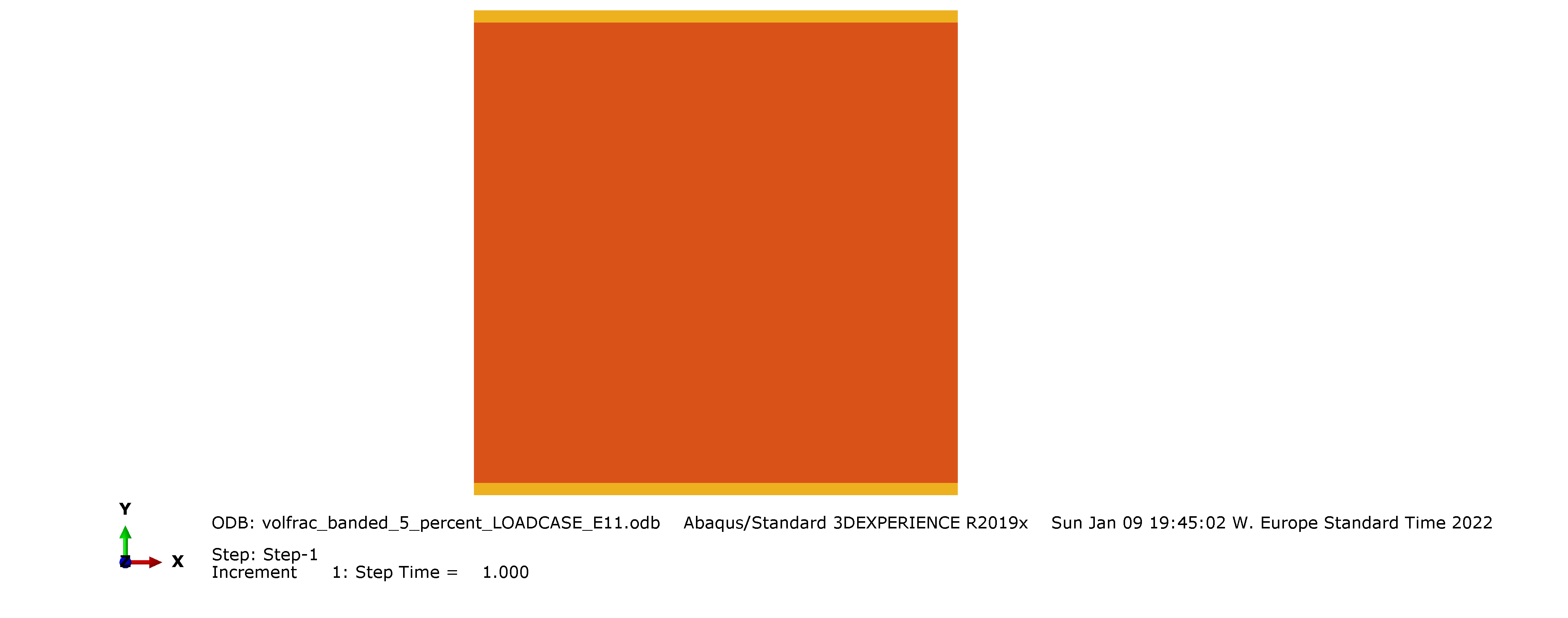}}}
\end{minipage}
\caption{Some of the investigated banded microstructures. In computations, the  volume fraction interval $[0,1]$ is traversed with volume fraction steps of $0.05$ corresponding to 21 models.}
\label{F:2D_banded}
\end{figure*}

\begin{figure*}[htb!]
% \vspace{5pt}
\centering
\subfigure[]{\begin{tikzpicture}
    \node[inner sep=0pt] (B11) at (0,0)
    {\includegraphics[height=0.37\textwidth]{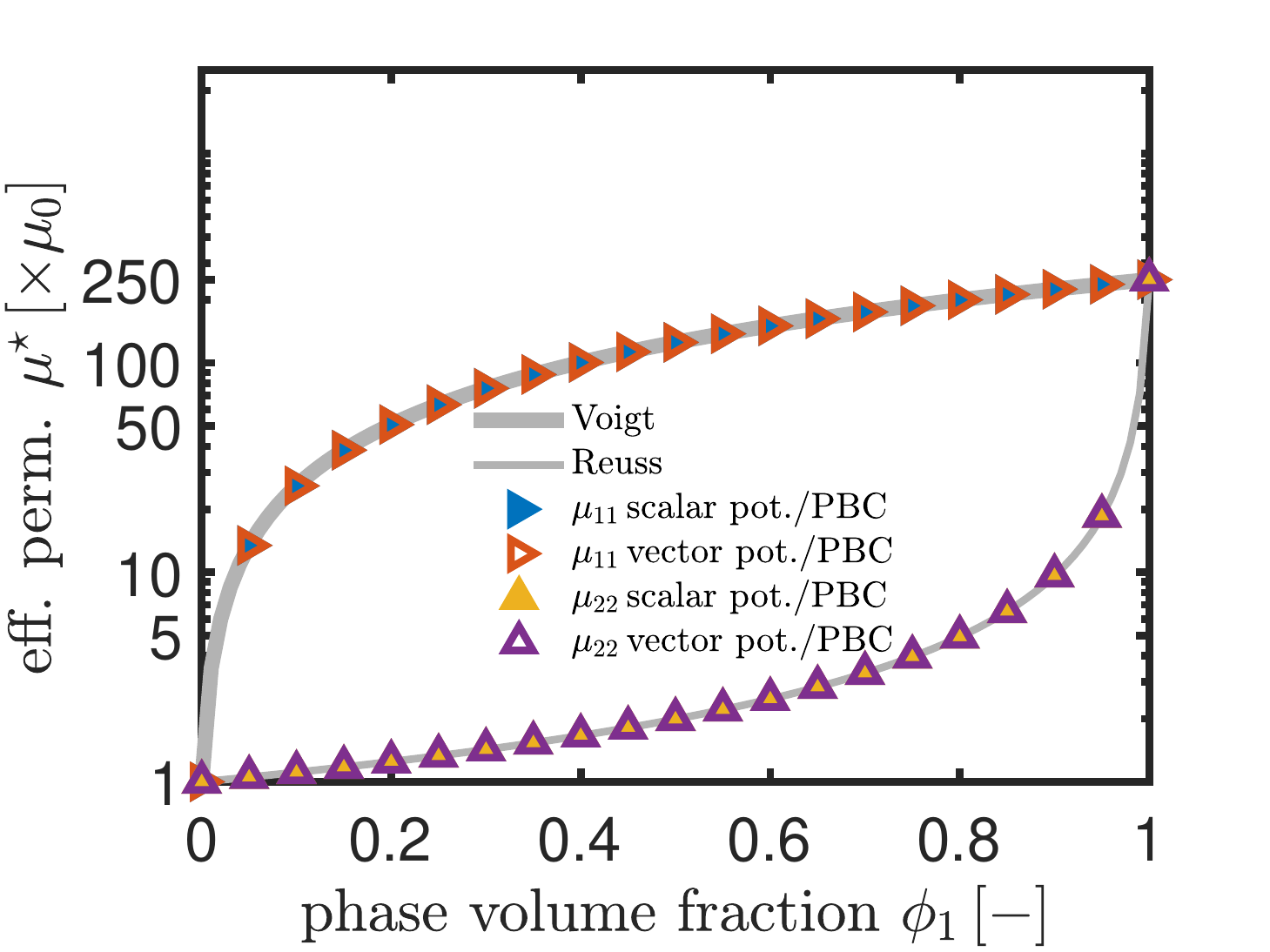}};
    \node[inner sep=0pt] (C1) at (-3.70+1.55,2.0)
    {{\includegraphics[height=0.055\textwidth,
trim=1238 324 1595 27, clip]{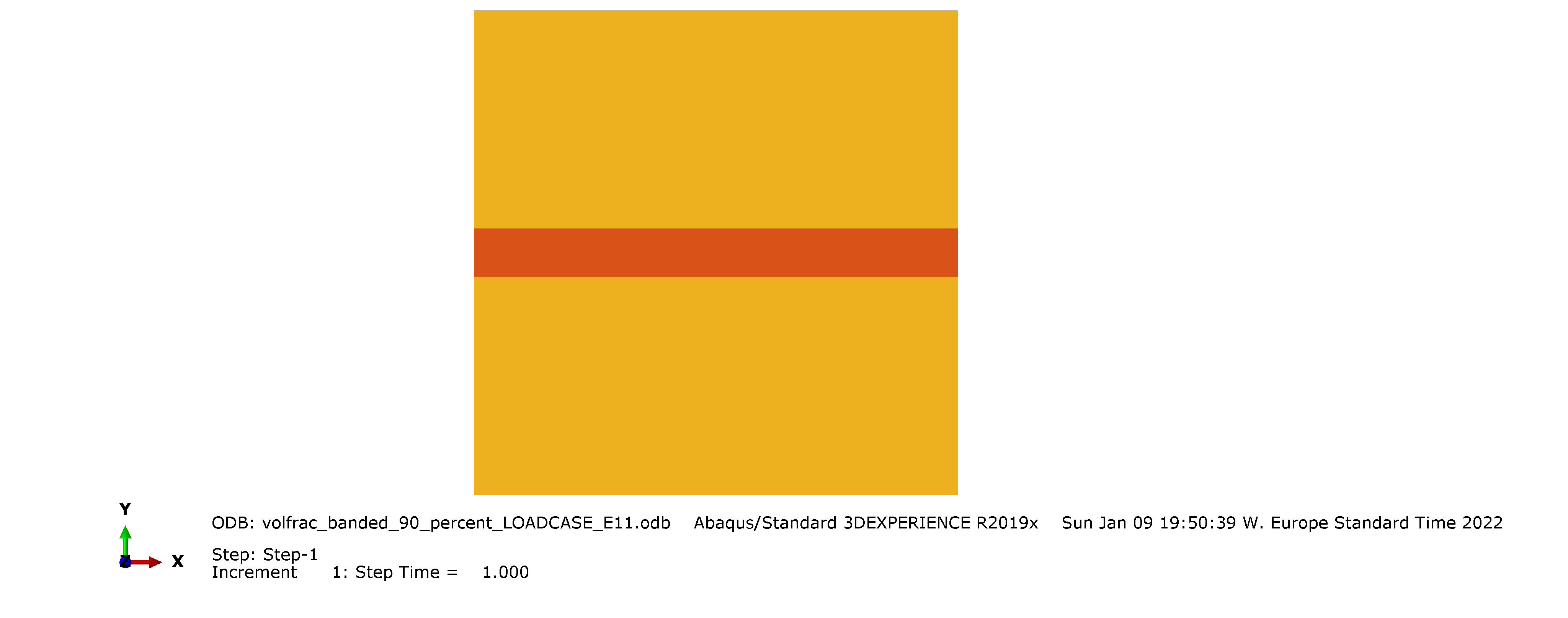}}};
    \node[inner sep=0pt] (C2) at (-3.70+1.55+1*1.2,2.0)
    {{\includegraphics[height=0.055\textwidth,
trim=1238 324 1595 27, clip]{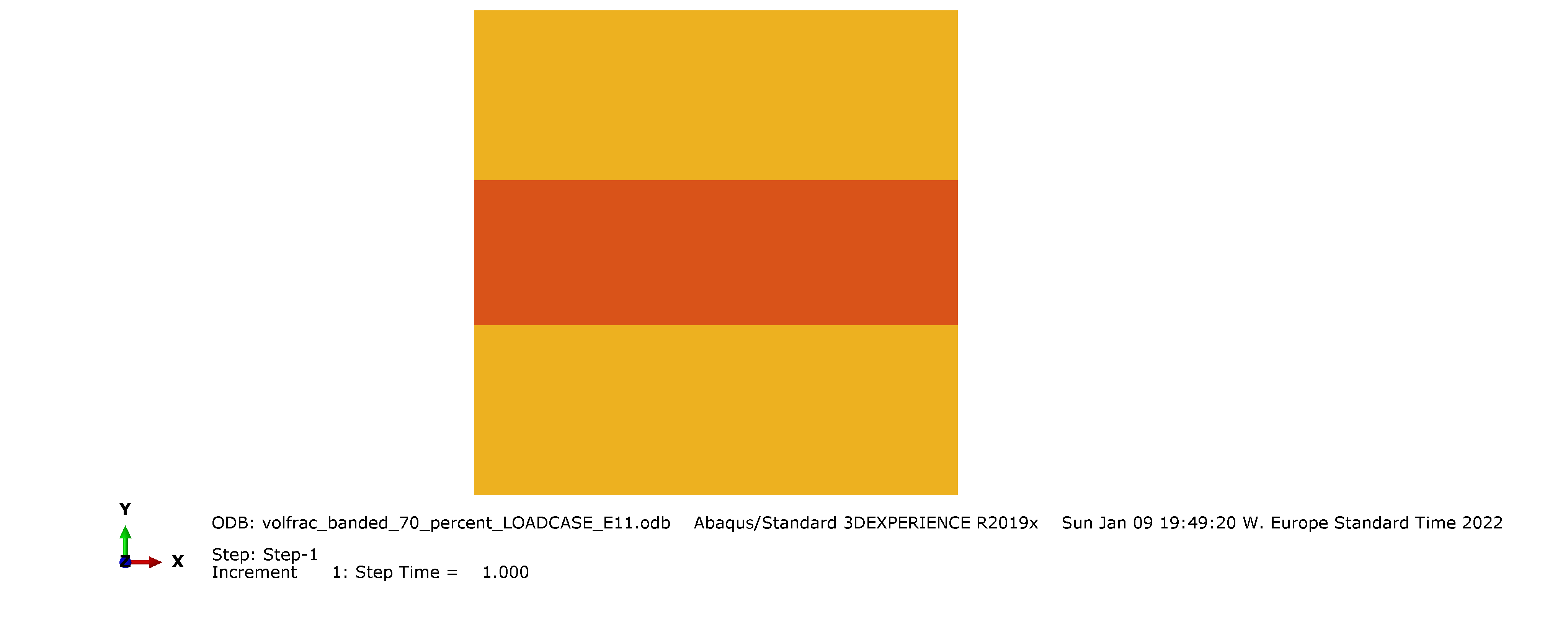}}};
    \node[inner sep=0pt] (C3) at (-3.70+1.55+2*1.2,2.0)
    {{\includegraphics[height=0.055\textwidth,
trim=1238 324 1595 27, clip]{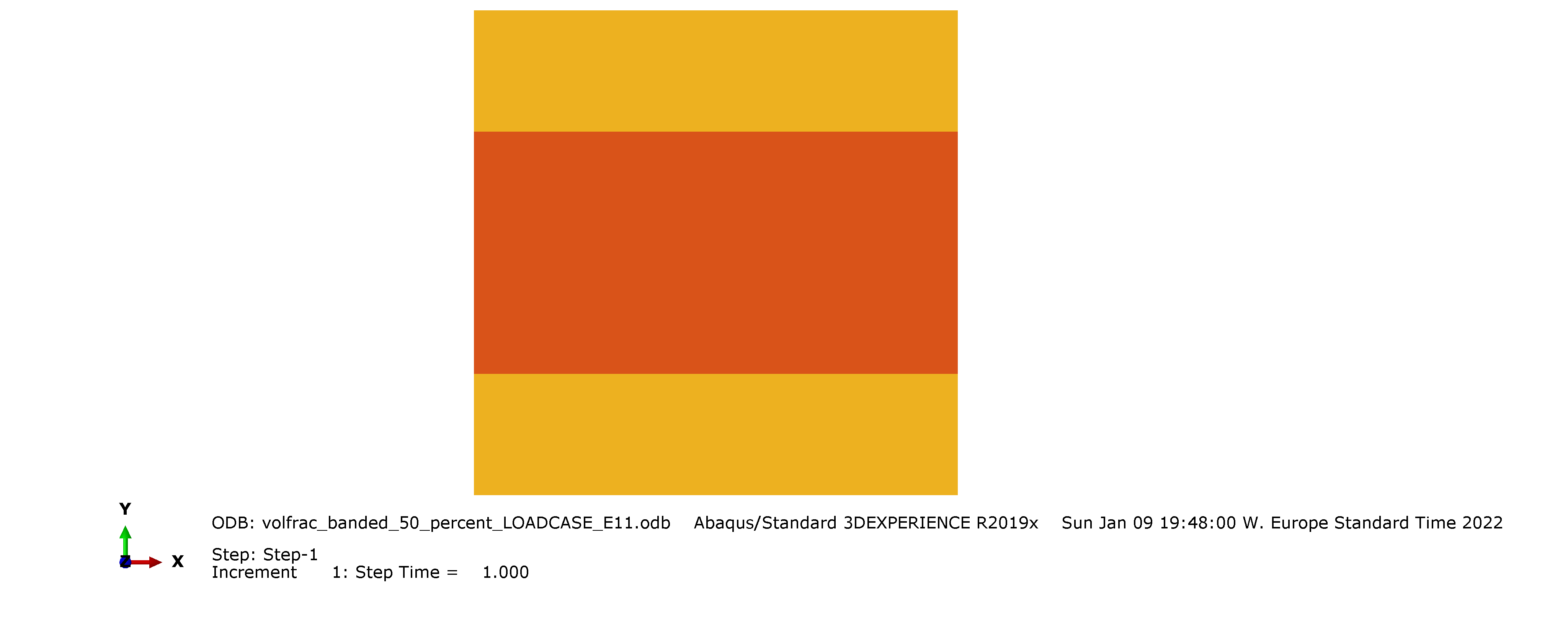}}};
    \node[inner sep=0pt] (C4) at (-3.70+1.55+3*1.2,2.0)
    {{\includegraphics[height=0.055\textwidth,
trim=1238 324 1595 27, clip]{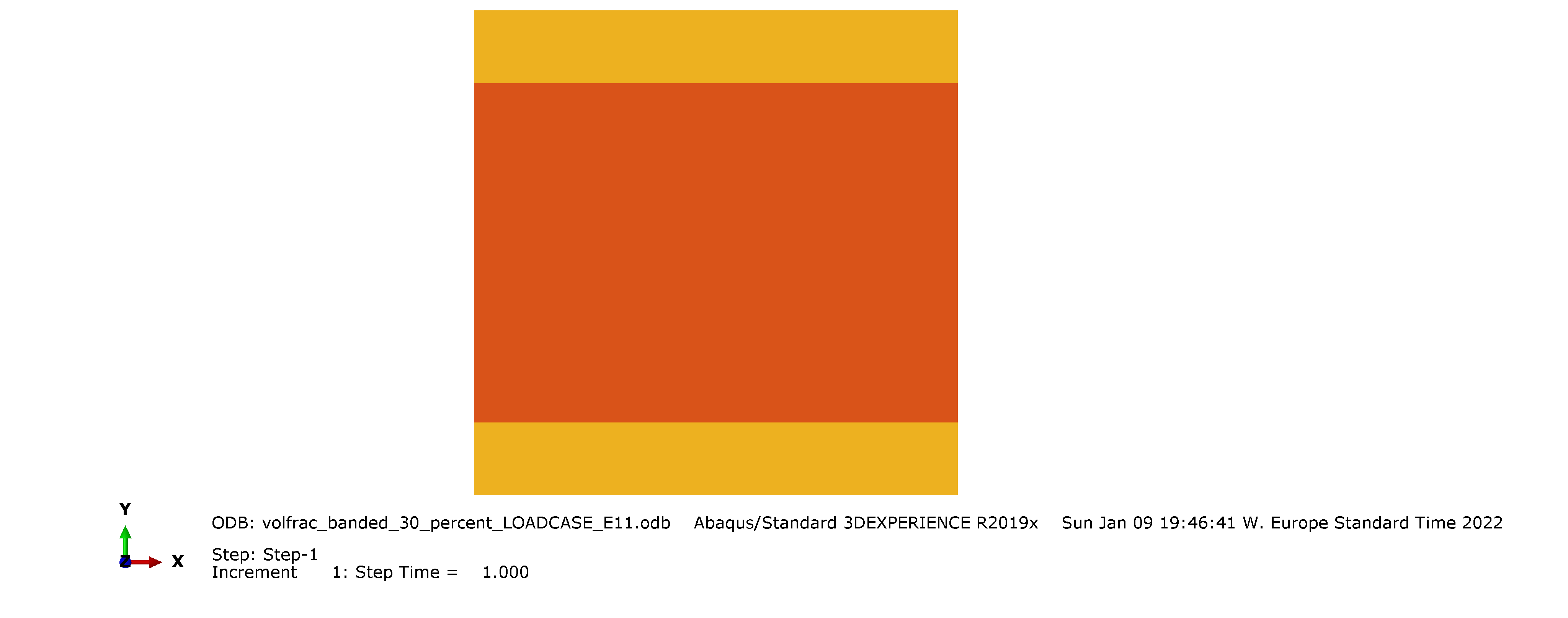}}};
    \node[inner sep=0pt] (C5) at (-3.70+1.55+4*1.2,2.0)
    {{\includegraphics[height=0.055\textwidth,
trim=1238 324 1595 27, clip]{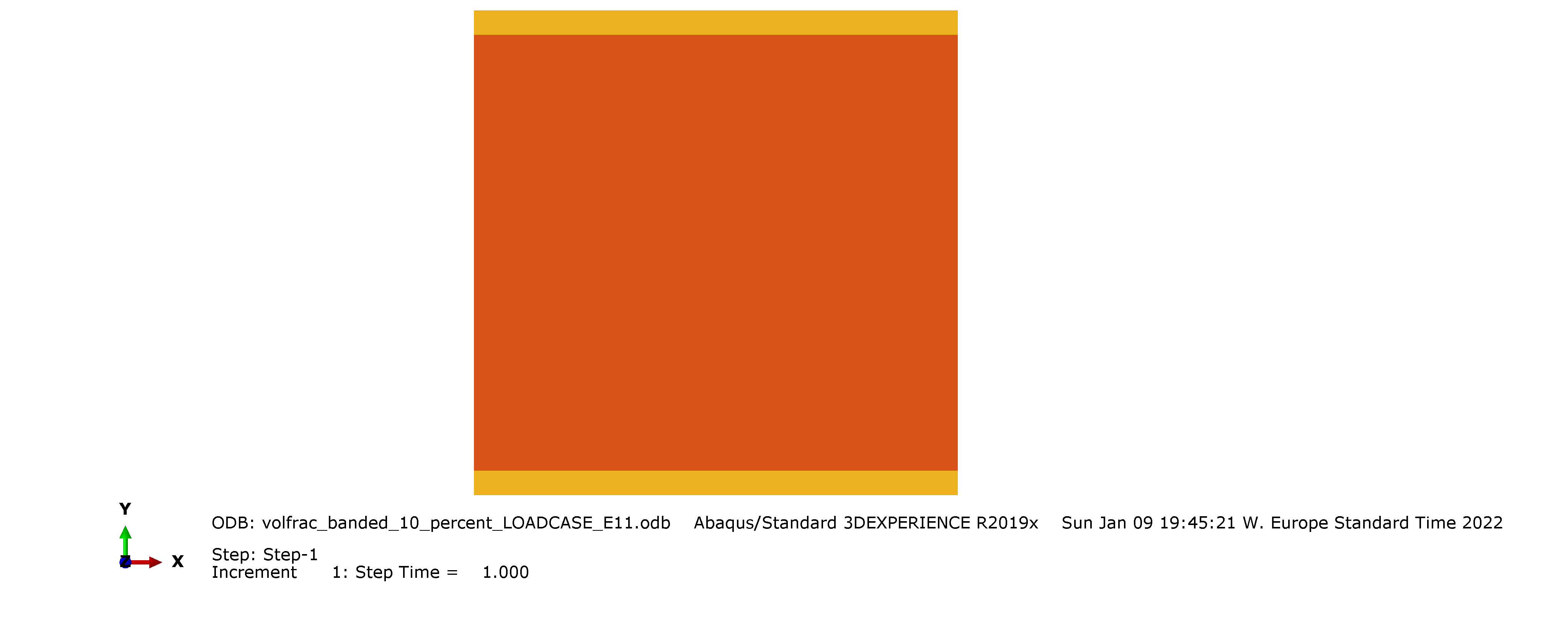}}};        \draw [-stealth](-3.75+1.55+2*1.225,2.0) -- (-3.75+1.55+0.25+2*1.225,2.0) node[below]{\scriptsize $\bs e_1$};
    \draw [-stealth](-3.75+1.55+2*1.225,2.0) -- (-3.75+1.55+2*1.225,2.00+0.25) node[left] {\scriptsize $\bs e_2$};
    \end{tikzpicture}}
\subfigure[]{\begin{tikzpicture}
    \node[inner sep=0pt] (B11) at (0,0)
    {\includegraphics[height=0.37\textwidth]{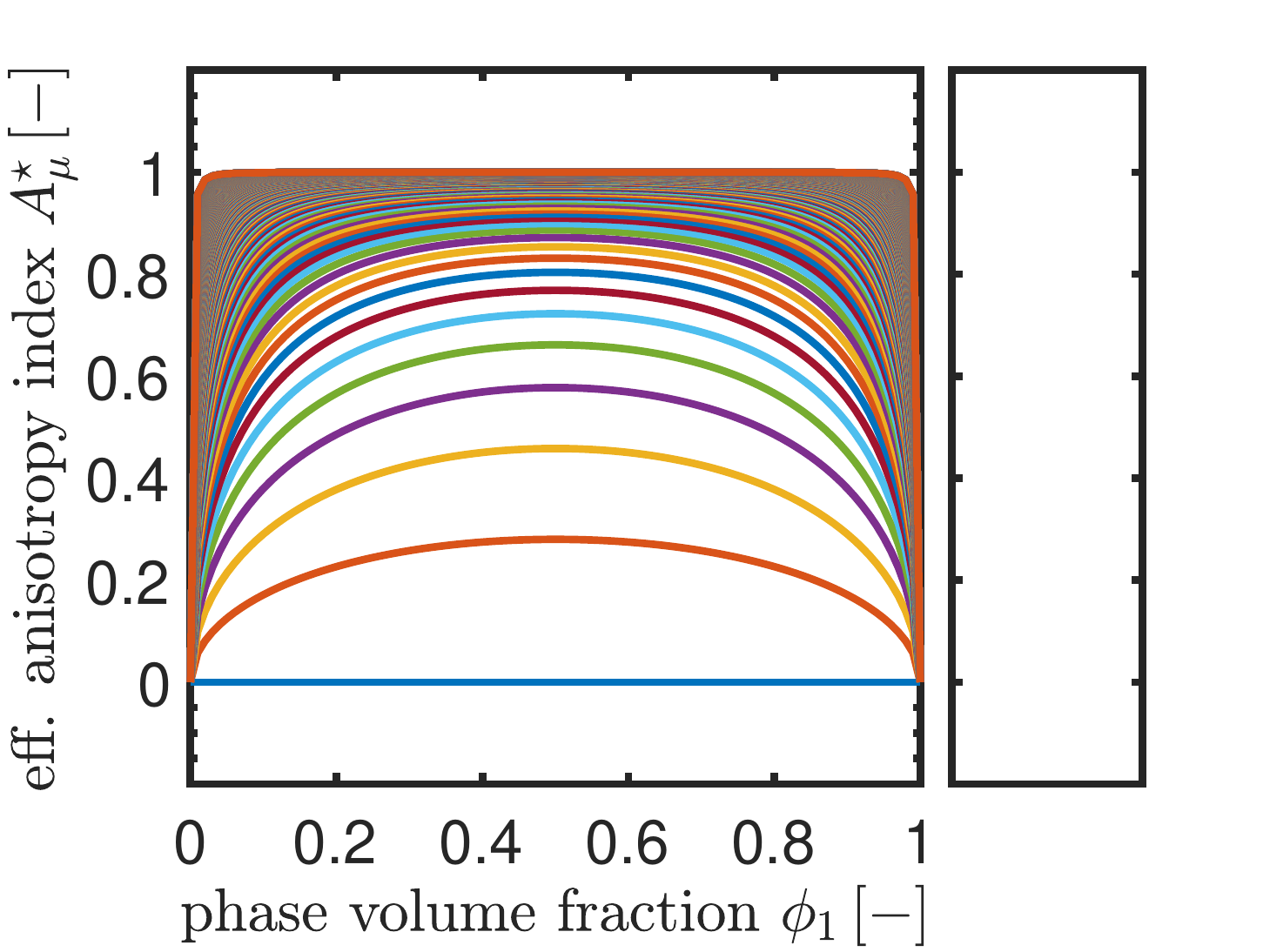}};
    \node (Start) at (-0.5,2.3) {\scriptsize $\mu_1/\mu_2=250$};
    \node (End) at (-0.5,-1.7) {\scriptsize $\mu_1/\mu_2=1$};
    \draw [-latex] (End) to (Start);
    \draw [line width=0.25mm](2.65,-1.3) ellipse (0.15cm and 0.15cm);
    \draw [line width=0.25mm](2.65,-1.3+1*0.65) ellipse (0.1531cm and 0.15cm);
    \draw [line width=0.25mm](2.65,-1.3+2*0.65) ellipse (0.1637cm and 0.15cm);
    \draw [line width=0.25mm](2.65,-1.3+3*0.65) ellipse (0.1875cm and 0.15cm);
    \draw [line width=0.25mm](2.65,-1.3+4*0.65) ellipse (0.25cm and 0.15cm);
    \draw [line width=0.25mm] (2.2,-1.3+5*0.65)--(3.1,-1.3+5*0.65);
    \end{tikzpicture}}
\caption{(a) Effective relative magnetic permeability and its comparison to Voigt and Reuss bounds for various phase volume fractions in stacked-layer microstructures. In  banded structures, an orthotropic magnetostatic response is observed, where the effective permeabilities along $x-$ and $y-$directions are those of the principal values of the corresponding magnetic permeability tensor. As depicted, these agree well with the (upper) Voigt and the (lower) Reuss bounds, respectively. (b)
Anisotropy index plots computed for the stacked-layer microstructures for various phase contrasts. The index is symmetric concerning the phase volume fraction. As anticipated, the phase contrast increases the degree of directionality in the magnetic response.}
\label{F:2D_banded_results}
\end{figure*}
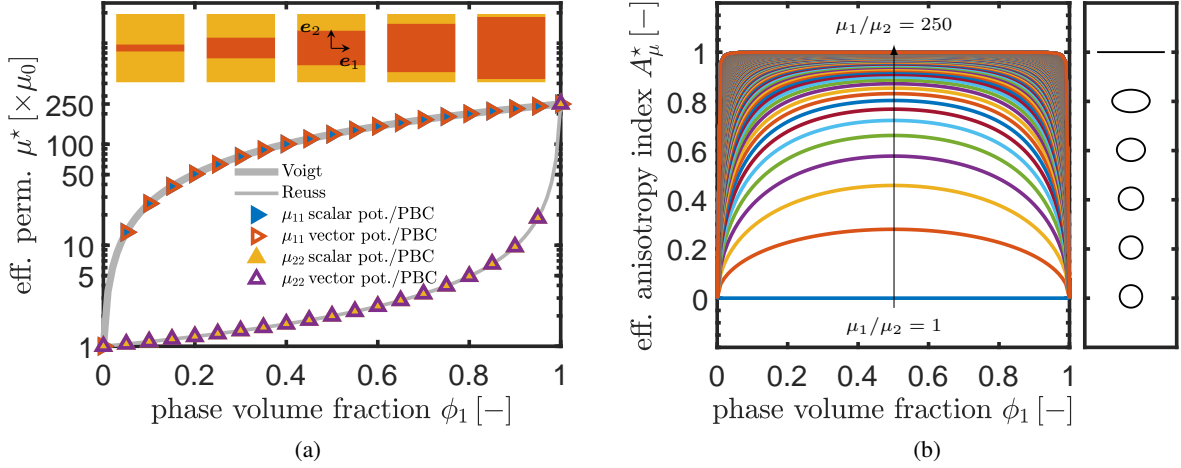

As depicted in Figure\ \ref{F:2D_banded_results}(a), our computations for scalar and vector potential formulations with the application of periodic boundary conditions are in excellent agreement with each other, as well as the analytical solutions. More specifically, the magnetic permeability computed along the direction that matches that of the band agrees with the (upper) Voigt bound. In contrast, the transverse direction magnetic permeability corresponds to the (lower) Reuss bound. This directionality in the magnetic properties can be quantified using the proposed anisotropy index, whose plots computed for the  stacked-layer microstructures are depicted in Figure\ \ref{F:2D_banded_results}(b) for various phase contrast magnitudes as a function of phase volume fraction. As anticipated, the index plots are symmetric concerning the point of phase balance, that is, $\phi_{1}=0.5$. As the phase property contrast increases, rapid growth in the anisotropy index occurs even for small phase volume fraction differences from zero.

For the considered two-dimensional banded system as well, the phase-interchange relation encapsulated in the  expression $\mu^\star(\mu_1,\mu_2)\,\mu^\star(\mu_2,\mu_1)=\mu_1\mu_2$ is satisfied in agreement with  Mendelson's generalization of Keller's relation \cite{Mendelson1975} to two-dimensional two-phase composites and considering the principal axes of the permeability tensor for which the macroscopic tensor components are computed.

\section{Two-point Probability Function}
\label{S:twopointprobabilityfunction}
Defining a phase indicator function $\mathcal{I}^{(i)}(\bs x)$ for phase $i$ with
\begin{equation}
\mathcal{I}^{(i)}(\bs x)=\left\{
\begin{array}{ll}
      1, & \text{if }x\in \mathcal{V}_i\,, \\
      0, & \text{otherwise\,,}\\
\end{array}
\right.
\label{E:n-pointprob}
\end{equation}
we define the $n$-point probability function for phase $i$ with \cite[p. 27]{Torquato2002}%
\begin{equation}
S^{(i)}_n(\bs x_1, \bs x_2, \ldots, \bs x_n)=\prec \mathcal{I}^{(i)}(\bs x_1) \mathcal{I}^{(i)}(\bs x_2)\cdots \mathcal{I}^{(i)}(\bs x_n) \succ\,,
\end{equation}
where $\prec \bullet \succ$ denotes ensemble average of $\bullet$. $n-$point probability function provide a mathematical description for the distribution and morphology of constituents in random composites. For $n\to 1$, the function gives the phase volume fraction,  whereas as $n$ increases, so does the corresponding information the function provides. Complete statistical information is attained at the limit $n\to\infty$ \cite[p.\ 203]{Beran1968}. From a practical point of view, higher-order probability functions are rarely achievable. For statistically homogeneous systems, the two-point probability function $S_2^{(i)}(r)$ gives the likelihood that the selected two points with $r$ distance away from each other belong to the same phase indexed by $i$. The information provided by the two-point probability function allows testing the  microstructures randomly generated using Monte Carlo simulations. For statistically isotropic material systems, the probability functions are rotational invariance; thus, one can use the representation $S^{(i)}_2(\bs x_1, \bs x_2)\to S^{(i)}_2(r)$ with $r=|\bs x_1-\bs x_2|$. Moreover, if the medium does not possess any long-range order, $S^{(i)}_2(r)\to\phi_i$ as $r\to 0$ and $S^{(i)}_2(r)\to\phi_i^2$ as $r\to \infty$. Oscillations in $S^{(i)}_2(r)$ for small $r$ represent short-range order and spatial correlations. Exponential monotonic decay to the asymptotic value signals the absence of spatial correlation.

\begin{figure}[t!]
    \centering
    \includegraphics[height=0.30\textwidth]{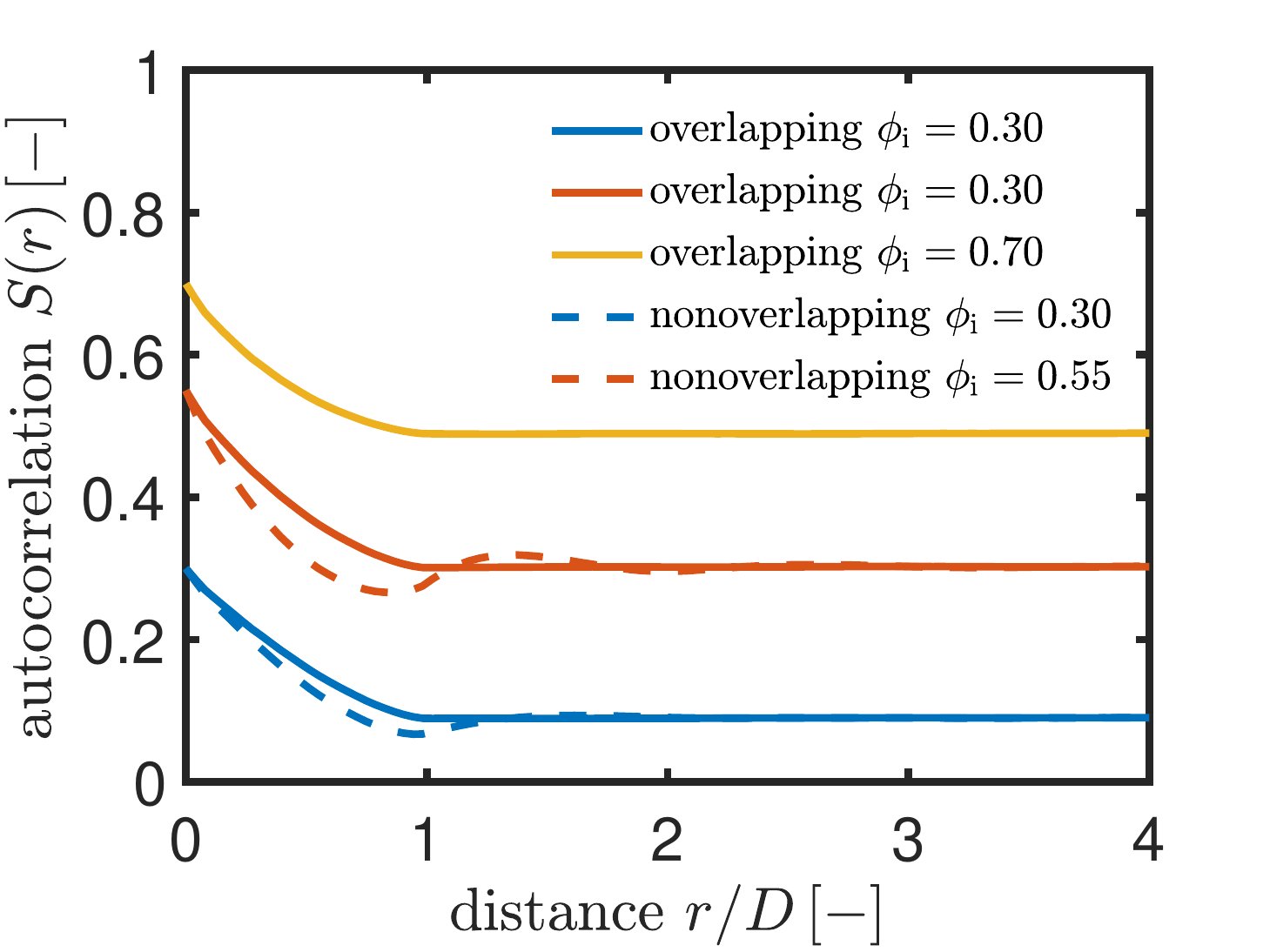}
\caption{The two-point probability function $S_2(r)$ plots for the selected microstructure generations with random disks. The VE size is 2048 pixels.}
\label{F:autocorrelations}
\end{figure}

Figure\ \ref{F:autocorrelations}  depicts the two-point probability function $S_2(r)$ plots for the inclusions for the selected overlapping and nonoverlapping random disk microstructures. For the computation of the $S_2(r)$, the  radially averaged autocorrelation function macro of ImageJ \cite{Schneider2012} is used, see, e.g., \cite{Berryman1986}.
Our computations agree with the findings given in Torquato \cite{Torquato2002}. Starting with the curves, satisfy  the indicated limits at $r\to 0$ and $r\to \infty$. Moreover, for the nonoverlapping disk systems, there occur oscillations in $S^{(i)}_2(r)$ for small $r$, denoting spatial correlations. In contrast, $S^{(i)}_2(r)$ exponentially decays to its asymptotic value at a distance equal to the disk diameter $r=D$. Thus, the overlapping disk systems are spatially uncorrelated.

\bibliography{bibliography}

\end{document}